\PassOptionsToPackage{unicode}{hyperref}
\PassOptionsToPackage{hyphens}{url}
\documentclass[
  12pt]{article}
\usepackage{xcolor}
\usepackage{amsmath,amssymb}
\usepackage{iftex}
\ifPDFTeX
  \usepackage[T1]{fontenc}
  \usepackage[utf8]{inputenc}
  \usepackage{textcomp} 
\else 
  \usepackage{unicode-math} 
  \defaultfontfeatures{Scale=MatchLowercase}
  \defaultfontfeatures[\rmfamily]{Ligatures=TeX,Scale=1}
\fi
\usepackage{lmodern}
\ifPDFTeX\else
\fi
\IfFileExists{upquote.sty}{\usepackage{upquote}}{}
\IfFileExists{microtype.sty}{
  \usepackage[]{microtype}
  \UseMicrotypeSet[protrusion]{basicmath} 
}{}
\makeatletter
\@ifundefined{KOMAClassName}{
  \IfFileExists{parskip.sty}{%
    \usepackage{parskip}
  }{
    \setlength{\parindent}{0pt}
    \setlength{\parskip}{6pt plus 2pt minus 1pt}}
}{
  \KOMAoptions{parskip=half}}
\makeatother
\makeatletter
\ifx\paragraph\undefined\else
  \let\oldparagraph\paragraph
  \renewcommand{\paragraph}{
    \@ifstar
      \xxxParagraphStar
      \xxxParagraphNoStar
  }
  \newcommand{\xxxParagraphStar}[1]{\oldparagraph*{#1}\mbox{}}
  \newcommand{\xxxParagraphNoStar}[1]{\oldparagraph{#1}\mbox{}}
\fi
\ifx\subparagraph\undefined\else
  \let\oldsubparagraph\subparagraph
  \renewcommand{\subparagraph}{
    \@ifstar
      \xxxSubParagraphStar
      \xxxSubParagraphNoStar
  }
  \newcommand{\xxxSubParagraphStar}[1]{\oldsubparagraph*{#1}\mbox{}}
  \newcommand{\xxxSubParagraphNoStar}[1]{\oldsubparagraph{#1}\mbox{}}
\fi
\makeatother

\usepackage{longtable,booktabs,array}
\usepackage{calc} 
\usepackage{etoolbox}
\makeatletter
\patchcmd\longtable{\par}{\if@noskipsec\mbox{}\fi\par}{}{}
\makeatother
\IfFileExists{footnotehyper.sty}{\usepackage{footnotehyper}}{\usepackage{footnote}}
\makesavenoteenv{longtable}
\usepackage{graphicx}
\makeatletter
\newsavebox\pandoc@box
\newcommand*\pandocbounded[1]{
  \sbox\pandoc@box{#1}%
  \Gscale@div\@tempa{\textheight}{\dimexpr\ht\pandoc@box+\dp\pandoc@box\relax}%
  \Gscale@div\@tempb{\linewidth}{\wd\pandoc@box}%
  \ifdim\@tempb\p@<\@tempa\p@\let\@tempa\@tempb\fi
  \ifdim\@tempa\p@<\p@\scalebox{\@tempa}{\usebox\pandoc@box}%
  \else\usebox{\pandoc@box}%
  \fi%
}
\def\fps@figure{htbp}
\makeatother

\providecommand{\tightlist}{%
  \setlength{\itemsep}{0pt}\setlength{\parskip}{0pt}}

\usepackage[]{natbib}
\usepackage{booktabs}
\usepackage{longtable}
\usepackage{array}
\usepackage{multirow}
\usepackage{wrapfig}
\usepackage{float}
\usepackage{colortbl}
\usepackage{pdflscape}
\usepackage{tabularx}
\usepackage{xltabular}
\usepackage{threeparttable}
\usepackage{threeparttablex}
\usepackage[normalem]{ulem}
\usepackage{makecell}
\usepackage{xcolor}
\usepackage{etoolbox}
\usepackage{pdflscape}
\usepackage{needspace}
\renewcommand{\topfraction}{0.9}
\renewcommand{\bottomfraction}{0.8}
\renewcommand{\textfraction}{0.07}
\renewcommand{\floatpagefraction}{0.8}
\usepackage{setspace}
\renewcommand{\singlespacing}{\renewcommand{\baselinestretch}{1}\selectfont}
\AtBeginEnvironment{longtable}{\footnotesize\setlength{\tabcolsep}{4pt}\singlespacing}
\AtBeginEnvironment{table}{\footnotesize\setlength{\tabcolsep}{3pt}\singlespacing}
\usepackage{colortbl}
\AtBeginDocument{\let\originaltoprule\toprule\renewcommand{\toprule}{\originaltoprule\rowcolor{gray!15}}}
\AtBeginEnvironment{thebibliography}{\singlespacing}
\AtBeginEnvironment{CSLReferences}{\singlespacing}
\usepackage{newunicodechar}
\newunicodechar{β}{\ensuremath{\beta}}
\newunicodechar{ε}{\ensuremath{\varepsilon}}
\newunicodechar{ᵢ}{\textsubscript{i}}
\newunicodechar{ᶜ}{\textsuperscript{c}}
\newunicodechar{≤}{\ensuremath{\le}}
\newunicodechar{≥}{\ensuremath{\ge}}
\newunicodechar{ρ}{\ensuremath{\rho}}
\newunicodechar{γ}{\ensuremath{\gamma}}
\AtBeginDocument{\newunicodechar{′}{\ensuremath{{}'}}}
\makeatletter
\@ifpackageloaded{caption}{}{\usepackage{caption}}
\AtBeginDocument{%
\ifdefined\contentsname
  \renewcommand*\contentsname{Table of contents}
\else
  \newcommand\contentsname{Table of contents}
\fi
\ifdefined\listfigurename
  \renewcommand*\listfigurename{List of Figures}
\else
  \newcommand\listfigurename{List of Figures}
\fi
\ifdefined\listtablename
  \renewcommand*\listtablename{List of Tables}
\else
  \newcommand\listtablename{List of Tables}
\fi
\ifdefined\figurename
  \renewcommand*\figurename{Figure}
\else
  \newcommand\figurename{Figure}
\fi
\ifdefined\tablename
  \renewcommand*\tablename{Table}
\else
  \newcommand\tablename{Table}
\fi
}
\@ifpackageloaded{float}{}{\usepackage{float}}
\floatstyle{ruled}
\@ifundefined{c@chapter}{\newfloat{codelisting}{h}{lop}}{\newfloat{codelisting}{h}{lop}[chapter]}
\floatname{codelisting}{Listing}

\makeatother
\makeatletter
\@ifpackageloaded{caption}{}{\usepackage{caption}}
\@ifpackageloaded{subcaption}{}{\usepackage{subcaption}}
\makeatother
\usepackage{bookmark}
\IfFileExists{xurl.sty}{\usepackage{xurl}}{} 
\makeatletter
\@ifundefined{xmpquote}{\newcommand{\xmpquote}[1]{#1}}{}
\makeatother
\hypersetup{
  pdftitle={Interpretable subgroup identification across survival and generalized-linear outcomes},
  pdfauthor={Larry F. León; Keaven M. Anderson},
  pdfkeywords={\xmpquote{interpretable subgroup}, \xmpquote{multiplier
resampling}, \xmpquote{post-selection inference}, \xmpquote{stability
selection}, \xmpquote{subgroup identification}},
  hidelinks,
  pdfcreator={LaTeX via pandoc}}

\begin{document}

\def\spacingset#1{\renewcommand{\baselinestretch}%
{#1}\small\normalsize} \spacingset{1}


\date{October 13, 2026}
\title{\bf Interpretable subgroup identification across survival and
generalized-linear outcomes}
\author{
Larry F. León\\
and\\Keaven M. Anderson\\
}
\makeatletter
\let\arxivSaveMaketitle\maketitle
\let\arxivSaveAtMaketitle\@maketitle
\let\arxivSaveTitle\title
\let\arxivSaveAuthor\author
\let\arxivSaveDate\date
\let\arxivSaveAnd\and
\let\arxivSaveThanks\thanks
\let\arxivSaveBibsection\bibsection
\makeatother
\maketitle

\bigskip
\bigskip
\begin{abstract}
A subgroup analysis searching a trial's covariates or many prespecified
subgroups often singles out a subgroup for its extremity, interpreted as
if prespecified. Selection inflates the estimate; for a given selection
rule, a correction removes the inflation. Two questions belong to the
rule: which candidate to return, and whether to declare none. A third is
whether inference extends to the patients left outside. We treat all
three for interpretable subgroups, defined by conjunctions of covariate
cuts, generated by forest search, a causal forest or a cross-fit surface
estimator under one set of rules, each a modelling choice trading size
against effect. Forest search screens candidates on splitting
consistency, the rate at which effects replicate across half-splits; the
screen is complementary-pairs stability selection in closed form, and
the law of the maximum over the family gives the family-wise
false-declaration rate when every candidate's effect is at least a
clinically specified benefit. Refit-free multiplier resampling gives
adjusted one-sided bounds (subgroup and complement), joined in a
two-subgroup claim, with coverage asymptotically exact under separated
selection, characterized under local competition, and conditional on a
family proposed from the outcomes. Two trial applications, and
simulations built from them, cover survival, binary and continuous
endpoints.
\end{abstract}

\noindent%
\vfill

\newpage

\section{Introduction}\label{sec-intro}

A confirmatory trial is powered for an effect in its whole population,
and the indication it supports describes a population in which efficacy
and a favourable benefit--risk balance have been concluded. Regulatory
guidance asks that this conclusion be examined across relevant subgroups
rather than assumed, and that a sub-population in which it does not hold
be left out of the indication \citep{ema2019}. It credits a subgroup
finding only when credible, which it ties to prior plausibility and to
replication, and it warns that an estimate singled out for its extremity
is biased by the selection. Approvals have been restricted to a subgroup
after a positive result in the whole population \citep{amatya2021}.

A subgroup of concern is reached by one of two routes: by searching the
trial's covariates after the data are seen, or by examining many
prespecified subgroups and singling one out. Either way it is chosen for
the extremity of its estimate, and its effect is reported through the
analysis the trial would have run had that subgroup been the only one:
the treatment coefficient of the trial's working model fitted within the
subgroup, for a time-to-event outcome a Cox log hazard ratio. So chosen,
the subgroup carries an estimate that overstates its effect, and the
same selection distorts the estimate for the patients outside it
\citep{ema2019, leon2024}. The overstatement is corrected here on that
scale rather than through alternative estimands, because it is the scale
on which the trial was powered and on which regulators and clinicians
interpret its results.

The three identifiers considered here --- forest search
\citep{leon2024}, the causal forest fitted by generalized random forests
(GRF) \citep{wagerathey2018, atheytibshiraniwager2019, cui2023}, and the
difference-in-natural-parameters estimator (DINA) of \citet{gao2025} ---
share one form. Each scores a family of candidate subgroups by an
estimated treatment effect, admits the candidates whose
standard-analysis estimate clears a threshold common to all three, and
reports one admitted candidate as a selection rule directs, or none if
none is admitted. The candidates are conjunctions of at most two
covariate cuts, so whether a patient belongs to one is read from the
patient's own measurements: the subgroups are interpretable, and a
model-based identifier is held to the same form rather than to its
fitted surface. Write \(\beta(g)\) for the treatment coefficient of the
trial's working model in the population restricted to candidate \(g\).
It is the model's least-false value: the value the coefficient estimates
whether or not the model is correct. Its estimate by the standard
analysis of \(g\) is \(\hat\beta(g)\). Forest search enumerates the
candidates exhaustively, scores each by \(\hat\beta(g)\) itself, and
requires in addition that its effect replicate across random
half-splits. DINA fits a cross-fit surface of per-patient effects, the
difference in natural parameters of \citet{gao2025}, and proposes as
candidates the covariate-cut regions whose mean fitted effect clears its
own floor. GRF fits a causal forest, floors the doubly robust contrast
of the forest's own region to decide whether it proposes at all, and
offers the regions of a covariate grid as candidates. DINA and GRF enter
as generators --- the identifiers are the generators, the candidates
what they generate --- and every candidate is re-ranked on
\(\hat\beta(g)\) and held to the same threshold; Supplementary Section
S5 gives each construction in full. The region returned is \(\hat H\),
and the targets of inference are \(\beta(\hat H)\) and, for the patients
outside it, \(\beta(\hat H^c)\).

Which region is returned depends on the selection rule, and the
correction follows it: removing the optimism of selection from
\(\beta(\hat H)\) is a necessary step toward reliable inference about
the region the rule returns and about the patients outside it. Whether
and how to act on such a finding is a broader judgement
\citep{ema2019, amatya2021}. Criteria proposed for extending a trial's
conclusion of benefit to a biomarker-negative subgroup ask that the
subgroup's own data estimate its effect reliably and that the estimate
be clinically relevant \citep{rothmann2012, mcshane2023}; the evidence
here must rest on valid inference to inform analogous evaluations of
benefit and risk.

Three lines of work bear on these questions. The first identifies
subgroups, by recursive partitioning and related searches, reviewed by
\citet{lipkovich2017,lipkovich2024}, or by effect surfaces
\citep{wagerathey2018, atheytibshiraniwager2019, cui2023, gao2025},
trading region size against effect explicitly only through a clinically
meaningful threshold \citep{cai2022} or a utility \citep{zhao2024}. The
second corrects the effect of a selected subgroup, and of that subgroup
alone: by bootstrap de-biasing of the largest estimated effect
\citep{guohe2021}, extended to electronic health records
\citep{guo2023}, by cross-validation \citep{zhao2023}, and by inference
conditional on the winner \citep{andrewskitagawamccloskey2024}. The
third controls errors of selection: by stability selection over
subsamples \citep{meinshausen2010} and its complementary-pairs form
\citep{shahsamworth2013}, by false discovery rate control over
prespecified tree-structured subgroups \citep{luoguo2023}, and by
controlled subgroup selection, which gives a guarantee for every member
of the region rather than for the trial's within-subgroup analysis
\citep{reeve2023, muller2025}.

This paper develops the inference that a corrected subgroup effect
leaves undone: whether a subgroup should be declared at all, what may be
claimed about the patients outside it, and which region a rule should
return. Its starting point is the post-selection framework of
\citet{leonanderson2026}, which treats the three identifiers as
generators of candidates re-ranked on \(\hat\beta(g)\) and corrects the
effect by refit-free multiplier resampling with an
infinitesimal-jackknife interval; that framework fixes one selection
rule, and its intervals are two-sided and separate, with no claim about
both groups. Three contributions follow.

\textbf{Family-wise control of the declaration rate.} To our knowledge,
no existing method governs how often a subgroup search declares a
subgroup when the treatment in fact benefits every candidate, and a
correction does not govern it either. Forest search declares a candidate
when its estimate clears a screening threshold and its effect exceeds
the claim threshold in both halves of a random split of the trial at a
stated rate. That screen is shown to be complementary-pairs stability
selection applied to a smooth estimating-function coefficient. Because
the coefficient is smooth, a candidate's replication rate follows in
closed form, to first order, from one fit of the working model within
it. In stability selection as originally developed, the selected objects
are variables chosen by a procedure such as the lasso, whose selection
probabilities have no closed form, so error rates are bounded rather
than computed. The law of the maximum over the enumerated family then
gives the family-wise rate of falsely declaring a subgroup: the rate at
which one is declared when every candidate subgroup's effect is at least
as beneficial as a clinically specified level. It serves as a diagnostic
of the rate that clinically chosen thresholds imply and, where a capped
rate is wanted, gives the cutoff that holds it, asymptotically, at a
nominal level.

\textbf{Bounds for the discovered subgroup and for its complement, and a
claim about both.} Post-selection methods correct the selected subgroup
and generally stop there. We develop the field, which simulates the law
of the residual the correction leaves and inverts it into one-sided
bounds for the discovered subgroup and for the patients outside it,
joined into a two-subgroup claim, with coverage asymptotically exact
under separated selection and, under local competition, characterized
with a bound on its error. The closest outside work, \citet{guohe2021},
corrects the argmax-selected subgroup through the maximum over the
candidate family; the complement is not the argmax of any functional, so
that framework does not address the complement, or a joint claim about
both.

\textbf{The selection rule as a modelling choice.} The three identifiers
are treated as admitting families of one candidate space of
interpretable conjunctions, with the same five rules acting on whichever
family is admitted. The rule moves the discovered region along a
frontier of size against effect, and so determines what identification
and inference are about.

The framework applies to survival and generalized-linear outcomes on
their working-model scales; the operating characteristics of
Section~\ref{sec-oc} are for survival, with binary and continuous
outcomes evaluated in the Supplementary Materials.

Sections 2--4 develop the identification, the bounds and the declaration
calibration; Sections 5 and 6 report operating characteristics and the
breast cancer application; Section 7 concludes; proofs and further
detail are in the Supplementary Materials.

\section{Interpretable identification across outcome
families}\label{interpretable-identification-across-outcome-families}

\subsection{Identifiers as admissible
families}\label{sec-identification}

Let \(\mathcal{S}\) denote the candidate space: every subgroup defined
by a conjunction of at most two covariate cuts that meets size
eligibility (Supplementary Section S5). The target for every candidate
\(g\) is the same: its working-model coefficient \(\beta(g)\) on its
natural scale, a log hazard ratio for the Cox model and a log odds ratio
or a mean difference for a generalized linear model. It is oriented so
that harm corresponds to \(\beta(g) > c_{\mathrm{cons}}\), where
\(c_{\mathrm{cons}}\) is the \emph{claim threshold}, the boundary of a
sub-optimal effect on that scale (\(\log 1 = 0\) for a ratio; the
difference itself for a mean or risk difference). An identifier does not
select from \(\mathcal{S}\) directly. It admits a subset
\(\mathcal{A} \subseteq \mathcal{S}\), its \emph{admissible family}, and
applies a selection rule to it. Forest search (FS) enumerates
\(\mathcal{S}\) exhaustively, scoring candidates by \(\hat\beta(g)\)
itself and admitting only those whose estimate clears a screening
threshold \(c\) (its floor). It requires in addition a
splitting-consistency rate \(p_{\mathrm{cons}}(g) \ge p^\star\), by
default \(0.90\): the probability over random half-splits that the
candidate's effect clears \(c_{\mathrm{cons}}\) in both halves, which
Section~\ref{sec-screen} gives in closed form. The cross-fit surface
(DINA) proposes candidates read off a fitted surface, floors them on its
own statistic, and has those that clear it re-ranked on \(\hat\beta(g)\)
against the same \(c\). The causal forest (GRF) floors its doubly robust
contrast only to decide whether it returns a region at all; after that,
every size-eligible region of its covariate grid is re-ranked on
\(\hat\beta(g)\) against \(c\). Supplementary Section S5 sets the three
side by side with every setting this paper uses. Four constants recur:
the screening threshold \(c\) a candidate's estimate must clear, the
claim threshold \(c_{\mathrm{cons}}\) at which an effect stops being
beneficial, the consistency level \(p^\star\) of the screen, and, from
Section~\ref{sec-declaration}, the protected level \(c_0\) against which
false declarations are controlled.

\begin{figure}

\centering{

\pandocbounded{\includegraphics[keepaspectratio,alt={A scatter plot with subgroup size on the horizontal axis, from about 60 to 90 subjects, and estimated hazard ratio on the vertical axis, from about 1.6 to 2.6. Nine points show the screened candidates, with a horizontal dashed line at 2.03 marking the epsilon band and the region above it shaded. Three are marked: a black circle at size 82 and hazard ratio 1.95, the largest subgroup; a blue square at 75 and 2.22, the largest subgroup above the line; and an orange triangle at 61 and 2.54, the maximum effect. A dotted step joins the three, showing the frontier of the nine.}]{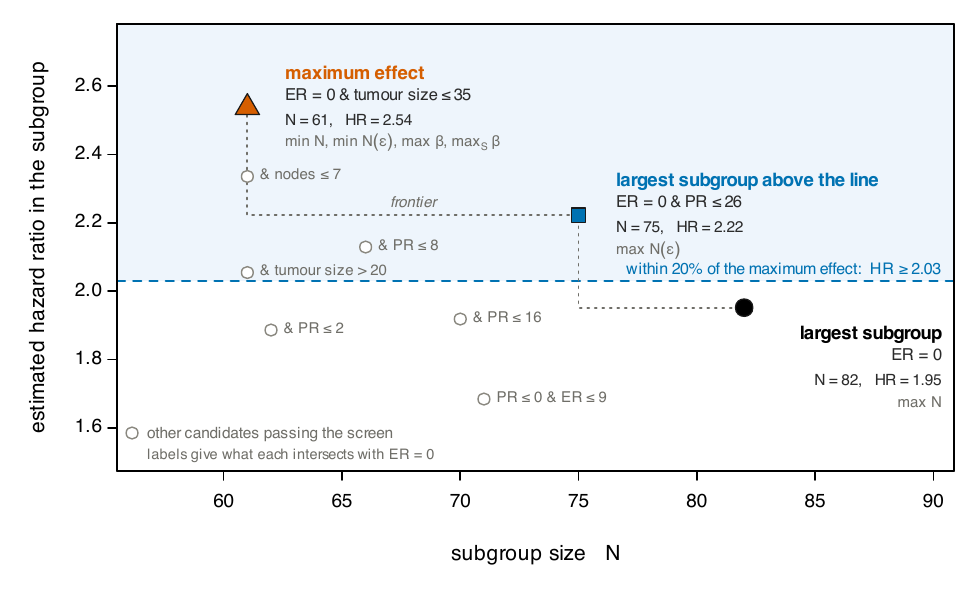}}

}

\caption{\label{fig-criteria}Selection criteria as positions on a
candidate frontier (German Breast Cancer Study Group trial, Section 6).
Each point is one of the nine candidates passing forest search's
consistency screen at 0.90, by its size and estimated hazard ratio:
\(\max_{\mathcal{A}} N\) takes the largest,
\(\max_{\mathcal{A}} N(\varepsilon)\) the largest within 20\% of the
maximum effect (the dashed line, HR \(\ge\) 2.03), and
\(\max_{\mathcal{A}} \beta\), \(\min_{\mathcal{A}} N\),
\(\min_{\mathcal{A}} N(\varepsilon)\) and the unscreened
\(\max_{\mathcal{S}} \beta\) all take the maximum effect. All nine
restrict the estrogen receptor; the dotted step is their frontier.}

\end{figure}

The selection rules are common to the three and act on the admissible
family. Writing \(N(g)\) for the size of \(g\): \(\max_{\mathcal{A}} N\)
takes the largest admissible candidate;
\(\max_{\mathcal{A}} N(\varepsilon)\) takes the largest whose effect is
at least a fraction \(1-\varepsilon\) of the maximum over
\(\mathcal{A}\), on the ratio scale (hazard, odds or rate ratio) or of
the effect itself for a difference; \(\max_{\mathcal{A}} \beta\) takes
the maximum effect; and \(\min_{\mathcal{A}} N\) and
\(\min_{\mathcal{A}} N(\varepsilon)\) are the smallest-subgroup
counterparts. The band of the two \(\varepsilon\) rules is taken
relative to a maximum in the harmful direction, so where no admissible
candidate shows harm those rules select nothing. Forest search alone
admits the unscreened argmax \(\max_{\mathcal{S}} \beta\), the
effect-maximiser over the whole candidate space, which is the selection
primitive of \citet{guohe2021}; it has no counterpart for GRF or DINA
because neither enumerates \(\mathcal{S}\). Figure~\ref{fig-criteria}
shows the rules as positions on one trial's candidate frontier.

\subsection{The splitting-consistency screen}\label{sec-screen}

The screen applies to a candidate that has cleared the floor
\(\hat\beta(g) > c\) on the full fit. As originally formulated, it draws
a fair coin \(Z_i\) for each subject \(i \in g\), refits the working
model on the two halves to obtain \(\hat\beta^{(Z=1)}(g)\) and
\(\hat\beta^{(Z=0)}(g)\), calls the split \emph{consistent} when both
exceed \(c_{\mathrm{cons}}\), and flags \(g\) when the proportion of
consistent splits, \(p_{\mathrm{cons}}(g)\), reaches \(p^\star\).

Repeatedly subsampling the data and retaining the objects that reappear
across most subsamples is stability selection \citep{meinshausen2010},
and drawing the subsamples as complementary halves is its
complementary-pairs form \citep{shahsamworth2013}. The consistency
screen is exactly this construction, with ``object'' read as candidate
subgroup and ``reappears'' as showing harm on both halves:
\(p_{\mathrm{cons}}(g)\) is candidate \(g\)'s selection frequency. Where
the screen departs from stability selection as originally developed is
decisive. There the selected object is a lasso active set, whose
selection probability is intractable, so error rates are bounded rather
than computed; here the base statistic is a smooth estimating-function
coefficient, and its smoothness replaces the bound by an equality.

\emph{The stability probability in closed form.} To first order the two
half-sample estimates are a mirror pair about the full fit,
\(\hat\beta^{(Z=1)}(g) \approx \hat\beta(g) + D_g\) and
\(\hat\beta^{(Z=0)}(g) \approx \hat\beta(g) - D_g\), with
\(D_g = \sum_{i \in g} G_i\,\mathrm{db}_{g,i}\) built from the influence
values \(\mathrm{db}_{g,i}\) of the fit and Rademacher multipliers
\(G_i = 2Z_i - 1\); its conditional variance is the robust variance
\(\sigma_{D,g}^2 = \sum_{i \in g} \mathrm{db}_{g,i}^2\). Both halves
exceed \(c_{\mathrm{cons}}\) exactly when
\(|D_g| < \hat\beta(g) - c_{\mathrm{cons}}\), and a central limit
theorem in \(n_g\) gives
\(D_g \mid \text{data} \approx \mathcal{N}(0, \sigma_{D,g}^2)\), so that
to first order \[
p_{\mathrm{cons}}(g)
\approx
\max\!\Big\{0,\; 2\,\Phi\!\big((\hat\beta(g) - c_{\mathrm{cons}})/\sigma_{D,g}\big) - 1\Big\}.
\] A single fit of \(g\) determines the rate; no splitting or refitting
enters, and the closed form is what the implementation evaluates.
Inverting it, the candidate passes at level \(p^\star\) if and only if
\[
\hat\beta(g) - z_{(1+p^\star)/2}\,\sigma_{D,g} \;\ge\; c_{\mathrm{cons}} :
\] the screen is a one-sided \(z\)-test of
\(H_0 : \beta(g) \le c_{\mathrm{cons}}\) with the split-matched robust
standard error, at level \((1-p^\star)/2\), \(0.05\) at the default. It
tightens with subgroup size only through \(\sigma_{D,g}\), and forest
search admits \(g\) exactly when \(\hat\beta(g) > c\) and
\(\hat\beta(g) \ge c_{\mathrm{cons}} + z_{(1+p^\star)/2}\,\sigma_{D,g}\);
at this paper's default \(c = c_{\mathrm{cons}}\) the floor is implied
by the screen.

The per-candidate false-flag probability is therefore not bounded but
calibrated. What is not is the family-wise rate: forest search flags a
subgroup when \emph{any} candidate survives, so the governing quantity
is the law of the maximum, \[
\mathbb{P}\!\left(\max_{g \in \mathcal{S}} T_g > \kappa \,\Big|\, \beta(g) \le c_{\mathrm{cons}}\ \text{for all } g\right),
\qquad
T_g := \frac{\hat\beta(g) - c_{\mathrm{cons}}}{\sigma_{D,g}},
\] over a family that forest search enumerates once and holds fixed.
Calibrating the cutoff \(\kappa\) so that this probability equals a
target \(\alpha\), and thereby promoting the per-candidate calibration
to family-wise control, is the subject of Section~\ref{sec-declaration},
which also moves the boundary of that null set from the claim threshold
to a clinically specified level of uniform benefit, where the control is
stated. The calibration is forest search's because forest search alone
applies the screen; that its family is enumerated once from the
covariates is what makes the control unconditional
(Section~\ref{sec-declaration-diagnostic}). Removing the screen reduces
forest search's selection to the argmax primitive of \citet{guohe2021}
and \citet{guo2023}; what the screen adds is the option to declare
nothing, and it is the family-wise error of that added structure that
Section~\ref{sec-declaration} calibrates.

\subsection{What the rules trade}\label{sec-rules-tradeoff}

\begin{figure}

\centering{

\pandocbounded{\includegraphics[keepaspectratio,alt={Four panels. A, B and C plot sensitivity, specificity and the identified-to-planted size ratio against the six selection rules for three identifiers, points coloured by rule and shaped by identifier and joined by grey lines; D plots harm captured against benefit retained for the same rules and identifiers.}]{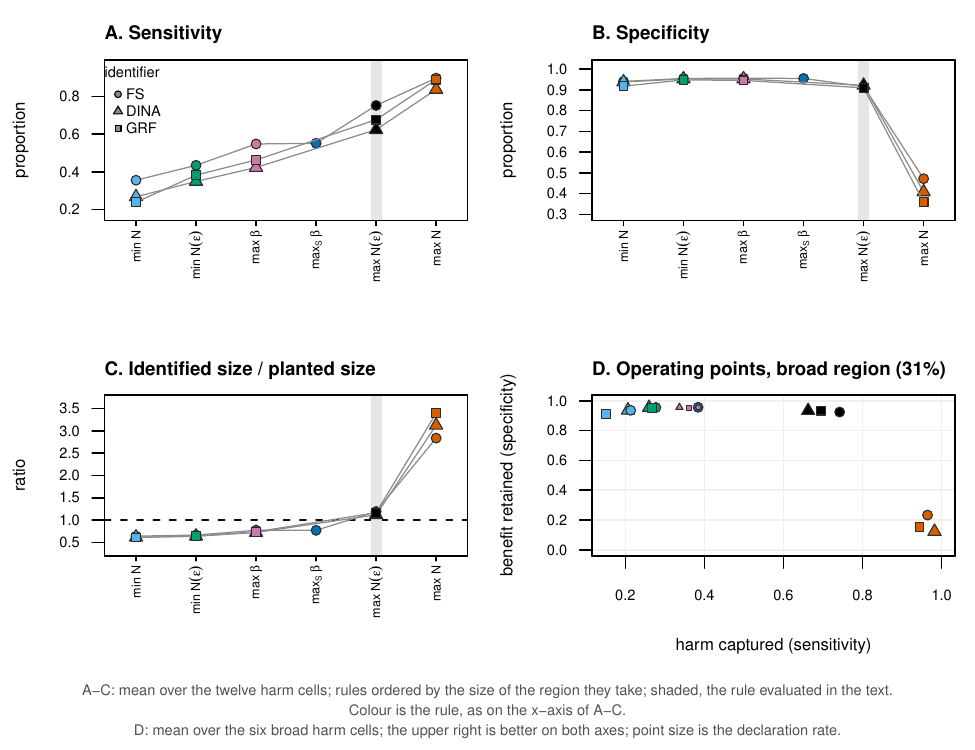}}

}

\caption{\label{fig-rules-sweep}Identification under the selection rules
for forest search (FS), DINA and GRF. A--C: sensitivity, specificity and
identified size as a multiple of the planted size, averaged over the
twelve harm cells, the rules ordered by the size of the region they
take, six positions in all; D: harm captured against benefit retained,
for the six harm cells at the larger prevalence. Each value in A--C is
the unweighted mean over the twelve cells of a per-cell summary,
sensitivity and specificity scoring a replicate that returns no region
as 0 and 1; D's point size is the declaration rate.}

\end{figure}

Identification is judged as classification of patients into the region
and its complement. Sensitivity of \(\hat H\) for the planted region is
harm captured, its specificity is benefit retained, and the rule sets
the trade between them because it sets how much of the trial the region
takes. Figure~\ref{fig-rules-sweep} shows this trade on the simulation
design of Section~\ref{sec-oc-design}. That design plants a region of
known effect in simulated trials built from real trial data. Each cell
of the design fixes the region's prevalence, its hazard ratio and the
sample size. Figure~\ref{fig-rules-sweep} uses the twelve harm cells,
where the planted hazard ratio is 1.50 or 1.75, at two prevalences and
three sample sizes. It covers all three identifiers under every rule, at
500 replicates per cell, and evaluates identification only, with no
inference. The six attenuated-benefit cells, where the planted region
carries no harm (its hazard ratio is 1.0), are shown the same way in
Supplementary Section S2.1. Sensitivity rises monotonically along the
order of region size, from 0.24--0.36 at min N to 0.84--0.90 at max N.
Specificity holds between 0.91 and 0.96 for every rule but max N and
then falls to 0.36--0.47. The third panel is the reason. Every rule but
two returns a region 0.61 to 0.77 times the planted size; max N(ε)
returns 1.12 to 1.19 times; max N returns 2.8 to 3.4 times, and captures
most of the planted region by capturing most of the trial. Everything
below max N(ε) is therefore dominated, with the same benefit retained
and less harm captured, and max N pays a third to two thirds of the
complement for its extra sensitivity. The pattern holds for all three
identifiers, so it is a property of the rule rather than of any one
method. The rule evaluated in the rest of the paper is max N(ε) at ε =
0.20. It captures the most harm among the rules that retain benefit at
0.91 or above on average over the twelve cells, and it is the only rule
whose identified region is approximately the size of the planted one.
The choice is an operating-characteristic one; no optimality is claimed
for it.

\section{Multiplier resampling}\label{sec-mr}

This section sets out the multiplier resampling, its conditions and
local limit law, and the selection-adjusted one-sided bounds for the
identified region, its complement and both jointly. Proofs are in
Supplementary Section S1.

\subsection{Re-selection by multiplier
resampling}\label{sec-reselection}

Multiplier resampling re-runs the search on randomly reweighted copies
of the data without refitting any model; every candidate is perturbed by
the same weights, so the copies carry the overlap among candidates. Each
candidate \(g \in \mathcal S\) receives the trial's standard analysis, a
Cox or generalized linear model fitted to its members, giving the
treatment coefficient \(\hat\beta(g)\) and influence values
\(\mathrm{db}_{g,i}\) (Section~\ref{sec-screen}), zero for
\(i \notin g\). Selection is re-applied over a fixed family
\(\mathcal F \subseteq \mathcal S\) of \(M\) candidates: the candidate
space itself for forest search, GRF's covariate grid, and DINA's
proposed candidates. Their influence values form the \(n \times M\)
matrix \(\mathbf D\). Selection is a map \(h_n\) from candidate effects
\(v\) to \(\mathcal F \cup \{\varnothing\}\). A candidate \(g\) is
admitted when \(v_g\) clears its floor \(t_g\) (Supplementary Section S5
gives each identifier's comparison) and, for forest search, when it also
clears the screen of Section~\ref{sec-screen}; the rule of
Section~\ref{sec-identification} chooses among the admitted, and
\(h_n(v) = \varnothing\) when none is. The identified region is
\(\hat H = h_n(\hat\beta)\).

For \(b = 1, \dots, B\) let \(G^{(b)} \in \mathbb R^n\) have independent
centred entries of unit variance, Poisson(1) − 1 by default, which
linearizes the nonparametric bootstrap \citep{efron2014}. With
\(D(b) = \mathbf D^\top G^{(b)}\) and \(\mathcal B\) the draws on which
\(\hat H^*_b = h_n(\hat\beta + D(b))\) is not empty, the two-term
correction of \citet{leonanderson2026} is
\begin{equation}\protect\phantomsection\label{eq-two-term}{
\tilde\beta(\hat H) = \hat\beta(\hat H) - \frac{1}{|\mathcal B|}\sum_{b\in\mathcal B}\bigl\{D_{\hat H^*_b}(b) + D_{\hat H}(b)\bigr\}.
}\end{equation} One draw perturbs every candidate and so carries their
overlap covariance, and on a fixed family re-selection reproduces the
bootstrap of the whole search to first order (their Theorem 1). The
first term in braces estimates the optimism of selection. The second
supplies the residuals of the infinitesimal-jackknife variance
\(\hat V\) on the same draws, and
\(\tilde\beta(\hat H) \pm z_{1-\alpha/2}\hat V^{1/2}\) is the two-sided
reference interval of Supplementary Section S2.5. The complement is
corrected likewise.

\subsection{Conditions and the local limit law}\label{sec-local-limit}

Let \(\beta(g)\) be the working-model parameter of candidate \(g\), its
least-false value (Section 1); the estimand is \(\beta(\hat H)\). The
data are independent and identically distributed, and limits are taken
under five conditions, stated in full in Supplementary Section S1.1.

(A1) Treatment is randomized, \(P(A = 1) \in (0, 1)\); for a
time-to-event outcome, censoring is independent and follow-up bounded.
(A2) The family \(\mathcal F\) is finite, with \(M\) fixed, generated
from the covariates alone and held fixed under resampling; each region
has positive probability and non-singular information. (A3) Each
\(\hat\beta(g)\) is asymptotically linear with influence values
consistently estimated by \(\mathrm{db}_{g,i}\), so that
\(n\mathbf D^\top\mathbf D \to_p \Sigma\), the overlap covariance of the
family. (A4) The multipliers are exchangeable, with mean zero and unit
variance, and satisfy the conditions of \citet{praestgaardwellner1993};
no single subject dominates any candidate's variance. (A5)
\(\Sigma \succ 0\), and every comparison that settles the selection,
each candidate's admission against its floor and the rule's choice among
the admitted, is either \emph{separated} or \emph{local}.
\emph{Separated} means decided at the population values by a margin
\(\delta_n\) with \(\sqrt n\,\delta_n \to \infty\) ((A5-i) for a fixed
margin, (A5-i′) in general); \emph{local}, (A5-ii), means
\(\beta(g) = \beta_0 + n^{-1/2}\mu_g\) on the competing candidates. The
limiting roots below are continuous at the quantiles inverted.

Under (A5-ii) selection remains random:
\(\sqrt n(\hat\beta - \beta_0\mathbf 1)\) converges to the Gaussian
field \(W = \mu + \zeta\), \(\zeta \sim N(0,\Sigma)\), and \(h_n\)
converges on this scale to a limit rule \(h\) whose discontinuities have
probability zero.

\textbf{Theorem 1} \citep{leonanderson2026}. Assume (A1)--(A5) with
(A5-ii), suppose that a subgroup is declared with probability tending to
one, on the data and on the draws, and let \(B \to \infty\). Then \[
\sqrt n\{\tilde\beta(\hat H) - \beta(\hat H)\} \Rightarrow \Lambda = \zeta_{h(W)} - m(W), \qquad m(w) = \mathbb E\,\zeta^*_{h(w+\zeta^*)},
\] where \(\zeta^*\) is an independent copy of \(\zeta\).

This is Theorem 2(ii) of \citet{leonanderson2026} for the implemented
rules (Supplementary Section S1.1, which also gives the limit when
declaration is uncertain). Because the correction evaluates the optimism
at \(W\) rather than \(\mu\), \(\Lambda\) is not centred: its mean is
positive near ties and, for the argmax, negative under separation (their
Theorem 2(iii)); for the size rules no sign reversal is established
(Supplementary Section S1.1).

\subsection{The field constructions}\label{sec-field-main}

Each of the three bounds simulates the law of Theorem 1 at a plug-in
estimate of its arguments and inverts its quantiles. For a point \(w\)
let \(\Lambda_w = \zeta^*_{h(w+\zeta^*)} - m(w+\zeta^*)\) be the root
centred at \(w\), so that \(\Lambda\) has the law of \(\Lambda_\mu\).

\emph{The field.} The field is the array of simulated candidate effects
over the family. The plug-in replaces \(N(0,\Sigma)\) by
\(N(0,\hat\Sigma)\), \(\hat\Sigma = n\mathbf D^\top\mathbf D\), and the
centre by the shrunk field \(w\), which is \(\hat\beta\) with its
\(\hat H\) entry replaced by \(\tilde\beta(\hat H)\). On the local scale
\(w\) converges to \(\tilde W\), which equals \(W\) except in the
selected coordinate, where it is \(W_{h(W)} - m(W)\). With independent
\(\xi_r, \xi'_j \sim N(0, I_n)\), \(r \le R_{\mathrm{out}}\),
\(j \le R_{\mathrm{in}}\), \(\zeta^*_r = \mathbf D^\top\xi_r\) and
\(\zeta'_j = \mathbf D^\top\xi'_j\), each outer draw that declares a
subgroup gives
\begin{equation}\protect\phantomsection\label{eq-field-root}{
g_r = h_n(w + \zeta^*_r), \qquad g_{rj} = h_n(w + \zeta^*_r + \zeta'_j), \qquad \Lambda^*_r = \zeta^*_{r,g_r} - \frac{1}{|\mathcal J_r|}\sum_{j \in \mathcal J_r} \zeta'_{j,g_{rj}},
}\end{equation} with \(\mathcal J_r\) the inner draws that declare one.
The field lower bound is
\(L = \tilde\beta(\hat H) - q_{1-\alpha}(\Lambda^*)\), \(q\) the
quantile over outer draws.

\textbf{Proposition 1.} Under the conditions of Theorem 1, as
\(R_{\mathrm{out}}, R_{\mathrm{in}} \to \infty\),
\(\lim_{n}\Pr\{\beta(\hat H) \ge L\} = \Pr\{\Lambda \le q_{1-\alpha}(\Lambda_{\tilde W})\}\),
which equals \(1-\alpha\) when \(\tilde W = \mu\), in particular under
(A5-i′).

The field reads the quantile at \(\tilde W\) where asymptotically exact
coverage needs it at \(\mu\), which is not estimable under local
asymptotics; the error vanishes as the competition separates.

\emph{The complement.} With \(\mathbf D^c\) holding in column \(g\) the
influence values of the fit to \(g^c\), the complement field reuses the
multipliers, \(\zeta^{c*}_r = (\mathbf D^c)^\top\xi_r\), and forms
\(\Lambda^{c*}_r\) as in (\ref{eq-field-root}) at the same \(g_r\) and
\(g_{rj}\), without re-selecting. Let \(s_g\) be the robust standard
error of the fit to \(g^c\). The studentized root rescales each reading
to the realized selection's scale,
\begin{equation}\protect\phantomsection\label{eq-field-s}{
\Lambda^{cs*}_r = \frac{s_{\hat H}}{s_{g_r}}\,\zeta^{c*}_{r,g_r} - \frac{1}{|\mathcal J_r|}\sum_{j \in \mathcal J_r}\frac{s_{\hat H}}{s_{g_{rj}}}\,\zeta^{c\prime}_{j,g_{rj}},
}\end{equation} and the field-s upper bound is
\(U = \tilde\beta(\hat H^c) - q_\alpha(\Lambda^{cs*})\).

\textbf{Proposition 2.} Under the conditions of Theorem 1, with (A2) and
(A3) also for the complements,
\(\sqrt n\{\tilde\beta(\hat H^c) - \beta(\hat H^c)\} \Rightarrow \Lambda^c = \zeta^c_{h(W)} - m^c(W)\),
where \(\zeta^c\) is the limiting complement field, jointly Gaussian
with \(\zeta\), and \(m^c(w) = \mathbb E\,\zeta^{c*}_{h(w+\zeta^*)}\);
no condition is placed on the covariance of \(\zeta^c\).

\textbf{Proposition 3.} Under the conditions of Proposition 2, write
\(\zeta^c_g = \sigma^c_g Z_g\) with \(\sigma^c_g = \lim \sqrt n\, s_g\),
let \(\tilde\Lambda^c_w\) be the root of the unit-scale field \(Z\)
centred at \(w\) and \(m^Z\) its optimism, and let
\(R_{\mathrm{out}}, R_{\mathrm{in}} \to \infty\). Then (i)
\(q_\alpha(\Lambda^{cs*}) = s_{\hat H}\,q_\alpha(\tilde\Lambda^{c*})\)
exactly; (ii)
\(\lim_{n}\Pr\{\beta(\hat H^c) \le U\} = \Pr\{T \ge q_\alpha(\tilde\Lambda^c_{\tilde W})\}\)
with \(T = \{\zeta^c_{h(W)} - m^c(W)\}/\sigma^c_{h(W)}\), which equals
\(1-\alpha\) when \(\tilde W = \mu\) and
\(m^c(W) = \sigma^c_{h(W)}\,m^Z(W)\); (iii) \(\Lambda^{cs}\) and
\(\Lambda^c\) coincide when \(\sigma^c_g\) is constant over the
candidates carrying re-selection probability; and (iv) under (A5-i′) the
coverage in (ii) is \(1-\alpha\).

\emph{The joint claim.} The two roots share their outer draws, and the
Bonferroni pair
\begin{equation}\protect\phantomsection\label{eq-bonf-pair}{
L_{\alpha/2} = \tilde\beta(\hat H) - q_{1-\alpha/2}(\Lambda^*), \qquad U_{\alpha/2} = \tilde\beta(\hat H^c) - q_{\alpha/2}(\Lambda^{cs*})
}\end{equation} supports the claim \(\beta(\hat H) \ge L_{\alpha/2}\)
and \(\beta(\hat H^c) \le U_{\alpha/2}\) at joint level at least
\(1-\alpha\) whenever each bound attains its marginal level.

\textbf{Proposition 4.} For every candidate \(g\),
\(\operatorname{Cov}(\zeta_g, \zeta^c_g) = 0\). Under (A1)--(A5), with
(A2) and (A3) also for the complements and declaration with probability
tending to one: under (A5-i′), \(\Lambda\) and \(\Lambda^c\) are
asymptotically independent and the pair (\ref{eq-bonf-pair}) has
asymptotic joint coverage \((1-\alpha/2)^2 = 1-\alpha+\alpha^2/4\);
under (A5-ii), conditionally on the selection the leading terms of the
two roots are uncorrelated, and their dependence enters only through the
common selection and the optimism terms.

\emph{Asymptotic exactness under separation.} Once the selection
settles, \(\tilde W\) and \(\mu\) coincide in the limit.

\textbf{Theorem 2.} Assume (A1)--(A4) and (A5-i′), with \(g^\star\) the
candidate the separated comparisons select, and let
\(B, R_{\mathrm{out}}, R_{\mathrm{in}} \to \infty\). Then
\(\sqrt n\{\tilde\beta(\hat H) - \beta(\hat H)\} \Rightarrow N(0, \Sigma_{g^\star g^\star})\),
the simulated root has the same limit conditionally on the data, and
\(\lim_n \Pr\{\beta(\hat H) \ge L\} = 1 - \alpha\). With (A2) and (A3)
for the complements,
\(\lim_n \Pr\{\beta(\hat H^c) \le U\} = 1 - \alpha\), the two limiting
errors are independent, and the pair (\ref{eq-bonf-pair}) has asymptotic
joint coverage \((1 - \alpha/2)^2\).

For a family proposed from the outcomes, the constructions are applied
with the realized family held fixed: every statement of this section is
conditional on that family, and none accounts for the variability of the
proposal step (Supplementary Section S1.9).

Coverage is asymptotically exact under (A5-i′), every comparison that
settles the selection being decided by a margin \(\delta_n\) with
\(\sqrt n\,\delta_n \to \infty\); where one is not, Propositions 1--4
describe the bounds, the coverage error being bounded by the
concentration of the root near its nominal quantile and the tail of the
quantile-location error between \(Q(\tilde W)\) and \(Q(\mu)\)
(Supplementary Section S1.2). Supplementary Section S1.7 evaluates both
regimes and the envelope in closed form. Bounds are one-sided because
Theorem 1's residual changes sign across regimes (Supplementary Section
S1.2). Where declaration is uncertain the field root omits the
same-draws term of (\ref{eq-two-term}); the omission is conservative on
the harm side under association, and the complement's sign is open
(Supplementary Section S1.2). No nominal guarantee is claimed under
local competition. The bounds are pointwise in \(\mu\)
\citep{leebpotscher2005}.

\section{Calibrating the declaration rate}\label{sec-declaration}

Section~\ref{sec-screen} fixes the level at which a single candidate is
flagged, and Section~\ref{sec-mr} corrects the effect of whichever
candidate the rule returns, but forest search declares a subgroup when
\emph{any} member of its enumerated family survives the screen. That
rate is governed by the law of the maximum of the standardized candidate
statistics, and what counts as a false declaration depends on the
benefit the trial is taken to deliver where no harm exists. The
screening and consistency thresholds are set on clinical grounds. The
calibration serves first as a diagnostic of the family-wise
false-declaration rate those thresholds imply and, where a capped rate
is wanted, as a rule that holds it at a nominal level. This section
calibrates the cutoff against a level of that benefit fixed in advance,
using the field Section~\ref{sec-field-main} already draws: one run of
the search yields the enumerated family, each candidate's coefficient
and robust standard error \(\sigma_{D,g}\), and the shared multiplier
draws, and nothing else is needed. Proofs are in Supplementary Section
S1.6.

\subsection{The declaration event}\label{sec-declaration-event}

The family \(\mathcal F\) of Section~\ref{sec-reselection} is what
selection runs over. The \emph{admitted set}
\begin{equation}\protect\phantomsection\label{eq-declaration-admitted}{
\hat{\mathcal A}(\kappa)
= \bigl\{ g \in \mathcal F : \hat\beta(g) > t_g,\; T_g > \kappa \bigr\},
\qquad
T_g = \frac{\hat\beta(g) - c_{\mathrm{cons}}}{\sigma_{D,g}},
}\end{equation} collects the candidates clearing their floor and the
screen at cutoff \(\kappa\); the conventional screen is
\(\kappa = z_{(1+p^\star)/2}\). The \emph{declaration event} is
\(\{\hat{\mathcal A}(\kappa) \ne \varnothing\}\); only then does the
rule act, and \(\hat H = h_n(\hat\beta)\) is its choice among the
admitted. None of the screen, the rule and the bounds of
Section~\ref{sec-field-main} governs the declaration event, and the
admitted set is not in general a singleton.

\subsection{The calibrated cutoff}\label{sec-declaration-cutoff}

The outer draws of Section~\ref{sec-field-main},
\(\zeta^*_r = \mathbf D^\top\xi_r\) with \(\xi_r \sim N(0, I_n)\), give
the standardized field
\begin{equation}\protect\phantomsection\label{eq-declaration-field}{
Z^*_{r,g} = \zeta^*_{r,g}\,/\,\sigma_{D,g}, \qquad g \in \mathcal F,
}\end{equation} which conditionally on the data is exactly
\(N(0, \hat R)\), \(\hat R\) the correlation matrix of the columns of
\(\mathbf D\).

Let \(c_0 \le c_{\mathrm{cons}}\) be the \emph{protected level}: the
uniform benefit, chosen on clinical grounds before the trial is read,
against which a false declaration is to be protected. A \emph{uniform
benefit} of \(c_0\) means \(\beta(g) = c_0\) for every candidate, each
on its region's own working-model scale; a treatment effect that is the
same for every patient gives regional coefficients close to one another
but not equal, the Cox coefficient not being collapsible, which is why
the null set of Section~\ref{sec-declaration-strong} is an inequality.
Write \[
\delta_g = \frac{c_{\mathrm{cons}} - c_0}{\sigma_{D,g}} \ge 0,
\qquad
M^*_r(c_0) = \max_{g \in \mathcal F}\bigl\{ Z^*_{r,g} - \delta_g \bigr\},
\] the law of \(\max_g T_g\) when every candidate's effect is exactly
\(c_0\). The calibrated cutoff and the diagnostic read that law in
opposite directions:
\begin{equation}\protect\phantomsection\label{eq-declaration-kappa}{
\hat\kappa_\alpha(c_0) = \text{the empirical } (1-\alpha) \text{ quantile of }
\bigl\{ M^*_r(c_0) : r \le R_{\mathrm{out}} \bigr\},
}\end{equation}
\begin{equation}\protect\phantomsection\label{eq-declaration-diagnostic}{
\hat\alpha_{\mathrm{FW}}(c_0;\kappa)
= \frac{1}{R_{\mathrm{out}}}\sum_{r}
\mathbf 1\bigl\{ M^*_r(c_0) > \kappa \bigr\}.
}\end{equation} The \emph{calibrated rule} declares when
\(\max_g T_g > \hat\kappa_\alpha(c_0)\), that is when
\(\hat\beta(g) > t_g\) and
\(\hat\beta(g) > c_{\mathrm{cons}} + \hat\kappa_\alpha(c_0)\sigma_{D,g}\)
for some \(g\); at \(c_0 = c_{\mathrm{cons}}\) every \(\delta_g\) is
zero and (\ref{eq-declaration-kappa}) is the cutoff calibrated at the
claim threshold. Because \(\hat R\) carries the overlap of the regions,
the cutoff prices the multiplicity the family actually has rather than
its nominal size \(M\); \(2\Phi\{\hat\kappa_\alpha(c_0)\} - 1\) is the
screening level it implies (Supplementary Section S1.8).

\textbf{Where the cutoff sits.} Let \(g^\dagger\) be the least
informative candidate, the one with the largest \(\sigma_{D,g}\), and
\(\Delta_g = \delta_g - \delta_{g^\dagger} \ge 0\). On every draw
\(Z^*_{r,g^\dagger} \le \max_g(Z^*_{r,g} - \Delta_g) \le \max_g Z^*_{r,g}\),
and \(Z^*_{r,g^\dagger}\) is exactly standard normal given the data, so
the centred cutoff
\(\tilde\kappa_\alpha(c_0) = \hat\kappa_\alpha(c_0) + \delta_{g^\dagger}\)
satisfies
\begin{equation}\protect\phantomsection\label{eq-declaration-sandwich}{
z_{1-\alpha} \;\lesssim\; \tilde\kappa_\alpha(c_0) \;\le\; \hat\kappa_\alpha(c_{\mathrm{cons}}),
}\end{equation} the right-hand inequality exactly, the draws being
shared, and the left up to the error of an empirical quantile. Read on
the scale of the estimates, the calibrated rule admits \(g\) when \[
\hat\beta(g) \;>\; c_{\mathrm{cons}} - (c_{\mathrm{cons}} - c_0)\,
\frac{\sigma_{D,g}}{\sigma_{D,g^\dagger}} + \tilde\kappa_\alpha(c_0)\,\sigma_{D,g}:
\] each candidate is tested against a threshold of its own --- \(c_0\)
for the least informative, and closer to the claim threshold the more
informative a candidate is --- at one critical value between a single
candidate's and the whole family's. Candidates whose shift exceeds
\(g^\dagger\)'s by more than a few units no longer reach the maximum.
Because \(\Delta_g\) is a fixed clinical distance times a difference of
inverse standard errors, their number falls as the trial's information
grows.

\textbf{Theorem 3.} Assume (A1)--(A4), with
\(\liminf_n n\sigma_{D,g}^2 > 0\) for every \(g \in \mathcal F\), and
let \(\hat R \to_p R\) and \(R_{\mathrm{out}} \to \infty\). Then
\(\hat\kappa_\alpha(c_{\mathrm{cons}}) \to_p \kappa_\alpha(R)\), the
\((1-\alpha)\) quantile of \(\max_g Z_g\) for \(Z \sim N(0,R)\). If in
addition \(\beta(g) = c_{\mathrm{cons}}\) for every \(g \in \mathcal F\)
and the floors \(t_g\) are inactive, then
\(\Pr\{\hat{\mathcal A}\{\hat\kappa_\alpha(c_{\mathrm{cons}})\} \ne \varnothing\} \to \alpha\).

The quantile behind (\ref{eq-declaration-kappa}) exists and is unique
whatever the rank of \(R\) (Supplementary Lemma S4).

\subsection{Control at the protected
level}\label{sec-declaration-strong}

The error that matters in a trial is flagging a subgroup as harmful when
the treatment's effect in it is in fact at least as beneficial as a
clinically meaningful level, and that is what the protected level names.
Write \(\mathcal I_0(c_0) = \{g \in \mathcal F : \beta(g) \le c_0\}\)
for the candidates at least as beneficial as \(c_0\), and
\(U_g = \{\hat\beta(g) - \beta(g)\}/\sigma_{D,g}\) for the studentized
error.

\textbf{Proposition 5.} Let \(c_0 \le c_{\mathrm{cons}}\) and
\(M(c_0) = \max_{g \in \mathcal F}(U_g - \delta_g)\). (i) On every
realization \(\max_{g \in \mathcal I_0(c_0)} T_g \le M(c_0)\), so that
for every \(n\),
\(\Pr\{\exists\, g \in \mathcal I_0(c_0) : T_g > \hat\kappa_\alpha(c_0)\} \le \Pr\{M(c_0) > \hat\kappa_\alpha(c_0)\}\),
and the same bound covers the event that the reported region \(\hat H\)
lies in \(\mathcal I_0(c_0)\). (ii) Under the working model that treats
the estimated covariance as known, \(U \sim N(0, \hat R)\) given
\(\mathbf D\), the right-hand side equals \(\alpha\) for every \(n\) and
every \(c_0\), up to the error of an empirical quantile.

Part (i) needs no model and part (ii) holds because
\(\hat\kappa_\alpha(c_0)\) is the quantile of that same law. What holds
without the working model, as the trial grows, is Theorem 3 at the claim
threshold and the following at any fixed protected level below it.

\textbf{Theorem 4.} Assume the conditions of Theorem 3, let
\(c_0 < c_{\mathrm{cons}}\) be fixed, and suppose the limiting standard
errors \(\sigma_g = \lim_n \sqrt n\,\sigma_{D,g}\) have a unique
maximizer \(g^\dagger\) on \(\mathcal F\) (the data-defined
\(g^\dagger\) of Section 4.2 coincides with it with probability tending
to one). Then (i) \(\tilde\kappa_\alpha(c_0) \to_p z_{1-\alpha}\); (ii)
whatever the effects of the candidates outside \(\mathcal I_0(c_0)\), \[
\limsup_n \Pr\bigl\{\, \exists\, g \in \mathcal I_0(c_0) : T_g > \hat\kappa_\alpha(c_0) \,\bigr\} \;\le\; \alpha ,
\] with equality when \(\beta(g^\dagger) = c_0\) and the floors are
inactive; (iii) if some candidate has \(\beta(g) \ge c_{\mathrm{cons}}\)
--- a region in which the treatment is not beneficial --- the
probability that a subgroup is declared tends to one with the floors
inactive, and with the floors at \(c_{\mathrm{cons}}\) whenever
\(\beta(g) > c_{\mathrm{cons}}\).

Theorems 3 and 4 are the two ends of (\ref{eq-declaration-sandwich}): at
the claim threshold the calibration pays for the whole family's
multiplicity through \(R\), and at any fixed level below it the shifts
separate the candidates as information accumulates, until the rule is a
level-\(\alpha\) one-sided test of \(\beta(g^\dagger) \le c_0\) for the
least informative candidate. A trial sits between the ends, where
\(\Delta_g\) puts it (Supplementary Section S1.8). Part (ii) is strong
control, pointwise in the data-generating law; (iii) is its other side.
A region in which the treatment delivers at least a clinically
meaningful benefit is declared with probability at most \(\alpha\); one
in which it delivers none is declared with certainty as the trial grows;
between the two, the rate rises with the shortfall. That is the screen's
purpose: to find where the treatment is sub-optimal against a complement
where it is consistently beneficial, with the boundary set by the
clinician rather than at no benefit. What a declaration licenses is
separate: the region goes to the bounds of Section~\ref{sec-field-main},
and calling it harmful requires the corrected lower bound to reach the
claim threshold, which on the attenuated-benefit cells of
Section~\ref{sec-oc} it did on at most 0.010 of declaring replicates.
Calibrating at the claim threshold instead (Theorem 3) protects even
regions with no benefit at all, at a cost in power that Supplementary
Section S1.8 quantifies.

Power at a fixed alternative is immediate at every protected level,
\(\hat\kappa_\alpha(c_0)\) being bounded above by
(\ref{eq-declaration-sandwich}) while \(T_g \to \infty\) for a candidate
with \(\beta(g) > c_{\mathrm{cons}}\); this is the condition
Section~\ref{sec-local-limit} assumes when it takes declaration as
certain. On the local scale
\(\beta(g) = c_{\mathrm{cons}} + \mu_g/\sqrt n\) the declaration
probability at the claim threshold tends to
\(\Pr[\max_g\{Z_g + \mu_g/\sigma_g\} > \kappa_\alpha(R)]\).

\subsection{Floors, scope and the
diagnostic}\label{sec-declaration-diagnostic}

Only the second of the two thresholds in (\ref{eq-declaration-admitted})
is calibrated. The floor \(t_g\) is deterministic given the fit and can
only remove candidates, so Theorem 3's convergence and Proposition 5's
bound hold with the floors active, the first as an inequality; at this
paper's default, \(t_g = c_{\mathrm{cons}}\), the floor is implied by
the calibrated screen at the claim threshold, since
\(\hat\kappa_\alpha(c_{\mathrm{cons}}) > 0\) for \(\alpha < 1/2\), while
below it the floor binds and Theorem 4(ii) holds conservatively
(Supplementary Section S1.6). The calibration rests on (A2): forest
search's family is enumerated once from the covariates, so conditioning
on it fixes the regions without conditioning on the outcome; for DINA
the family is proposed using the outcomes (GRF's grid is fixed, its one
outcome-dependent step deciding only whether a region is returned), and
(\ref{eq-declaration-kappa}) then prices selection within the realized
family without pricing which family would have been proposed, the
conditional scope Supplementary Section S1.9 sets out. A family growing
with \(n\) would need a high-dimensional Gaussian approximation for the
maximum \citep{cck2017}, and is not pursued.

The same draws give the conventional screen's family-wise size on this
family under a uniform benefit of \(c_0\),
(\ref{eq-declaration-diagnostic}) at the screen's own cutoff: it
quantifies the cost of the default without controlling it, and is a
boundary size; at the claim threshold it is large and not any trial's
error rate.

In practice, we recommend a cap of 0.99 for the consistency level.
Consistency is a floor, not a ranking: every candidate that clears it
enters the trade-off between size and effect that the rule makes, and
rates above 0.99 are not practically distinguishable. Raising the level
further could admit a single candidate while excluding more clinically
meaningful ones just below it. Where the calibrated cutoff implies a
level above 0.99, we would apply the cap and let the diagnostic
(\ref{eq-declaration-diagnostic}) report the family-wise rate the capped
level carries; the control of Theorem 4 applies to the uncapped cutoff.

\section{Operating characteristics}\label{sec-oc}

\subsection{Design}\label{sec-oc-design}

The evaluation uses a survival design built from the German Breast
Cancer Study Group data, so that covariate structure, censoring and
event rates are those of a real trial. The design plants a region of
known effect in each simulated trial. Eighteen cells cross two
prevalences of the planted region, 12.4\% (narrow) and 31\% (broad);
three planted hazard ratios, 1.50 and 1.75 in the twelve harm cells and
one in the six attenuated-benefit cells; and three sample sizes, 500,
1000 and 1500. Each cell is run for 2,000 replicates. The three
identifiers are applied to the same simulated trials, so every contrast
between them is paired.

In the attenuated-benefit cells the planted region has a hazard ratio of
one, while its complement benefits, at a hazard ratio of 0.657 where the
region is narrow and 0.721 where it is broad. The region is not harmful,
but it is meaningfully worse than the rest of the trial. In every cell
the trial as a whole benefits: its marginal hazard ratio runs from 0.685
to 0.710 where the planted region is narrow and from 0.792 to 0.894
where it is broad (Supplementary Section S2). Every simulated trial is
therefore of the kind the method is built for, one whose overall result
would support the treatment.

There is no strict-null cell. The question is therefore never whether a
method finds structure in noise, but whether it correctly characterises
structure that is there. The rule throughout is
\(\max_{\mathcal{A}} N(\varepsilon)\) at \(\varepsilon = 0.20\), chosen
in Section~\ref{sec-rules-tradeoff}; nothing below speaks to the other
rules. The corrections and bounds of Section~\ref{sec-mr} are valid for
any choice of screening and consistency thresholds; the simulations use
the conventional settings, and the false-declaration rate those settings
imply is assessed with the diagnostic of Section~\ref{sec-declaration}
in Supplementary Section S1.8. Supplementary Section S2.8 evaluates the
same rule and bounds on binary and continuous outcomes, in designs built
from the ACTG 175 trial.

Three quantities are reported and they are not interchangeable. The
\textbf{declaration rate} is the share of replicates returning a
subgroup at all. \textbf{Classification} is reported unconditionally, a
replicate that declares nothing being scored as making no assignment,
because that is what happens to a trial. \textbf{Bias, coverage and
bound location} are over declaring replicates, since none exists where
no region was returned. On the attenuated-benefit cells the declaration
rate is also an error rate of the kind Section~\ref{sec-declaration}
controls: there every candidate has \(\beta(g) \le c_{\mathrm{cons}}\),
so \(\mathcal I_0(c_{\mathrm{cons}}) = \mathcal F\) and every
declaration admits a candidate with no true harm. The recommended
products are the field one-sided lower bound on \(\beta(\hat H)\) and
the field-s one-sided upper bound on \(\beta(\hat H^c)\), each at
nominal 0.95, with the Bonferroni pair joining them at joint 0.95. Two
references appear throughout: the \textbf{unadjusted} analysis, which
proceeds as though the region had been fixed in advance, with the
two-sided Wald limits a standard analysis reports; and the
\textbf{oracle}, which refits on the true region and marks what an
analyst who knew the answer could attain.

\subsection{The correction and its bounds}\label{sec-oc-bounds}

Unadjusted, the effect in the identified region is overstated in every
cell, by a relative bias of 0.10 at the mildest and 1.10 at the most
extreme. The correction brings it to between \(-0.03\) and \(+0.33\),
and to between \(-0.023\) and \(+0.161\) for forest search. On the
complement the unadjusted effect is understated by 0.02 to 0.13, and the
correction brings it within roughly 0.03 of zero. A region chosen
because its effect looked extreme leaves behind a complement that looks
better than it is. What remains after correction is the residual of
Section~\ref{sec-local-limit}, positive near ties and negative under
separation. Supplementary Section S2.2 shows the bias cell by cell.

\begin{figure}

\centering{

\pandocbounded{\includegraphics[keepaspectratio,alt={Four stacked panels by cell for forest search, DINA and GRF: the declaration rate, the coverage of the field lower bound, of the field-s upper bound and of the Bonferroni pair jointly, each point with a Wilson interval and a dashed line at 0.95.}]{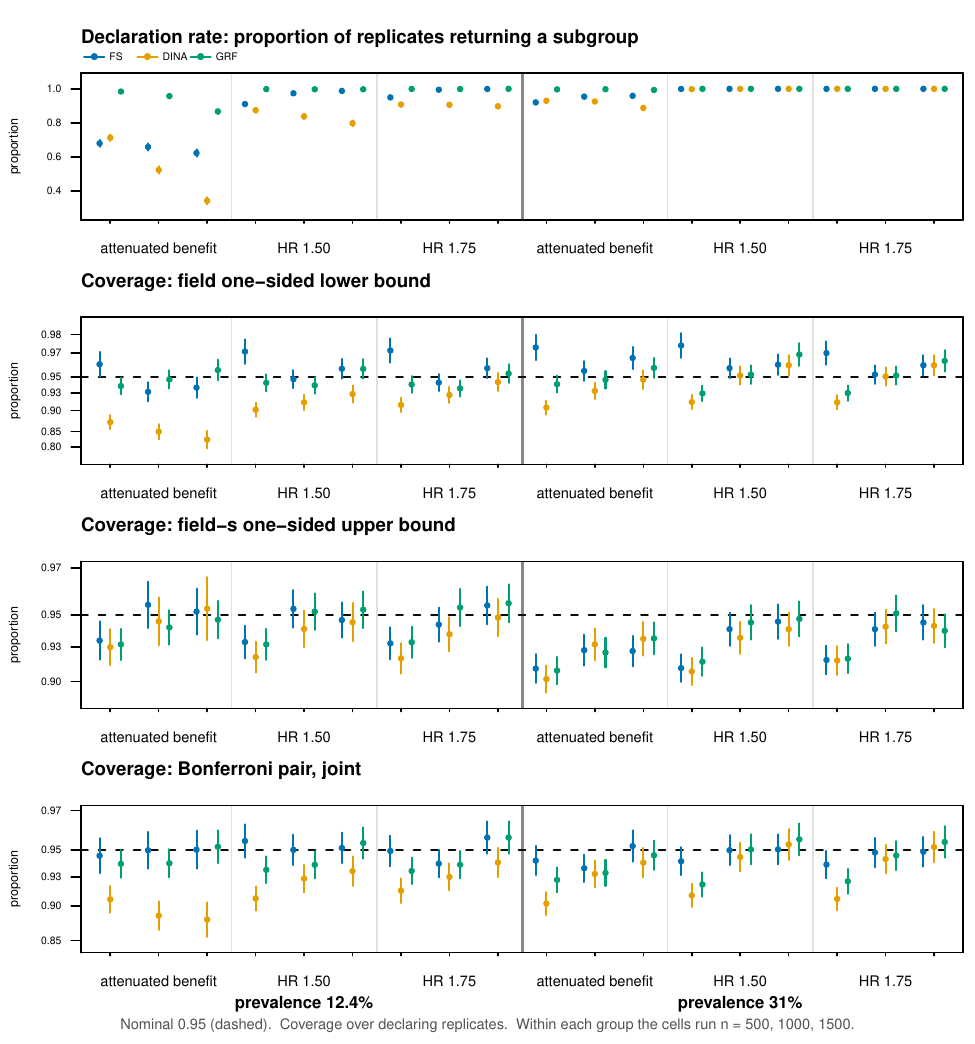}}

}

\caption{\label{fig-oc-coverage}Declaration rate and coverage by cell
for forest search (FS), DINA and GRF: the proportion of replicates
returning a subgroup and, over declaring replicates, the coverage of the
field lower bound on \(\beta(\hat H)\), the field-s upper bound on
\(\beta(\hat H^c)\) and the Bonferroni pair, against its nominal 0.95
(dashed), on a logit scale with Wilson 95\% intervals. Within each cell
group the sample size runs 500, 1000, 1500.}

\end{figure}

The unadjusted lower bound covers between 0.144 and 0.888 against a
nominal 0.95, lowest where no harm is planted and the sample small. The
corrected bound covers between 0.825 and 0.974: forest search 0.932 to
0.974, GRF 0.929 to 0.969, DINA 0.825 to 0.961. DINA lies below the
other two at seventeen of the eighteen cells, including three of the
four at which all three identifiers declare on every replicate. In those
cells coverage conditions on nothing, so the difference is not an
artefact of computing coverage over declaring replicates; it is the
bound itself covering less on the family DINA proposes. On the
complement the unadjusted upper bound covers 0.827 to 0.965 and the
corrected bound 0.903 to 0.956; the studentized Bonferroni pair covers
0.908 to 0.958 across the harm cells. Figure~\ref{fig-oc-coverage} gives
coverage cell by cell beneath the declaration rate that is its
denominator, with Wilson intervals. Supplementary Section S2.2 reads it
by cell type. The attenuated-benefit cells are read separately: there a
region is declared on only part of the replicates, and where declaration
is uncertain the corrected estimate carries the same-draws term that the
field root omits (Section~\ref{sec-field-main}), so those cells are
where that omission would show.

\subsection{Where the bounds sit}\label{sec-oc-location}

Coverage establishes that a bound contains its target at the stated
rate, not that the bound is useful; the question is how often a bound
reaches a value that would support a claim. In the harm cells the
corrected lower bound reaches the claim threshold, a hazard ratio of
one, on 0.023 to 0.817 of replicates, climbing steeply with sample size
and effect strength; much of what limits it is the size of the trial
rather than the price of not knowing the region (Supplementary Section
S2.6 compares it with an oracle that knows the region). In the
attenuated-benefit cells the share stays between 0.004 and 0.030 and
does not climb with sample size: it grows with information where harm is
planted and not where it is absent, so it responds to the effect rather
than to the sample.

Read that way, the attenuated-benefit cells report both errors of the
conventional screen at \(p^\star = 0.90\). Forest search declares a
region on 0.624 to 0.959 of replicates across the six cells, and its
corrected lower bound reaches a hazard ratio of one on 0.004 to 0.010 of
those that declare: the first is the family-wise false-declaration rate
that Section~\ref{sec-declaration} assesses, the second the claim error
that the bound's coverage governs. Used together, the two bounds support
a two-subgroup claim at a joint level: both land simultaneously on 0.58
of replicates at the strongest harm cell and on at most 0.013 in the
attenuated-benefit cells. Classification under the chosen rule is
reported in Supplementary Section S2.4.

\section{Applications}\label{sec-app}

\subsection{The trial and the specification}\label{sec-app-trial}

The search is run once, with all three identifiers on identical data.
The German Breast Cancer Study Group trial compared hormonal therapy
with none in node-positive breast cancer, with recurrence-free survival
as the endpoint. The analysis set is the 686 patients with complete
prognostic data, out of 720 randomized; 246 received hormonal therapy
and 440 did not, with 299 recurrences or deaths, and over the whole
trial the hazard ratio is 0.695 (0.544, 0.888). Randomization status is
masked in the distributed data, roughly a third of patients having
received hormonal therapy outside randomization \citep{schmoor1996}, and
hormonal therapy is analysed here as randomized. Seven covariates enter
the search: age, menopausal status, tumour size, grade, positive lymph
nodes, and the progesterone and estrogen receptor concentrations, for
which the value zero, no measured receptor, is a cut of its own.

Candidates are conjunctions of at most two covariate cuts meeting the
eligibility minimum of sixty patients, with forest search additionally
requiring ten events in each arm; the rule is
\(\max_{\mathcal{A}} N(\varepsilon)\) at \(\varepsilon = 0.20\), and the
screening and per-split thresholds are both a hazard ratio of one, with
the consistency screen at 0.90 for forest search and DINA and GRF
re-ranked on \(\hat\beta(g)\) without it (Supplementary Section S5).
Each analysis reports the unadjusted estimate with its two-sided Wald
interval, the corrected estimate (\ref{eq-two-term}), the nominal 95\%
field lower bound on \(\beta(\hat H)\), the field-s upper bound on
\(\beta(\hat H^c)\), and the Bonferroni pair at joint 0.95, from 5,000
multiplier draws over every size-eligible candidate the identifier
proposes. Every region was declared at the identifiers' default
settings, forest search's by the conventional screen, and the close of
Section~\ref{sec-app-regions} reads forest search's declaration against
the calibrated cutoff. The corrected estimates carry the reading. Each
is read by its location against the trial's own effect; the bounds are
reported beside them and are not read as tests at a hazard ratio of one.
A corrected estimate above 0.90 is read as suggestive of limited
benefit, and where its lower bound lies below 1.0 a benefit cannot be
ruled out. Supplementary Section S3 gives every supporting quantity.

\subsection{The regions and their adjusted
effects}\label{sec-app-regions}

Forest search returns the 75 patients with no measured estrogen receptor
and progesterone receptor at most 26 fmol/l, 11\% of the trial, the
region the rule chooses among the nine candidates of
Figure~\ref{fig-criteria}. GRF returns the 82 patients with no measured
estrogen receptor, a region containing forest search's. DINA returns the
113 patients who are grade 3 with progesterone receptor at most 20
fmol/l, 16\% of the trial, a different classification. Supplementary
Section S3 gives each identifier's candidates and how the three regions
overlap.

Table~\ref{tbl-app-gbsg} reports the analysis, and forest search's
column shows how to read it. The unadjusted hazard ratio in its region
is 2.222, opposite in direction to the trial's 0.695. The correction
removes 0.543 on the log scale and leaves 1.276. That estimate is above
0.90, so it is suggestive of limited benefit; with a field lower bound
of 0.634, below 1.0, a benefit cannot be ruled out. The complement, 89\%
of the trial, has a corrected estimate of 0.645, a larger benefit than
the trial's 0.695, with a field-s upper bound of 0.804. The DINA and GRF
columns read the same way: corrected estimates of 1.076 and 1.111 in
their regions, with lower bounds below 1.0, and of 0.658 and 0.663 in
their complements.

\begin{table}[tbp]

\caption{\label{tbl-app-gbsg}Selection-adjusted analysis of the three
identified regions on the German Breast Cancer Study Group trial,
hazard-ratio scale (ER, PR: estrogen and progesterone receptor, fmol/l).
Unadjusted: two-sided 95\% Wald interval; field and field-s: one-sided
at nominal 0.95; Bonferroni pair: region / complement, joint 0.95. Whole
trial: 686 patients, 299 events, hazard ratio 0.695 (0.544, 0.888).}

\centering{

\begin{tabular}[t]{lrrr}
\toprule
  & FS & DINA & GRF\\
\midrule
Region & ER 0 and PR ≤ 26 & grade 3 and PR ≤ 20 & ER 0\\
N (events) & 75 (41) & 113 (61) & 82 (45)\\
Unadjusted, 95\% CI & 2.222 (1.169, 4.221) & 1.315 (0.746, 2.317) & 1.951 (1.037, 3.673)\\
Corrected & 1.276 & 1.076 & 1.111\\
Field, lower bound & 0.634 & 0.638 & 0.510\\
Complement N (events) & 611 (258) & 573 (238) & 604 (254)\\
Unadjusted, 95\% CI & 0.605 (0.465, 0.788) & 0.641 (0.489, 0.839) & 0.615 (0.472, 0.801)\\
Corrected & 0.645 & 0.658 & 0.663\\
Field-s, upper bound & 0.804 & 0.842 & 0.846\\
Bonferroni pair & 0.509 / 0.840 & 0.572 / 0.887 & 0.404 / 0.874\\
\bottomrule
\end{tabular}

}

\end{table}%

Every region above was declared at the identifiers' default settings,
forest search's by the conventional screen at 0.90.
Section~\ref{sec-declaration} asks whether a screen calibrated to
protect a clinically meaningful uniform benefit would have declared it,
and the answer is read for forest search alone, whose family is
enumerated (Section~\ref{sec-screen}). At a protected hazard ratio of
0.75 and \(\alpha = 0.10\), a level common for screening, the calibrated
cutoff on this family is \(\hat\kappa_{0.10}(c_0) = 2.726\), an implied
screening level of 0.994, while the region's statistic is \(T = 2.438\)
and the largest in the family 2.564: the calibrated screen admits no
candidate, at every protected level from 0.70 to 0.85 and at the claim
threshold, where the cutoff is 3.503. The diagnostic
(\ref{eq-declaration-diagnostic}) puts its boundary size at 0.667 under
a uniform benefit of 0.75; it is computed under the working model with
every candidate's coefficient exactly at \(c_0\), which the Cox model's
non-collapsibility makes an approximation, and it is a rough guide to
the family's multiplicity rather than a rate a trial would show.

The realized rate is the evidence to weigh: Supplementary Section S1.8.4
holds this trial's 686 patients, their covariates and their observed
treatment assignment fixed and generates outcomes under a uniform
benefit at the protected level, and the screen as executed declares a
subgroup on 0.432 of such trials at a consistency level \(p^\star\) of
0.90, on 0.074 at 0.99, and on 0.022 at 1.00, against diagnostics of
0.667 and 0.186 at the first two. Under the recommended cap of
Section~\ref{sec-declaration-diagnostic} the level is 0.99, at which the
region's statistic, 2.438, clears by the narrowest of margins the cutoff
of 2.432 at which the screen as executed, with consistency rates rounded
to two decimals, admits at 0.99 (Supplementary Section S1.8.4), with a
realized false-declaration rate of 0.074 on this design, below
\(\alpha\); the uncapped calibrated cutoff, 2.726, would not have
admitted it. The evidence for declaring the region is marginal on either
reading.

The substantive reading is the same under all three partitions: a small
region whose unfavourable unadjusted estimate the correction largely
removes, and a complement, most of the trial, with a corrected benefit
larger than the trial's. With 41 to 61 events in the regions, a claim of
harm at a conventional level is not realistic at this trial's size; the
bounds are reported, and how much statistical assurance to require of
them is the analyst's decision in their own setting. Forest search on a
binary and a continuous endpoint of ACTG 175 gives the same shape of
result (Supplementary Section S3).

\section{Discussion}\label{sec-discussion}

Section~\ref{sec-mr} proves the one-sided bounds asymptotically nominal
under separation, fixed or increasing. Under local competition it bounds
their discrepancy from nominal by a root anti-concentration term and the
tail of a quantile-location error; on a two-candidate example with every
limit object in closed form (Supplementary Section S1.7), the location
error is small and of either sign, and the envelope is substantially
wider than the exact discrepancy. Against the post hoc design of
\citet{guohe2021}, run as published (Supplementary Section S4), both
corrections restore coverage near the nominal level where the unadjusted
bound covers below it, and only the field construction bounds the
complement and pairs the two bounds. On the survival design built from
the breast cancer trial, the correction removes most of the
overstatement in the identified region, and the bounds cover close to
nominal for forest search and the causal forest and below it for DINA.
The methods are not confined to survival outcomes. Supplementary Section
S3 applies forest search, with the same corrections and bounds, to a
binary and a continuous endpoint of the ACTG 175 trial, on the
odds-ratio and mean-difference scales, with the same shape of result as
on the breast cancer trial; Supplementary Section S2.8 evaluates the
same rule and bounds for all three identifiers by simulation on binary
and continuous designs built from that trial.

Declaration and estimation carry separate guarantees.
Section~\ref{sec-declaration} supplies the first at a level of benefit
the clinician chooses: the family-wise rate of declaring a harm subgroup
is asymptotically at most the nominal level whenever every candidate
subgroup's effect is at least as beneficial as \(c_0\), while a region
with no benefit is declared with certainty as the trial grows. Whether a
declared region is then called harmful is governed by the corrected
bounds of Section~\ref{sec-field-main}. The guarantee is a property of
an enumerated family; for a family proposed by a fitted model, as DINA's
is, it holds conditionally on the family proposed, and for GRF with the
forest's decision to return a region held fixed (Supplementary Section
S1.9). The declaration theory is fixed-dimensional; a family growing
with the sample would need a high-dimensional Gaussian approximation for
the maximum (Supplementary Section S1.10). A bound on the
quantile-location term for a general family, and the treatment of the
same-draws term where declaration is uncertain, remain open.

\medskip
\begin{singlespace}
\setlength{\parindent}{0pt}

\textbf{Supplementary materials.} The supplement contains the proofs
(S1), the operating characteristics in detail (S2), the applications in
full, including binary and continuous analyses of ACTG 175 (S3), the
comparison with \citet{guohe2021} (S4) and the identifiers' settings
(S5). {The methods are implemented in the R package forestsearch
(version 0.4.0, available from CRAN); the causal forest is fitted by the
grf package.}

\textbf{Disclosure statement.} Both authors are employees of Merck \&
Co., Inc., Rahway, NJ, USA, and hold stock in the company; they report
no other competing interests.

\textbf{Data availability.} The German Breast Cancer Study Group and
ACTG 175 data are publicly available in the R packages survival and
speff2trial.

\textbf{AI use.} The authors used Anthropic's Claude models (2025--2026
releases, including Claude Opus 5.5) and Claude Code to draft and revise
text, to draft and check mathematical arguments, and to write and check
code and reported numbers. The authors reviewed all output, verified the
mathematical results, every reported number and the references, and take
full responsibility for the content.

\end{singlespace}

\section*{References}\label{references}
\addcontentsline{toc}{section}{References}

\renewcommand{\bibsection}{}

\clearpage
\makeatletter
\let\maketitle\arxivSaveMaketitle
\let\@maketitle\arxivSaveAtMaketitle
\let\title\arxivSaveTitle
\let\author\arxivSaveAuthor
\let\date\arxivSaveDate
\let\and\arxivSaveAnd
\let\thanks\arxivSaveThanks
\let\bibsection\arxivSaveBibsection
\makeatother
\renewcommand{\topfraction}{0.7}
\renewcommand{\bottomfraction}{0.3}
\renewcommand{\textfraction}{0.2}
\renewcommand{\floatpagefraction}{0.5}
\setcounter{topnumber}{2}
\setcounter{totalnumber}{3}
\setcounter{section}{0}
\setcounter{subsection}{0}
\setcounter{subsubsection}{0}
\setcounter{figure}{0}
\setcounter{table}{0}
\setcounter{equation}{0}
\setcounter{footnote}{0}
\renewcommand{\thesection}{S\arabic{section}}
\renewcommand{\thesubsection}{S\arabic{section}.\arabic{subsection}}
\renewcommand{\thesubsubsection}{S\arabic{section}.\arabic{subsection}.\arabic{subsubsection}}
\renewcommand{\thetable}{S\arabic{table}}
\renewcommand{\thefigure}{S\arabic{figure}}
\renewcommand{\theequation}{S\arabic{equation}}
\renewcommand{\theHsection}{S\arabic{section}}
\renewcommand{\theHfigure}{S\arabic{figure}}
\renewcommand{\theHtable}{S\arabic{table}}
\renewcommand{\theHequation}{S\arabic{equation}}

\def\spacingset#1{\renewcommand{\baselinestretch}%
{#1}\small\normalsize} \spacingset{1}


\date{October 13, 2026}
\title{\bf Supplementary Materials for ``Interpretable subgroup
identification across survival and generalized-linear outcomes''}
\author{
Larry F. León\\
and\\Keaven M. Anderson\\
}
\maketitle

\bigskip
\bigskip
\begin{abstract}
Proofs of all results (Section S1), the operating characteristics in
detail (S2), the applications in full, including the ACTG 175
illustrations (S3), the comparison with \citet{guohe2021} (S4), and the
identifiers' settings and constructions (S5). Section, equation, figure
and table numbers without an S prefix refer to the main paper.
\end{abstract}

\newpage

This supplement accompanies the main paper. It carries the conditions
and proofs behind the paper's theory, the simulation and application
detail behind its results, a comparison with the closest prior method,
and the identifiers' settings and constructions by outcome type. Each
section is self-contained. References to the main paper are by section
number in plain text; no cross-reference crosses the boundary between
the two documents.

Section S1 gives the conditions, limit laws and proofs behind Sections 3
and 4 of the main paper. Its Section S1.7 works the field constructions
out exactly on the smallest selection problem --- two candidate regions,
each effect estimated by ordinary least squares, the region with the
larger estimate selected --- and then runs the implementation on data
drawn from that model: the field lower bound's coverage rates track
their limits closely. A short summary at the head of Section S1.7 gives
the comparison in brief.

Section S2 carries the operating characteristics of Section 5 in full:
the selection rules as operating points, coverage cell by cell, bias
against the planted target, classification under both conventions, the
two-sided interval, where the bounds sit against fixed thresholds, Monte
Carlo error, the generalized-linear paths, and what the design does not
recover. Section S3 gives the applications of Section 6 in the detail
the main text omits: the searches in full, the candidate funnels, the
breast cancer regions, the ACTG 175 analyses, declaration at a protected
level, and the quantities that support each bound.

Section S4 sets the paper's post-selection bounds beside those of
\citet{guohe2021}, on the post hoc design of their Section 5.2: a nested
continuum of cutpoints, identified by the argmax, where neighbouring
candidates differ by a single subject. It reports what the argmax
selects on such a family, the lower bounds on the selected subgroup's
effect, the complement bound and the joint two-subgroup claim, and the
limits of what the design can show.

Section S5 sets out the identifiers' settings and constructions --- how
each generates its candidates, its size and event minimums, its floors
at proposal and at re-ranking, its replicability screen, and the family
over which multiplier resampling re-selects --- for survival and binary
outcomes, with a note on continuous and count outcomes.

\section{Multiplier resampling: conditions, limits and
proofs}\label{sec-field}

This section supports Sections 3 and 4 of the main paper; section,
equation and table numbers without an S prefix refer to them.
Section~\ref{sec-field-conditions} states conditions (A1)--(A5), which
adapt those of \citet{leonanderson2026} to the implemented selection
rule, and proves the local limit law in a form that also covers
uncertain declaration; Section~\ref{sec-field-prop1} proves Proposition
1 and records the Monte Carlo regime; Section~\ref{sec-field-prop2} and
Section~\ref{sec-field-prop3} prove Propositions 2 and 3;
Section~\ref{sec-field-prop4} proves Proposition 4 and, with Lemma S7,
Theorem 2; Section~\ref{sec-field-declaration} proves Theorems 3 and 4
and Proposition 5 of Section 4; Section~\ref{sec-field-example}
evaluates the coverage functionals on a two-candidate example in closed
form and sets the implementation's finite-sample results beside them;
Section~\ref{sec-field-declcal} sets out calibration at a protected
level, its implementation and evidence;
Section~\ref{sec-field-conditional} states the conditional scope for
model-generated families; Section~\ref{sec-field-open} lists what
remains open. The notation is that of Section 3: \(\mathcal F\) is the
family re-selection runs over, \(\mathbf D\) the \(n \times M\) matrix
of influence values, \(h_n\) the selection map, \(t_g\) the admission
floors and \(\hat H = h_n(\hat\beta)\). Results stated in the main paper
are restated here in the more general form in which they are proved.

\subsection{Conditions and the local limit
law}\label{sec-field-conditions}

The data \(O_i = (Y_i, A_i, X_i)\), \(i = 1, \dots, n\), are independent
and identically distributed.

\textbf{(A1) Sampling and randomization.} \(A \perp X\) with
\(\pi = P(A = 1) \in (0, 1)\). For a time-to-event outcome, censoring is
independent and follow-up is bounded by \(\tau\) with
\(P(\tilde T \ge \tau) > 0\), \(\tilde T\) the uncensored time.

\textbf{(A2) Fixed family.} \(\mathcal F = \{g_1, \dots, g_M\}\) with
\(M\) fixed. Cut locations are prespecified or computed from the
observed covariates alone and then held fixed under resampling, so that
each \(g\) is a fixed measurable region with \(p_g = P(X \in g) > 0\)
and per-observation information
\(n^{-1}\mathcal I_g \to \bar{\mathcal I}_g \succ 0\). Forest search's
family satisfies this: it is the size-eligible enumeration, generated
from the covariates alone, and the size minimum of Section 2.1 gives
\(p_g > 0\), from which the information condition follows in the limit.
The finite-sample search fits only candidates that meet its per-arm
event minimum, which is outcome-dependent. The event minimum is not
re-applied at multiplier re-selection or at the calibration of Section
4; those constructions run over the larger size-eligible family
generated from the covariates. Under the assumed positive event rates
the minimum ceases to bind with probability tending to one, so the
fitted family and the larger family coincide asymptotically; where they
differ in a finite sample, the correction is computed over the larger
family. A family proposed from the outcomes, as DINA's is, is not
covered by the covariate-measurability of (A2). In the implementation
the realized proposed family is held fixed while multiplier re-selection
reruns admission, re-ranking and final selection within it; this does
not reproduce the randomness of the proposal step. The fixed-family
results below therefore describe the constructions applied to the
realized family; they establish neither unconditional validity nor a
selective conditional-coverage theorem given the proposal. GRF's
covariate grid is fixed, but its fitted-model decision to return a
region is likewise not replayed. Section S1.9 states this scope.

\textbf{(A3) Regular asymptotically linear estimators.} For each
\(g \in \mathcal F\), \(\beta(g)\) is the unique root of
\(\mathbb E[\psi(\beta; g)\mathbf 1\{X \in g\}] = 0\), the working-model
parameter, and \[
\hat\beta(g) - \beta(g) = \sum_{i=1}^n \mathrm{db}^0_{g,i} + o_p(n^{-1/2}),
\] where \(\mathrm{db}^0_{g,i}\) is the influence of subject \(i\) on
the treatment coefficient of the fit to \(g\) alone, zero for
\(i \notin g\): the M-estimation form
\(\mathcal I_g^{-1}\psi_i(\beta(g); g)\mathbf 1\{i \in g\}\) for a
generalized linear model \citep{white1980} and the martingale-residual
form of \citet{linwei1989} for the Cox model, under the conditions of
\citet{andersen1993}. The empirical influence values
\(\mathrm{db}_{g,i}\) of Section 2.2 are consistent for them in the
sense that \(n\mathbf D^\top\mathbf D \to_p \Sigma\), where
\(\Sigma_{jk} = \lim_n n\sum_i \mathrm{db}^0_{g_j,i}\,\mathrm{db}^0_{g_k,i}\)
is the overlap covariance of the family.

\emph{(A3) for the working models used here.} For a generalized linear
model fitted to the members of \(g\), the treatment coefficient is an
M-estimator, and (A3) holds under the standard M-estimation conditions
--- a unique root, a non-singular expected derivative of the score at
it, and finite second moments of the score and its derivative --- which
a canonical-link model with bounded covariates and an outcome of finite
variance satisfies whether or not its mean is correctly specified, the
population quasi-likelihood being strictly concave under a non-singular
design; the influence is
\(\mathcal I_g^{-1}\psi_i(\beta(g); g)\mathbf 1\{i \in g\}\) and the
sandwich estimate is consistent \citep{white1980}. For the Cox model the
linearization with the martingale-residual influence and the consistency
of the robust variance are \citet{linwei1989}, under the conditions of
\citet{andersen1993}. The complements \(g^c\) are regions of the same
kind, so the same results give (A3) for them. Nothing in the field or in
the calibration of Section 4 uses the working model beyond
\(\mathbf D\), so the statements of Sections 3 and 4 hold on the
mean-difference and log-odds paths under these conditions;
Section~\ref{sec-supp-glm-oc} evaluates their operating characteristics
by simulation.

\textbf{(A4) Multipliers.} The multipliers \(G^{(b)}_i\) are
exchangeable with mean zero and unit variance and satisfy the conditions
of \citet{praestgaardwellner1993}; Poisson(1) − 1, Rademacher and
Gaussian multipliers all do. For each \(g\),
\(\max_i \mathrm{db}^2_{g,i}/\sigma^2_{D,g} \to 0\), where
\(\sigma^2_{D,g} = \sum_i \mathrm{db}^2_{g,i}\).

\textbf{(A5) Selection regularity.} \(\Sigma \succ 0\). The selection is
settled by finitely many comparisons, admission of each candidate
against its floor and the rule's choice among the admitted, and (A5)
asks that none of them be tied in the limit. Two regimes are
distinguished.

\begin{enumerate}
\def\labelenumi{(\roman{enumi})}
\tightlist
\item
  \emph{Separated.} With \(t^\infty_g\) the limit in probability of
  \(t_g\), the population admitted set
  \(\mathcal F^+ = \{g : \beta(g) > t^\infty_g\}\) is fixed, every
  admitted effect is at least \(\eta > 0\) above its floor and every
  other effect at least \(\eta\) below, and the rule's choice over
  \(\mathcal F^+\) is unique: for max \(\beta\), a unique maximizer of
  \(\beta\) over \(\mathcal F^+\) separated from the others by \(\eta\);
  for the four size rules, distinct population shares \(p_g\) over
  \(\mathcal F^+\).
\end{enumerate}

(i′) \emph{Separation, fixed or increasing.} There are a candidate
\(g^\star \in \mathcal F\) and margins \(\delta_n > 0\) with
\(\sqrt n\,\delta_n \to \infty\) such that, at the population values,
every comparison that settles the selection is decided by at least
\(\delta_n\): (a) admission, \(|\beta(g) - t^\infty_g| \ge \delta_n\)
for every \(g \in \mathcal F\), with
\(\beta(g^\star) > t^\infty_{g^\star}\); (b) for max \(\beta\),
\(\beta(g^\star) - \beta(g) \ge \delta_n\) for every other admitted
\(g\); (c) for the two \(\varepsilon\) rules, membership in the band,
\(|\beta(g) - \{\max_{g' \in \mathcal F^+}\beta(g') - b_\varepsilon\}| \ge \delta_n\)
for every admitted \(g\), with \(g^\star\) in the band; and (d) for the
four size rules, distinct shares \(p_g\) over the admitted (and, for the
\(\varepsilon\) rules, in-band) candidates, with \(g^\star\) the rule's
choice among them. (A5-i) is the case \(\delta_n = \eta\) fixed. When
\(\delta_n \to 0\) with \(\sqrt n\,\delta_n \to \infty\) the competition
tightens, but not at the \(n^{-1/2}\) rate of (A5-ii): (i′) and (ii) are
separated by exactly the margins decided at that rate.

\begin{enumerate}
\def\labelenumi{(\roman{enumi})}
\setcounter{enumi}{1}
\item
  \emph{Local.} A subset \(\mathcal F_0 \subseteq \mathcal F\) of local
  competitors has \(\beta(g) = \beta_0 + n^{-1/2}\mu_g\) and rescaled
  floors \(\sqrt n(t_g - \beta_0) \to \tau_g \in [-\infty, \infty)\),
  with distinct shares \(p_g\) over \(\mathcal F_0\); every
  \(g \notin \mathcal F_0\) has \(\beta(g) \le t_g - \eta\) and is
  admitted with probability tending to zero. Limits are taken on
  \(\mathcal F_0\), and \(\zeta \sim N(0, \Sigma)\) is indexed by it.
\item
  \emph{Quantiles.} The laws of the limiting roots are continuous at the
  quantiles that Propositions 1--4 invert, so that the coverage events
  there are continuity sets.
\end{enumerate}

\emph{The limit rule.} For \(w \in \mathbb R^{\mathcal F_0}\) let
\(A(w) = \{g : w_g \ge \tau_g\}\). The limit rule is
\(h(w) = \varnothing\) when \(A(w)\) is empty and otherwise the
maximizer of \(w_g\) over \(A(w)\) for max \(\beta\), the maximizer of
\(p_g\) for max N and max N(ε), and the minimizer of \(p_g\) for min N
and min N(ε). The band of the two ε rules is
\(\{g : \exp(v_g) \ge (1 - \varepsilon)\max_{g'}\exp(v_{g'})\}\), the
maximum over the admitted candidates, on the ratio scale, that is
\(v_g \ge \max v - b_\varepsilon\) with
\(b_\varepsilon = -\log(1 - \varepsilon)\) on the working scale, of
fixed width; on the \(n^{-1/2}\) scale its width is
\(\sqrt n\, b_\varepsilon \to \infty\), so for large \(n\) every
admitted candidate lies in the band and the two ε rules reduce to max N
and min N over the admitted. The sizes \(N_g\) are fixed under
resampling, so every draw orders them as the data do, and with distinct
shares that order is the population order with probability tending to
one. The discontinuities of \(h\) lie on the floor surfaces
\(\{w_g = \tau_g\}\) and, for max \(\beta\), on the tie surfaces
\(\{w_j = w_k\}\), a finite union of hyperplanes to which
\(\Sigma \succ 0\) assigns probability zero under \(W\) and under
\(w + \zeta^*\) for every \(w\). \citet{leonanderson2026} state (A5) for
the thresholded argmax; the version here covers the rules of Section
2.1. Their (A6), which controls the variance estimate \(\hat V\), and
their Theorem 2(iv) are not used, since the bounds of Section 3.3 invert
simulated quantiles rather than \(\hat V\).

\emph{Two remarks on Section 3.2.} A band of fixed width on the working
scale contains every admitted candidate on the \(n^{-1/2}\) scale, so
max N(ε) and min N(ε) converge to max N and min N over the admitted
candidates; their selection is random only through admission, and no
sign law is established for their residual. The bounds of Section 3.3
apply \(h_n\) itself, band included. And the null hypothesis of the
change-plane tests cited in Section 1 of the main paper, under which the
change-plane is not identified, ties every candidate at the common
effect: the exact tie of the local regime.

\emph{The local-scale objects.} Write
\(W_n = \sqrt n(\hat\beta - \beta_0\mathbf 1)\) on \(\mathcal F_0\) and,
for draw \(b\),
\(\zeta^*_n(b) = \sqrt n\,\mathbf D^\top G^{(b)} = \sqrt n\, D(b)\).
Under (A1)--(A3) and (A5-ii), \(W_n \Rightarrow W = \mu + \zeta\) by the
multivariate central limit theorem, (A3) supplying the linearization and
(A4) the Lindeberg condition. Under (A4), conditionally on the data,
\(\zeta^*_n(b) \Rightarrow \zeta^* \sim N(0, \Sigma)\) in probability,
the conditional multiplier central limit theorem \citep[Chapter
10]{praestgaardwellner1993, kosorok2008}, as in \citet[Lemma
2]{leonanderson2026}; for Gaussian multipliers the conditional law is
\(N(0, \hat\Sigma)\) exactly,
\(\hat\Sigma = n\mathbf D^\top \mathbf D\). On this scale
\(h_n(\beta_0\mathbf 1 + n^{-1/2}w)\) admits \(g\) if and only if
\(w_g \ge \tau_{g,n}\) with
\(\tau_{g,n} = \sqrt n(t_g - \beta_0) \to_p \tau_g\), the fluctuation of
\(t_g\) through the estimated \(\sigma_{D,g}\) being \(O_p(n^{-1})\),
and for large \(n\) the band contains every admitted candidate; hence
\(h_n(\beta_0\mathbf 1 + n^{-1/2}w_n) \to h(w)\) for every sequence
\(w_n \to w\) with \(w\) off the discontinuity set of \(h\). This is the
convergence of \(h_n\) to \(h\) that the extended continuous mapping
theorem requires \citep[Theorem 1.11.1]{vandervaartwellner1996}, and it
is used repeatedly below.

\emph{Declaration and the two optimism functionals.} For
\(w \in \mathbb R^{\mathcal F_0}\) let
\(q(w) = P\{h(w + \zeta^*) \ne \varnothing\}\) and, where \(q(w) > 0\),
\begin{equation}\protect\phantomsection\label{eq-s-optimism}{
m(w) = \mathbb E\{\zeta^*_{h(w+\zeta^*)} \mid h(w+\zeta^*) \ne \varnothing\},
\qquad
m_0(w) = \mathbb E\{\zeta^*_{h(w)} \mid h(w+\zeta^*) \ne \varnothing\}
\quad (h(w) \ne \varnothing).
}\end{equation} Here \(m\) is the optimism of re-selection and \(m_0\)
the conditional mean of the same-draws perturbation, the fixed-index
term of equation (1) of the main paper, given that the draw declares a
subgroup. Both are conditional because the implementation averages both
terms of (1) over the draws that declare a subgroup. When \(q(w) = 1\),
\(m_0(w) = 0\) and \(m\) is the unconditional optimism
\(\mathbb E\,\zeta^*_{h(w+\zeta^*)}\) of Section 3.2.

\textbf{Lemma S1} (regularity of the optimism functionals). Let
\(\zeta^* \sim N(0, \Sigma)\) with \(\Sigma \succ 0\) and let
\(h : \mathbb R^{M} \to \mathcal F \cup \{\varnothing\}\) be measurable.
Then
\(\nu(w) = \mathbb E[\zeta^*_{h(w+\zeta^*)}\mathbf 1\{h(w+\zeta^*) \ne \varnothing\}]\),
\(q(w)\) and
\(\nu_k(w) = \mathbb E[\zeta^*_k \mathbf 1\{h(w+\zeta^*) \ne \varnothing\}]\)
are continuously differentiable on \(\mathbb R^M\), with
\begin{equation}\protect\phantomsection\label{eq-s-gradm}{
\nabla \nu(w) = \Sigma^{-1}\mathbb E\bigl[\zeta^*\,\zeta^*_{h(w+\zeta^*)}\mathbf 1\{h(w+\zeta^*) \ne \varnothing\}\bigr] - p(w),
\qquad p_g(w) = P\{h(w+\zeta^*) = g\},
}\end{equation} and
\(\nabla q(w) = \Sigma^{-1}\mathbb E[\zeta^*\mathbf 1\{h(w+\zeta^*) \ne \varnothing\}]\).
Consequently \(m = \nu/q\) is continuously differentiable where
\(q > 0\), \(m_0 = \nu_{h(w)}/q\) is continuously differentiable on each
open set where \(h\) is constant, and every \(p_g\) is continuous.

\emph{Proof.} Substitute \(u = w + \zeta^*\):
\(\nu(w) = \int (u - w)_{h(u)}\mathbf 1\{h(u) \ne \varnothing\}\,\phi_\Sigma(u - w)\,du\).
The rule enters only through the sets \(\{u : h(u) = g\}\), which do not
move with \(w\); the \(w\)-dependence sits in the coordinate
\((u - w)_{h(u)}\) and in the Gaussian density. The integrand and its
\(w\)-derivative are bounded by a polynomial times a Gaussian, uniformly
on compacts, so differentiation under the integral is justified by
dominated convergence, and
\(\partial_{w_k}\phi_\Sigma(u - w) = \phi_\Sigma(u - w)[\Sigma^{-1}(u - w)]_k\)
gives (\ref{eq-s-gradm}); the derivatives are continuous in \(w\) by the
same domination. The other two functions are treated identically.
\(\square\)

No continuity of \(h\) is used: the differentiation lands on the
density, not on the rule. Lemma S1 supplies the continuity the proofs
below need, and it shows that the optimism functional is smooth for any
measurable rule, so the rule enters the limit law through the value of
\(m\) and not through the argument. It does not by itself extend the
limit law to the implemented rule; that extension is the convergence
\(h_n \to h\) above, which rests on (A5) for the rule in force.

\textbf{Theorem S1} (local limit law, general form). Assume (A1)--(A5)
with (A5-ii), and let \(B \to \infty\). Conditionally on
\(\hat H \ne \varnothing\),
\begin{equation}\protect\phantomsection\label{eq-s-limit}{
\sqrt n\{\tilde\beta(\hat H) - \beta(\hat H)\} \Rightarrow \Lambda = \zeta_{h(W)} - m(W) - m_0(W),
}\end{equation} the law of the right side being conditional on
\(h(W) \ne \varnothing\), and the convergence is joint with
\(W_n \Rightarrow W\). When \(q(W) = 1\) almost surely and
\(P\{h(W) \ne \varnothing\} = 1\), (\ref{eq-s-limit}) is Theorem 1.

\emph{Proof.} Step 1, the fields, is the paragraph above. Step 2,
re-selection. Fix \(w\) off the discontinuity set of \(h\) and a
sequence of data sets with \(W_n \to w\) and \(\hat\Sigma \to \Sigma\).
Conditionally on the data,
\(\hat H^*_b = h_n(\hat\beta + D(b)) = h_n\{\beta_0\mathbf 1 + n^{-1/2}(W_n + \zeta^*_n(b))\}\),
and the map \((z, w) \mapsto h(w + z)\) is continuous at every
\((z, w)\) with \(w + z\) off the discontinuity set, a set of
probability zero under \(N(0, \Sigma)\) for every \(w\). By Step 1 and
the extended continuous mapping theorem,
\((\zeta^*_n(b), \hat H^*_b) \Rightarrow (\zeta^*, h(w + \zeta^*))\)
conditionally on the data; the same theorem gives
\(\hat H = h_n(\hat\beta) \to h(w)\). Since
\(\mathbb E^*|\zeta^*_{n,g}(b)|^2\) is bounded, equal to
\(\hat\Sigma_{gg}\) for independent multipliers, the conditional
expectations converge: \(P^*(\hat H^*_b \ne \varnothing) \to q(w)\),
\(\mathbb E^*[\zeta^*_{n,\hat H^*_b}(b)\mathbf 1\{\hat H^*_b \ne \varnothing\}] \to \nu(w)\)
and
\(\mathbb E^*[\zeta^*_{n,\hat H}(b)\mathbf 1\{\hat H^*_b \ne \varnothing\}] \to \nu_{h(w)}(w)\).
Step 3, the draws. With \(q(w) > 0\), the averages over the declaring
draws in (1) converge as \(B \to \infty\), by the law of large numbers
over independent draws, to the ratios of these conditional expectations,
so along the sequence \[
\sqrt n\{\hat\beta(\hat H) - \tilde\beta(\hat H)\} \to m(w) + m_0(w).
\] Step 4, assembly. The limits in Steps 2--3 are continuous functions
of \(w\) (Lemma S1), so by the subsequence criterion
\(\sqrt n\{\hat\beta(\hat H) - \tilde\beta(\hat H)\} - (m + m_0)(W_n) \to_p 0\)
on the declaring event and, since \(W_n \Rightarrow W\), the correction
converges to \((m + m_0)(W)\) jointly with \(W_n\). The raw error is
\(\sqrt n\{\hat\beta(\hat H) - \beta(\hat H)\} = W_{n,\hat H} - \mu_{\hat H}\),
and \(w \mapsto w_{h(w)} - \mu_{h(w)}\) is continuous off the
discontinuity set, so it converges to \(\zeta_{h(W)}\) jointly.
Subtracting gives the joint convergence of the error and \(W_n\);
restricting to the event \(\{\hat H \ne \varnothing\}\), whose limit
\(\{h(W) \ne \varnothing\}\) has its boundary in the floor surfaces and
is therefore a continuity set, gives (\ref{eq-s-limit}). \(\square\)

Two consequences are recorded. First, the residual bias. Under certain
declaration,
\(\mathbb E\Lambda = m^{(\Sigma)}(\mu) - \tfrac12 m^{(2\Sigma)}(\mu)\),
with \(m^{(S)}\) the optimism under noise covariance \(S\): the
projection identity
\(\mathbb E\,m(\mu + \zeta) = \tfrac12 m^{(2\Sigma)}(\mu)\) uses only
that \(\zeta\) and \(\zeta^*\) are independent copies and so holds for
every rule \citep[Theorem 2(iii)]{leonanderson2026}. The sign law of
Section 3.2, positive near ties with the fraction \(1 - 2^{-1/2}\) of
the optimism retained at an exact tie and negative under separation,
uses in addition that the argmax is invariant to positive rescaling of
the field. The size rules are invariant only when every floor is at
zero, \(\tau_g = -\infty\) or \(0\); there the tie value carries over,
and no sign reversal is established. For their size-dominant limit
\citet[Section S4.5]{leonanderson2026} report the residual positive
across the separations they examine.

Second, uncertain declaration. When the local competitors lie within
\(O(n^{-1/2})\) of their floors, \(q(W) < 1\) and \(m_0(W) \ne 0\) in
general. Declaration is a coordinatewise increasing event in
\(\zeta^*\), so \(m_0(w) \ge 0\) whenever the limiting field is
associated; entrywise non-negative covariance suffices for a Gaussian
field \citep{pitt1982}. That covariance condition is an additional
assumption for the directional conclusion: overlap of candidate regions
is what makes it plausible, but overlap alone does not fix the sign of
the covariance between two candidates' influence functions. The
implemented correction then subtracts, beyond the optimism, the
conditional mean of the same-draws term, and the corrected estimate is
shifted downward by it on the \(n^{-1/2}\) scale. The field root of
Section 3.3 does not carry this term; Section~\ref{sec-field-prop1}
records the consequence.

\subsection{Proposition 1: the field}\label{sec-field-prop1}

On the local scale the shrunk field is
\(\tilde W_n = \sqrt n(w - \beta_0\mathbf 1)\), with
\(\tilde W_{n,g} = W_{n,g}\) for \(g \ne \hat H\) and
\(\tilde W_{n,\hat H} = W_{n,\hat H} - \sqrt n\{\hat\beta(\hat H) - \tilde\beta(\hat H)\}\);
by Theorem S1, jointly with \(W_n \Rightarrow W\),
\begin{equation}\protect\phantomsection\label{eq-s-shrunk}{
\tilde W_n \Rightarrow \tilde W, \qquad \tilde W_g = W_g \ (g \ne h(W)), \qquad \tilde W_{h(W)} = W_{h(W)} - m(W) - m_0(W).
}\end{equation} The outer draws are \(\zeta^*_r = \mathbf D^\top\xi_r\)
with \(\xi_r \sim N(0, I_n)\), so that, given the data,
\(\sqrt n\,\zeta^*_r \sim N(0, \hat\Sigma)\) exactly; likewise the inner
draws. For a point \(w\) define the simulated root at \(w\),
\begin{equation}\protect\phantomsection\label{eq-s-simroot}{
\Lambda_w = \zeta^*_{h(w+\zeta^*)} - m(w + \zeta^*) \quad \text{on } \{h(w + \zeta^*) \ne \varnothing\},
}\end{equation} the root of Section 3.3 with the conditional optimism of
(\ref{eq-s-optimism}). Its law at \(w = \mu\) is that of
\(\zeta_{h(W)} - m(W)\) given declaration: (\ref{eq-s-limit}) without
the same-draws term.

\textbf{Proposition 1} (general form). Assume (A1)--(A5) with (A5-ii),
and let \(R_{\mathrm{out}}, R_{\mathrm{in}} \to \infty\). Then
\begin{equation}\protect\phantomsection\label{eq-s-prop1}{
\lim_{n \to \infty} \Pr\{\beta(\hat H) \ge L \mid \hat H \ne \varnothing\}
= \Pr\bigl\{\zeta_{h(W)} - m(W) - m_0(W) \le q_{1-\alpha}(\Lambda_{\tilde W}) \,\big|\, h(W) \ne \varnothing\bigr\},
}\end{equation} where \(q_{1-\alpha}(\Lambda_w)\) is the quantile of the
law of (\ref{eq-s-simroot}). When \(q \equiv 1\) this is Proposition 1,
and the right side equals \(1 - \alpha\) when \(\tilde W = \mu\).

\emph{Proof.} Multiplying by \(\sqrt n\), the coverage event is
\(\{\sqrt n(\tilde\beta(\hat H) - \beta(\hat H)) \le q_{1-\alpha}(\sqrt n\Lambda^*)\}\),
with the quantile that of the outer sample. Given the data,
\(\sqrt n\Lambda^*_r\) is, up to the inner Monte Carlo error, the root
of (\ref{eq-s-simroot}) formed with \((\hat\Sigma, h_n)\) in place of
\((\Sigma, h)\) and centred at \(\tilde W_n\); write \(\hat q_n\) for
the \(1 - \alpha\) quantile of that law. (a) As
\(R_{\mathrm{in}} \to \infty\) the inner average converges to the
conditional optimism under \((\hat\Sigma, h_n)\) uniformly in its
argument: the selection cells \(\{z : h_n(v + z) = g\}\) are finite
Boolean combinations of half-spaces whose normals lie in a finite set, a
Vapnik--Červonenkis class \citep[Lemma 2.6.17]{vandervaartwellner1996},
so the class of functions \(z \mapsto z_g\mathbf 1\{h_n(v + z) = g\}\)
indexed by \(v\) is Glivenko--Cantelli under the integrable envelope
\(|z_g|\) (their Theorem 2.4.3); the inner draws may therefore be shared
across outer draws. (b) As \(R_{\mathrm{out}} \to \infty\) the empirical
quantile converges to \(\hat q_n\) given the data, the limit law having
a unique quantile by (A5-iii). (c) As \(n \to \infty\),
\(\hat\Sigma \to_p \Sigma\), \(\tilde W_n \Rightarrow \tilde W\) and
\(h_n \to h\) in the sense of Section~\ref{sec-field-conditions}, so by
the extended continuous mapping theorem the law of the simulated root
converges weakly to that of \(\Lambda_{\tilde W}\), and
\(\hat q_n \Rightarrow q_{1-\alpha}(\Lambda_{\tilde W})\) by continuity
of the quantile map at a law with a unique quantile \citep[Lemma
21.2]{vandervaart1998}, jointly with the convergence of Theorem S1, both
being continuous functions of the same converging sequence. (d) The
coverage event is a continuity set by (A5-iii), and the portmanteau
theorem gives (\ref{eq-s-prop1}). \(\square\)

\textbf{Proposition S1} (coverage-error envelope). Assume the conditions
of Theorem 1 with declaration certain, so that
\(\Lambda \stackrel{d}{=} \Lambda_\mu\), and for a centre \(w\) let
\(Q(w) = q_{1-\alpha}(\Lambda_w)\). Suppose that \(Q(\mu)\) is the
unique \((1-\alpha)\)-quantile of \(\Lambda_\mu\) and that the law of
\(\Lambda_\mu\) is continuous at \(Q(\mu)\), as (A5) provides. Write
\(C(\mu) = \Pr\{\Lambda \le Q(\tilde W)\}\), the limit of Proposition 1.
Then for every \(\delta > 0\), \[
\bigl|\,C(\mu) - (1-\alpha)\,\bigr| \;\le\; \Pr\bigl\{\,|\Lambda - Q(\mu)| \le \delta\,\bigr\}
\;+\; \Pr\bigl\{\,|Q(\tilde W) - Q(\mu)| > \delta\,\bigr\}.
\] The result needs neither smoothness of \(Q\) nor independence of
\(\Lambda\) and \(\tilde W\); it uses only that \(\Lambda\) and
\(Q(\tilde W)\) are defined on one probability space.

\emph{Proof.} Write \(q_0 = Q(\mu)\) and \(\tilde q = Q(\tilde W)\).
Since \(\Lambda \stackrel{d}{=} \Lambda_\mu\) and the law of
\(\Lambda_\mu\) is continuous at \(q_0\),
\(\Pr\{\Lambda \le q_0\} = 1-\alpha\), so
\(C(\mu) - (1-\alpha) = \mathbb E[\mathbf 1\{\Lambda \le \tilde q\} - \mathbf 1\{\Lambda \le q_0\}]\).
The difference of indicators takes values in \(\{-1,0,1\}\) and is
nonzero only on
\(E = \{\min(q_0,\tilde q) < \Lambda \le \max(q_0,\tilde q)\}\), so
\(|C(\mu) - (1-\alpha)| \le \Pr(E)\). Let
\(A_\delta = \{|\tilde q - q_0| \le \delta\}\). On \(A_\delta\) every
point between \(q_0\) and \(\tilde q\) lies within \(\delta\) of
\(q_0\), so
\(E \cap A_\delta \subseteq \{|\Lambda - q_0| \le \delta\}\). Hence
\(\Pr(E) \le \Pr(E \cap A_\delta) + \Pr(A_\delta^c) \le \Pr\{|\Lambda - q_0| \le \delta\} + \Pr\{|\tilde q - q_0| > \delta\}\).
\(\square\)

\emph{What the envelope says.} The first term is the concentration of
the true limiting root near its nominal quantile, a property of one
one-dimensional law. The second is the tail of the quantile-location
error, the displacement of the quantile the field reads at \(\tilde W\)
from the one asymptotically exact coverage needs at \(\mu\). The
proposition does not imply nominal coverage under local competition: for
fixed \(\delta\) the second term need not vanish there. It isolates the
two contributors to the discrepancy.

\textbf{Corollary S1} (local-density form). Let \(q_0 = Q(\mu)\) and
suppose \(\Lambda\) has a density \(f_\Lambda\) with
\(\sup_{|x - q_0| \le \delta_0} f_\Lambda(x) \le c_\Lambda\) for some
\(\delta_0 > 0\) and finite \(c_\Lambda\). Then for every
\(0 < \delta \le \delta_0\), \[
\bigl|\,C(\mu) - (1-\alpha)\,\bigr| \;\le\; 2c_\Lambda\delta + \Pr\bigl\{|Q(\tilde W) - Q(\mu)| > \delta\bigr\},
\] and hence
\(|C(\mu) - (1-\alpha)| \le \inf_{0<\delta\le\delta_0}\bigl[2c_\Lambda\delta + \Pr\{|Q(\tilde W) - Q(\mu)| > \delta\}\bigr]\).

\emph{Proof.} By Proposition S1 it suffices to bound the first term; for
\(0 < \delta \le \delta_0\),
\(\Pr\{|\Lambda - q_0| \le \delta\} = \int_{q_0-\delta}^{q_0+\delta} f_\Lambda \le 2c_\Lambda\delta\).
Taking the infimum over admissible \(\delta\) gives the second display.
\(\square\)

The constant \(c_\Lambda\) is an assumed local density bound; the
sufficient condition of the next remark gives one route to it, and is
not established uniformly over the implemented rules and families.

\emph{Remark (a sufficient condition for the density constant).} The
following is an additional sufficient condition, not implied by
(A1)--(A5). On the cell \(\{h(W) = g\}\) the root is
\(\zeta_g - m(\mu+\zeta)\), and \(m\) is \(C^1\) by Lemma S1. Suppose
that, almost everywhere on each cell, the partial derivative of \(m\) in
the selected coordinate, the other coordinates held fixed, satisfies
\(1 - \partial_g m(\mu+\zeta) \ge 1 - L > 0\). Conditionally on the
other coordinates the map \(\zeta_g \mapsto \zeta_g - m(\mu+\zeta)\) is
then strictly increasing with inverse derivative at most \((1-L)^{-1}\);
the conditional Gaussian density of \(\zeta_g\) is at most
\((2\pi v_g)^{-1/2}\) with \(v_g = 1/(\Sigma^{-1})_{gg}\); and since
cell boundaries have probability zero (Section S1.1), integrating over
the other coordinates and summing over the at most \(M\) cells gives
\(c_\Lambda \le M/\{(1-L)\sqrt{2\pi v_{\min}}\}\),
\(v_{\min} = \min_g v_g\). On the two-candidate example of Section S1.7,
\(m\) is \(\tau\phi(\delta/\tau)\) in the standardized gap, so
\(L = \phi(1) \approx 0.242\).

\emph{Remark (separated selection).} Under (A5-i′) the envelope is
compatible with Theorem 2 rather than a route to it. By Lemma S7(ii) the
simulated root at the corrected centre collapses conditionally to the
\(g^\star\) coordinate, and by Theorem 2 its conditional
\((1-\alpha)\)-quantile and the \((1-\alpha)\)-quantile of the true root
both converge to \(z_{1-\alpha}\sigma_\star\); the finite-sample
analogue of the second term therefore vanishes for every fixed
\(\delta\), and the first vanishes as \(\delta \downarrow 0\) by
continuity of the Gaussian limit.

\emph{Remark (uncertain declaration).} The proposition is stated for the
root of Theorem 1 with declaration certain. Where declaration is
uncertain the true root carries the same-draws term \(-m_0\) (Theorem
S1) and the field root omits it; the envelope then applies to the root
with the term included, and the omission is the separate shift recorded
above.

Proposition 1 is a characterization, not a guarantee. The substitution
of \(\hat\Sigma\) for \(\Sigma\) is consistent, but \(\tilde W\) is the
observed field with its selected coordinate corrected, not the centre
\(\mu\), and under local asymptotics \(\mu\) is not consistently
estimable: the field reads the quantile of the root at \(\tilde W\)
where asymptotically exact coverage needs it at \(\mu\). The results
here characterize and bound the consequence of that substitution; they
do not rule out other constructions under additional structure. The gap
\(q_{1-\alpha}(\Lambda_{\tilde W}) - q_{1-\alpha}(\Lambda_\mu)\) is a
location error, \(O_p(1)\) on the local scale, with the same source as
the residual of Theorem 1, the optimism evaluated at the observed field
rather than at its centre. It vanishes as the competition separates,
since under (A5-i′) both quantiles collapse to the Gaussian one (Theorem
2), and it is not reduced by more draws. When declaration is uncertain a
second discrepancy appears: the true root carries \(-m_0(W)\) and the
simulated root does not. If the limiting field is associated,
\(m_0(W) \ge 0\) and the omitted term shifts the selected subgroup's
estimation error in the conservative direction for the lower bound. This
is a statement about the omitted component alone; it does not make the
lower bound conservative under local competition, where the
quantile-location discrepancy of Proposition S1 remains. A version of
the root carrying the same-draws term would simulate (\ref{eq-s-limit})
itself and is noted in Section~\ref{sec-field-open}.

\emph{The Monte Carlo regime.} The two draw counts do different work.
\(R_{\mathrm{in}}\) estimates the optimism at each outer point; its
error is \(O_p(R_{\mathrm{in}}^{-1/2})\) and enters \(\Lambda^*_r\)
additively. \(R_{\mathrm{out}}\) is the sample from which the quantile
is read, with error \(O_p(R_{\mathrm{out}}^{-1/2})\) under a positive
density at the quantile. Neither needs a rate tied to \(n\), because
both approximate a limit object rather than a sampling law; the
correction's \(B\) approximates an expectation inside an \(n^{-1/2}\)
quantity and does carry a joint regime \citep[Section
S4.4]{leonanderson2026}.

\emph{Why no two-sided interval is proposed.} The residual of Theorem 1
changes sign across the competition regime. A one-sided bound is exposed
to one sign of that residual only, and in the direction that widens it;
a two-sided interval is exposed to both, so whichever end is not
protected by the residual's sign is the end that fails. The asymmetry is
a property of the shape of \(\Lambda\), not of any design, and it is why
the one-sided products are the recommended ones.

\subsection{Proposition 2: the complement}\label{sec-field-prop2}

Let \(\mathbf D^c\) be the \(n \times M\) matrix whose column \(g\)
holds the influence values of the fit to \(g^c\),
\(\zeta^{c*}_r = (\mathbf D^c)^\top\xi_r\) and
\(\zeta^{c\prime}_j = (\mathbf D^c)^\top\xi'_j\) the complement fields
drawn from the harm field's multipliers, and \(\Sigma^c\) and
\(\Sigma_\times\) the limits of \(n(\mathbf D^c)^\top\mathbf D^c\) and
\(n\mathbf D^\top\mathbf D^c\). The complement's two-term correction is
(1) with \(D\) replaced by \(D^c(b) = (\mathbf D^c)^\top G^{(b)}\), read
at the complement of each draw's winner and at \(\hat H^c\); both
averages run over the draws that declare a subgroup, and a draw whose
winner's complement could not be fitted is dropped from them. With
\(\zeta^{c*}\) the limiting complement copy, define as in
(\ref{eq-s-optimism}) \[
m^c(w) = \mathbb E\{\zeta^{c*}_{h(w+\zeta^*)} \mid h(w+\zeta^*) \ne \varnothing\}, \qquad
m^c_0(w) = \mathbb E\{\zeta^{c*}_{h(w)} \mid h(w+\zeta^*) \ne \varnothing\},
\] the expectations taken over the harm-side rule.

\textbf{Proposition 2} (general form). Assume (A1)--(A5) with (A5-ii),
with (A2) and (A3) holding also for the complements \(g^c\),
\(g \in \mathcal F\). Conditionally on \(\hat H \ne \varnothing\), and
jointly with (\ref{eq-s-limit}),
\begin{equation}\protect\phantomsection\label{eq-s-comp}{
\sqrt n\{\tilde\beta(\hat H^c) - \beta(\hat H^c)\} \Rightarrow \Lambda^c = \zeta^c_{h(W)} - m^c(W) - m^c_0(W),
}\end{equation} where \((\zeta, \zeta^c)\) is jointly Gaussian with
covariance
\(\begin{pmatrix}\Sigma & \Sigma_\times \\ \Sigma_\times^\top & \Sigma^c\end{pmatrix}\).
No condition is placed on \(\Sigma^c\).

\emph{Proof.} The complement family \(\{g^c : g \in \mathcal F\}\) is
fixed and finite, complementation being a fixed operation on regions
generated by one cut vector, so (A2) for it asks only that each
complement have positive share and non-degenerate information, which is
the assumption. (A3) for it gives the linearization with influence
\(\mathrm{db}^{c,0}_{g,i}\). The harm and complement estimates are
linear in the same sample, so the multivariate central limit theorem
applies to the stacked \(2M\)-vector and gives the joint Gaussian limit
with the cross-covariance \(\Sigma_\times\) carried rather than assumed;
the same holds for the multiplier fields conditionally on the data, and
the shared \(\xi\) of the implementation draws from precisely this
stacked law. A degenerate \(\Sigma^c\) is admissible in both statements.
No selection acts on the complements: the index read is \(h(w + z)\), a
function of the candidate coordinates alone, so the map
\((z, w) \mapsto (h(w + z), z^c_{h(w+z)})\) is continuous off the
discontinuity set of \(h\), the same null set as before. Steps 2--4 of
the proof of Theorem S1 then apply verbatim with \(\zeta^{c*}\) read at
the harm-side index, and the joint convergence follows because both
errors are continuous functions of the same stacked sequence.
\(\square\)

The proposition is cheap because nothing is selected among complements.
That matters in practice: small candidates have nearly identical
complements, so \(\Sigma^c\) is typically near-singular, the
configuration that would defeat an argmax over the complements. The
complement is read through the harm-side rule, whose non-degeneracy is
(A5)'s business. The estimand is \(\beta(\hat H^c)\) on the same terms
as Theorem 1, with coverage averaged over the selection.

Propositions 1--4 have no counterpart in \citet{leonanderson2026}.

\subsection{Proposition 3: studentizing the
complement}\label{sec-field-prop3}

\textbf{Lemma S2.}
\(s_g^2 = \lVert\mathbf D^c_{\cdot g}\rVert^2 = \sum_i (\mathrm{db}^c_{g,i})^2\)
is the robust variance estimate of \(\hat\beta(g^c)\), and
\(\sqrt n\, s_g \to_p \sigma^c_g = (\Sigma^c_{gg})^{1/2}\).

\emph{Proof.} The robust variance of an asymptotically linear estimator
is the sum of its squared empirical influences
\citep{white1980, linwei1989}, which is \(s_g^2\) by definition, and the
convergence is the \((g, g)\) entry of
\(n(\mathbf D^c)^\top\mathbf D^c \to_p \Sigma^c\). \(\square\)

So the scale re-attached by the studentized root is the standard error
the analysis already reports for the complement. Write the limiting
complement field in scale-and-shape form, \(\zeta^c_g = \sigma^c_g Z_g\)
and \(\zeta^{c*}_g = \sigma^c_g Z^*_g\), and let
\(m^Z(w) = \mathbb E\{Z^*_{h(w+\zeta^*)} \mid h(w+\zeta^*) \ne \varnothing\}\)
be the unit-scale optimism. On the local scale the two simulated
complement roots centred at \(w\) are
\begin{equation}\protect\phantomsection\label{eq-s-two-roots}{
\Lambda^c_w = \sigma^c_{G}Z^*_{G} - m^c(w + \zeta^*),
\qquad
\Lambda^{cs}_w = \sigma^c_{\hat H}\bigl(Z^*_{G} - m^Z(w + \zeta^*)\bigr) = \sigma^c_{\hat H}\,\tilde\Lambda^c_w,
}\end{equation} with \(G = h(w + \zeta^*)\) the outer re-selection,
\(m^c(w + \zeta^*) = \mathbb E\{\sigma^c_{G'}Z'_{G'} \mid \zeta^*\}\)
the inner mean over \(G' = h(w + \zeta^* + \zeta')\) on declaring inner
draws, and \(\tilde\Lambda^c_w\) the root of the unit-scale field. The
first is a scale mixture over the complements that re-selection favours;
the second refers every reading to the scale of the selected complement.
The studentized error of the complement estimate is \[
T = \frac{\zeta^c_{h(W)} - m^c(W) - m^c_0(W)}{\sigma^c_{h(W)}} = Z_{h(W)} - \frac{m^c(W) + m^c_0(W)}{\sigma^c_{h(W)}}.
\]

\textbf{Proposition 3} (general form). Under the conditions of
Proposition 2, with \(R_{\mathrm{out}}, R_{\mathrm{in}} \to \infty\):

\begin{enumerate}
\def\labelenumi{(\roman{enumi})}
\item
  \(q_\alpha(\Lambda^{cs*}) = s_{\hat H}\,q_\alpha(\tilde\Lambda^{c*})\)
  exactly, with \(\tilde\Lambda^{c*}_r = \Lambda^{cs*}_r/s_{\hat H}\)
  the unit-scale draws, for every finite number of draws;
\item
  \(\lim_n \Pr\{\beta(\hat H^c) \le U \mid \hat H \ne \varnothing\} = \Pr\{T \ge q_\alpha(\tilde\Lambda^c_{\tilde W}) \mid h(W) \ne \varnothing\}\),
  which equals \(1 - \alpha\) when \(\tilde W = \mu\),
  \(m^c_0 \equiv 0\) and \(m^c(W) = \sigma^c_{h(W)}\,m^Z(W)\);
\item
  \(\Lambda^{cs}_w - \Lambda^c_w = (\sigma^c_{\hat H} - \sigma^c_G)Z^*_G - \mathbb E\{(\sigma^c_{\hat H} - \sigma^c_{G'})Z'_{G'} \mid \zeta^*\}\),
  which vanishes when \(\sigma^c_g\) is constant over the candidates
  whose selection cell has positive measure;
\end{enumerate}

and (iv) under (A5-i′), the coverage in (ii) is \(1 - \alpha\) (Theorem
2).

\emph{Proof.} (i) is (\ref{eq-s-two-roots}) read at the quantile,
\(s_{\hat H}\) being a positive constant given the data; it holds at
finite \(n\) with \(s_{\hat H}\) in place of \(\sigma^c_{\hat H}\). (ii)
Dividing the coverage event by \(s_{\hat H}\) and multiplying by
\(\sqrt n\) gives
\(\{\sqrt n(\tilde\beta(\hat H^c) - \beta(\hat H^c))/(\sqrt n s_{\hat H}) \ge q_\alpha(\tilde\Lambda^{c*})\}\);
by Proposition 2 and Lemma S2 the left side converges to \(T\), and the
proof of Proposition 1 applies to the unit-scale root, whose Monte Carlo
quantile converges to \(q_\alpha(\tilde\Lambda^c_{\tilde W})\). Under
the three conditions the quantile is the constant
\(q_\alpha(\tilde\Lambda^c_\mu)\) and \(T = Z_{h(W)} - m^Z(W)\) has the
law of \(\tilde\Lambda^c_\mu\), since \(\zeta\) and \(\zeta^*\) are
copies, so the probability is \(1 - \alpha\). (iii) is
(\ref{eq-s-two-roots}) subtracted term by term. Both \(G\) and \(G'\)
take, with positive probability, exactly the candidates whose selection
cell has positive measure, whatever the noise covariance, so constancy
of \(\sigma^c_g\) on that set makes both terms vanish. (iv) is Theorem
2(iii), proved in Section S1.5. \(\square\)

The equality in (ii) uses the first-moment relation
\(m^c(W) = \sigma^c_{h(W)}\, m_Z(W)\) stated there; studentization
removes scale heterogeneity across re-selected complements but does not
itself establish that relation. The asymptotically exact result for the
complement is Theorem 2, under separated selection.

Studentization is the bootstrap-\(t\) device of resampling an
approximately pivotal root rather than the raw statistic
\citep{hall1992}, and its force is greatest exactly when the
non-pivotality is driven by an estimable scale, here \(s_g\), which by
Lemma S2 is the reported standard error. It has no tuning parameter,
since \(s_g\) is fixed by the linearization before any coverage is
consulted, and by part (iii) it changes nothing where complements nearly
coincide, which is where admitted candidates are small. What it does not
repair is named in part (ii): the unstudentized root mixes the optimism
over the scales of the re-selected complements, whereas the studentized
root evaluates it at \(\sigma^c_{\hat H}\); the two agree when scale
heterogeneity across re-selections is uncorrelated with the standardized
noise at the re-selected complement. Studentization thus makes the
spread of the root refer to the realized selection, the leading effect,
while the optimism term keeps its mixture form, a shift rather than a
spread.

\subsection{Proposition 4 and Theorem 2: the joint claim, and asymptotic
exactness under separation}\label{sec-field-prop4}

\textbf{Lemma S3.} For every candidate \(g\),
\(\sum_i \mathrm{db}_{g,i}\,\mathrm{db}^c_{g,i} = 0\); hence
\((\Sigma_\times)_{gg} = 0\) and
\(\operatorname{Cov}(\zeta_g, \zeta^c_g) = 0\).

\emph{Proof.} The fit to \(g\) uses only its members, so
\(\mathrm{db}_{g,i} = 0\) for \(i \notin g\), and the fit to \(g^c\)
uses only its own, so \(\mathrm{db}^c_{g,i} = 0\) for \(i \notin g^c\).
Since \(g \cap g^c = \emptyset\), every product vanishes. \(\square\)

The lemma concerns the diagonal of \(\Sigma_\times\) only:
\(\operatorname{Cov}(\zeta_g, \zeta^c_k)\) for \(k \ne g\) is in general
non-zero, since \(g\) and \(k^c\) overlap, and those entries are what
the stacked law of Proposition 2 carries. The Bonferroni pair, equation
(4) of the main paper, supports its two-subgroup claim at joint level at
least \(1 - \alpha\) by the union bound; its joint coverage exceeds
\(1 - \alpha\) by the probability that neither bound covers.

\textbf{Proposition 4.} Assume (A1)--(A5), with (A2) and (A3) holding
also for the complements, and that a subgroup is declared with
probability tending to one on the data and on the draws.

\begin{enumerate}
\def\labelenumi{(\roman{enumi})}
\item
  Under (A5-i′), \(\Lambda\) and \(\Lambda^c\) are asymptotically
  independent Gaussians, each bound of (4) is asymptotically exact at
  level \(1 - \alpha/2\), and the pair has asymptotic joint coverage
  \((1 - \alpha/2)^2 = 1 - \alpha + \alpha^2/4\).
\item
  Under (A5-ii), conditionally on \(h(W) = g\) the leading terms
  \(\zeta_g\) and \(\zeta^c_g\) of the two roots are uncorrelated, and
  dependence between \(\Lambda\) and \(\Lambda^c\) enters only through
  the common selection \(h(W)\) and the optimism terms \(m(W)\) and
  \(m^c(W)\).
\end{enumerate}

\emph{Proof.} (i) Under (A5-i′) the selection stabilizes on \(g^\star\)
(Lemma S7) and both roots collapse to its coordinate with the optimism
terms \(o(n^{-1/2})\) (Theorem 2, Steps 2--3; under fixed margins this
is Theorem 2(i) of \citet{leonanderson2026}), so
\(\Lambda = \zeta_{g^\star}\) and \(\Lambda^c = \zeta^c_{g^\star}\),
jointly Gaussian with covariance zero by Lemma S3, hence independent.
The simulated roots are then Gaussian with the consistently estimated
variances, each bound inverts the correct quantile, and the joint event
factorizes. (ii) In the local regime \(\Lambda = \zeta_{h(W)} - m(W)\)
and \(\Lambda^c = \zeta^c_{h(W)} - m^c(W)\) share the index \(h(W)\);
conditionally on \(h(W) = g\) the leading terms are uncorrelated by
Lemma S3, so all dependence is carried by the variation of \(h(W)\) and
by \(m\) and \(m^c\), which are functions of the whole field.
\(\square\)

Part (i) says that under separation Bonferroni's conservatism is
\(\alpha^2/4\), six parts in ten thousand at \(\alpha = 0.05\). Part
(ii) says that in the local regime the departure from \(\alpha^2/4\)
comes only through the common selection and the optimism terms, not
through the noise at the selected pair. The aligned draws
\((\Lambda^*_r, \Lambda^{cs*}_r)\) make a calibrated split of \(\alpha\)
available, the largest \(\gamma \in [\alpha/2, \alpha]\) whose simulated
joint coverage still reaches \(1 - \alpha\); Proposition 4 says there is
little for it to recover, and it serves as a check on the union bound
rather than as a second product. It is ordinary simultaneous inference
on a joint law the field simulates directly, and it allocates error
between two bounds that are already selection-adjusted; it does not
calibrate the correction itself. Everything downstream of Lemma S3 rests
on \(\hat H\) and \(\hat H^c\) partitioning the sample: a second region
defined otherwise than as the complement, or overlapping the first,
would carry a non-zero covariance at the selected pair, and there the
calibrated split would be the better construction.

\textbf{Lemma S7} (the comparisons settle). Assume (A1)--(A4) and
(A5-i′). Then (i) \(\Pr\{\hat H = g^\star\} \to 1\); (ii) conditionally
on the data, at each centre the construction evaluates --- \(\hat\beta\)
in equation (1) of the main paper, and \(w\), \(w + \zeta^*_r\) and
\(w + \zeta^*_r + \zeta'_j\) in equation (2) of the main paper --- the
probability that the re-selection is not \(g^\star\) tends to zero in
probability; (iii) a subgroup is declared with probability tending to
one on the data and on the draws.

\emph{Proof.} Each comparison is of the form ``an estimated quantity
exceeds a threshold'', in which the estimated quantity differs from its
population value by \(O_p(n^{-1/2})\): by (A3) for \(\hat\beta(g)\) and
for the differences \(\hat\beta(g) - \hat\beta(g')\); for the
thresholds, because \(t_g - t^\infty_g = O_p(n^{-1/2})\), the screen's
floor being \(c_{\mathrm{cons}} + z_{(1+p^\star)/2}\,\sigma_{D,g}\) with
\(\sigma_{D,g} = O_p(n^{-1/2})\), and because the band edge is
\(b_\varepsilon\) below a maximum of finitely many such estimates; for
the correction that forms \(w\) from \(\hat\beta\), which is an average
of draw coordinates and so at most
\(|\mathcal B|^{-1}\sum_b \max_g|D_g(b)| = O_p(n^{-1/2})\) by the
second-moment bound
\(\mathbb E^*\max_g D_g(b)^2 \le \sum_g\hat\Sigma_{gg}/n\); and for the
draws, by (A4) and Chebyshev's inequality, since \(D_g(b)\) has
conditional variance \(\sigma^2_{D,g} = \hat\Sigma_{gg}/n\), so that
\(P^*\{|D_g(b)| \ge \delta_n/2\} \le 4\hat\Sigma_{gg}/(n\delta_n^2) \to_p 0\),
the sum of an outer and an inner draw having twice the variance. Since
\(\sqrt n\,\delta_n \to \infty\), each comparison is decided as at the
population values with probability tending to one, and there are
finitely many, so (i) and (ii) follow by the union bound; the sizes are
fixed under resampling, and by (d) their order is the population order
with probability tending to one. (iii) holds because \(g^\star\) is
admitted. \(\square\)

\textbf{Theorem 2} (asymptotically exact coverage under separation).
Assume (A1)--(A4) and (A5-i′), let
\(B, R_{\mathrm{out}}, R_{\mathrm{in}} \to \infty\), and write
\(\sigma_\star^2 = \Sigma_{g^\star g^\star}\). Then

\begin{enumerate}
\def\labelenumi{(\roman{enumi})}
\item
  \(\sqrt n\{\tilde\beta(\hat H) - \beta(\hat H)\} \Rightarrow N(0, \sigma_\star^2)\),
  and conditionally on the data the law of \(\sqrt n\,\Lambda^*\)
  converges in probability to \(N(0, \sigma_\star^2)\);
\item
  \(\lim_{n\to\infty} \Pr\{\beta(\hat H) \ge L\} = 1 - \alpha\);
\item
  if (A2) and (A3) hold also for the complements, then with
  \(\sigma_{c\star}^2 = \Sigma^c_{g^\star g^\star}\),
  \(\sqrt n\{\tilde\beta(\hat H^c) - \beta(\hat H^c)\} \Rightarrow N(0, \sigma_{c\star}^2)\),
  \(\lim_n \Pr\{\beta(\hat H^c) \le U\} = 1 - \alpha\) for the field-s
  bound and for its unstudentized form, the two limiting errors are
  independent, and the pair, equation (4) of the main paper, has
  asymptotic joint coverage \((1 - \alpha/2)^2\).
\end{enumerate}

Under (A5-i) the hypotheses hold with \(\delta_n\) fixed, so Theorem 2
contains the separated case of Propositions 1, 3(iv) and 4(i).

\emph{Proof of Theorem 2.} Step 1, stabilization: Lemma S7. Step 2, the
correction collapses. On the draws that declare, write
\(D_{\hat H^*_b}(b) = D_{g^\star}(b) + \{D_{\hat H^*_b}(b) - D_{g^\star}(b)\}\,\mathbf 1\{\hat H^*_b \ne g^\star\}\).
The average of the first term over \(B\) draws has conditional mean zero
and conditional variance \(\hat\Sigma_{g^\star g^\star}/(nB)\), so it is
\(o_p(n^{-1/2})\) as \(B \to \infty\). The average of the second is
bounded by the Cauchy--Schwarz inequality: its conditional absolute mean
is at most
\(2\{\mathbb E^*\max_g D_g(b)^2\}^{1/2}\,P^*\{\hat H^*_b \ne g^\star\}^{1/2} \le 2\{\sum_g \hat\Sigma_{gg}/n\}^{1/2}\,P^*\{\hat H^*_b \ne g^\star\}^{1/2}\),
the bounded term being a difference of two coordinates of the draw,
which is \(O_p(n^{-1/2})\,o_p(1)\) by Lemma S7(ii). Restricting both
averages to the declaring draws divides by
\(P^*\{\hat H^*_b \ne \varnothing\} \to 1\) and removes the
non-declaring draws, whose contribution is bounded the same way with
\(P^*\{\hat H^*_b = \varnothing\} \to 0\) in place of the re-selection
error. The same-draws average, of \(D_{\hat H}(b)\) over the declaring
draws, is the first term's average on \(\{\hat H = g^\star\}\), whose
probability tends to one, with the same restriction, so it too is
\(o_p(n^{-1/2})\). Hence
\(\tilde\beta(\hat H) = \hat\beta(g^\star) + o_p(n^{-1/2})\) on
\(\{\hat H = g^\star\}\), and by (A3)
\(\sqrt n\{\tilde\beta(\hat H) - \beta(\hat H)\} = \sqrt n\{\hat\beta(g^\star) - \beta(g^\star)\} + o_p(1) \Rightarrow N(0, \sigma_\star^2)\).

Step 3, the simulated root. In equation (2) of the main paper,
\(g_r = g^\star\) and \(g_{rj} = g^\star\) except on events whose
conditional probability vanishes (Lemma S7(ii)), so
\(\sqrt n\,\Lambda^*_r = \sqrt n\,\zeta^*_{r,g^\star} - |\mathcal J_r|^{-1}\sum_{j \in \mathcal J_r}\sqrt n\,\zeta'_{j,g^\star} + \epsilon_r\)
with \(\mathbb E^*|\epsilon_r| \to 0\), by the bound of Step 2 applied
to the outer and to the inner draws; the inner average has conditional
variance \(\hat\Sigma_{g^\star g^\star}/|\mathcal J_r| \to 0\) as
\(R_{\mathrm{in}} \to \infty\), since
\(|\mathcal J_r|/R_{\mathrm{in}} \to 1\). Given the data,
\(\sqrt n\,\zeta^*_{r,g^\star} \sim N(0, \hat\Sigma_{g^\star g^\star})\)
exactly, and \(\hat\Sigma \to_p \Sigma\) by (A3), so the conditional law
of \(\sqrt n\,\Lambda^*\) converges to \(N(0, \sigma_\star^2)\) in
probability. This is (i).

Step 4, the quantile. The empirical \((1-\alpha)\) quantile of
\(\{\sqrt n\,\Lambda^*_r : r \le R_{\mathrm{out}}\}\) converges in
probability to \(z_{1-\alpha}\sigma_\star\) as
\(n, R_{\mathrm{out}} \to \infty\): the limit is Gaussian, so its
distribution function is continuous and its quantile unique, and Step 2
of the proof of Theorem 3 --- Hoeffding's inequality on either side of
the limiting quantile, with the conditional distribution function
converging by Step 3 --- applies unchanged.

Step 5, inversion. The coverage event is
\(\{\sqrt n(\tilde\beta(\hat H) - \beta(\hat H)) \le \hat q_{1-\alpha}\}\).
By Steps 2 and 4 the pair converges jointly to
\((Z, z_{1-\alpha}\sigma_\star)\) with \(Z \sim N(0, \sigma_\star^2)\);
\(\{(x, q) : x \le q\}\) is a continuity set of the limit, the Gaussian
having no atom, and the portmanteau theorem gives
\(\Pr\{Z \le z_{1-\alpha}\sigma_\star\} = 1 - \alpha\), which is (ii).

Step 6, the complement and the pair. With (A2) and (A3) for the
complements, Steps 2--5 apply verbatim to the complement field read at
the same indices, giving the Gaussian limit and
\(\lim_n \Pr\{\beta(\hat H^c) \le U\} = 1 - \alpha\) for the
unstudentized bound; the studentized root coincides with it to
\(o_p(1)\), since \(s_{\hat H}/s_{g_r} = 1\) on
\(\{g_r = \hat H = g^\star\}\) (Proposition 3(iii)). The two limiting
errors are the \(g^\star\) coordinates of \((\zeta, \zeta^c)\),
uncorrelated by Lemma S3 and jointly Gaussian, hence independent; the
pair (4) inverts each at level \(\alpha/2\), so its joint coverage is
\((1 - \alpha/2)^2\). \(\square\)

(A5-i′) enters only through Lemma S7: any condition under which observed
selection, declaration and the re-selection at the centres of (1) and
(2) stabilize on one candidate gives the same conclusions. What it
excludes is a comparison decided within \(O(n^{-1/2})\) --- a tie among
effects, a candidate at its floor, a candidate at the band edge ---
which is the local regime (A5-ii), where Propositions 1--4 characterize
the coverage and do not claim the nominal level.

\subsection{Theorems 3 and 4 and Proposition 5: the declaration
calibration}\label{sec-field-declaration}

Section 4 calibrates the cutoff of the consistency screen against the
maximum of the standardized candidate field, using the outer draws of
Section~\ref{sec-field-prop1}. On \(\mathcal F\) write
\(\hat\sigma_g = \hat\Sigma_{gg}^{1/2} = \sqrt n\,\sigma_{D,g}\),
\(\sigma_g = \Sigma_{gg}^{1/2}\), and \(\hat R\) and \(R\) for the
correlation matrices
\(\mathrm{diag}(\hat\Sigma)^{-1/2}\hat\Sigma\,\mathrm{diag}(\hat\Sigma)^{-1/2}\)
and
\(\mathrm{diag}(\Sigma)^{-1/2}\Sigma\,\mathrm{diag}(\Sigma)^{-1/2}\).
The standardized statistic of Section 4 is then
\(T_g = \sqrt n\{\hat\beta(g) - c_{\mathrm{cons}}\}/\hat\sigma_g\), and
the standardized field \(Z^*_{r,g} = \zeta^*_{r,g}/\sigma_{D,g} =
\sqrt n\,\zeta^*_{r,g}/\hat\sigma_g\). For a correlation matrix \(R\)
let \(F_R(t) = \Pr\{\max_g Z_g \le t\}\) with \(Z \sim N(0,R)\), and let
\(\kappa_\alpha(R)\) be its \((1-\alpha)\) quantile. Three lemmas carry
the argument; the first is what lets the calibration dispense with
\(\Sigma \succ 0\).

\textbf{Lemma S4} (the maximum of a Gaussian vector with non-degenerate
margins). Let \(R\) be an \(M \times M\) correlation matrix, positive
semidefinite with unit diagonal and of any rank, and \(Z \sim N(0, R)\).
Then (i) \(F_R\) is continuous on \(\mathbb R\); (ii) \(F_R(t) < 1\) for
every \(t\), and \(F_R\) is strictly increasing on
\(\{t : F_R(t) > 0\}\), which is an interval \((t_0, \infty)\) with
\(t_0 \le 0\); (iii) for every \(\alpha \in (0,1)\) the quantile
\(\kappa_\alpha(R)\) is unique, and \(\kappa_\alpha(R) > 0\) when
\(\alpha < 1/2\). The same holds for the maximum \(\max_g (Z_g - a_g)\)
shifted by a fixed \(a \in \mathbb R^M\), with \(t_0\) the infimum of
that maximum over the support of \(Z\) in place of \(t_0 \le 0\) and
without the sign statement; write \(\kappa_\alpha(R, a)\) for its
\((1-\alpha)\) quantile, so that
\(\kappa_\alpha(R) = \kappa_\alpha(R, 0)\).

\emph{Proof.} (i)
\(\{\max_g Z_g = t\} \subseteq \bigcup_g \{Z_g = t\}\), a finite union
of null events since each \(Z_g\) is a non-degenerate Gaussian. (ii)
\(1 - F_R(t) \ge
\Pr\{Z_1 > t\} > 0\). Let \(L\) be the column space of \(R\), the
support of \(Z\), of dimension \(k \ge 1\), and \(\gamma\) the law of
\(Z\), which has a positive continuous density with respect to
\(k\)-dimensional Lebesgue measure on \(L\). The map
\(m(z) = \max_g z_g\) is continuous on \(L\), so \(m(L)\) is an
interval; it is unbounded above, since \(Z_1\) is non-degenerate and
\(L\) therefore contains a \(v\) with \(v_1 > 0\), along which
\(m(sv) \ge s v_1 \to \infty\); and it contains \(m(0) = 0\). Write
\(t_0 = \inf m(L) \le 0\), so that \(m(L) \supseteq (t_0, \infty)\). If
\(F_R(s) > 0\) then \(\{z \in L : m(z) \le s\}\) has positive
\(\gamma\)-measure and so contains a relatively open subset \(U\) of
\(L\). A point \(z^0 \in U\) has \(m(z^0) < s\): were
\(z^0_g = m(z^0) = s\) for some \(g\), the coordinate \(z \mapsto
z_g\) is non-constant on \(L\) because \(Z_g\) is non-degenerate, so
some \(v \in L\) has \(v_g > 0\) and \(z^0 + \epsilon v \in U\) has
\(m > s\) for small \(\epsilon\), contradicting
\(U \subseteq \{m \le s\}\). Hence \(t_0 < s\), and for any \(t > s\)
the set \(\{z \in L : s < m(z) < t\}\) is a non-empty relatively open
subset of \(L\), of positive \(\gamma\)-measure, so \(F_R(t) > F_R(s)\).
That \(\{F_R > 0\} =
(t_0, \infty)\) follows in the same way: \(\{m < t\}\) is non-empty and
open in \(L\) for \(t > t_0\), while
\(\{m \le t\} \subseteq \{m = t_0\}\) for \(t \le t_0\), a null event by
(i). (iii) By (i) and (ii), \(F_R\) maps \((t_0, \infty)\) continuously
and strictly increasingly onto \((0, 1)\), so every level in \((0,1)\)
has exactly one preimage. Finally
\(F_R(0) = \Pr\{Z_g \le 0 \text{ for all } g\} \le \Pr\{Z_1 \le
0\} = 1/2 < 1 - \alpha\) when \(\alpha < 1/2\), so
\(\kappa_\alpha(R) > 0\). For a fixed shift \(a\) the argument is the
same with \(m_a(z) = \max_g (z_g - a_g)\) in place of \(m\): \(m_a\) is
continuous on \(L\), unbounded above along the same \(v\), and
non-constant in each coordinate, which is all that (i) and (ii) use;
only the sign statement in (iii) rests on \(m(0) = 0\), and it is not
claimed for \(a \ne 0\). \(\square\)

\textbf{Lemma S5} (continuity in the correlation matrix). If correlation
matrices \(R_n \to R\) entrywise, then \(F_{R_n} \to F_R\) uniformly on
\(\mathbb R\) and \(\kappa_\alpha(R_n) \to \kappa_\alpha(R)\) for every
\(\alpha \in (0,1)\).

\emph{Proof.} \(N(0, R_n) \Rightarrow N(0, R)\) by Lévy's continuity
theorem, the characteristic functions being \(\exp(-u^\top R_n u/2)\);
the maximum is continuous, so
\(\max_g Z^{(n)}_g \Rightarrow \max_g Z_g\); the limit distribution
function is continuous by Lemma S4(i), so the convergence is uniform
(Pólya); and the limit quantile is unique by Lemma S4(iii), so the
quantiles converge \citep[Lemma 21.2]{vandervaart1998}. \(\square\)

\textbf{Lemma S6} (the standardized field). Conditionally on the data,
and on the event that every \(\sigma_{D,g} > 0\), the vectors \(Z^*_r\),
\(r \le
R_{\mathrm{out}}\), are independent and exactly \(N(0, \hat R)\), and
\(M^*_r =
\max_g Z^*_{r,g}\) are independent with distribution function
\(F_{\hat R}\).

\emph{Proof.} Section~\ref{sec-field-conditions} records that
\(\sqrt n\,\zeta^*_r \sim N(0,
\hat\Sigma)\) exactly given the data, the multipliers being Gaussian;
\(Z^*_r =
\mathrm{diag}(\hat\Sigma)^{-1/2}\sqrt n\,\zeta^*_r\) is a fixed linear
map of it, and \(\hat R\) is a correlation matrix when its diagonal is
positive. Independence across \(r\) is that of the \(\xi_r\).
\(\square\)

\emph{Proof of Theorem 3.} Three steps.

\emph{Step 1: the correlation estimate.} Under (A3),
\(\hat\Sigma = n\mathbf
D^\top\mathbf D \to_p \Sigma\). The hypothesis
\(\liminf_n n\sigma_{D,g}^2 > 0\) is \(\Sigma_{gg} > 0\) for every
\(g\), since \(n\sigma_{D,g}^2 = \hat\Sigma_{gg}\); under it \(R\) is a
correlation matrix, every \(\sigma_{D,g} > 0\) with probability tending
to one, and \(\hat R \to_p R\) by the continuous mapping theorem. The
theorem states \(\hat R \to_p R\) as a hypothesis for the record; it is
a consequence of (A3).

\emph{Step 2: the calibrated cutoff.} By Lemma S6 the \(M^*_r\) are,
given the data, independent with distribution function \(F_{\hat R}\),
whose \((1-\alpha)\) quantile \(\kappa_\alpha(\hat R)\) is unique by
Lemma S4, so that conditionally on the data
\(\hat\kappa_\alpha \to \kappa_\alpha(\hat R)\) almost surely as
\(R_{\mathrm{out}}
\to \infty\) \citep[Lemma 21.2, with the Glivenko--Cantelli
theorem]{vandervaart1998}. For the joint limit in \(n\) and
\(R_{\mathrm{out}}\), fix \(\delta > 0\). The event
\(\{\hat\kappa_\alpha > \kappa_\alpha(\hat R) + \delta\}\) requires the
empirical distribution function of the \(M^*_r\) to fall below
\(1 - \alpha\) at \(\kappa_\alpha(\hat R) + \delta\), a point where
\(F_{\hat R}\) equals \(1 - \alpha +
\eta_n\) with \(\eta_n > 0\); Hoeffding's inequality bounds its
conditional probability by \(\exp(-2R_{\mathrm{out}}\eta_n^2)\), and
likewise below with
\(\eta'_n = 1 - \alpha - F_{\hat R}(\kappa_\alpha(\hat R) - \delta) > 0\),
both positive by Lemma S4(ii). By Step 1 and Lemma S5, \(\eta_n \to_p
F_R(\kappa_\alpha(R) + \delta) - (1-\alpha) > 0\) and likewise
\(\eta'_n\), so both bounds tend to zero in probability and, being
bounded by one, in expectation; hence
\(\hat\kappa_\alpha - \kappa_\alpha(\hat R) \to_p 0\) as
\(n, R_{\mathrm{out}}
\to \infty\). Lemma S5 with Step 1 gives \(\kappa_\alpha(\hat R) \to_p
\kappa_\alpha(R)\), and so \(\hat\kappa_\alpha \to_p \kappa_\alpha(R)\).

\emph{Step 3: every candidate at the claim threshold.} Under
\(\beta(g) = c_{\mathrm{cons}}\) for every \(g\), which is the local
regime of Section~\ref{sec-field-conditions} with \(\beta_0 =
c_{\mathrm{cons}}\) and \(\mu = 0\),
\(W_n = \sqrt n(\hat\beta - c_{\mathrm{cons}}
\mathbf 1) \Rightarrow \zeta \sim N(0, \Sigma)\) by the multivariate
central limit theorem there, which uses (A3) for the linearization and
(A4) for the Lindeberg condition and does not use the non-degeneracy of
\(\Sigma\). With Step 1 and Slutsky's lemma,
\(T = \mathrm{diag}(\hat\Sigma)^{-1/2}W_n \Rightarrow
\mathrm{diag}(\Sigma)^{-1/2}\zeta =: Z \sim N(0, R)\). With the floors
inactive,
\(\{\hat{\mathcal A}(\hat\kappa_\alpha) \ne \varnothing\} = \{\max_g T_g >
\hat\kappa_\alpha\}\). By Step 2 and Slutsky's lemma, \((\max_g T_g,
\hat\kappa_\alpha) \Rightarrow (\max_g Z_g, \kappa_\alpha(R))\) jointly;
the set \(\{(x, c) : x > c\}\) has boundary \(\{x = c\}\), to which the
limit assigns probability \(\Pr\{\max_g Z_g = \kappa_\alpha(R)\} = 0\)
by Lemma S4(i); so by the portmanteau theorem \[
\Pr\{\max_g T_g > \hat\kappa_\alpha\} \to \Pr\{\max_g Z_g > \kappa_\alpha(R)\}
= 1 - F_R(\kappa_\alpha(R)) = \alpha. \qquad\square
\]

\emph{Strong control at the claim threshold.} The case
\(c_0 = c_{\mathrm{cons}}\) of Theorem 4(ii), which that theorem
excludes, holds under the conditions of Theorem 3 alone. Fix a law \(P\)
satisfying them, so that
\(\mathcal I_0 = \{g : \beta(g) \le c_{\mathrm{cons}}\}\) is a fixed
subset of \(\mathcal F\), and split it into
\(\mathcal I_{00} = \{g : \beta(g) =
c_{\mathrm{cons}}\}\) and
\(\mathcal I_{0-} = \{g : \beta(g) < c_{\mathrm{cons}}\}\). For
\(g \in \mathcal I_{0-}\), \[
T_g = \frac{\sqrt n\{\hat\beta(g) - \beta(g)\}}{\hat\sigma_g}
+ \frac{\sqrt n\{\beta(g) - c_{\mathrm{cons}}\}}{\hat\sigma_g},
\] the first term \(O_p(1)\) by (A3) and Step 1 and the second tending
to \(-\infty\), while \(\hat\kappa_\alpha = O_p(1)\) by Step 2; so
\(\Pr\{T_g > \hat\kappa_\alpha\}
\to 0\) for each such \(g\), and \(\mathcal I_{0-}\) contributes nothing
to the limit. On \(\mathcal I_{00}\) the argument of Step 3 applies to
the subvector: \((T_g)_{g
\in \mathcal I_{00}} \Rightarrow (Z_g)_{g \in \mathcal I_{00}}\) with
\(Z \sim N(0,
R)\), whatever the values of \(\beta(g)\) off \(\mathcal I_{00}\), and
\(\hat\kappa_\alpha \to_p \kappa_\alpha(R)\) by Step 2, which uses
nothing about \(P\) beyond (A1)--(A4). The boundary event
\(\{\max_{g \in \mathcal I_{00}} Z_g =
\kappa_\alpha(R)\}\) is null by Lemma S4(i) applied to the principal
submatrix, so \[
\Pr\Bigl\{\max_{g \in \mathcal I_{00}} T_g > \hat\kappa_\alpha\Bigr\}
\to \Pr\Bigl\{\max_{g \in \mathcal I_{00}} Z_g > \kappa_\alpha(R)\Bigr\}
\le \Pr\Bigl\{\max_{g \in \mathcal F} Z_g > \kappa_\alpha(R)\Bigr\} = \alpha,
\] the inequality because the maximum over a subset is dominated on
every realization by the maximum over the family, against which
\(\hat\kappa_\alpha\) is calibrated. This is the bound. The floors, when
active, remove candidates from \(\hat{\mathcal A}\) and so lower the
probability further. The same bound covers the reported region:
\(\hat H\) is admitted, so
\(\{\hat H \in \mathcal I_0\} \subseteq \{\exists\,
g \in \mathcal I_0 : g \in \hat{\mathcal A}(\hat\kappa_\alpha)\} \subseteq
\{\exists\, g \in \mathcal I_0 : T_g > \hat\kappa_\alpha\}\), whatever
rule chooses \(\hat H\). \(\square\)

The control is pointwise in \(P\) because the rate at which
\(\mathcal I_{0-}\) drops out depends on
\(\min_{g \in \mathcal I_{0-}}\{c_{\mathrm{cons}} - \beta(g)\}\), and
because Step 1 is a convergence in probability under the fixed \(P\);
both would have to hold uniformly over a class of laws for uniform
control.

\emph{The floors.} Two facts stated in Section 4.4 are recorded here.
First, at the paper's default the floor coincides with the claim
threshold, \(t_g = c =
c_{\mathrm{cons}}\), and is then implied by the calibrated screen: for
\(\alpha <
1/2\), \(\kappa_\alpha(R) > 0\) by Lemma S4(iii), so
\(\hat\kappa_\alpha > 0\) with probability tending to one by Step 2, and
\(T_g > \hat\kappa_\alpha\) gives
\(\hat\beta(g) > c_{\mathrm{cons}} = t_g\). The equality of Theorem 3
therefore holds with the floors in force. Second, a fixed floor
\(c > c_{\mathrm{cons}}\) makes the calibration conservative when every
candidate sits at the claim threshold: \(\hat\beta(g) > c\) requires
\(\sqrt n\{\hat\beta(g) - \beta(g)\} \ge \sqrt n(c - \beta(g)) \to
\infty\) when \(\beta(g) = c_{\mathrm{cons}}\), while the left side is
\(O_p(1)\), so each candidate clears the floor with probability tending
to zero and, \(M\) being finite, so does the declaration event. The
floor has an \(O(1)\) effect on declaration only when it is local,
\(t_g = c_{\mathrm{cons}} + \tau_g/\sqrt n\), which is the regime of
Section~\ref{sec-field-conditions}; there the declaration probability
tends to
\(\Pr\{\exists\, g : \sigma_g Z_g \ge \tau_g,\ Z_g > \kappa_\alpha(R)\}\),
between zero and \(\alpha\).

\emph{Power.} If \(\beta(g^\star) - c_{\mathrm{cons}} \ge \Delta > 0\)
and \(\beta(g^\star)
> t_{g^\star}\) for some \(g^\star \in \mathcal F\), then
\(\hat\beta(g^\star) > t_{g^\star}\) with probability tending to one by
(A3), and \(T_{g^\star} \ge
\sqrt n\,\Delta/\hat\sigma_{g^\star} + O_p(1) \to \infty\) while
\(\hat\kappa_\alpha = O_p(1)\), so \(\Pr\{g^\star \in \hat{\mathcal A}
(\hat\kappa_\alpha)\} \to 1\). Under local alternatives \(\beta(g) =
c_{\mathrm{cons}} + \mu_g/\sqrt n\) with the floors inactive, Step 3
gives \(T
\Rightarrow \mathrm{diag}(\Sigma)^{-1/2}(\mu + \zeta)\), whose
coordinates are \(Z_g + \mu_g/\sigma_g\); the maximum has a continuous
distribution function, each coordinate being non-degenerate, and the
portmanteau argument of Step 3 gives
\(\Pr\{\max_g T_g > \hat\kappa_\alpha\} \to \Pr\{\max_g(Z_g + \mu_g/\sigma_g) >
\kappa_\alpha(R)\}\). At any protected level the fixed-alternative
statement holds a fortiori,
\(\hat\kappa_\alpha(c_0) \le \hat\kappa_\alpha(c_{\mathrm{cons}})\) by
equation (9) of the main paper.

\emph{Proof of Proposition 5.} Let \(c_0 \le c_{\mathrm{cons}}\) be
fixed, so that \(\delta_g = (c_{\mathrm{cons}} - c_0)/\sigma_{D,g}\) is
a fixed vector given the data. Part (i) is the pathwise inequality
\(T_g \le U_g - \delta_g\) on \(\mathcal I_0(c_0)\), the maximum over a
subset being at most the maximum over \(\mathcal F\); nothing is
assumed. For part (ii), Lemma S6 gives \(M^*_r(c_0)\),
\(r \le R_{\mathrm{out}}\), independent conditionally on the data with
distribution function \(F_{\hat R, \delta}\), whose \((1-\alpha)\)
quantile is unique by Lemma S4(iii); under the working model
\(U \sim N(0, \hat R)\) given \(\mathbf D\), \(M(c_0)\) has the same
conditional law, so \(\Pr\{M(c_0) > \kappa_\alpha(\hat R, \delta)\} =
\alpha\) exactly, and Step 2's Hoeffding bound controls the distance
between \(\hat\kappa_\alpha(c_0)\) and \(\kappa_\alpha(\hat R, \delta)\)
as before, uniformly in the shift since the bound does not depend on it.
Monotonicity in the delivered benefit is the same pathwise inequality
read at a uniform benefit \(c' \ne c_0\): every \(T_g\) then equals
\(U_g - \delta'_g\) with \(\delta'_g \gtrless \delta_g\) as
\(c' \lessgtr
c_0\), so under the working model the rate is below or above \(\alpha\)
accordingly, and rises as \(c'\) rises toward \(c_{\mathrm{cons}}\).

\emph{The ordering (9).} For \(g \ne g^\dagger\), \(\Delta_g \ge 0\), so
on every outer draw
\(Z^*_{r,g^\dagger} \le \max_g (Z^*_{r,g} - \Delta_g) \le \max_g Z^*_{r,g}\).
Empirical quantiles over the same draws preserve the order, which gives
the right-hand inequality exactly; the left compares the empirical
quantile of one column, exactly \(N(0,1)\) given the data by Lemma S6,
with \(z_{1-\alpha}\), and Step 2 of the proof of Theorem 3 bounds that
difference. \(\square\)

\emph{Proof of Theorem 4.} Write
\(\hat\sigma_g = \sqrt n\,\sigma_{D,g}\), so that
\(\hat\sigma_g \to_p \sigma_g\) under (A3) and
\(\delta_g = \sqrt n\,(c_{\mathrm{cons}} -
c_0)/\hat\sigma_g\). Because the maximizer \(g^\dagger\) of \(\sigma_g\)
is unique, the maximizer of \(\hat\sigma_g\) equals it with probability
tending to one; on that event, for \(g \ne g^\dagger\),
\(\Delta_g = \sqrt n\,(c_{\mathrm{cons}} - c_0)(1/\hat\sigma_g -
1/\hat\sigma_{g^\dagger}) \to_p +\infty\), since
\(1/\sigma_g - 1/\sigma_{g^\dagger} > 0\). Write
\(\underline\Delta = \min_{g \ne g^\dagger} \Delta_g\).

\begin{enumerate}
\def\labelenumi{(\roman{enumi})}
\item
  Let \(\tilde F\) be the conditional distribution function of
  \(\max_g(Z^*_{r,g} - \Delta_g)\), whose \((1-\alpha)\) quantile is
  \(\kappa_\alpha(\hat R, \delta) + \delta_{g^\dagger}\). From the
  ordering above and a union bound,
  \(\Phi(t) - (M-1)\{1 - \Phi(t + \underline\Delta)\} \le \tilde F(t) \le \Phi(t)\)
  for every \(t\). As \(\underline\Delta \to_p \infty\),
  \(\tilde F \to \Phi\) uniformly on compact sets, in probability, and
  since \(\Phi\) is continuous and strictly increasing, the quantile
  converges in probability to \(z_{1-\alpha}\). Step 2 of the proof of
  Theorem 3 bounds the distance between the empirical quantile over
  \(R_{\mathrm{out}}\) draws and the exact one by the Hoeffding bound of
  that step, whose exponent \(\eta_n\), at that step's tolerance
  \(\delta' > 0\), converges in probability to
  \(\Phi(z_{1-\alpha} + \delta') - (1-\alpha) > 0\) by the convergence
  of \(\tilde F\) just shown, so that the bound vanishes as
  \(n, R_{\mathrm{out}} \to \infty\) as there. This is (i).
\item
  For \(g \in \mathcal I_0(c_0)\),
  \(T_g + \delta_{g^\dagger} = U_g - \Delta_g - \{c_0 - \beta(g)\}/\sigma_{D,g} \le U_g - \Delta_g\),
  so the probability in (ii) is at most
  \(\Pr\{\max_{g \in \mathcal F}(U_g - \Delta_g) > \tilde\kappa_\alpha(c_0)\}\).
  By (A3) and Slutsky's lemma \(U \Rightarrow N(0, R)\), so every
  \(U_g\) is \(O_p(1)\) and
  \(\max_{g \ne g^\dagger}(U_g - \Delta_g) \to_p -\infty\). The event is
  the union of \(\{U_{g^\dagger} > \tilde\kappa_\alpha(c_0)\}\) and
  \(\{\max_{g \ne g^\dagger}(U_g - \Delta_g) > \tilde\kappa_\alpha(c_0)\}\),
  the second of probability tending to zero since
  \(\tilde\kappa_\alpha(c_0)\) is tight by (i). The bound is therefore
  \(\Pr\{U_{g^\dagger} > \tilde\kappa_\alpha(c_0)\} + o(1)\), which
  tends to \(1 - \Phi(z_{1-\alpha}) = \alpha\) by (i), \(U_{g^\dagger}\)
  converging to a standard normal. No property of the candidates outside
  \(\mathcal I_0(c_0)\) was used. When \(\beta(g^\dagger) = c_0\) and
  the floors are inactive,
  \(T_{g^\dagger} + \delta_{g^\dagger} = U_{g^\dagger}\) exactly, so the
  probability is at least
  \(\Pr\{U_{g^\dagger} > \tilde\kappa_\alpha(c_0)\} \to \alpha\), and
  the limit is \(\alpha\). The data-dependent \(\delta_{g^\dagger}\)
  appears on both sides of every comparison and cancels; nothing is
  assumed about the fluctuation of the variance estimates.
\item
  If some candidate has \(\beta(g) \ge c_{\mathrm{cons}}\), then
  \(T_g = U_g + \{\beta(g) -
  c_{\mathrm{cons}}\}/\sigma_{D,g} \ge U_g\), and by (i)
  \(\hat\kappa_\alpha(c_0) = z_{1-\alpha} -
  \delta_{g^\dagger} + o_p(1) \to_p -\infty\), since
  \(\delta_{g^\dagger} = \sqrt n\,(c_{\mathrm{cons}} -
  c_0)/\hat\sigma_{g^\dagger} \to_p \infty\); so
  \(\Pr\{\max_h T_h > \hat\kappa_\alpha(c_0)\} \ge
  \Pr\{U_g > \hat\kappa_\alpha(c_0)\} \to 1\) with the floors inactive.
  With every floor at \(c_{\mathrm{cons}}\) the candidate is admitted
  only if also \(\hat\beta(g) > c_{\mathrm{cons}}\), which has
  probability tending to one when \(\beta(g) > c_{\mathrm{cons}}\),
  since then
  \(\sqrt n\{\hat\beta(g) - c_{\mathrm{cons}}\}/\hat\sigma_g \to_p \infty\),
  and to one half at \(\beta(g) = c_{\mathrm{cons}}\) exactly, where the
  floor decides. \(\square\)
\end{enumerate}

\subsection{The field on a two-candidate example in closed
form}\label{sec-field-example}

Proposition 1 characterizes the field lower bound's limiting coverage as
a functional of the limit objects \((\mu, \Sigma, h)\), Theorem 2 proves
it nominal once the comparison that settles the selection is decided by
a margin large against \(n^{-1/2}\), and Proposition 4 gives the pair's
joint coverage in the two regimes. This section evaluates all three
statements exactly on the smallest configuration in which selection is a
live contest, two candidates, where every object of Section 3.3 has a
closed form and the coverage functionals are one-dimensional Gaussian
integrals. The limit objects are not simulated: their integrals are
evaluated by deterministic quadrature. The implementation is then run on
the example's own data, and its finite-sample quantities are set beside
those limits. The example is, for the field, what the closed forms of
\citet[Section S4.5]{leonanderson2026} are for the two-term estimate's
residual, which it recovers as a check.

\textbf{Summary.}

\begin{figure}

\centering{

\pandocbounded{\includegraphics[keepaspectratio,alt={Two panels of coverage against the standardized gap, limiting values as lines and simulated coverage rates as points with interval bars at n = 400 and 1,600. In the disjoint panel the field lower bound\textquotesingle s line stays between 0.946 and 0.950, the unadjusted lower bound\textquotesingle s falls to 0.90 at a tie, and the pair\textquotesingle s runs from 0.945 to 0.951; the simulated points scatter around the lines. In the nested panel the field lower bound\textquotesingle s line runs from 0.950 to 0.957 and the unadjusted lower bound\textquotesingle s falls to 0.92 near a tie, again with the simulated points around the lines.}]{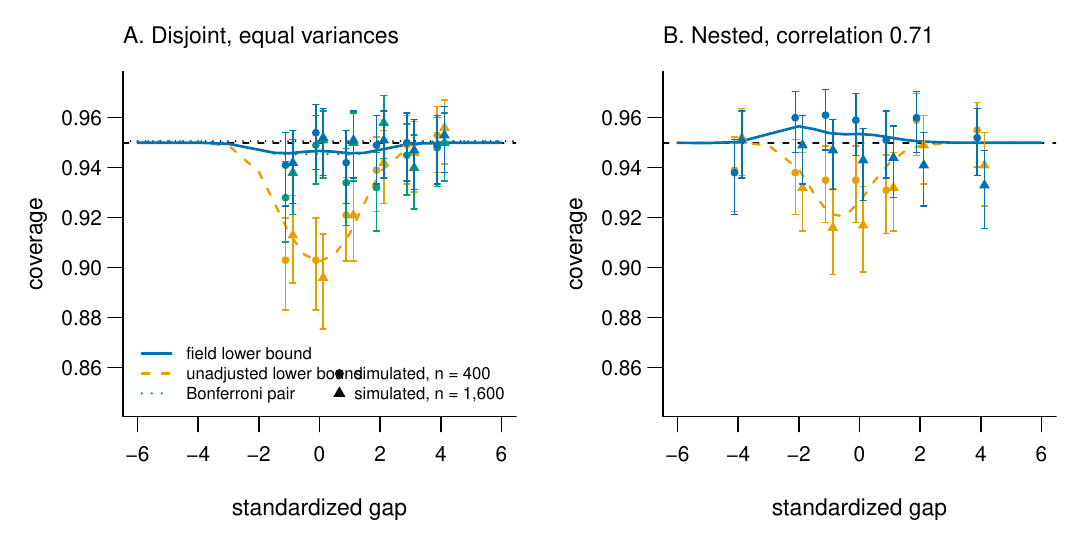}}

}

\caption{\label{fig-supp-two-candidate}Coverage of the field lower
bound, the unadjusted lower bound and, on the disjoint configuration,
the Bonferroni pair, against the standardized gap \(d/\tau\) on the
two-candidate example. Lines: the limiting coverage, by quadrature.
Points: the coverage rate of the implementation on the example's own
data, 1,000 replicates per point, with Wilson 95\% intervals; circles
\(n = 400\), triangles \(n = 1{,}600\), offset slightly in the gap so
the intervals separate. Dashed: nominal \(0.95\); dotted:
\((1-\alpha/2)^2\).}

\end{figure}

\begin{itemize}
\tightlist
\item
  \emph{The setup.} Two candidate regions; in each, the treatment effect
  is estimated by ordinary least squares of the outcome on treatment, so
  the estimate is the difference in arm means. The region with the
  larger estimate is selected --- the argmax, with no screening floor
  --- so a subgroup is always declared. In configuration A the two
  regions are disjoint halves of the sample; in B a region holding a
  quarter of the sample is nested in one holding half, and their
  estimates are correlated \(0.71\). The gap \(d\) between the two
  regions' effects is measured in units of \(\tau\), the standard error
  of their contrast.
\item
  \emph{The limits, exactly.} The limiting coverage of the field lower
  bound, of the unadjusted lower bound and, in A, of the Bonferroni pair
  are one-dimensional Gaussian integrals, evaluated by quadrature
  (Table~\ref{tbl-supp-two-candidate},
  Figure~\ref{fig-supp-two-candidate}). The field's lies between
  \(0.946\) and \(0.957\) across the gaps of
  Table~\ref{tbl-supp-two-candidate} and is \(0.9500\) once the regions
  are four contrast standard errors apart, six in B's slow direction;
  the unadjusted bound's is \(0.9025\) (A) and \(0.926\) (B) at a tie;
  the pair's lies between its lowest value, \(0.9452\), and its
  separated value \((1-\alpha/2)^2 = 0.9506\), which it approaches from
  below.
\item
  \emph{The implementation on the example's data.} The procedure was run
  on data drawn from the same model at \(n = 400\) and \(1{,}600\),
  \(1{,}000\) replicates per cell, with its default draws
  (Table~\ref{tbl-supp-two-candidate-sim}, the points of
  Figure~\ref{fig-supp-two-candidate}).
\item
  \emph{Selection and bias.} The first region is selected with a
  frequency within \(0.024\) of its limit \(\Phi(d/\tau)\) in every
  cell; the corrected estimate's bias is within two Monte Carlo standard
  errors of its limit in 23 of the 26 cells.
\item
  \emph{Coverage.} The field lower bound covers between \(0.933\) and
  \(0.961\), its limiting value inside the Wilson 95\% interval in 25 of
  the 26 cells; at a tie in A it covers \(0.954\) and \(0.952\) where
  the unadjusted lower bound covers \(0.903\) and \(0.896\). Across all
  one hundred coverage intervals of the check, seven exclude their
  limiting value, against five expected by chance.
\item
  \emph{The one visible departure.} In B at \(n = 400\) the quantile the
  field uses lies above its limit, by up to \(0.142\tau\), which lowers
  the bound; at \(n = 1{,}600\) it is within \(0.033\tau\). Run end to
  end, forest search returns the argmax's region in every replicate.
\end{itemize}

\emph{The setting.} Two candidate regions \(g_1\) and \(g_2\) receive
the mean-difference working model,
\(Y = \alpha_g + \beta(g)A + \epsilon\), so that \(\hat\beta(g)\) is the
difference in arm means, with the influence values \(e_i/n_{g,1}\) for a
treated member and \(-e_i/n_{g,0}\) for a control
(Section~\ref{sec-field-conditions}). With residual variance
\(\sigma^2\) and arms of equal size, a region of share \(p_g\) has
\(\operatorname{Var}\{\sqrt n\,\hat\beta(g)\} \to 4\sigma^2/p_g\); two
disjoint regions have independent estimates, and for nested regions
\(g_1 \subset g_2\) the covariance is the larger region's variance, so
the correlation is \((p_1/p_2)^{1/2}\). Two configurations are
evaluated, both at \(\sigma = 1\): (A) disjoint regions of equal share,
\(\Sigma\) proportional to \(I_2\); (B) \(g_1\) of share \(1/4\) nested
in \(g_2\) of share \(1/2\),
\(\Sigma = \bigl(\begin{smallmatrix}16 & 8\\ 8 & 8\end{smallmatrix}\bigr)\),
correlation \(0.71\) --- a subgroup competing with its own superset, the
configuration of the nested family of Section~\ref{sec-guohe} in
miniature. The rule is the pure argmax with no floor, so a subgroup is
always declared and the one comparison that settles the selection is the
effect order, condition (A5-i′)(b); the local parameter is the gap
\(d = \mu_1 - \mu_2\) of the local means
\(\beta(g) = \beta_0 + n^{-1/2}\mu_g\). Write
\(\tau = \operatorname{sd}(\zeta_1 - \zeta_2)\), the limiting standard
error of the contrast; \(d\) is reported in units of \(\tau\) so that
the two configurations share an axis, and \(d > 0\) makes \(g_1\) the
population winner.

\emph{The objects in closed form.} Let
\(U = \zeta_1 - \zeta_2 \sim N(0, \tau^2)\) and write each coordinate as
its regression on the contrast plus a residual, \(\zeta_g = a_g U + R\),
with \(a_1 = (\Sigma_{11} - \Sigma_{12})/\tau^2\),
\(a_2 = (\Sigma_{12} - \Sigma_{22})/\tau^2\), so that \(a_1 - a_2 = 1\),
and one residual \(R \sim N(0, \Sigma_{11} - a_1^2\tau^2)\) common to
both coordinates and independent of \(U\) (in A, \(a_1 = -a_2 = 1/2\)
and \(R = (\zeta_1 + \zeta_2)/2\); in B, \(a_1 = 1\), \(a_2 = 0\) and
\(R = \zeta_2\): the larger region's estimate is uncorrelated with the
contrast). The argmax applied to a field of gap \(\delta\) perturbed by
an independent copy \(U^*\) selects \(g_1\) exactly when
\(U^* > -\delta\). Hence the optimism functional of
Section~\ref{sec-field-conditions}, the selected coordinate, and the
root of Theorem 1 are

\begin{equation}\protect\phantomsection\label{eq-example-root}{
\begin{gathered}
m(\delta) = \tau\,\phi(\delta/\tau), \qquad
\zeta_{h(W)} = R + a_{G}U, \\
\Lambda = R + h_d(U), \quad h_\delta(u) = a_{G(\delta,u)}\,u - m(\delta + u),
\end{gathered}
}\end{equation}

where \(G(\delta, u) = 1\) if \(u + \delta > 0\) and \(2\) otherwise,
and \(m\) depends on a field only through its gap: for \(M = 2\) the
optimism is \(\tau\) times the standard normal density at the
standardized gap, maximal at a tie. The shrunk field of Section 3.3
replaces the winner's coordinate by its corrected value, so its gap is

\begin{equation}\protect\phantomsection\label{eq-example-shrunk}{
\tilde\delta(\delta_W) = \delta_W - \operatorname{sgn}(\delta_W)\,m(\delta_W), \qquad \delta_W = d + U,
}\end{equation}

the observed gap shrunk toward zero by the optimism. The simulated root
at a centre of gap \(\delta\), equation (2) of the main paper with
\(R_{\mathrm{in}} \to \infty\), is
\(\Lambda_\delta = R^* + h_\delta(U^*)\) with \((R^*, U^*)\) an
independent copy, so its distribution function is one Gaussian integral,

\begin{equation}\protect\phantomsection\label{eq-example-cdf}{
F_\delta(x) = \mathbb E_U\,\Phi\bigl\{(x - h_\delta(U))/\operatorname{sd}(R)\bigr\},
}\end{equation}

its \((1-\alpha)\) quantile \(q_{1-\alpha}(\delta)\) follows by
root-finding, and Proposition 1's coverage functional is

\begin{equation}\protect\phantomsection\label{eq-example-coverage}{
C(d) = \Pr\{\Lambda \le q_{1-\alpha}(\Lambda_{\tilde W})\}
     = \mathbb E_U\,\Phi\Bigl[\bigl\{q_{1-\alpha}\bigl(\tilde\delta(d+U)\bigr) - h_d(U)\bigr\}\big/\operatorname{sd}(R)\Bigr].
}\end{equation}

The unadjusted lower bound
\(\hat\beta(\hat H) - z_{1-\alpha}\,\sigma_{\hat H}/\sqrt n\) covers
with probability
\(\mathbb E_U\,\Phi\{(z_{1-\alpha}\Sigma_{GG}^{1/2} - a_G U)/\operatorname{sd}(R)\}\).
In configuration A the complement of the winner is the loser, its
optimism is \(-m\), its root is the mirror image \(R - h_d(U)\) of
\(\Lambda\), and the joint coverage of the Bonferroni pair, equation (4)
of the main paper, is again one integral over \(U\), both roots being
affine in \(R\) given \(U\); the studentization of Proposition 3 is
inert, the two complements having equal variance. Only in A is the law
of \(\Lambda_\delta\) symmetric in \(\delta\), so the quantile is
tabulated on a signed grid of gaps. Two checks tie the example to the
record:
\(\mathbb E[\Lambda]/\tau = \phi(d/\tau) - 2^{-1/2}\phi(d/\sqrt2\tau)\)
depends on \(d/\tau\) alone, and at \(\tau = \sqrt2\) it is the closed
form of \citet[Section S4.5]{leonanderson2026}, recovered to seven
decimals; and as \(|d|\) grows every quantile tends to
\(z_{1-\alpha}\Sigma_{g^\star g^\star}^{1/2}\), the value Theorem 2
inverts.

{

\begin{longtable}[]{@{}
  >{\raggedright\arraybackslash}p{(\linewidth - 16\tabcolsep) * \real{0.0800}}
  >{\raggedleft\arraybackslash}p{(\linewidth - 16\tabcolsep) * \real{0.0700}}
  >{\raggedleft\arraybackslash}p{(\linewidth - 16\tabcolsep) * \real{0.1100}}
  >{\raggedleft\arraybackslash}p{(\linewidth - 16\tabcolsep) * \real{0.1200}}
  >{\raggedleft\arraybackslash}p{(\linewidth - 16\tabcolsep) * \real{0.1400}}
  >{\raggedleft\arraybackslash}p{(\linewidth - 16\tabcolsep) * \real{0.1000}}
  >{\raggedleft\arraybackslash}p{(\linewidth - 16\tabcolsep) * \real{0.1200}}
  >{\raggedleft\arraybackslash}p{(\linewidth - 16\tabcolsep) * \real{0.1400}}
  >{\raggedleft\arraybackslash}p{(\linewidth - 16\tabcolsep) * \real{0.1200}}@{}}

\caption{\label{tbl-supp-two-candidate}Proposition 1's coverage
functional and its neighbours on the two-candidate example, evaluated by
quadrature, at \(\alpha = 0.05\). The gap \(d = \mu_1 - \mu_2\) is in
units of the contrast's limiting standard error \(\tau\); \(d > 0\)
makes the first region the population winner.
\(\mathbb E[\Lambda]/\tau\) is the residual of Theorem 1, the same
function of \(d/\tau\) in both configurations. Coverage is of the field
one-sided lower bound at nominal \(0.95\), of the unadjusted lower bound
at the same nominal level, and, in A, of the Bonferroni pair at joint
nominal \(0.95\), whose separated-selection value is
\((1-\alpha/2)^2 = 0.9506\). Envelope: the bound of Proposition S1 on
the field's coverage error, minimized over its free width.}

\tabularnewline

\toprule\noalign{}
\begin{minipage}[b]{\linewidth}\raggedright
\end{minipage} & \begin{minipage}[b]{\linewidth}\raggedleft
\(d/\tau\)
\end{minipage} & \begin{minipage}[b]{\linewidth}\raggedleft
\(\mathbb E[\Lambda]/\tau\)
\end{minipage} & \begin{minipage}[b]{\linewidth}\raggedleft
\(q_{1-\alpha}(d)/\tau\)
\end{minipage} & \begin{minipage}[b]{\linewidth}\raggedleft
\(\mathbb E\,q_{1-\alpha}(\tilde\delta)/\tau\)
\end{minipage} & \begin{minipage}[b]{\linewidth}\raggedleft
field
\end{minipage} & \begin{minipage}[b]{\linewidth}\raggedleft
envelope
\end{minipage} & \begin{minipage}[b]{\linewidth}\raggedleft
unadjusted
\end{minipage} & \begin{minipage}[b]{\linewidth}\raggedleft
pair
\end{minipage} \\
\midrule\noalign{}
\endhead
\bottomrule\noalign{}
\endlastfoot
A & -6.0 & -0.0000 & 1.163 & 1.163 & 0.9500 & 0.0026 & 0.9500 &
0.9506 \\
A & -4.0 & -0.0050 & 1.163 & 1.166 & 0.9500 & 0.0203 & 0.9500 &
0.9506 \\
A & -3.0 & -0.0253 & 1.167 & 1.177 & 0.9496 & 0.0190 & 0.9489 &
0.9499 \\
A & -2.0 & -0.0498 & 1.194 & 1.197 & 0.9472 & 0.0111 & 0.9383 &
0.9470 \\
A & -1.5 & -0.0312 & 1.213 & 1.209 & 0.9460 & 0.0142 & 0.9263 &
0.9457 \\
A & -1.0 & 0.0223 & 1.226 & 1.218 & 0.9458 & 0.0179 & 0.9137 & 0.9452 \\
A & -0.5 & 0.0871 & 1.231 & 1.224 & 0.9464 & 0.0194 & 0.9052 & 0.9454 \\
A & 0.0 & 0.1168 & 1.233 & 1.226 & 0.9467 & 0.0197 & 0.9025 & 0.9456 \\
A & 0.5 & 0.0871 & 1.231 & 1.224 & 0.9464 & 0.0194 & 0.9052 & 0.9454 \\
A & 1.0 & 0.0223 & 1.226 & 1.218 & 0.9458 & 0.0179 & 0.9137 & 0.9452 \\
A & 1.5 & -0.0312 & 1.213 & 1.209 & 0.9460 & 0.0142 & 0.9263 & 0.9457 \\
A & 2.0 & -0.0498 & 1.194 & 1.197 & 0.9472 & 0.0111 & 0.9383 & 0.9470 \\
A & 3.0 & -0.0253 & 1.167 & 1.177 & 0.9496 & 0.0190 & 0.9489 & 0.9499 \\
A & 4.0 & -0.0050 & 1.163 & 1.166 & 0.9500 & 0.0203 & 0.9500 & 0.9506 \\
A & 6.0 & -0.0000 & 1.163 & 1.163 & 0.9500 & 0.0026 & 0.9500 & 0.9506 \\
B & -6.0 & -0.0000 & 1.645 & 1.644 & 0.9500 & 0.0033 & 0.9500 & --- \\
B & -4.0 & -0.0050 & 1.640 & 1.645 & 0.9503 & 0.0373 & 0.9500 & --- \\
B & -3.0 & -0.0253 & 1.629 & 1.678 & 0.9534 & 0.1109 & 0.9490 & --- \\
B & -2.0 & -0.0498 & 1.700 & 1.798 & 0.9566 & 0.1009 & 0.9394 & --- \\
B & -1.5 & -0.0312 & 1.835 & 1.890 & 0.9554 & 0.0740 & 0.9296 & --- \\
B & -1.0 & 0.0223 & 1.989 & 1.986 & 0.9538 & 0.0546 & 0.9216 & --- \\
B & -0.5 & 0.0871 & 2.099 & 2.073 & 0.9534 & 0.0719 & 0.9204 & --- \\
B & 0.0 & 0.1168 & 2.173 & 2.143 & 0.9535 & 0.0827 & 0.9263 & --- \\
B & 0.5 & 0.0871 & 2.230 & 2.196 & 0.9530 & 0.0739 & 0.9348 & --- \\
B & 1.0 & 0.0223 & 2.274 & 2.238 & 0.9521 & 0.0543 & 0.9421 & --- \\
B & 1.5 & -0.0312 & 2.302 & 2.270 & 0.9511 & 0.0401 & 0.9467 & --- \\
B & 2.0 & -0.0498 & 2.316 & 2.294 & 0.9505 & 0.0325 & 0.9489 & --- \\
B & 3.0 & -0.0253 & 2.325 & 2.319 & 0.9500 & 0.0250 & 0.9499 & --- \\
B & 4.0 & -0.0050 & 2.326 & 2.325 & 0.9500 & 0.0098 & 0.9500 & --- \\
B & 6.0 & -0.0000 & 2.326 & 2.326 & 0.9500 & 0.0007 & 0.9500 & --- \\

\end{longtable}

}

\emph{What the numbers say.} Three things, one for each statement the
example evaluates. First, Theorem 2: on the disjoint pair the field's
coverage is \(0.9500\) by four contrast standard errors of separation in
either direction; on the nested pair it is \(0.9500\) by four when the
smaller region is the population winner, and in the slow direction, the
larger region the winner, \(0.9503\) at four and \(0.9500\) by six. The
quantile the field uses converges to
\(z_{0.95}\Sigma_{g^\star g^\star}^{1/2}\) --- \(1.163\tau\) in A, and
in B \(2.326\tau\) when the smaller region wins and \(1.645\tau\) when
the larger does, reached to three decimals by six --- the convergence in
Lemma S7 and Step 4 of the proof made visible. Second, Proposition 1:
the location error is small and is not signed. In A the field covers
\(0.9467\) at an exact tie and \(0.9458\) near one contrast standard
error, its lowest value, and \(0.9496\) at three; the shrunk gap
\(|\delta_W| - m\) over-states the separation on average, so the field
reads the quantile of a competition more settled than the truth
(\(1.226\tau\) against \(1.233\tau\) at the tie). In B the field covers
\(0.9535\) at the tie and up to \(0.9566\) when the larger region is the
population winner by two contrast standard errors, and \(0.9505\) at two
in the other direction: there the whole of the optimism sits in the
smaller region's coordinate, and a realization in which the larger
region is observed to win reads a quantile that is generous for the root
it faces. The sign of the error depends on the covariance geometry
because coverage is a joint functional --- the quantile the field reads
and the root it is compared with are evaluated on the same realization
and co-move --- which is what Section~\ref{sec-field-prop1} says in
words. The envelope of Proposition S1 evaluated at each gap lies between
\(0.0007\) and \(0.1109\); the exact error sits inside it at every gap,
and, at the gaps where the exact error is not zero to four decimals, the
envelope is looser than the exact error by a factor of \(3.5\) to
\(136\), which is the cost of a bound that holds for a general family
against a closed form. Third, Proposition 4: the pair covers \(0.9456\)
at the tie and \(0.9452\) near one contrast standard error, its lowest
value, the dependence the common selection induces, then \(0.9470\) at
two and \(0.9499\) at three, reaching the separated value
\((1-\alpha/2)^2 = 0.9506\) from below by four. For reference, the
unadjusted lower bound covers \(0.9025\) (A) and \(0.926\) (B) at the
tie and at least \(0.949\) on both by three.

\emph{The implementation on the example's data.} The limit objects are
what the procedure converges to; the procedure itself was run on data
drawn from the example's own model --- treatment assigned with
probability one half, unit-variance normal errors, no prognostic effect;
in A a binary covariate that splits the sample into the two regions, in
B a covariate uniform on four levels whose two upper levels carry an
effect a fixed margin below the two regions' --- at \(n = 400\) and
\(1{,}600\), 1,000 replicates for each configuration, gap and sample
size, with the implementation's defaults: 2,000 multiplier draws for the
correction and a field of 1,000 outer by 500 inner draws. The pair
\(\{g_1, g_2\}\) is the family and the argmax is applied to it.
Table~\ref{tbl-supp-two-candidate-sim} sets the selection frequency, the
corrected estimate's bias and the mean quantile the field uses beside
their limits, and Figure~\ref{fig-supp-two-candidate} draws the coverage
rates as points. The selection frequency is within \(0.024\) of
\(\Phi(d/\tau)\) in every cell. The bias of the corrected estimate,
\(\sqrt n\{\tilde\beta(\hat H) - \beta(\hat H)\}/\tau\), is within two
Monte Carlo standard errors of \(\mathbb E[\Lambda]/\tau\) in 23 of the
26 cells. The quantile the field uses is within \(0.009\tau\) of its
limit in A already at \(n = 400\); in B it lies above its limit at
\(n = 400\), by \(0.015\tau\) to \(0.142\tau\), the largest gap where
the larger region wins by four contrast standard errors, and within
\(0.033\tau\) of it at \(n = 1{,}600\) --- the finite-sample approach of
the quantile in the configuration where a region holds a quarter of the
sample. The field lower bound covers between \(0.933\) and \(0.961\),
its limiting value inside the Wilson 95\% interval in 25 of the 26
cells; at the tie in A it covers \(0.954\) and \(0.952\) where the
unadjusted lower bound covers \(0.903\) and \(0.896\), and the
unadjusted bound's limiting value is inside the interval in all 26
cells. In A the field-s complement bound's and the pair's limiting
values are each inside the interval in 10 of 12 cells. Across all one
hundred intervals, seven exclude their limiting value, against five
expected of 95\% intervals by chance. Run end to end --- forest search
on the binary covariate of A, the unscreened argmax with every floor
inert, at gaps 0, 1 and 3 --- the identifier returned the argmax's
region in every replicate, and the field lower bound's limiting coverage
is inside the interval in all six cells.

{

\begin{longtable}[]{@{}
  >{\raggedright\arraybackslash}p{(\linewidth - 16\tabcolsep) * \real{0.0333}}
  >{\raggedleft\arraybackslash}p{(\linewidth - 16\tabcolsep) * \real{0.1000}}
  >{\raggedleft\arraybackslash}p{(\linewidth - 16\tabcolsep) * \real{0.0667}}
  >{\raggedleft\arraybackslash}p{(\linewidth - 16\tabcolsep) * \real{0.1000}}
  >{\raggedleft\arraybackslash}p{(\linewidth - 16\tabcolsep) * \real{0.1667}}
  >{\raggedleft\arraybackslash}p{(\linewidth - 16\tabcolsep) * \real{0.0778}}
  >{\raggedleft\arraybackslash}p{(\linewidth - 16\tabcolsep) * \real{0.2889}}
  >{\raggedleft\arraybackslash}p{(\linewidth - 16\tabcolsep) * \real{0.1000}}
  >{\raggedleft\arraybackslash}p{(\linewidth - 16\tabcolsep) * \real{0.0667}}@{}}

\caption{\label{tbl-supp-two-candidate-sim}The implementation on the
two-candidate example's own data, beside the limits of
Table~\ref{tbl-supp-two-candidate}: 1,000 replicates per row, the two
regions \(g_1\) and \(g_2\) supplied as the family and the argmax
applied to it. Selected: the share of replicates selecting \(g_1\),
beside its limit \(\Phi(d/\tau)\). Bias: the mean of
\(\sqrt n(\tilde\beta(\hat H) - \beta(\hat H))/\tau\), Monte Carlo
standard errors \(0.02\) to \(0.05\), beside
\(\mathbb E[\Lambda]/\tau\). Quantile: the mean of the quantile the
field uses, over \(\tau\), beside its limit, the mean of
\(q_{1-\alpha}(\tilde\delta)/\tau\).}

\tabularnewline

\toprule\noalign{}
\begin{minipage}[b]{\linewidth}\raggedright
\end{minipage} & \begin{minipage}[b]{\linewidth}\raggedleft
\(d/\tau\)
\end{minipage} & \begin{minipage}[b]{\linewidth}\raggedleft
\(n\)
\end{minipage} & \begin{minipage}[b]{\linewidth}\raggedleft
selected
\end{minipage} & \begin{minipage}[b]{\linewidth}\raggedleft
\(\Phi(d/\tau)\)
\end{minipage} & \begin{minipage}[b]{\linewidth}\raggedleft
bias
\end{minipage} & \begin{minipage}[b]{\linewidth}\raggedleft
\(\mathbb E[\Lambda]/\tau\)
\end{minipage} & \begin{minipage}[b]{\linewidth}\raggedleft
quantile
\end{minipage} & \begin{minipage}[b]{\linewidth}\raggedleft
limit
\end{minipage} \\
\midrule\noalign{}
\endhead
\bottomrule\noalign{}
\endlastfoot
A & -1 & 400 & 0.182 & 0.159 & 0.082 & 0.022 & 1.227 & 1.218 \\
A & -1 & 1,600 & 0.177 & 0.159 & -0.027 & 0.022 & 1.216 & 1.218 \\
A & 0 & 400 & 0.506 & 0.500 & 0.102 & 0.117 & 1.233 & 1.226 \\
A & 0 & 1,600 & 0.500 & 0.500 & 0.089 & 0.117 & 1.225 & 1.226 \\
A & 1 & 400 & 0.838 & 0.841 & -0.013 & 0.022 & 1.227 & 1.218 \\
A & 1 & 1,600 & 0.840 & 0.841 & 0.050 & 0.022 & 1.218 & 1.218 \\
A & 2 & 400 & 0.979 & 0.977 & -0.034 & -0.050 & 1.204 & 1.197 \\
A & 2 & 1,600 & 0.975 & 0.977 & -0.063 & -0.050 & 1.194 & 1.197 \\
A & 3 & 400 & 0.995 & 0.999 & -0.038 & -0.025 & 1.180 & 1.177 \\
A & 3 & 1,600 & 0.999 & 0.999 & -0.031 & -0.025 & 1.176 & 1.177 \\
A & 4 & 400 & 1.000 & 1.000 & -0.052 & -0.005 & 1.171 & 1.166 \\
A & 4 & 1,600 & 1.000 & 1.000 & -0.039 & -0.005 & 1.161 & 1.166 \\
B & -4 & 400 & 0.000 & 0.000 & -0.018 & -0.005 & 1.787 & 1.645 \\
B & -4 & 1,600 & 0.000 & 0.000 & -0.031 & -0.005 & 1.677 & 1.645 \\
B & -2 & 400 & 0.021 & 0.023 & -0.087 & -0.050 & 1.864 & 1.798 \\
B & -2 & 1,600 & 0.022 & 0.023 & -0.008 & -0.050 & 1.808 & 1.798 \\
B & -1 & 400 & 0.150 & 0.159 & -0.021 & 0.022 & 2.030 & 1.986 \\
B & -1 & 1,600 & 0.158 & 0.159 & 0.045 & 0.022 & 1.992 & 1.986 \\
B & 0 & 400 & 0.499 & 0.500 & 0.061 & 0.117 & 2.157 & 2.143 \\
B & 0 & 1,600 & 0.512 & 0.500 & 0.169 & 0.117 & 2.149 & 2.143 \\
B & 1 & 400 & 0.826 & 0.841 & 0.027 & 0.022 & 2.268 & 2.238 \\
B & 1 & 1,600 & 0.842 & 0.841 & 0.063 & 0.022 & 2.244 & 2.238 \\
B & 2 & 400 & 0.983 & 0.977 & -0.092 & -0.050 & 2.321 & 2.294 \\
B & 2 & 1,600 & 0.978 & 0.977 & -0.039 & -0.050 & 2.293 & 2.294 \\
B & 4 & 400 & 1.000 & 1.000 & -0.031 & -0.005 & 2.357 & 2.325 \\
B & 4 & 1,600 & 1.000 & 1.000 & 0.004 & -0.005 & 2.326 & 2.325 \\

\end{longtable}

}

\emph{What the example does and does not verify.} The example evaluates
the limiting coverage functional exactly for one two-candidate Gaussian
selection problem and checks that the implementation approaches that
limit in the configurations run. It does not establish nominal local
coverage for a general family: it has no admission floor, declaration is
certain, the same-draws term is absent, the selection map is the pure
argmax, and the candidate dimension is two, so it isolates the
quantile-location discrepancy from the other features of the implemented
procedures. The complement and the pair are evaluated in configuration A
only, where the complement of the winner is the loser. The script that
produces every number in this section is distributed with the paper; it
runs in base R in seconds and asserts the closed-form check and the two
limits.

\textbf{Declaration on the example.} The rule above is the argmax with
no screen, so a subgroup is always declared. Adding the screen of
Section 4 of the main paper --- a candidate admitted when
\(T_g = \{\hat\beta(g) - c_{\mathrm{cons}}\}/\sigma_{D,g}\) exceeds a
cutoff \(\kappa\) --- makes declaration a live event on the same data,
and the rule then plays no part in it, since a subgroup is declared
exactly when some candidate is admitted. Conditionally on the data the
standardized draws of equation (6) of the main paper are exactly
\(N(0, \hat R)\), and in the limit the off-diagonal of \(R\) is the
correlation \(\rho\) of the two estimates: \(\rho = 0\) in configuration
A and \(\rho = (p_1/p_2)^{1/2} = 2^{-1/2}\) in B. Writing
\(\Phi_2(\cdot, \cdot; \rho)\) for the standard bivariate normal
distribution function, the diagnostic of equation (8) of the main paper
tends to \[
1 - \Phi_2\bigl(z_{(1+p^\star)/2},\, z_{(1+p^\star)/2};\, \rho\bigr),
\] the calibrated cutoff of equation (7) tends to the root
\(\kappa_\alpha\) of \(\Phi_2(\kappa, \kappa; \rho) = 1 - \alpha\), and
at local effects \(\beta(g) = c_{\mathrm{cons}} + n^{-1/2}\mu_g\), with
\(h_g = \mu_g/\sigma_g\), the limiting declaration probability of
Section 4.3 of the main paper under the calibrated screen is \[
1 - \Phi_2(\kappa_\alpha - h_1,\, \kappa_\alpha - h_2;\, \rho).
\] With both candidates at the claim threshold, \(h_1 = h_2 = 0\), this
is \(\alpha\), as Theorem 3 states; with both on the beneficial side,
\(h_1, h_2 < 0\), it is smaller; and when one candidate sits at the
claim threshold and the other is clearly harmful, the first is admitted
with probability \(1 - \Phi(\kappa_\alpha)\), below \(\alpha\), as
strong control at the claim threshold requires (Section S1.6). At
\(\alpha = 0.05\) and \(p^\star = 0.90\) the conventional screen's
family-wise size is 0.0975 in A and 0.0801 in B; the calibrated cutoff
is 1.955 in A and 1.875 in B, an implied screening level
\(2\Phi(\kappa_{0.05}) - 1\) of 0.949 and 0.939; and a candidate at the
claim threshold beside a clearly harmful one is admitted with
probability 0.0253 in A and 0.0304 in B. The overlap of the nested pair
lowers the cutoff: B's two candidates count for less than two
independent tests. Because the draws are exactly Gaussian given the
data, an implementation of equations (7) and (8) can be checked on a
single dataset against these expressions evaluated at \(\hat\rho\), up
to the Monte Carlo error of its draws. At a protected level \(c_0\) the
same forms carry the shift. With
\(\delta_j = (c_{\mathrm{cons}} - c_0)/\sigma_{D,j}\) for the two
candidates, the diagnostic of equation (8) at a screen cutoff
\(\kappa_{\mathrm{scr}}\) is, up to the Monte Carlo error of the draws
and at the realized \(\delta_j\) and \(\hat\rho\),
\(1 - \Phi_2(\kappa_{\mathrm{scr}} + \delta_1,\, \kappa_{\mathrm{scr}} + \delta_2;\, \hat\rho)\),
and the cutoff of equation (7) the root of
\(\Phi_2(\kappa + \delta_1,\, \kappa + \delta_2;\, \hat\rho) = 1 - \alpha\)
--- no limit in \(n\) is taken, \(c_0\) being a constant; the unshifted
forms above are the case \(\delta_1 = \delta_2 = 0\). The two candidates
of configuration B differ in share by a factor of two and so in
\(\sigma_D\) by \(\sqrt 2\), which makes \(\delta_1\) and \(\delta_2\)
differ by the same factor: the shift is heterogeneous even here, and the
larger region is pushed down furthest. The closed form also shows
Theorem 4 of the main paper at work. The smaller candidate is the least
informative; write \(\delta_s\) and \(\delta_l\) for the smaller and
larger candidates' shifts. As \(\delta_l - \delta_s \to \infty\) the
equation for the cutoff becomes
\(\Phi(\kappa + \delta_s) = 1 - \alpha\), so
\(\kappa + \delta_s \to z_{1-\alpha}\): the pair is calibrated as a
single test of the smaller region.

\subsection{Calibrating at a protected level: implementation and
evidence}\label{sec-field-declcal}

\subsubsection{What one run produces}\label{what-one-run-produces}

One run of the search yields everything the calibration needs, and
nothing in it depends on the screening level the run was made at: the
enumerated family, each candidate's coefficient and robust standard
error, and the shared multiplier draws. The calibration is therefore
read off the run post hoc, and applying the calibrated constant to
\(T_g\) is the same declaration as re-running the screen at the implied
level, provided the re-run applies that constant and not a rounded
version of it. In applications the rule is applied on \(T_g\).

\emph{Two computational steps sit between the enumeration and the
executed screen.} The search fits only candidates with at least ten
events in each arm, and removes duplicate members from the family it
ranks. Both are clean-up rather than method: the calibration and the
diagnostic are computed over the enumerated family --- the same family
multiplier re-selection has always used, in which the event minimum is
not re-applied (Section S5) --- while the executed screen evaluates the
cleaned one, and the difference is a few candidates. Its footprint in
the campaign below is 16 declarations in 20,000 replicates.

\emph{The executed screen rounds, and the implied level is not always
settable.} The screen as implemented compares the consistency rate
rounded to two places against the screening level, so a nominal level
\(p^\star\) executes as a rate of at least \(p^\star - 0.005\) and a
cutoff of \(\Phi^{-1}\{(1 + p^\star - 0.005)/2\}\): 1.621 at a nominal
0.90 rather than 1.645, 2.432 at 0.99, 2.807 at 1.00. The diagnostic is
therefore evaluated at the cutoff the screen executes, and the
calibrated screen never passes through that comparison at all. The same
rounding decides whether the implied level can be set. On the three
applications (Section 6 and Section S3) a level resolved to four digits
comes within 0.003 of the calibrated cutoff, while at the default two
digits only one of the three is reachable at all, at a level of 1.00
that sits 0.081 above the cutoff it is meant to reproduce. The implied
level is therefore an interpretation of \(\hat\kappa\) and not a route
to it, which is why the rule is applied on \(T_g\).

\emph{A protected level must be fixed in advance.} If the level were set
at the trial's own observed effect the constant would become
outcome-dependent and the field would not propagate its variability: for
a candidate holding a share \(\pi\) of the trial the true standard
deviation of the contrast between the candidate and the whole-trial
effect is about \(\sqrt{1-\pi}\) of what a field centred at the observed
effect assumes, so the calibration would be conservative by that factor
and conservative by an amount that depends on the candidate. A level
fixed on clinical grounds avoids this entirely. The principled repair,
were an estimated centre ever wanted, is a contrast field --- perturbing
the difference between the candidate's coefficient and the whole-trial
coefficient with the difference of their influence values on the same
draws, standardized by the contrast's own robust standard error ---
which is not developed here.

\subsubsection{The schema}\label{the-schema}

\begin{figure}

\centering{

\includegraphics[width=0.8\linewidth,height=\textheight,keepaspectratio,alt={A schematic in three tiers. Top: two boxes, the claim at the claim threshold and the protection at a protected level. Middle: one forest-search run feeding a standardized statistic and a standardized field, which feed two screens side by side, the conventional screen with its family-wise-size diagnostic and the calibrated screen with the shift, the shifted maximum, its quantile as cutoff and the implied level. Bottom: a hazard-ratio number line marked at the protected level and the claim threshold, with the rate at most alpha to the left of the protected level, above alpha between the two, and harm per the claim to the right.}]{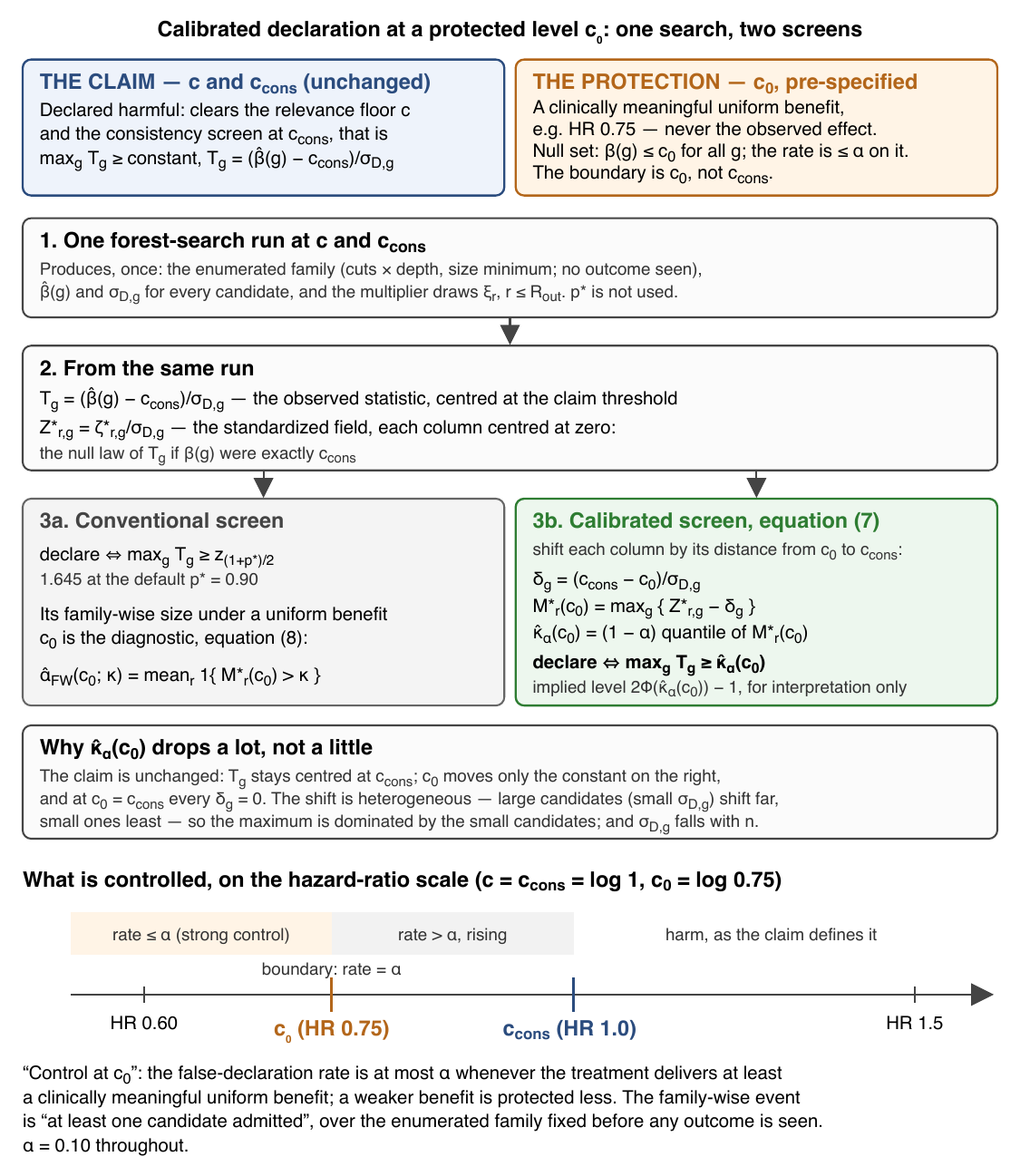}

}

\caption{\label{fig-supp-declcal-schema}The calibrated declaration
threshold at a protected level, in one picture. Top: the claim at the
claim threshold, and the protection at a pre-specified uniform benefit
\(c_0\). Middle: one forest-search run gives the standardized statistic
and the standardized field; the conventional screen's family-wise size
at \(c_0\) is the diagnostic of equation (8), and the calibrated screen
shifts each column by that candidate's own distance from \(c_0\) to the
claim threshold and takes the \((1-\alpha)\) quantile of the shifted
maximum as its cutoff, equation (7). Bottom: what is controlled where on
the hazard-ratio scale. Section 4 gives the construction in full.}

\end{figure}

Figure~\ref{fig-supp-declcal-schema} draws the construction on one page:
the three thresholds, the one run that serves both screens, and what is
controlled where on the hazard-ratio scale.

\subsubsection{Evidence}\label{evidence}

The construction was evaluated on the covariate design of Section 5,
with planted-harm cells of its own, forest search alone, with the
screening and claim thresholds both at a hazard ratio of one, 2,000
replicates per cell, on the same seeds and draws as the
operating-characteristics campaign. Uniform-benefit cells put every
candidate on the beneficial side of the claim threshold, at two
uniform-benefit designs with marginal Cox hazard ratios of 0.657 and
0.721, at sample sizes 500, 1,000 and 1,500; planted-harm cells put a
region at hazard ratio 1.5 or 2.0 at sample sizes 1,000 and 1,500.

The two uniform-benefit levels are not free choices. Each is the
marginal Cox hazard ratio in the complement of the planted region in the
campaign's other cells, at its two planted-region prevalences --- 0.657
at a prevalence of 12.4\% and 0.721 at 31\% --- and it is the same value
whether the planted region carries harm or sits at a hazard ratio of
one. A uniform-benefit cell therefore delivers to the whole trial
exactly the benefit those cells deliver outside their planted region,
and, the benefit being uniform, that level is also the cell's own
trial-wide estimand. Setting the two side by side shows how much a
planted region dilutes the trial-wide effect: where that region sits at
a hazard ratio of one, the trial-wide marginal hazard ratio is 0.685 at
the smaller prevalence and 0.792 at the larger, against complements of
0.657 and 0.721. The declaration rate a uniform-benefit cell produces is
therefore the rate the same design would show with no planted region
anywhere. (The uniform-benefit designs were solved to the three-decimal
levels, which sit within 0.0005 of the complement values they match.)
Rates are over 2,000 replicates, with Wilson 95\% intervals on the
false-declaration column.

{

\begin{longtable}[]{@{}
  >{\raggedright\arraybackslash}p{(\linewidth - 6\tabcolsep) * \real{0.2800}}
  >{\raggedright\arraybackslash}p{(\linewidth - 6\tabcolsep) * \real{0.2400}}
  >{\raggedright\arraybackslash}p{(\linewidth - 6\tabcolsep) * \real{0.2400}}
  >{\raggedright\arraybackslash}p{(\linewidth - 6\tabcolsep) * \real{0.2400}}@{}}

\caption{\label{tbl-supp-declcal-conventional}The conventional screen
under uniform benefit, cell by cell: the declaration rate of the screen
as executed at a screening level of 0.90, as a proportion of 2,000
replicates, with Wilson 95\% intervals.}

\tabularnewline

\toprule\noalign{}
\begin{minipage}[b]{\linewidth}\raggedright
Uniform benefit (marginal Cox HR)
\end{minipage} & \begin{minipage}[b]{\linewidth}\raggedright
\(n\) = 500
\end{minipage} & \begin{minipage}[b]{\linewidth}\raggedright
\(n\) = 1,000
\end{minipage} & \begin{minipage}[b]{\linewidth}\raggedright
\(n\) = 1,500
\end{minipage} \\
\midrule\noalign{}
\endhead
\bottomrule\noalign{}
\endlastfoot
0.657 & 0.0950 (0.0829, 0.1086) & 0.0325 (0.0256, 0.0412) & 0.0070
(0.0042, 0.0117) \\
0.721 & 0.2325 (0.2145, 0.2515) & 0.1110 (0.0980, 0.1255) & 0.0325
(0.0256, 0.0412) \\

\end{longtable}

}

The rate falls with the benefit delivered and with the sample size, and
falls by more than an order of magnitude across the six cells. It is a
family-wise false-declaration rate in the sense of Section 4.3, every
candidate in these designs being at least as beneficial as the claim
threshold, and it is far below the boundary size the diagnostic reports
at that threshold. The screen discriminates well against a clinically
meaningful uniform benefit; what it lacks is family-wise control, not
discrimination.

No uniform-benefit cell is worse than the weaker benefit at the smallest
sample size in any calibrated row, and at the smallest rates several
cells tie with it; the tables give that worst cell's rate. Every
declaration in a uniform-benefit cell is false as a claim of harm;
whether it also falls within the guarantee depends on the protected
level. At a protected hazard ratio of 0.70 the weaker design's
trial-wide level, 0.721, lies above the protected level and not every
candidate need lie in the protected set, so the rate in those cells,
though below \(\alpha\), is not one the guarantee covers.

\needspace{0.36\textheight}

{

\begin{longtable}[]{@{}
  >{\raggedright\arraybackslash}p{(\linewidth - 6\tabcolsep) * \real{0.3000}}
  >{\raggedright\arraybackslash}p{(\linewidth - 6\tabcolsep) * \real{0.3000}}
  >{\raggedright\arraybackslash}p{(\linewidth - 6\tabcolsep) * \real{0.2000}}
  >{\raggedright\arraybackslash}p{(\linewidth - 6\tabcolsep) * \real{0.2000}}@{}}

\caption{\label{tbl-supp-declcal-alpha10}The calibrated screen at
\(\alpha = 0.10\), by protected level, against the conventional screen
and against a screening level tuned by simulation. Every entry is a
proportion of 2,000 replicates; the primary protected level is set in
bold. The level each cutoff implies is in
Table~\ref{tbl-supp-declcal-cutoff-n}.}

\tabularnewline

\toprule\noalign{}
\begin{minipage}[b]{\linewidth}\raggedright
Screen
\end{minipage} & \begin{minipage}[b]{\linewidth}\raggedright
Worst uniform-benefit false declaration (Wilson 95\%)
\end{minipage} & \begin{minipage}[b]{\linewidth}\raggedright
Power at HR 1.5, \(n\) = 1,000 / 1,500
\end{minipage} & \begin{minipage}[b]{\linewidth}\raggedright
Power at HR 2.0, \(n\) = 1,000 / 1,500
\end{minipage} \\
\midrule\noalign{}
\endhead
\bottomrule\noalign{}
\endlastfoot
Conventional, \(p^\star = 0.90\) & 0.2325 (0.2145, 0.2515) & 0.699 /
0.791 & 0.964 / 0.988 \\
Tuned fixed \(p^\star = 0.9545\) & 0.0725 (0.0619, 0.0847) & 0.526 /
0.648 & 0.910 / 0.971 \\
Calibrated, \(c_0 = \log 0.70\) & 0.0705 (0.0601, 0.0826) & 0.617 /
0.824 & 0.943 / 0.992 \\
\textbf{Calibrated, \(c_0 = \log 0.75\)} & \textbf{0.0350 (0.0278,
0.0440)} & \textbf{0.501 / 0.723} & \textbf{0.896 / 0.982} \\
Calibrated, \(c_0 = \log 0.80\) & 0.0165 (0.0118, 0.0231) & 0.370 /
0.599 & 0.833 / 0.964 \\
Calibrated, \(c_0 = \log 0.85\) & 0.0055 (0.0031, 0.0098) & 0.265 /
0.459 & 0.756 / 0.921 \\
Calibrated, \(c_0 = c_{\mathrm{cons}}\) & 0.0005 (0.0001, 0.0028) &
0.083 / 0.145 & 0.448 / 0.726 \\

\end{longtable}

}

\needspace{0.30\textheight}

{

\begin{longtable}[]{@{}
  >{\raggedright\arraybackslash}p{(\linewidth - 6\tabcolsep) * \real{0.3000}}
  >{\raggedright\arraybackslash}p{(\linewidth - 6\tabcolsep) * \real{0.3000}}
  >{\raggedright\arraybackslash}p{(\linewidth - 6\tabcolsep) * \real{0.2000}}
  >{\raggedright\arraybackslash}p{(\linewidth - 6\tabcolsep) * \real{0.2000}}@{}}

\caption{\label{tbl-supp-declcal-alpha05}The same at \(\alpha = 0.05\).}

\tabularnewline

\toprule\noalign{}
\begin{minipage}[b]{\linewidth}\raggedright
Screen
\end{minipage} & \begin{minipage}[b]{\linewidth}\raggedright
Worst uniform-benefit false declaration (Wilson 95\%)
\end{minipage} & \begin{minipage}[b]{\linewidth}\raggedright
Power at HR 1.5, \(n\) = 1,000 / 1,500
\end{minipage} & \begin{minipage}[b]{\linewidth}\raggedright
Power at HR 2.0, \(n\) = 1,000 / 1,500
\end{minipage} \\
\midrule\noalign{}
\endhead
\bottomrule\noalign{}
\endlastfoot
Calibrated, \(c_0 = \log 0.70\) & 0.0315 (0.0247, 0.0401) & 0.503 /
0.747 & 0.896 / 0.983 \\
\textbf{Calibrated, \(c_0 = \log 0.75\)} & \textbf{0.0130 (0.0089,
0.0190)} & \textbf{0.373 / 0.620} & \textbf{0.834 / 0.966} \\
Calibrated, \(c_0 = \log 0.80\) & 0.0040 (0.0020, 0.0079) & 0.271 /
0.491 & 0.753 / 0.928 \\
Calibrated, \(c_0 = \log 0.85\) & 0.0010 (0.0003, 0.0036) & 0.188 /
0.363 & 0.676 / 0.882 \\
Calibrated, \(c_0 = c_{\mathrm{cons}}\) & 0.0005 (0.0001, 0.0028) &
0.050 / 0.094 & 0.355 / 0.641 \\

\end{longtable}

}

The cutoff is invariant to about 0.03 across data-generating laws at a
fixed sample size and protected level, that variation being systematic
and coming from the standard errors; it falls with the sample size at
every protected level below the claim threshold, and is flat in the
sample size at the claim threshold itself --- the two ends of equation
(9) of the main paper: flat where Theorem 3 gives a fixed limit, and
falling where Theorem 4 sends the centred cutoff toward a single
candidate's.

\needspace{0.42\textheight}

{

\begin{longtable}[]{@{}
  >{\raggedright\arraybackslash}p{(\linewidth - 14\tabcolsep) * \real{0.1500}}
  >{\raggedright\arraybackslash}p{(\linewidth - 14\tabcolsep) * \real{0.0700}}
  >{\raggedleft\arraybackslash}p{(\linewidth - 14\tabcolsep) * \real{0.1300}}
  >{\raggedleft\arraybackslash}p{(\linewidth - 14\tabcolsep) * \real{0.1300}}
  >{\raggedleft\arraybackslash}p{(\linewidth - 14\tabcolsep) * \real{0.1300}}
  >{\raggedleft\arraybackslash}p{(\linewidth - 14\tabcolsep) * \real{0.1300}}
  >{\raggedleft\arraybackslash}p{(\linewidth - 14\tabcolsep) * \real{0.1300}}
  >{\raggedleft\arraybackslash}p{(\linewidth - 14\tabcolsep) * \real{0.1300}}@{}}

\caption{\label{tbl-supp-declcal-cutoff-n}The calibrated cutoff by
sample size, and the screening level it implies. At each sample size the
cutoff is the median over cells of each cell's median cutoff, every cell
at that sample size; the implied level is \(2\Phi(\hat\kappa) - 1\), the
level at which an exact comparison on the standardized statistic would
reproduce the same constant, reported for interpretation, the rule being
applied on that statistic directly.}

\tabularnewline

\toprule\noalign{}
\begin{minipage}[b]{\linewidth}\raggedright
\(c_0\)
\end{minipage} & \begin{minipage}[b]{\linewidth}\raggedright
\(\alpha\)
\end{minipage} & \begin{minipage}[b]{\linewidth}\raggedleft
\(\hat\kappa\), \(n\) = 500
\end{minipage} & \begin{minipage}[b]{\linewidth}\raggedleft
\(n\) = 1,000
\end{minipage} & \begin{minipage}[b]{\linewidth}\raggedleft
\(n\) = 1,500
\end{minipage} & \begin{minipage}[b]{\linewidth}\raggedleft
implied \(p^\star\), \(n\) = 500
\end{minipage} & \begin{minipage}[b]{\linewidth}\raggedleft
\(n\) = 1,000
\end{minipage} & \begin{minipage}[b]{\linewidth}\raggedleft
\(n\) = 1,500
\end{minipage} \\
\midrule\noalign{}
\endhead
\bottomrule\noalign{}
\endlastfoot
\(\log 0.70\) & 0.10 & 2.19 & 1.82 & 1.51 & 0.972 & 0.932 & 0.868 \\
\(\log 0.70\) & 0.05 & 2.45 & 2.07 & 1.75 & 0.986 & 0.962 & 0.921 \\
\(\log 0.75\) & 0.10 & 2.39 & 2.09 & 1.82 & 0.983 & 0.963 & 0.931 \\
\(\log 0.75\) & 0.05 & 2.65 & 2.33 & 2.07 & 0.992 & 0.980 & 0.961 \\
\(\log 0.80\) & 0.10 & 2.59 & 2.34 & 2.12 & 0.990 & 0.981 & 0.966 \\
\(\log 0.80\) & 0.05 & 2.85 & 2.59 & 2.37 & 0.996 & 0.990 & 0.982 \\
\(\log 0.85\) & 0.10 & 2.78 & 2.59 & 2.42 & 0.995 & 0.990 & 0.985 \\
\(\log 0.85\) & 0.05 & 3.04 & 2.84 & 2.67 & 0.998 & 0.995 & 0.992 \\
\(c_{\mathrm{cons}}\) & 0.10 & 3.34 & 3.34 & 3.33 & 0.9992 & 0.9992 &
0.9991 \\
\(c_{\mathrm{cons}}\) & 0.05 & 3.60 & 3.59 & 3.58 & 0.9997 & 0.9997 &
0.9997 \\

\end{longtable}

}

\emph{Reading the tables.} Two things are visible beyond what
Table~\ref{tbl-supp-declcal-conventional} already shows. The calibrated
screen holds its rate at or below the nominal level in every
uniform-benefit cell and does so at a cost in power that the protected
level controls directly: at a protected hazard ratio of 0.75 and
\(\alpha = 0.10\) it declares on 0.501 and 0.723 of replicates in the
planted-harm cells at hazard ratio 1.5 against the conventional screen's
0.699 and 0.791, and 0.896 and 0.982 at hazard ratio 2.0 against 0.964
and 0.988, while cutting the worst false-declaration rate from 0.2325 to
0.0350. And the implied \(p^\star\)
(Table~\ref{tbl-supp-declcal-cutoff-n}) falls with the sample size at
every protected level, from 0.983 at 500 to 0.931 at 1,500 at a
protected hazard ratio of 0.75, which is the ordering (9) of the main
paper at work, in the quantity a reader of the screen recognizes. At a
protected hazard ratio of 0.70 and the largest sample size the implied
level falls below the conventional 0.90, to 0.868, where the calibrated
rule is the looser of the two --- correctly: at that information the
family's multiplicity has largely drained away, the regime of Theorem 4,
in which the rule approaches a single test of the least informative
candidate at the protected level.

\emph{A fixed level tuned by simulation is the calibration at one sample
size.} A screening level of 0.9545, chosen by simulation to hold
uniform-benefit false declaration below a tenth, sits almost exactly at
the calibration's implied level at a protected hazard ratio of 0.75,
\(\alpha = 0.10\) and \(n\) = 1,000, which is 0.963. The tuned rule was
the calibration at one sample size; the calibration generalizes it
across sample sizes and attaches a guarantee to it.

\emph{Power at the smaller planted effect is a property of the design.}
At hazard ratio 1.5 and \(n\) = 1,000 the planted region holds about 124
subjects, for which a Schoenfeld approximation at half events puts the
oracle standardized effect for the true region near 1.6; no
size-controlling rule has much power there, and the calibrated screen's
power runs between one and a half and two times that single-candidate
figure at the same cutoff.

\emph{Three free checks.} First, the diagnostic at a protected level
exceeds the realized rate in every uniform-benefit cell, which is what a
boundary size should do when the designs sit strictly inside the
protected level. Compared like for like --- the diagnostic at the exact
cutoff against the rate on the same pre-reduction family --- the margins
at their narrowest run from 0.040 to 0.061, at the loosest protected
level in the three cells of the weaker benefit. Compared as each is
actually used, the diagnostic at the cutoff the screen executes against
the rate the executed search realizes, the same three cells give 0.043
to 0.092, the widest at the cell with the highest rate. Two small
effects compound there --- the lower cutoff raises the diagnostic, and
the two clean-up steps remove a few candidates from the executed family
--- and they are the reason the two figures should not be quoted as one
range. Second, the calibrated screen can declare where the conventional
one did not: pooled over 20,000 replicates, 18 at \(\alpha = 0.05\) and
172 at \(\alpha = 0.10\), most of the latter through the lower cutoff at
the loosest protected level and largest sample size, and 16 of the 18
through a candidate the executed search's clean-up had removed --- the
footprint noted above. The inclusion of the calibrated admitted set in
the conventional one is therefore not a general fact once the protected
level sits below the claim threshold, though it held throughout at the
claim threshold itself. Third, replacing the per-replicate cutoff by a
single median value per sample size reproduces the exact rule's ordering
and magnitudes to within 0.0135, reading slightly low on size.

\subsubsection{The diagnostic against a realized
rate}\label{the-diagnostic-against-a-realized-rate}

The boundary size can be set against a realized rate on a real trial's
own covariate design. The breast cancer application's 686 patients and
their covariates were held fixed, forest search alone, 5,000 replicates
per cell, under a uniform benefit calibrated to a marginal Cox hazard
ratio of 0.75 --- the protected level itself. One cell re-randomizes the
arm assignment and one keeps the observed one. The consistency rates
were recorded as the screen compares them, so the realized rates are the
executed screen's and are directly comparable with the diagnostic.

\begin{longtable}[]{@{}
  >{\raggedright\arraybackslash}p{(\linewidth - 6\tabcolsep) * \real{0.2500}}
  >{\raggedright\arraybackslash}p{(\linewidth - 6\tabcolsep) * \real{0.2500}}
  >{\raggedright\arraybackslash}p{(\linewidth - 6\tabcolsep) * \real{0.2500}}
  >{\raggedright\arraybackslash}p{(\linewidth - 6\tabcolsep) * \real{0.2500}}@{}}
\caption{The diagnostic at the primary protected level against the
realized declaration rate, on the application's own design. The
diagnostic is \(\hat\alpha_{\mathrm{FW}}(c_0;\kappa)\) at
\(c_0 = \log 0.75\), \(\alpha = 0.10\), evaluated at each screening
level's executed cutoff. Realized rates are proportions of 5,000
replicates with Wilson 90\%
intervals.}\label{tbl-supp-declcal-realized}\tabularnewline
\toprule\noalign{}
\begin{minipage}[b]{\linewidth}\raggedright
Screening level
\end{minipage} & \begin{minipage}[b]{\linewidth}\raggedright
Diagnostic at \(c_0 = \log 0.75\)
\end{minipage} & \begin{minipage}[b]{\linewidth}\raggedright
Realized rate, re-randomized
\end{minipage} & \begin{minipage}[b]{\linewidth}\raggedright
Realized rate, observed assignment
\end{minipage} \\
\midrule\noalign{}
\endfirsthead
\toprule\noalign{}
\begin{minipage}[b]{\linewidth}\raggedright
Screening level
\end{minipage} & \begin{minipage}[b]{\linewidth}\raggedright
Diagnostic at \(c_0 = \log 0.75\)
\end{minipage} & \begin{minipage}[b]{\linewidth}\raggedright
Realized rate, re-randomized
\end{minipage} & \begin{minipage}[b]{\linewidth}\raggedright
Realized rate, observed assignment
\end{minipage} \\
\midrule\noalign{}
\endhead
\bottomrule\noalign{}
\endlastfoot
0.90 & 0.6672 & 0.448 (0.437, 0.460) & 0.432 (0.421, 0.444) \\
0.99 & 0.186 & 0.081 (0.075, 0.087) & 0.074 (0.068, 0.080) \\
1.00 & not evaluated & 0.026 & 0.022 \\
\end{longtable}

\emph{Reading it.} The diagnostic exceeds the realized rate at both
levels at which both are available, by 0.22 at the conventional level
and by 0.11 at the tightest settable one, and the gap narrows as the
screen tightens. That is what a boundary size should do, and two
separate things make it do so here. The truth sits inside the protected
level for most of the family: the candidates' own marginal hazard ratios
run from 0.674 to 0.798 with a median of 0.733, where the diagnostic is
computed as though every candidate sat exactly at 0.75. And the
diagnostic is computed over the pre-reduction family of 1,744
candidates, while the realized rate is produced by the 1,507 that
survive the search's clean-up steps, so it is computed over slightly
more candidates than produce the rate. The comparison is not a pure
boundary reading: about a quarter of the family, 364 of the 1,507,
carries a true marginal hazard ratio above the protected level, and a
declaration of one of those is not a false one. Those candidates raise
the realized rate rather than lower it, so the margin between the two
columns is if anything understated.

\emph{The calibration and the level the realized rate would require
agree closely.} At \(\alpha =
0.10\) and the primary protected level the calibrated cutoff on this
design is 2.726. The realized rate first falls to a tenth or below at a
screening level of 0.99, the finest step the default two-digit
resolution allows, whose executed cutoff is 2.432. The calibrated rule
is therefore a little stricter than the tuned level the design would
need, in the conservative direction and by about what the boundary
reading predicts --- the same reconciliation the simulated tuned level
gives, now on a trial's own covariate design rather than a simulated
one.

\emph{What this evidence does not establish.} The campaign measured
declaration and not identification: no rate here says that the declared
subgroup is the planted one, and which region the calibrated admitted
set yields under a selection rule was not measured --- a live question,
since under planted harm 96.6\% of calibrated declarations at
\(\alpha = 0.05\) (20,242 of 20,944, pooled over the four protected
levels) admit more than one candidate. The strict null, in which no
region anywhere carries harm, is not among these cells, and the
uniform-benefit cells are a composite null for declaration rather than a
strict one. A treatment effect varying smoothly with the covariates was
not run; the alternatives here are a planted discrete region.

\subsection{Scope when the candidate family is generated from the
outcomes}\label{sec-field-conditional}

This section distinguishes fixed-family inference from inference applied
after an outcome-dependent proposal step. The theory of Sections S1.1 to
S1.8 is stated for a finite family \(\mathcal F\) satisfying (A2). When
an identifier first uses the outcomes to construct a random family
\(\hat{\mathcal F}\), holding the realized \(\hat{\mathcal F}\) fixed
during multiplier re-selection does not replay the variability of that
proposal step. The construction still adjusts re-ranking and final
selection among the regions in the realized family, but the present
theory does not establish unconditional validity over repeated proposals
or a selective conditional-coverage theorem given \(\hat{\mathcal F}\).

\begin{enumerate}
\def\labelenumi{\arabic{enumi}.}
\item
  \emph{What transfers mechanically.} Once a finite realized family is
  specified, the standard-analysis coefficients, the influence-value
  covariance field, the multiplier perturbations, the re-ranking rule
  and the subgroup and complement corrections are computed as in
  Sections S1.1 to S1.8.
\item
  \emph{What is not reproduced.} If the outcomes were used to decide
  which regions enter the family, the multiplier construction does not
  replay that step; holding the family fixed during re-selection does
  not account for the families that could have been proposed.
\item
  \emph{Scope of the present theory.} The fixed-family results apply to
  families satisfying (A2). For an outcome-generated realized family the
  procedure may be reported as an adjustment for the subsequent
  selection within that family; this paper does not establish
  unconditional validity or a selective conditional-coverage theorem
  given the proposal.
\item
  \emph{Routes to broader validity.} Independent proposal and inference
  samples would make the family independent of the inference sample;
  re-selection over a covariate-measurable superset containing every
  region the proposal could return would price the proposal, at a cost
  in power. Neither is evaluated here.
\item
  \emph{The three identifiers.} Forest search is closest to (A2),
  subject to the treatment of the finite-sample event minimum in Section
  S1.1. GRF's grid is covariate-generated, but the fitted forest's
  decision to return a region is outcome-dependent and is not replayed;
  the field perturbs the standard-analysis coefficients over the
  realized grid without refitting the forest. DINA's fitted surface
  determines which regions enter the proposed family. The inferential
  scope therefore differs across the three.
\end{enumerate}

\emph{The mechanics transfer.} For a family proposed by a fitted model
the boundary is the screening threshold rather than the claim threshold:
the standardized statistic is \(\{\hat\beta(g) - c\}/\sigma_{D,g}\) and
the shift \(\delta_g = (c - c_0)/\sigma_{D,g}\), with the field, the
shifted maximum, the cutoff and the diagnostic formed exactly as in
Section 4. Nothing in the construction uses how the candidates arose.

\emph{The diagnostic and the cutoff are not alike here.} The family-wise
size over a proposed family transfers as a conditional multiplicity
diagnostic, and reading it as such is sound: it says what the proposed
family's overlap costs. The cutoff does not carry the forest search
guarantee over, because the quantile it inverts is the conditional one.

\emph{Cuts at covariate quantiles.} Cuts computed from the empirical
covariate distribution make the realized regions, and hence their
working-model estimands, conditional on the observed cuts unless their
variability is shown to be asymptotically negligible. The cited
change-point results give faster-than-\(n^{-1/2}\) estimation under
specific discontinuity and model conditions and do not imply first-order
equivalence for every empirical quantile grid used here. The present
theory therefore treats the realized cuts as fixed during resampling
\citep{chan1993, hansen2000, deng2022}.

\subsection{What remains open}\label{sec-field-open}

\begin{itemize}
\item
  The tail probability \(\Pr\{|Q(\tilde W) - Q(\mu)| > \delta\}\) of
  Proposition S1, the quantile-location term: no general rate or
  computable bound is given for it on a general family; Section S1.7
  evaluates it exactly on two candidates.
\item
  \emph{The same-draws term under uncertain declaration.} When
  declaration remains non-degenerate, Theorem S1 shows that the
  corrected estimator's limiting root contains \(-m_0(W)\) and the
  complement's \(-m_0^c(W)\); the field roots omit both. If the limiting
  field is associated, \(m_0(W) \ge 0\) and the omission shifts the
  lower bound in the conservative direction, as a componentwise effect;
  overlap alone does not establish association, and the
  quantile-location discrepancy remains. For the complement, zero
  covariance between the matched subgroup and complement coordinates
  does not fix the sign of \(m_0^c(W)\), because declaration depends on
  the whole subgroup field and unmatched coordinates may be correlated.
  A field root carrying both terms remains open.
\item
  Asymptotic exactness of the studentized bound, Proposition 3(ii),
  rests on the first-moment condition
  \(m^c(W) = \sigma^c_{h(W)}m^Z(W)\), which is a condition and not a
  theorem. A structural condition on the complement influences under
  which it holds would make the asymptotic exactness unconditional.
\item
  The two-sided interval has no result here.
  Section~\ref{sec-field-prop1} gives the reason it is the harder
  object; it does not give a construction.
\item
  The generalized-linear paths. Everything above is written for the
  survival estimand; Section~\ref{sec-field-conditions} records that
  (A3), and its complement form, hold on the mean-difference and
  log-odds paths under standard M-estimation conditions, so Sections 3
  and 4 apply to them as stated. Their operating characteristics are
  evaluated in Section~\ref{sec-supp-glm-oc} on one design per outcome
  family, built from ACTG 175; broader designs are not.
\item
  The declaration control of Theorems 3 and 4 is pointwise in the
  data-generating law. Uniform control over a class of laws would need
  (A3) and the convergence \(\hat R \to_p R\) to hold uniformly, and a
  uniform rate at which candidates strictly inside the protected set
  drop out of the maximum; Section~\ref{sec-field-declaration} records
  where each enters.
\item
  Theorem 4 assumes a unique least informative candidate. With ties
  among the limiting standard errors, the shifts of the tied candidates
  differ by amounts of order one that depend on the fluctuation of their
  variance estimates, and the limit of the declaration probability then
  depends on the joint law of those fluctuations and the estimates; it
  is \(\alpha\) when the two are asymptotically independent, which is
  not established in general. Between the two ends of equation (9) no
  limit is taken: where a given trial sits is decided by \(\Delta_g\),
  and Section S1.8 measures it on the evaluation's designs.
\item
  The family is fixed and finite, (A2). A family growing with \(n\)
  would need a high-dimensional Gaussian approximation with
  anti-concentration for the maximum \citep{cck2017, cck2015} and a
  growth restriction relating \(\log M\) to the influence tails; nothing
  here reaches that case.
\end{itemize}

\section{Operating characteristics: further detail}\label{sec-supp-oc}

This section supports the operating-characteristics section of the main
paper. It carries the material the main text refers to but does not
reproduce: the selection-rule comparison, coverage cell by cell, the
bias against the planted target, the classification tables under both
conventions, and the two-sided interval that serves throughout as a
reference rather than a recommendation.

Conventions are those of the main text and are not restated.
Classification is unconditional unless a table says otherwise; bias,
coverage and bound location are over declaring replicates; bounds are
read by their location against effect sizes and never as significance at
a hazard ratio of one; and comparisons across identifiers are
descriptive. Table~\ref{tbl-supp-design} gives the design's hazard
ratios, each a marginal Cox hazard ratio of the treatment on the
simulated population's potential outcomes, fixed by design and not
estimated.

{

\begin{longtable}[]{@{}rrrrr@{}}

\caption{\label{tbl-supp-design}Hazard ratios of the simulation design
by cell: in the planted region, in its complement, and over the whole
trial. The region's is set by calibration; the complement keeps the
treatment's base effect, so its value differs between prevalences only
because the complement is a different subpopulation.}

\tabularnewline

\toprule\noalign{}
Planted HR & Prevalence & Region & Complement & Whole trial \\
\midrule\noalign{}
\endhead
\bottomrule\noalign{}
\endlastfoot
1.50 & 12.4\% & 1.509 & 0.657 & 0.704 \\
1.75 & 12.4\% & 1.769 & 0.657 & 0.710 \\
1.00 & 12.4\% & 1.000 & 0.657 & 0.685 \\
1.50 & 31\% & 1.499 & 0.721 & 0.869 \\
1.75 & 31\% & 1.746 & 0.721 & 0.894 \\
1.00 & 31\% & 1.000 & 0.721 & 0.792 \\

\end{longtable}

}

\subsection{The selection rules as operating
points}\label{sec-supp-rules}

The main text closes its identification section with the comparison
behind the rule it evaluates: identification and classification under
the five rules of Section 2 of the main paper --- max N, max N(ε), max
β, min N and min N(ε) --- and, for forest search alone, under the
unscreened argmax \(\max_{\mathcal{S}}\beta\), which has no counterpart
for DINA or GRF because neither enumerates \(\mathcal S\); eighteen
cells, 500 replicates, no inference. That figure averages over the
twelve harm cells and shows the operating points where the planted
region is broad. The operating points where it is narrow are here.

\begin{figure}

\centering{

\pandocbounded{\includegraphics[keepaspectratio]{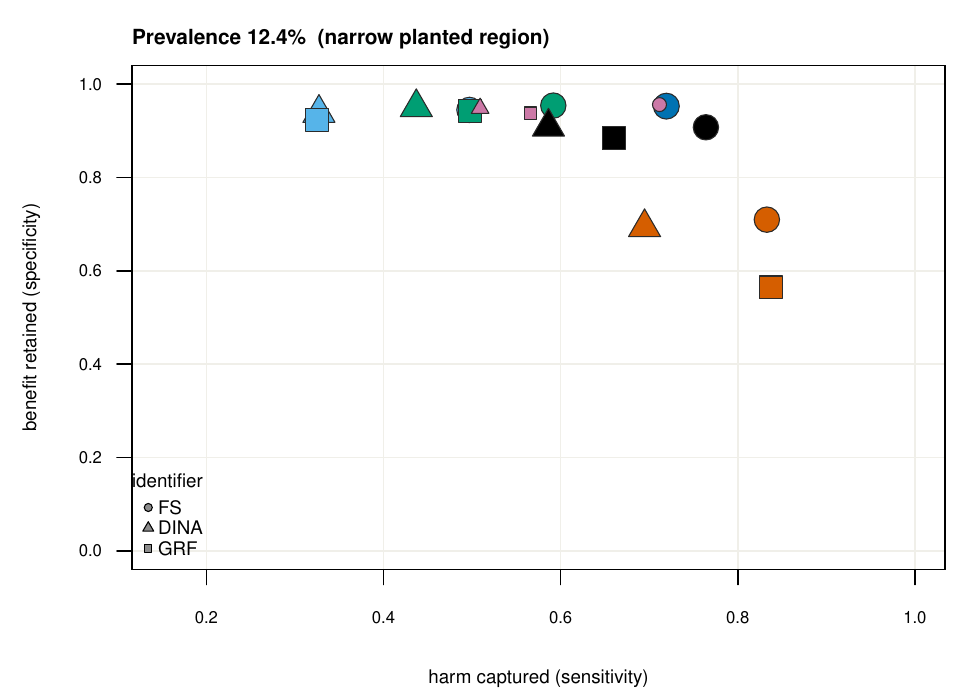}}

}

\caption{\label{fig-supp-rules-12}The selection rules as operating
points where the planted region is narrow. Harm captured against benefit
retained, averaged over the six harm cells at that prevalence; the upper
right is better on both. Symbol marks the identifier, colour the rule,
point size the rate at which a region is returned. The rules are not
joined: they are distinct rules, not thresholds on a common score.}

\end{figure}

Figure~\ref{fig-supp-rules-12} shows the same runs as operating points,
which makes the dominance argument direct --- a rule below and to the
left of another is worse on both objectives at once. Under DINA, max N
returns the entire trial on between 159 and 299 of 500 replicates at the
broad harm cells, leaving the complement empty and its negative
predictive value undefined.

Every rule but max N lies along the upper edge at benefit retained above
0.88, spread across harm captured from 0.15 to 0.76; max N alone leaves
that edge, and at the broad prevalence it leaves it decisively,
retaining between 0.12 and 0.23 of the complement. The rule the main
text evaluates, max N(ε), is the rightmost point on the upper edge at
both prevalences.

For forest search, \(\max_{\mathcal{A}} \beta\) and the unscreened
\(\max_{\mathcal{S}} \beta\) coincide at the broad prevalence and differ
by less than 0.01 at the narrow one. The consistency screen decides
whether a region is returned, not which: it declines on about 3\% of
replicates at the narrow harm cells and on none at the broad ones, and
it does not move the operating point.

\begin{figure}

\centering{

\pandocbounded{\includegraphics[keepaspectratio,alt={A two-by-two panel figure: sensitivity, specificity and identified-to-planted size ratio by selection rule for three identifiers on the attenuated-benefit cells, and their operating points where the planted region is broad.}]{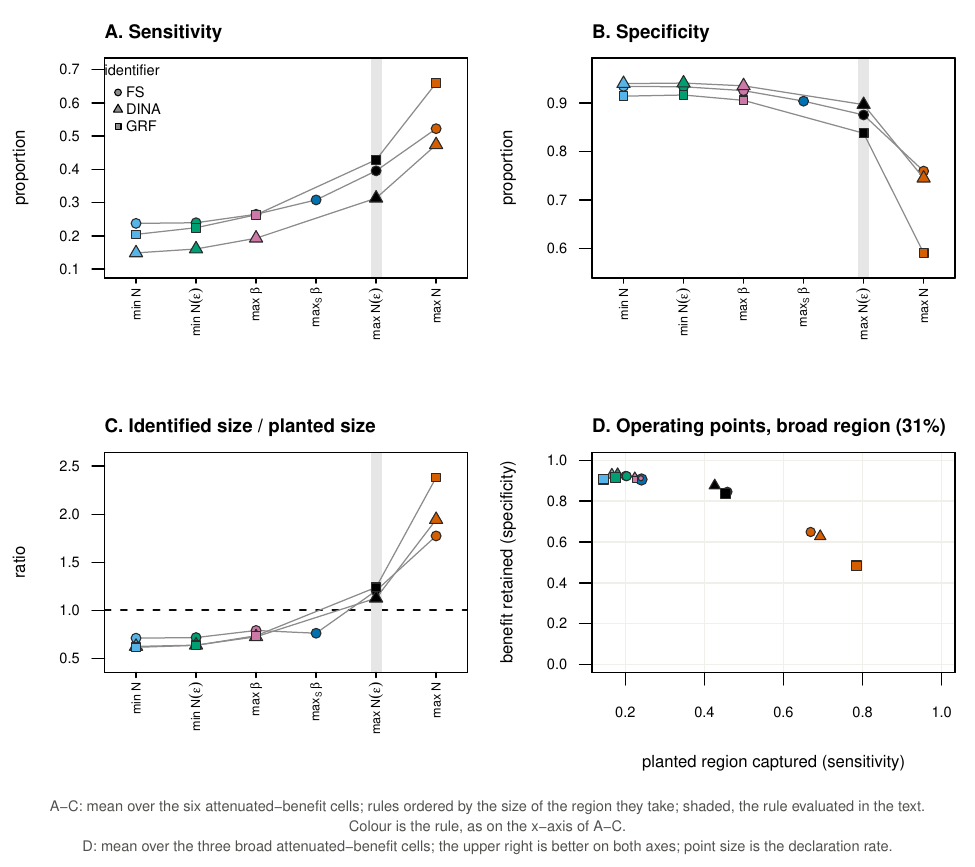}}

}

\caption{\label{fig-supp-rules-attenuated}Identification under the
selection rules on the six attenuated-benefit cells of the survival
design (Section 5.1 of the main paper), where the planted region carries
a hazard ratio of one against a complement that benefits, for forest
search (FS), DINA and GRF, 500 replicates per cell. Panels and encodings
as in Figure 2 of the main paper: A--C sensitivity, specificity and the
identified region as a multiple of the planted one, averaged over the
six cells, the rules in the same order; D the planted region captured
against benefit retained, averaged over the three cells where the
planted region is broad, prevalence 31\%, with point size the rate at
which a region is returned. Symbol marks the identifier, as keyed in A;
colour marks the rule, as on the x-axis of A--C. The rules are not
joined in D.}

\end{figure}

The main paper's Figure 2 shows the selection rules on the twelve harm
cells. Figure~\ref{fig-supp-rules-attenuated} shows them the same way on
the six attenuated-benefit cells, where the planted region carries a
hazard ratio of one against a complement that benefits, so that a region
returned there is not a harm region. A region is returned on 0.72 to
0.97 of replicates under min N, min N(ε), max β, max N(ε) and max N, the
rate the identifier's rather than the rule's, and on every replicate
under \(\max_{\mathcal{S}}\beta\), the unscreened rule: most often
there, least often under the five screened rules. Under the
unconditional convention, where no region scores sensitivity 0 and
specificity 1, sensitivity cannot exceed the declaration rate: it is
0.15--0.24 at min N, 0.16--0.24 at min N(ε), 0.19--0.26 at max β, 0.31
at \(\max_{\mathcal{S}}\beta\), 0.31--0.43 at max N(ε) and 0.47--0.66 at
max N, against specificity 0.91--0.94, 0.92--0.94, 0.91--0.94, 0.90,
0.84--0.90 and 0.59--0.76. The identified region is 0.62--0.71,
0.64--0.72, 0.73--0.79, 0.76, 1.13--1.24 and 1.78--2.38 times the
planted one. In D the points run from the upper left at min N to the
lower right at max N, spreading apart there; point size, the declaration
rate, runs 0.91 to 1.00.

\subsection{Coverage, cell by cell}\label{sec-supp-coverage}

\begin{figure}

\centering{

\pandocbounded{\includegraphics[keepaspectratio,alt={Relative bias of the within-subgroup hazard ratio by cell for forest search, DINA and GRF, on the identified region and on its complement; open and filled symbols for the unadjusted and corrected estimates, joined by a segment.}]{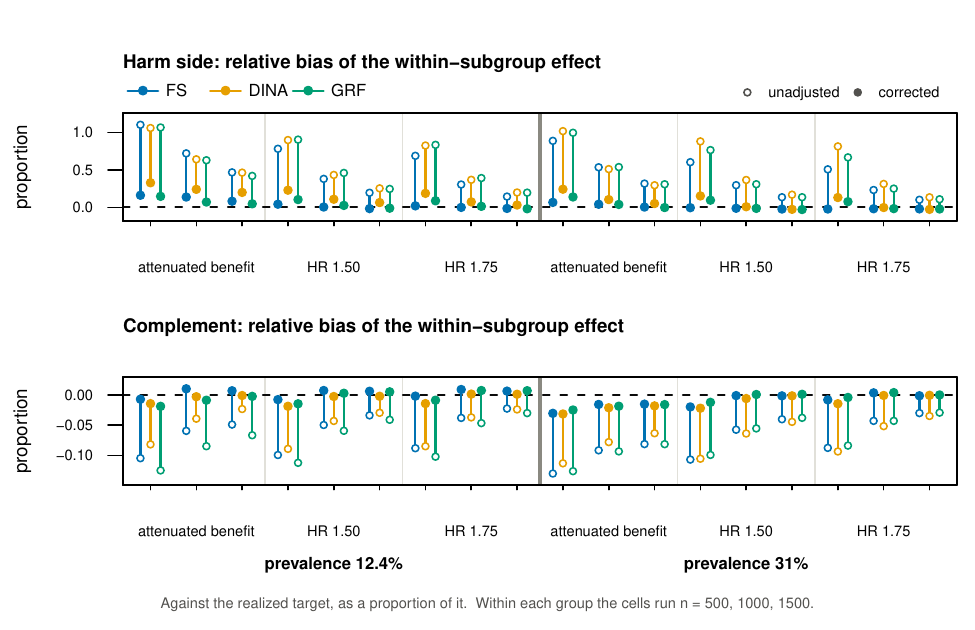}}

}

\caption{\label{fig-supp-oc-bias}Relative bias of the within-subgroup
hazard ratio against the realized targets, as a proportion of the
target, on the identified region \(\hat H\) (harm side) and its
complement \(\hat H^c\). Open symbols are the unadjusted analysis,
filled symbols the multiplier-resampling correction, joined so that the
length of the segment is the bias removed. The dashed line marks zero.
Summaries are over declaring replicates.}

\end{figure}

Figure 3 of the main paper gives coverage by cell, beneath the
declaration rate that is its denominator, for the two recommended
one-sided products and the Bonferroni pair; this section reads it.
Figure~\ref{fig-supp-oc-bias} shows the relative bias of the
within-subgroup effect cell by cell, the unadjusted and corrected values
joined so that the correction reads as a movement and its magnitude as a
length.

The declaration panel is placed first because it sets the denominator of
the three below it. On the harm cells forest search and GRF declare on
nearly every replicate once the trial is of moderate size, at 0.97 and
above. DINA does so at the higher prevalence, but at the lower
prevalence it declares on between 0.80 and 0.91 of replicates, falling
as the trial grows, so its coverage on those cells is also conditional
on a selected subset. At the attenuated-benefit cells at the lower
prevalence, where the planted region is not harmful and narrow, all
three declare less often as the trial grows: forest search from 0.68 to
0.62, GRF from 0.98 to 0.87, and DINA from 0.71 to 0.34. Every
conditional summary at those cells is therefore computed over a
shrinking subset, most severely for DINA, whose denominator falls by
more than a third.

The unadjusted lower bound covers between 0.144 and 0.888 against the
nominal 0.95, lowest where no harm is planted and the sample small; its
limits are the two-sided Wald limits a standard analysis reports, so the
comparison runs in its favour. The corrected bound covers between 0.825
and 0.974. Forest search covers 0.932 to 0.974, at or above nominal at
fourteen of the eighteen cells and easing toward it as the trial grows;
GRF covers 0.929 to 0.969, rising toward nominal; DINA covers 0.825 to
0.961 and lies below the other two at seventeen of the eighteen cells.
DINA's coverage rises with sample size in five of the six cell groups,
by 0.0269 to 0.0455, and falls only in the sixth --- the narrow
attenuated-benefit cells, where its declaration rate falls from 0.71 to
0.34 and a summary conditional on declaring is computed over a subset
that is both shrinking and increasingly selected.

On the complement the unadjusted upper bound covers between 0.827 and
0.965, understating the effect rather than overstating it, and the
corrected bound covers between 0.903 and 0.956. The studentized
Bonferroni pair covers between 0.908 and 0.958 across the harm cells
against a joint nominal of 0.95.

Section~\ref{sec-field-conditions} gives a reason to read both coverage
panels by cell type. Where a subgroup is declared with probability below
one, as at the attenuated-benefit cells above, the corrected estimates
carry a same-draws term that the field roots do not simulate: on the
identified region it is non-negative under association and makes the
lower bound conservative, and on the complement its sign is not settled,
a positive value moving the upper bound below its target. On the
identified region the corrected bound covers 0.944 to 0.974 on the
twelve harm cells and 0.932 to 0.973 on the six attenuated-benefit cells
for forest search, 0.902 to 0.961 and 0.825 to 0.947 for DINA, and 0.929
to 0.969 and 0.940 to 0.959 for GRF; on the complement the corrected
bound covers 0.913 to 0.955 and 0.913 to 0.955 for forest search, 0.910
to 0.949 and 0.903 to 0.953 for DINA, and 0.918 to 0.956 and 0.911 to
0.948 for GRF. On the identified region the attenuated-benefit coverage
is lower than the harm-cell coverage for forest search and DINA and
indistinguishable from it for GRF: the direction of the residual of
Theorem 1 of the main paper, which is positive near ties and moves the
lower bound up, rather than of the same-draws term, which moves it down;
the two act against each other in these cells, and the comparison does
not separate them. On the complement it is lower than the harm-cell
coverage for GRF, the direction a positive same-draws term would
produce, and indistinguishable for forest search and DINA. These cells
are also where the conditional summaries are computed over the smallest
subsets, so the comparison is read with that in mind; the term vanishes
where declaration becomes certain, so the harm cells are unaffected by
it.

\subsection{Bias against the planted target}\label{sec-supp-planted}

The main text reports bias against the realized effect in the identified
region, which is what the bounds are about. Bias against the
\emph{planted} effect is a different quantity, and the difference is
instructive rather than incidental.

\begin{figure}

\centering{

\pandocbounded{\includegraphics[keepaspectratio]{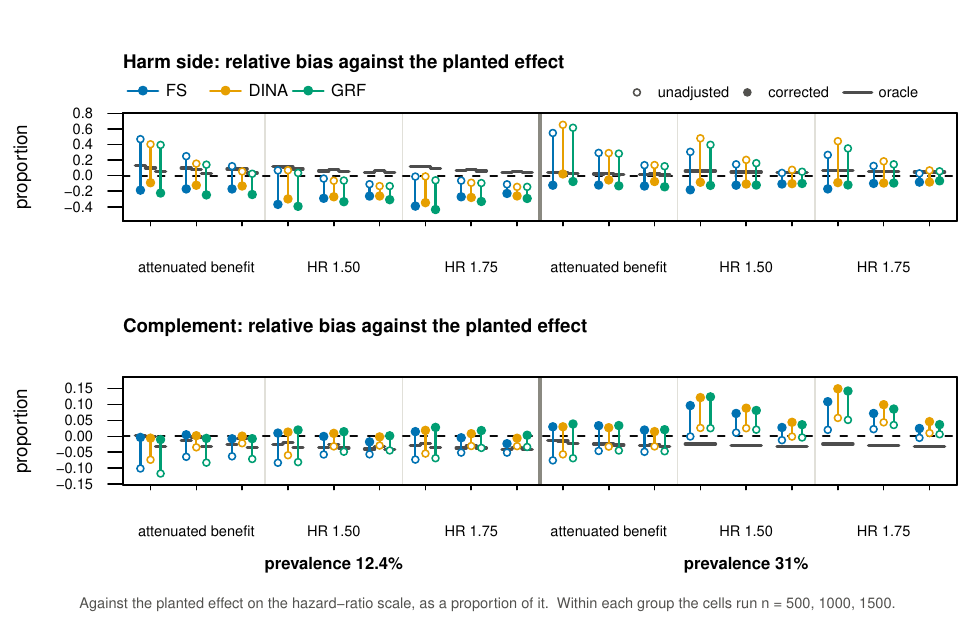}}

}

\caption{\label{fig-supp-bias-planted}Relative bias against the planted
effect, as a proportion of the target, harm side and complement. Grey
bars mark the oracle, which refits on the true region and so estimates
the planted target directly. Open symbols are the unadjusted analysis,
filled symbols the correction.}

\end{figure}%

Against the planted target the unadjusted analysis often sits closer to
zero than the corrected one. It is not the better estimator. Two errors
of opposite sign are cancelling: selection inflates the effect in the
identified region, and the identified region is not the planted one, so
the effect it contains is diluted relative to the planted effect. Where
the two are of similar magnitude the unadjusted estimate appears
unbiased for a quantity it does not estimate.

The oracle bar separates them. It refits on the planted region, so its
distance from zero is what an estimator free of both errors achieves;
the corrected estimator's distance below it is dilution alone, and the
unadjusted estimator's position reflects dilution offset by selection.
Reporting the planted-target panel without the oracle would invite
exactly the wrong conclusion, which is why the main text reports the
realized target and this section reports both.

One feature of the complement panel is worth stating because it looks
like an anomaly. The oracle's bias against the planted complement effect
is not zero at the broad prevalence. The oracle bounds the complement of
the \emph{planted} region; where the identified region is smaller than
the planted one --- as it is at all but one of the broad cells --- the
identified complement is larger than the planted complement, and the two
are different populations. The panel compares estimators of two related
but distinct quantities and should be read as such.

\subsection{Classification, both
conventions}\label{sec-supp-classification}

Every classification rate moves in the expected direction with sample
size and effect strength for every identifier; across the harm cells
sensitivity runs 0.46 to 0.93 and specificity 0.84 to 0.96. The
identifiers separate on sensitivity and not on specificity --- forest
search captures more of the planted region than DINA at every harm cell,
by 0.05 to 0.29, while their specificities agree to within 0.03 --- and
that is not an accuracy ordering, because it follows from region extent:
DINA returns the smallest region at every cell, and a larger region
captures more of the planted one while being less precise about it. The
full tables, under both conventions and with Monte Carlo standard
errors, follow.

The main text reports unconditional rates: a replicate that declares
nothing is scored as making no assignment, because that is what happens
to a trial. The conditional convention --- averaging only over
replicates that declared --- answers a different question, one no
investigator is in a position to ask of their own trial, but it isolates
the quality of the regions that were returned from how often they were
returned. Both are given here.

{

\begin{longtable}[]{@{}
  >{\raggedright\arraybackslash}p{(\linewidth - 20\tabcolsep) * \real{0.1494}}
  >{\raggedright\arraybackslash}p{(\linewidth - 20\tabcolsep) * \real{0.0714}}
  >{\raggedleft\arraybackslash}p{(\linewidth - 20\tabcolsep) * \real{0.0519}}
  >{\raggedleft\arraybackslash}p{(\linewidth - 20\tabcolsep) * \real{0.0909}}
  >{\raggedleft\arraybackslash}p{(\linewidth - 20\tabcolsep) * \real{0.0909}}
  >{\raggedleft\arraybackslash}p{(\linewidth - 20\tabcolsep) * \real{0.0909}}
  >{\raggedleft\arraybackslash}p{(\linewidth - 20\tabcolsep) * \real{0.0909}}
  >{\raggedleft\arraybackslash}p{(\linewidth - 20\tabcolsep) * \real{0.0909}}
  >{\raggedleft\arraybackslash}p{(\linewidth - 20\tabcolsep) * \real{0.0909}}
  >{\raggedleft\arraybackslash}p{(\linewidth - 20\tabcolsep) * \real{0.0909}}
  >{\raggedleft\arraybackslash}p{(\linewidth - 20\tabcolsep) * \real{0.0909}}@{}}

\caption{\label{tbl-supp-classification}Classification of the planted
region, harm cells, under both conventions, with Monte Carlo standard
errors computed across replicates. Declaration rate is given beside each
set of rates because it is the denominator of the conditional convention
and a summary in its own right.}

\tabularnewline

\toprule\noalign{}
\begin{minipage}[b]{\linewidth}\raggedright
Cell
\end{minipage} & \begin{minipage}[b]{\linewidth}\raggedright
Identifier
\end{minipage} & \begin{minipage}[b]{\linewidth}\raggedleft
Declare
\end{minipage} & \begin{minipage}[b]{\linewidth}\raggedleft
Sens
\end{minipage} & \begin{minipage}[b]{\linewidth}\raggedleft
Sens (c)
\end{minipage} & \begin{minipage}[b]{\linewidth}\raggedleft
Spec
\end{minipage} & \begin{minipage}[b]{\linewidth}\raggedleft
Spec (c)
\end{minipage} & \begin{minipage}[b]{\linewidth}\raggedleft
Ppv
\end{minipage} & \begin{minipage}[b]{\linewidth}\raggedleft
Ppv (c)
\end{minipage} & \begin{minipage}[b]{\linewidth}\raggedleft
Npv
\end{minipage} & \begin{minipage}[b]{\linewidth}\raggedleft
Npv (c)
\end{minipage} \\
\midrule\noalign{}
\endhead
\bottomrule\noalign{}
\endlastfoot
12.4\%, HR 1.50, n=500 & FS & 0.911 & 0.569 (0.008) & 0.625 (0.008) &
0.870 (0.002) & 0.858 (0.002) & 0.381 (0.006) & 0.418 (0.006) & 0.947
(0.001) & 0.942 (0.001) \\
12.4\%, HR 1.50, n=500 & DINA & 0.875 & 0.463 (0.008) & 0.530 (0.008) &
0.871 (0.002) & 0.852 (0.002) & 0.306 (0.006) & 0.349 (0.006) & 0.938
(0.001) & 0.929 (0.001) \\
12.4\%, HR 1.50, n=500 & GRF & 0.999 & 0.515 (0.007) & 0.516 (0.007) &
0.837 (0.002) & 0.836 (0.002) & 0.318 (0.005) & 0.319 (0.005) & 0.926
(0.001) & 0.925 (0.001) \\
12.4\%, HR 1.50, n=1000 & FS & 0.974 & 0.755 (0.007) & 0.776 (0.007) &
0.909 (0.002) & 0.906 (0.002) & 0.563 (0.006) & 0.578 (0.006) & 0.967
(0.001) & 0.966 (0.001) \\
12.4\%, HR 1.50, n=1000 & DINA & 0.839 & 0.545 (0.008) & 0.650 (0.007) &
0.912 (0.002) & 0.895 (0.002) & 0.412 (0.007) & 0.492 (0.006) & 0.956
(0.001) & 0.948 (0.001) \\
12.4\%, HR 1.50, n=1000 & GRF & 0.998 & 0.642 (0.007) & 0.644 (0.007) &
0.886 (0.002) & 0.885 (0.002) & 0.479 (0.006) & 0.481 (0.006) & 0.947
(0.001) & 0.947 (0.001) \\
12.4\%, HR 1.50, n=1500 & FS & 0.988 & 0.870 (0.005) & 0.881 (0.005) &
0.919 (0.001) & 0.918 (0.001) & 0.633 (0.005) & 0.640 (0.005) & 0.982
(0.001) & 0.982 (0.001) \\
12.4\%, HR 1.50, n=1500 & DINA & 0.798 & 0.580 (0.008) & 0.727 (0.007) &
0.923 (0.002) & 0.903 (0.002) & 0.436 (0.007) & 0.546 (0.006) & 0.968
(0.001) & 0.959 (0.001) \\
12.4\%, HR 1.50, n=1500 & GRF & 0.998 & 0.743 (0.005) & 0.745 (0.005) &
0.896 (0.002) & 0.896 (0.002) & 0.550 (0.005) & 0.552 (0.005) & 0.962
(0.001) & 0.962 (0.001) \\
12.4\%, HR 1.75, n=500 & FS & 0.950 & 0.653 (0.008) & 0.687 (0.007) &
0.881 (0.002) & 0.875 (0.002) & 0.455 (0.006) & 0.479 (0.006) & 0.954
(0.001) & 0.952 (0.001) \\
12.4\%, HR 1.75, n=500 & DINA & 0.908 & 0.538 (0.008) & 0.593 (0.008) &
0.879 (0.002) & 0.867 (0.002) & 0.370 (0.006) & 0.408 (0.006) & 0.945
(0.001) & 0.939 (0.001) \\
12.4\%, HR 1.75, n=500 & GRF & 1.000 & 0.574 (0.007) & 0.575 (0.007) &
0.850 (0.002) & 0.850 (0.002) & 0.366 (0.005) & 0.366 (0.005) & 0.935
(0.001) & 0.935 (0.001) \\
12.4\%, HR 1.75, n=1000 & FS & 0.995 & 0.835 (0.006) & 0.840 (0.006) &
0.928 (0.001) & 0.928 (0.001) & 0.658 (0.006) & 0.662 (0.006) & 0.976
(0.001) & 0.976 (0.001) \\
12.4\%, HR 1.75, n=1000 & DINA & 0.906 & 0.658 (0.008) & 0.726 (0.006) &
0.923 (0.001) & 0.916 (0.002) & 0.528 (0.007) & 0.582 (0.006) & 0.963
(0.001) & 0.960 (0.001) \\
12.4\%, HR 1.75, n=1000 & GRF & 0.999 & 0.697 (0.006) & 0.697 (0.006) &
0.908 (0.002) & 0.908 (0.002) & 0.563 (0.006) & 0.563 (0.006) & 0.955
(0.001) & 0.955 (0.001) \\
12.4\%, HR 1.75, n=1500 & FS & 0.999 & 0.928 (0.004) & 0.929 (0.004) &
0.943 (0.001) & 0.943 (0.001) & 0.728 (0.005) & 0.729 (0.005) & 0.990
(0.001) & 0.990 (0.001) \\
12.4\%, HR 1.75, n=1500 & DINA & 0.898 & 0.722 (0.007) & 0.804 (0.006) &
0.935 (0.001) & 0.928 (0.001) & 0.578 (0.006) & 0.643 (0.005) & 0.974
(0.001) & 0.971 (0.001) \\
12.4\%, HR 1.75, n=1500 & GRF & 1.000 & 0.777 (0.005) & 0.777 (0.005) &
0.921 (0.002) & 0.921 (0.002) & 0.639 (0.005) & 0.639 (0.005) & 0.967
(0.001) & 0.967 (0.001) \\
31\%, HR 1.50, n=500 & FS & 1.000 & 0.590 (0.006) & 0.590 (0.006) &
0.893 (0.002) & 0.893 (0.002) & 0.706 (0.005) & 0.706 (0.005) & 0.837
(0.002) & 0.837 (0.002) \\
31\%, HR 1.50, n=500 & DINA & 0.999 & 0.496 (0.006) & 0.497 (0.006) &
0.902 (0.002) & 0.902 (0.002) & 0.673 (0.005) & 0.674 (0.005) & 0.808
(0.002) & 0.808 (0.002) \\
31\%, HR 1.50, n=500 & GRF & 1.000 & 0.522 (0.006) & 0.522 (0.006) &
0.899 (0.002) & 0.899 (0.002) & 0.678 (0.005) & 0.678 (0.005) & 0.816
(0.002) & 0.816 (0.002) \\
31\%, HR 1.50, n=1000 & FS & 1.000 & 0.708 (0.006) & 0.708 (0.006) &
0.928 (0.002) & 0.928 (0.002) & 0.804 (0.004) & 0.804 (0.004) & 0.887
(0.002) & 0.887 (0.002) \\
31\%, HR 1.50, n=1000 & DINA & 1.000 & 0.642 (0.006) & 0.642 (0.006) &
0.933 (0.001) & 0.933 (0.001) & 0.799 (0.004) & 0.799 (0.004) & 0.863
(0.002) & 0.863 (0.002) \\
31\%, HR 1.50, n=1000 & GRF & 1.000 & 0.683 (0.006) & 0.683 (0.006) &
0.931 (0.002) & 0.931 (0.002) & 0.805 (0.004) & 0.805 (0.004) & 0.878
(0.002) & 0.878 (0.002) \\
31\%, HR 1.50, n=1500 & FS & 1.000 & 0.839 (0.005) & 0.839 (0.005) &
0.926 (0.002) & 0.926 (0.002) & 0.837 (0.003) & 0.837 (0.003) & 0.937
(0.002) & 0.937 (0.002) \\
31\%, HR 1.50, n=1500 & DINA & 1.000 & 0.788 (0.005) & 0.788 (0.005) &
0.939 (0.001) & 0.939 (0.001) & 0.850 (0.003) & 0.850 (0.003) & 0.918
(0.002) & 0.918 (0.002) \\
31\%, HR 1.50, n=1500 & GRF & 1.000 & 0.818 (0.005) & 0.818 (0.005) &
0.934 (0.002) & 0.934 (0.002) & 0.852 (0.003) & 0.852 (0.003) & 0.928
(0.002) & 0.928 (0.002) \\
31\%, HR 1.75, n=500 & FS & 1.000 & 0.660 (0.006) & 0.660 (0.006) &
0.919 (0.002) & 0.919 (0.002) & 0.778 (0.005) & 0.779 (0.005) & 0.866
(0.002) & 0.866 (0.002) \\
31\%, HR 1.75, n=500 & DINA & 1.000 & 0.538 (0.006) & 0.538 (0.006) &
0.921 (0.002) & 0.921 (0.002) & 0.733 (0.005) & 0.733 (0.005) & 0.824
(0.002) & 0.824 (0.002) \\
31\%, HR 1.75, n=500 & GRF & 1.000 & 0.582 (0.006) & 0.582 (0.006) &
0.922 (0.002) & 0.922 (0.002) & 0.749 (0.005) & 0.749 (0.005) & 0.840
(0.002) & 0.840 (0.002) \\
31\%, HR 1.75, n=1000 & FS & 1.000 & 0.769 (0.006) & 0.769 (0.006) &
0.948 (0.001) & 0.948 (0.001) & 0.860 (0.003) & 0.860 (0.003) & 0.911
(0.002) & 0.911 (0.002) \\
31\%, HR 1.75, n=1000 & DINA & 1.000 & 0.689 (0.006) & 0.689 (0.006) &
0.950 (0.001) & 0.950 (0.001) & 0.848 (0.004) & 0.848 (0.004) & 0.882
(0.002) & 0.882 (0.002) \\
31\%, HR 1.75, n=1000 & GRF & 1.000 & 0.733 (0.006) & 0.733 (0.006) &
0.950 (0.001) & 0.950 (0.001) & 0.860 (0.004) & 0.860 (0.004) & 0.899
(0.002) & 0.899 (0.002) \\
31\%, HR 1.75, n=1500 & FS & 1.000 & 0.874 (0.004) & 0.874 (0.004) &
0.947 (0.001) & 0.947 (0.001) & 0.883 (0.003) & 0.883 (0.003) & 0.951
(0.002) & 0.951 (0.002) \\
31\%, HR 1.75, n=1500 & DINA & 1.000 & 0.822 (0.005) & 0.822 (0.005) &
0.955 (0.001) & 0.955 (0.001) & 0.889 (0.003) & 0.889 (0.003) & 0.931
(0.002) & 0.931 (0.002) \\
31\%, HR 1.75, n=1500 & GRF & 1.000 & 0.844 (0.005) & 0.844 (0.005) &
0.956 (0.001) & 0.956 (0.001) & 0.897 (0.003) & 0.897 (0.003) & 0.940
(0.002) & 0.940 (0.002) \\

\end{longtable}

}

The two conventions agree closely where the declaration rate is near one
and separate where it is not, which is the low-prevalence
attenuated-benefit block and DINA in particular. A conditional rate
computed over a third of the replicates describes a selected subset, and
the declaration rate beside it is what tells a reader so.

\subsection{The two-sided interval}\label{sec-supp-twosided}

The two-term infinitesimal-jackknife interval is reported throughout as
a reference. It is not recommended, for the reason given in the main
text: the residual bias is positive near ties and negative under
separation, so a two-sided interval meets both signs and a one-sided
bound meets one.

\begin{figure}

\centering{

\pandocbounded{\includegraphics[keepaspectratio]{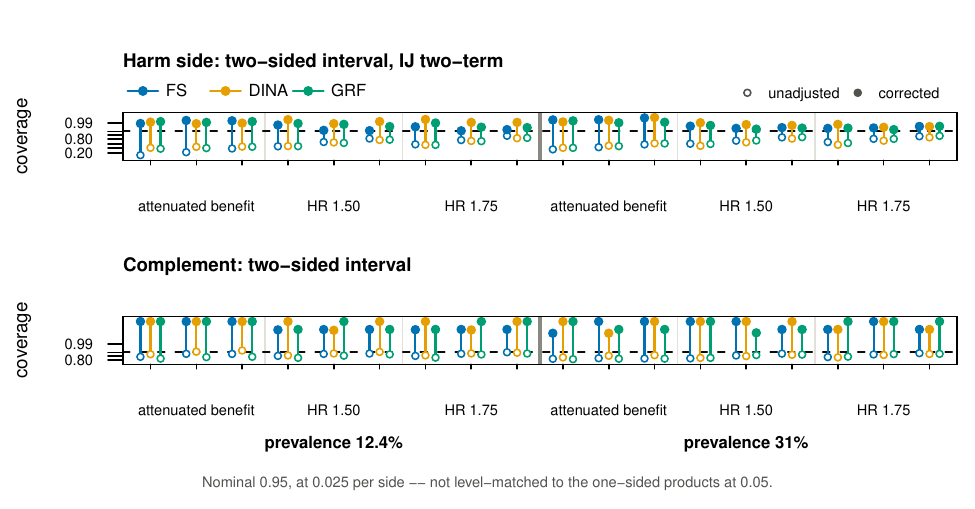}}

}

\caption{\label{fig-supp-twosided}Coverage of the two-sided interval on
\(\beta(\hat H)\), the identified region, and on \(\beta(\hat H^c)\),
its complement, against a nominal 0.95, beside the one-sided products.
Logit scale; Wilson intervals.}

\end{figure}%

Its behaviour is consistent with that argument. Its coverage is above
nominal at every identifier and cell. Read with the winner's
re-selection probability as the measure of separation, its largest
movements with sample size are where the competition separates, where it
falls toward nominal. Its location shares are low, and they are not
level-matched to the one-sided bounds in any case: its limits are at
0.025 per side against 0.05 for the recommended products.

\subsection{Bound location at fixed
thresholds}\label{sec-supp-thresholds}

The main text reports the share of bounds reaching a hazard ratio of
one, the claim threshold, at which a claim of harm or of retained
benefit becomes admissible at all. Shares at fixed thresholds further
from one are higher bars, and are reported here rather than in the main
text because a bound clearing them is not what a primary analysis would
be designed to achieve.

{

\begin{longtable}[]{@{}
  >{\raggedright\arraybackslash}p{(\linewidth - 12\tabcolsep) * \real{0.2267}}
  >{\raggedright\arraybackslash}p{(\linewidth - 12\tabcolsep) * \real{0.0733}}
  >{\raggedleft\arraybackslash}p{(\linewidth - 12\tabcolsep) * \real{0.1400}}
  >{\raggedleft\arraybackslash}p{(\linewidth - 12\tabcolsep) * \real{0.1400}}
  >{\raggedleft\arraybackslash}p{(\linewidth - 12\tabcolsep) * \real{0.1400}}
  >{\raggedleft\arraybackslash}p{(\linewidth - 12\tabcolsep) * \real{0.1400}}
  >{\raggedleft\arraybackslash}p{(\linewidth - 12\tabcolsep) * \real{0.1400}}@{}}

\caption{\label{tbl-supp-thresholds}Share of replicates whose bound
reaches each threshold, with Wilson intervals. Lower bounds on the
identified region at or above 1.00 and 1.25; upper bounds on the
complement at or below 1.00, 0.85 and 0.80.}

\tabularnewline

\toprule\noalign{}
\begin{minipage}[b]{\linewidth}\raggedright
Cell
\end{minipage} & \begin{minipage}[b]{\linewidth}\raggedright
Identifier
\end{minipage} & \begin{minipage}[b]{\linewidth}\raggedleft
lower \textgreater= 1.00
\end{minipage} & \begin{minipage}[b]{\linewidth}\raggedleft
lower \textgreater= 1.25
\end{minipage} & \begin{minipage}[b]{\linewidth}\raggedleft
upper \textless= 1.00
\end{minipage} & \begin{minipage}[b]{\linewidth}\raggedleft
upper \textless= 0.85
\end{minipage} & \begin{minipage}[b]{\linewidth}\raggedleft
upper \textless= 0.80
\end{minipage} \\
\midrule\noalign{}
\endhead
\bottomrule\noalign{}
\endlastfoot
12.4\%, attenuated benefit, n=500 & FS & 0.004 (0.002, 0.010) & 0.000
(0.000, 0.003) & 0.930 (0.915, 0.943) & 0.589 (0.562, 0.614) & 0.407
(0.381, 0.433) \\
12.4\%, attenuated benefit, n=500 & DINA & 0.019 (0.013, 0.027) & 0.004
(0.001, 0.008) & 0.940 (0.926, 0.951) & 0.608 (0.582, 0.633) & 0.426
(0.401, 0.452) \\
12.4\%, attenuated benefit, n=500 & GRF & 0.006 (0.003, 0.011) & 0.003
(0.001, 0.006) & 0.936 (0.925, 0.946) & 0.615 (0.594, 0.637) & 0.449
(0.427, 0.471) \\
12.4\%, attenuated benefit, n=1000 & FS & 0.005 (0.002, 0.010) & 0.001
(0.000, 0.004) & 0.996 (0.991, 0.998) & 0.822 (0.800, 0.842) & 0.632
(0.606, 0.658) \\
12.4\%, attenuated benefit, n=1000 & DINA & 0.016 (0.010, 0.026) & 0.001
(0.000, 0.005) & 0.999 (0.995, 1.000) & 0.829 (0.805, 0.851) & 0.655
(0.626, 0.683) \\
12.4\%, attenuated benefit, n=1000 & GRF & 0.004 (0.002, 0.008) & 0.001
(0.000, 0.004) & 0.998 (0.995, 0.999) & 0.861 (0.844, 0.875) & 0.686
(0.665, 0.707) \\
12.4\%, attenuated benefit, n=1500 & FS & 0.004 (0.002, 0.009) & 0.000
(0.000, 0.003) & 1.000 (0.997, 1.000) & 0.954 (0.941, 0.965) & 0.824
(0.802, 0.844) \\
12.4\%, attenuated benefit, n=1500 & DINA & 0.009 (0.004, 0.019) & 0.001
(0.000, 0.008) & 1.000 (0.994, 1.000) & 0.953 (0.935, 0.967) & 0.814
(0.783, 0.841) \\
12.4\%, attenuated benefit, n=1500 & GRF & 0.005 (0.002, 0.009) & 0.002
(0.001, 0.005) & 0.999 (0.997, 1.000) & 0.960 (0.950, 0.968) & 0.821
(0.802, 0.838) \\
12.4\%, HR 1.50, n=500 & FS & 0.026 (0.019, 0.034) & 0.007 (0.004,
0.011) & 0.906 (0.892, 0.919) & 0.551 (0.528, 0.574) & 0.384 (0.362,
0.407) \\
12.4\%, HR 1.50, n=500 & DINA & 0.055 (0.046, 0.067) & 0.014 (0.010,
0.021) & 0.901 (0.886, 0.914) & 0.561 (0.538, 0.584) & 0.390 (0.367,
0.413) \\
12.4\%, HR 1.50, n=500 & GRF & 0.023 (0.017, 0.031) & 0.005 (0.002,
0.009) & 0.894 (0.880, 0.907) & 0.535 (0.513, 0.557) & 0.373 (0.352,
0.394) \\
12.4\%, HR 1.50, n=1000 & FS & 0.093 (0.081, 0.107) & 0.030 (0.024,
0.039) & 0.994 (0.990, 0.997) & 0.831 (0.814, 0.847) & 0.640 (0.619,
0.661) \\
12.4\%, HR 1.50, n=1000 & DINA & 0.116 (0.101, 0.132) & 0.026 (0.020,
0.035) & 0.996 (0.992, 0.998) & 0.819 (0.800, 0.836) & 0.630 (0.607,
0.653) \\
12.4\%, HR 1.50, n=1000 & GRF & 0.061 (0.051, 0.073) & 0.016 (0.011,
0.023) & 0.994 (0.990, 0.997) & 0.800 (0.782, 0.817) & 0.595 (0.574,
0.617) \\
12.4\%, HR 1.50, n=1500 & FS & 0.147 (0.132, 0.164) & 0.039 (0.031,
0.048) & 0.999 (0.997, 1.000) & 0.950 (0.940, 0.959) & 0.825 (0.808,
0.841) \\
12.4\%, HR 1.50, n=1500 & DINA & 0.180 (0.162, 0.200) & 0.033 (0.025,
0.043) & 1.000 (0.998, 1.000) & 0.949 (0.937, 0.958) & 0.803 (0.783,
0.822) \\
12.4\%, HR 1.50, n=1500 & GRF & 0.108 (0.095, 0.123) & 0.030 (0.023,
0.038) & 0.999 (0.997, 1.000) & 0.935 (0.924, 0.945) & 0.769 (0.750,
0.787) \\
12.4\%, HR 1.75, n=500 & FS & 0.056 (0.046, 0.067) & 0.024 (0.018,
0.032) & 0.896 (0.882, 0.909) & 0.544 (0.521, 0.566) & 0.379 (0.357,
0.401) \\
12.4\%, HR 1.75, n=500 & DINA & 0.095 (0.083, 0.110) & 0.037 (0.029,
0.047) & 0.900 (0.886, 0.913) & 0.551 (0.528, 0.573) & 0.371 (0.349,
0.394) \\
12.4\%, HR 1.75, n=500 & GRF & 0.042 (0.034, 0.052) & 0.010 (0.006,
0.015) & 0.883 (0.868, 0.896) & 0.515 (0.493, 0.537) & 0.352 (0.331,
0.373) \\
12.4\%, HR 1.75, n=1000 & FS & 0.229 (0.211, 0.248) & 0.095 (0.083,
0.109) & 0.995 (0.991, 0.997) & 0.843 (0.826, 0.858) & 0.656 (0.635,
0.676) \\
12.4\%, HR 1.75, n=1000 & DINA & 0.265 (0.245, 0.286) & 0.088 (0.076,
0.102) & 0.996 (0.991, 0.998) & 0.817 (0.799, 0.835) & 0.628 (0.605,
0.650) \\
12.4\%, HR 1.75, n=1000 & GRF & 0.158 (0.143, 0.175) & 0.056 (0.047,
0.067) & 0.993 (0.988, 0.996) & 0.791 (0.772, 0.808) & 0.579 (0.557,
0.601) \\
12.4\%, HR 1.75, n=1500 & FS & 0.402 (0.381, 0.424) & 0.161 (0.145,
0.177) & 1.000 (0.998, 1.000) & 0.962 (0.953, 0.970) & 0.851 (0.835,
0.866) \\
12.4\%, HR 1.75, n=1500 & DINA & 0.400 (0.377, 0.423) & 0.136 (0.121,
0.153) & 1.000 (0.998, 1.000) & 0.955 (0.945, 0.964) & 0.813 (0.794,
0.830) \\
12.4\%, HR 1.75, n=1500 & GRF & 0.278 (0.258, 0.298) & 0.108 (0.095,
0.122) & 1.000 (0.997, 1.000) & 0.934 (0.923, 0.945) & 0.775 (0.756,
0.793) \\
31\%, attenuated benefit, n=500 & FS & 0.006 (0.003, 0.011) & 0.001
(0.000, 0.003) & 0.697 (0.676, 0.717) & 0.262 (0.242, 0.282) & 0.145
(0.130, 0.162) \\
31\%, attenuated benefit, n=500 & DINA & 0.030 (0.023, 0.039) & 0.009
(0.005, 0.014) & 0.710 (0.689, 0.730) & 0.275 (0.255, 0.295) & 0.156
(0.140, 0.173) \\
31\%, attenuated benefit, n=500 & GRF & 0.020 (0.014, 0.027) & 0.007
(0.004, 0.011) & 0.689 (0.668, 0.709) & 0.257 (0.238, 0.276) & 0.141
(0.127, 0.157) \\
31\%, attenuated benefit, n=1000 & FS & 0.010 (0.007, 0.016) & 0.002
(0.001, 0.005) & 0.909 (0.896, 0.921) & 0.397 (0.375, 0.419) & 0.187
(0.170, 0.205) \\
31\%, attenuated benefit, n=1000 & DINA & 0.016 (0.011, 0.022) & 0.003
(0.001, 0.007) & 0.920 (0.906, 0.931) & 0.416 (0.394, 0.439) & 0.200
(0.182, 0.219) \\
31\%, attenuated benefit, n=1000 & GRF & 0.011 (0.007, 0.017) & 0.003
(0.001, 0.006) & 0.919 (0.907, 0.930) & 0.383 (0.362, 0.405) & 0.191
(0.174, 0.209) \\
31\%, attenuated benefit, n=1500 & FS & 0.009 (0.006, 0.015) & 0.001
(0.000, 0.004) & 0.982 (0.975, 0.987) & 0.538 (0.516, 0.560) & 0.268
(0.249, 0.288) \\
31\%, attenuated benefit, n=1500 & DINA & 0.010 (0.006, 0.015) & 0.002
(0.001, 0.005) & 0.988 (0.981, 0.992) & 0.575 (0.552, 0.598) & 0.302
(0.281, 0.324) \\
31\%, attenuated benefit, n=1500 & GRF & 0.009 (0.006, 0.014) & 0.001
(0.000, 0.004) & 0.981 (0.974, 0.986) & 0.543 (0.521, 0.565) & 0.258
(0.239, 0.278) \\
31\%, HR 1.50, n=500 & FS & 0.088 (0.076, 0.101) & 0.028 (0.021, 0.036)
& 0.519 (0.497, 0.541) & 0.164 (0.148, 0.180) & 0.092 (0.080, 0.105) \\
31\%, HR 1.50, n=500 & DINA & 0.159 (0.144, 0.176) & 0.050 (0.041,
0.060) & 0.473 (0.451, 0.495) & 0.142 (0.127, 0.158) & 0.075 (0.064,
0.088) \\
31\%, HR 1.50, n=500 & GRF & 0.137 (0.122, 0.152) & 0.048 (0.039, 0.058)
& 0.464 (0.442, 0.485) & 0.138 (0.124, 0.154) & 0.074 (0.064, 0.087) \\
31\%, HR 1.50, n=1000 & FS & 0.228 (0.210, 0.247) & 0.047 (0.039, 0.057)
& 0.792 (0.774, 0.810) & 0.292 (0.272, 0.312) & 0.144 (0.129, 0.160) \\
31\%, HR 1.50, n=1000 & DINA & 0.257 (0.238, 0.277) & 0.059 (0.049,
0.070) & 0.770 (0.751, 0.788) & 0.245 (0.227, 0.264) & 0.117 (0.103,
0.131) \\
31\%, HR 1.50, n=1000 & GRF & 0.216 (0.199, 0.235) & 0.060 (0.051,
0.072) & 0.781 (0.763, 0.799) & 0.267 (0.248, 0.287) & 0.132 (0.117,
0.147) \\
31\%, HR 1.50, n=1500 & FS & 0.399 (0.378, 0.421) & 0.080 (0.068, 0.092)
& 0.939 (0.927, 0.948) & 0.510 (0.488, 0.532) & 0.291 (0.272, 0.311) \\
31\%, HR 1.50, n=1500 & DINA & 0.419 (0.398, 0.441) & 0.076 (0.066,
0.089) & 0.927 (0.915, 0.938) & 0.457 (0.435, 0.478) & 0.248 (0.230,
0.268) \\
31\%, HR 1.50, n=1500 & GRF & 0.405 (0.384, 0.427) & 0.083 (0.071,
0.095) & 0.940 (0.929, 0.950) & 0.473 (0.451, 0.495) & 0.253 (0.234,
0.273) \\
31\%, HR 1.75, n=500 & FS & 0.203 (0.186, 0.221) & 0.070 (0.059, 0.082)
& 0.492 (0.470, 0.514) & 0.156 (0.140, 0.172) & 0.084 (0.072, 0.096) \\
31\%, HR 1.75, n=500 & DINA & 0.301 (0.282, 0.322) & 0.124 (0.111,
0.140) & 0.419 (0.398, 0.441) & 0.110 (0.097, 0.124) & 0.061 (0.051,
0.072) \\
31\%, HR 1.75, n=500 & GRF & 0.265 (0.246, 0.284) & 0.110 (0.097, 0.124)
& 0.418 (0.397, 0.440) & 0.119 (0.106, 0.134) & 0.065 (0.055, 0.077) \\
31\%, HR 1.75, n=1000 & FS & 0.499 (0.478, 0.521) & 0.183 (0.167, 0.201)
& 0.775 (0.757, 0.793) & 0.299 (0.279, 0.319) & 0.156 (0.141, 0.173) \\
31\%, HR 1.75, n=1000 & DINA & 0.530 (0.509, 0.552) & 0.187 (0.171,
0.205) & 0.727 (0.707, 0.746) & 0.235 (0.217, 0.255) & 0.118 (0.105,
0.133) \\
31\%, HR 1.75, n=1000 & GRF & 0.505 (0.484, 0.527) & 0.183 (0.167,
0.201) & 0.755 (0.736, 0.774) & 0.261 (0.242, 0.280) & 0.141 (0.126,
0.156) \\
31\%, HR 1.75, n=1500 & FS & 0.797 (0.779, 0.815) & 0.333 (0.312, 0.353)
& 0.937 (0.925, 0.947) & 0.529 (0.507, 0.551) & 0.313 (0.293, 0.334) \\
31\%, HR 1.75, n=1500 & DINA & 0.811 (0.793, 0.828) & 0.323 (0.303,
0.344) & 0.907 (0.894, 0.919) & 0.463 (0.441, 0.485) & 0.255 (0.236,
0.274) \\
31\%, HR 1.75, n=1500 & GRF & 0.817 (0.799, 0.833) & 0.351 (0.331,
0.373) & 0.921 (0.908, 0.932) & 0.475 (0.454, 0.497) & 0.276 (0.257,
0.296) \\

\end{longtable}

}

\begin{figure}

\centering{

\pandocbounded{\includegraphics[keepaspectratio]{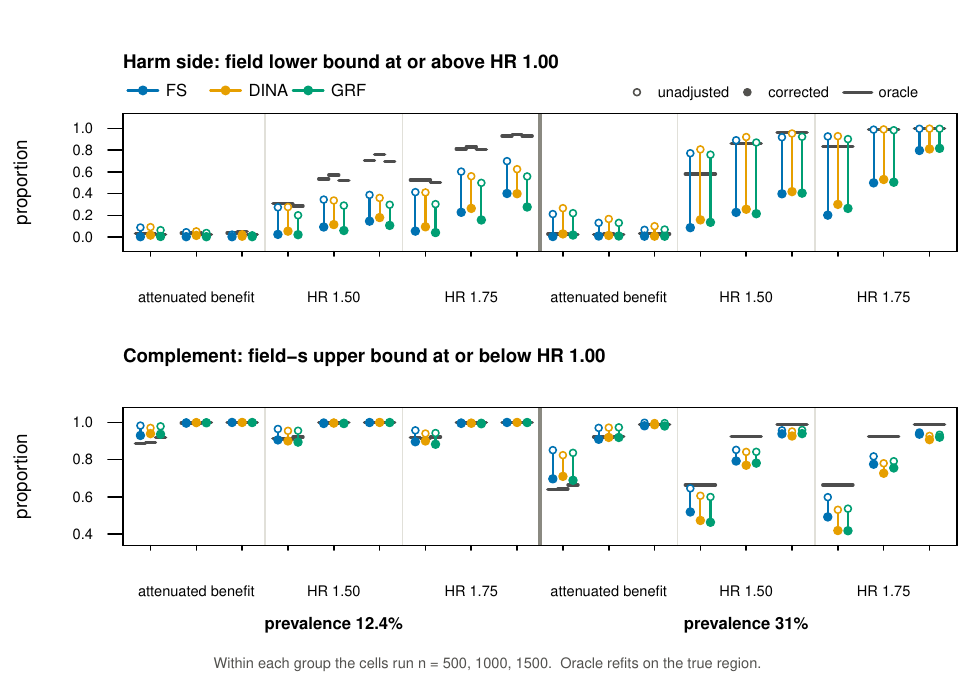}}

}

\caption{\label{fig-oc-location}Proportion of replicates whose bound
reaches a hazard ratio of one: the field lower bound on
\(\beta(\hat H)\), the identified region, at or above it, and the
field-s upper bound on \(\beta(\hat H^c)\), its complement, at or below
it, for forest search (FS), DINA and GRF. Open symbols are the
unadjusted bound and filled symbols the corrected bound, joined so that
the length of the segment is the change the correction makes. Grey bars
mark the oracle, which refits on the true region and so marks an ideal
ceiling rather than a competing method. Within each cell group the
sample size runs 500, 1000, 1500.}

\end{figure}%

The oracle bars of Figure~\ref{fig-oc-location} are the ceiling for the
shares the main text reports. In the harm cells the corrected lower
bound's share of replicates reaching a hazard ratio of one sits against
an oracle ceiling of 0.286 to 1.000; the ceiling is itself far from one
over much of the grid, and at the smallest sample size it runs between
0.286 and 0.837, so much of what limits a claim is the size of the trial
rather than the price of not knowing the region, though that price is
real. In the attenuated-benefit cells the unadjusted bound is most
clearly misplaced, reaching a hazard ratio of one on 0.017 to 0.267 of
replicates against an oracle ceiling of 0.024 to 0.052.

\begin{figure}

\centering{

\pandocbounded{\includegraphics[keepaspectratio,alt={Joint proportion by cell for forest search, DINA and GRF: the proportion of replicates in which the lower bound on the identified region is at or above a hazard ratio of one and the upper bound on the complement at or below one, with Wilson intervals.}]{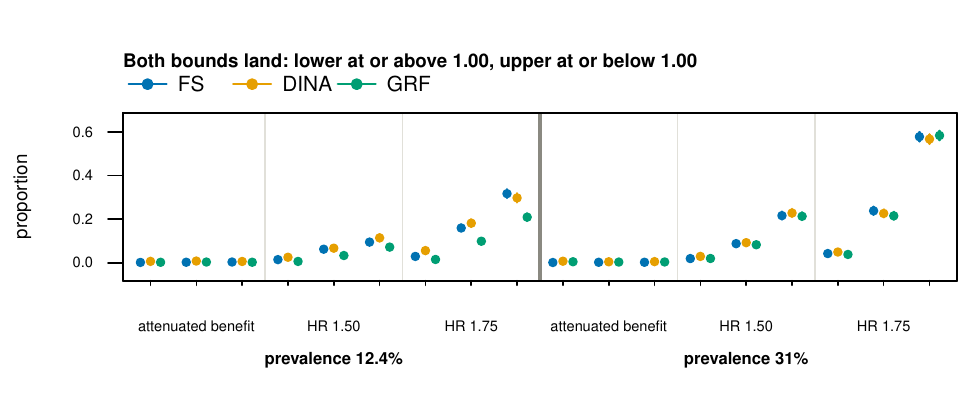}}

}

\caption{\label{fig-supp-joint}Proportion of replicates in which both
bounds of the studentized Bonferroni pair land: the lower bound on
\(\beta(\hat H)\), the identified region, at or above a hazard ratio of
one, and the upper bound on \(\beta(\hat H^c)\), its complement, at or
below one. Wilson intervals throughout.}

\end{figure}

Figure~\ref{fig-supp-joint} gives the share of replicates in which both
bounds of the studentized Bonferroni pair land at once, the event the
pair is constructed to protect.

\subsection{Monte Carlo error}\label{sec-supp-mcerror}

Every cell of the survival design is 2,000 replicates, except the
selection-rule comparison of Section S2.1, which is 500; the
generalized-linear designs of Section~\ref{sec-supp-glm-oc} run 1,000
replicates per binary cell and 2,000 per continuous cell. On 2,000
replicates a proportion near one half carries a Monte Carlo standard
error of 0.011 and one near 0.95 carries 0.005; on 500 the corresponding
values are 0.022 and 0.010. Rates are reported with Wilson intervals and
means with Monte Carlo standard errors computed across replicates rather
than from pooled counts, so that a cell in which few replicates
contribute is visibly less certain rather than silently so.

The identifiers are run on identical draws within each cell, so
contrasts between them are paired and carry less Monte Carlo error than
the individual rates suggest. Contrasts across cells are not paired.

\subsection{The generalized-linear paths}\label{sec-supp-glm-oc}

The rule and bounds of the main paper were also evaluated on the two
generalized-linear paths, in designs built from the ACTG 175 trial whose
binary and continuous endpoints Section~\ref{sec-supp-applications}
analyses. Each design plants a region of known effect in simulated
trials drawn from the trial's covariates, with the complement keeping
the treatment's fitted effect, and runs forest search, GRF and DINA on
identical draws under \(\max_{\mathcal{A}} N(\varepsilon)\) at
\(\varepsilon = 0.20\) with the field lower bound, the field-s upper
bound and their Bonferroni pair, as in Section 5 of the main paper.
DINA's figures are conditional on its proposed family and GRF's on its
returning a region.

\emph{The binary design.} The endpoint is the absence of CD4 improvement
at week 20, as in Section~\ref{sec-supp-applications}, and an odds ratio
above one is harm. The planted region is the patients with weight above
a quantile and baseline CD4 above a quantile, with prevalence 14.9\%.
Its odds ratio is planted at 0.75, 1.00 and 1.50 against 0.656 in the
complement, so the whole-trial odds ratio is 0.672 to 0.750 and every
simulated trial benefits overall. Each design point is run at \(n\) =
500 and 2,000, with 1,000 replicates per cell and the screening
thresholds the campaign template records: an effect threshold of 0.90
and a consistency threshold of 0.80 on the odds-ratio scale, with
consistency proportion \(p^\star = 0.90\).

\needspace{0.6\textheight}

{

\begin{longtable}[]{@{}
  >{\raggedright\arraybackslash}p{(\linewidth - 16\tabcolsep) * \real{0.1200}}
  >{\raggedleft\arraybackslash}p{(\linewidth - 16\tabcolsep) * \real{0.1200}}
  >{\raggedleft\arraybackslash}p{(\linewidth - 16\tabcolsep) * \real{0.0800}}
  >{\raggedleft\arraybackslash}p{(\linewidth - 16\tabcolsep) * \real{0.1000}}
  >{\raggedleft\arraybackslash}p{(\linewidth - 16\tabcolsep) * \real{0.1400}}
  >{\raggedleft\arraybackslash}p{(\linewidth - 16\tabcolsep) * \real{0.1000}}
  >{\raggedleft\arraybackslash}p{(\linewidth - 16\tabcolsep) * \real{0.1300}}
  >{\raggedleft\arraybackslash}p{(\linewidth - 16\tabcolsep) * \real{0.1100}}
  >{\raggedleft\arraybackslash}p{(\linewidth - 16\tabcolsep) * \real{0.1000}}@{}}

\caption{\label{tbl-supp-glm-binary}The binary design, by identifier,
planted odds ratio in the region and sample size. Declared is the share
of replicates returning a region; the other columns are over declaring
replicates. Bias is that of the field estimate against \(\beta(\hat H)\)
on the log-odds scale. Lower is the one-sided coverage of the field
lower bound on \(\beta(\hat H)\); Unadjusted the two-sided coverage of
the unadjusted interval on \(\beta(\hat H)\), a reference; Upper the
one-sided coverage of the field-s upper bound on \(\beta(\hat H^c)\);
Joint the joint coverage of their Bonferroni pair. GRF's and DINA's rows
are conditional on the family each proposed.}

\tabularnewline

\toprule\noalign{}
\begin{minipage}[b]{\linewidth}\raggedright
Identifier
\end{minipage} & \begin{minipage}[b]{\linewidth}\raggedleft
Planted OR
\end{minipage} & \begin{minipage}[b]{\linewidth}\raggedleft
\(n\)
\end{minipage} & \begin{minipage}[b]{\linewidth}\raggedleft
Declared
\end{minipage} & \begin{minipage}[b]{\linewidth}\raggedleft
Bias, log OR
\end{minipage} & \begin{minipage}[b]{\linewidth}\raggedleft
Lower
\end{minipage} & \begin{minipage}[b]{\linewidth}\raggedleft
Unadjusted
\end{minipage} & \begin{minipage}[b]{\linewidth}\raggedleft
Upper
\end{minipage} & \begin{minipage}[b]{\linewidth}\raggedleft
Joint
\end{minipage} \\
\midrule\noalign{}
\endhead
\bottomrule\noalign{}
\endlastfoot
FS & 0.75 & 500 & 0.781 & 0.130 & 0.974 & 0.010 & 0.950 & 0.960 \\
FS & 0.75 & 2,000 & 0.910 & 0.066 & 0.978 & 0.022 & 0.942 & 0.963 \\
FS & 1.00 & 500 & 0.852 & 0.092 & 0.978 & 0.110 & 0.948 & 0.964 \\
FS & 1.00 & 2,000 & 0.961 & 0.042 & 0.974 & 0.117 & 0.938 & 0.955 \\
FS & 1.50 & 500 & 0.931 & 0.030 & 0.977 & 0.284 & 0.928 & 0.957 \\
FS & 1.50 & 2,000 & 0.998 & \(-0.048\) & 0.964 & 0.439 & 0.945 &
0.959 \\
GRF & 0.75 & 500 & 0.999 & 0.136 & 0.931 & 0.272 & 0.934 & 0.936 \\
GRF & 0.75 & 2,000 & 0.999 & 0.105 & 0.945 & 0.179 & 0.934 & 0.942 \\
GRF & 1.00 & 500 & 0.999 & 0.130 & 0.936 & 0.302 & 0.939 & 0.934 \\
GRF & 1.00 & 2,000 & 1.000 & 0.092 & 0.945 & 0.212 & 0.931 & 0.939 \\
GRF & 1.50 & 500 & 1.000 & 0.109 & 0.941 & 0.343 & 0.941 & 0.933 \\
GRF & 1.50 & 2,000 & 1.000 & \(-0.017\) & 0.948 & 0.448 & 0.944 &
0.949 \\
DINA & 0.75 & 500 & 0.944 & 0.358 & 0.870 & 0.244 & 0.916 & 0.881 \\
DINA & 0.75 & 2,000 & 0.870 & 0.333 & 0.869 & 0.298 & 0.940 & 0.906 \\
DINA & 1.00 & 500 & 0.968 & 0.328 & 0.881 & 0.272 & 0.908 & 0.893 \\
DINA & 1.00 & 2,000 & 0.934 & 0.298 & 0.878 & 0.273 & 0.928 & 0.898 \\
DINA & 1.50 & 500 & 0.979 & 0.264 & 0.903 & 0.325 & 0.919 & 0.904 \\
DINA & 1.50 & 2,000 & 0.987 & 0.172 & 0.903 & 0.402 & 0.939 & 0.923 \\

\end{longtable}

}

\emph{The continuous design.} The outcome is the change in CD4 count
from baseline to week 20, and the design takes didanosine alone as the
treated arm, so the treatment's fitted effect, which the complement
keeps, is a deficit of 26.26 cells per cubic millimetre against the
combination; mean differences are oriented so that a positive value is
harm. The planted region, defined by age and the length of prior
antiretroviral therapy, has prevalence 34.5\% and a planted mean
difference of 40 or 120; a third design point plants no region, so that
the difference is 26.26 throughout. The first design point is run at
\(n\) = 500 and 700, the other two at \(n\) = 500, with 2,000 replicates
per cell, an effect threshold of 30 and a consistency threshold of 10
cells, and consistency proportion \(p^\star = 0.90\). Where no region is
planted every declared \(\hat H\) has \(\beta(\hat H)\) equal to that
common difference, so those rows are coverage of a homogeneous effect,
not a false-declaration rate.

\needspace{0.4\textheight}

{

\begin{longtable}[]{@{}
  >{\raggedright\arraybackslash}p{(\linewidth - 16\tabcolsep) * \real{0.1200}}
  >{\raggedleft\arraybackslash}p{(\linewidth - 16\tabcolsep) * \real{0.1200}}
  >{\raggedleft\arraybackslash}p{(\linewidth - 16\tabcolsep) * \real{0.0800}}
  >{\raggedleft\arraybackslash}p{(\linewidth - 16\tabcolsep) * \real{0.1000}}
  >{\raggedleft\arraybackslash}p{(\linewidth - 16\tabcolsep) * \real{0.1400}}
  >{\raggedleft\arraybackslash}p{(\linewidth - 16\tabcolsep) * \real{0.1000}}
  >{\raggedleft\arraybackslash}p{(\linewidth - 16\tabcolsep) * \real{0.1300}}
  >{\raggedleft\arraybackslash}p{(\linewidth - 16\tabcolsep) * \real{0.1100}}
  >{\raggedleft\arraybackslash}p{(\linewidth - 16\tabcolsep) * \real{0.1000}}@{}}

\caption{\label{tbl-supp-glm-continuous}The continuous design, by
identifier, planted mean difference in the region and sample size; null
marks the design point with no planted region. Columns as in
Table~\ref{tbl-supp-glm-binary}, with bias in CD4 cells per cubic
millimetre, positive meaning harm.}

\tabularnewline

\toprule\noalign{}
\begin{minipage}[b]{\linewidth}\raggedright
Identifier
\end{minipage} & \begin{minipage}[b]{\linewidth}\raggedleft
Planted MD
\end{minipage} & \begin{minipage}[b]{\linewidth}\raggedleft
\(n\)
\end{minipage} & \begin{minipage}[b]{\linewidth}\raggedleft
Declared
\end{minipage} & \begin{minipage}[b]{\linewidth}\raggedleft
Bias, cells
\end{minipage} & \begin{minipage}[b]{\linewidth}\raggedleft
Lower
\end{minipage} & \begin{minipage}[b]{\linewidth}\raggedleft
Unadjusted
\end{minipage} & \begin{minipage}[b]{\linewidth}\raggedleft
Upper
\end{minipage} & \begin{minipage}[b]{\linewidth}\raggedleft
Joint
\end{minipage} \\
\midrule\noalign{}
\endhead
\bottomrule\noalign{}
\endlastfoot
FS & 40 & 500 & 0.999 & 7.067 & 0.948 & 0.293 & 0.914 & 0.928 \\
FS & 40 & 700 & 1.000 & 7.725 & 0.939 & 0.208 & 0.928 & 0.927 \\
FS & 120 & 500 & 1.000 & \(-6.403\) & 0.967 & 0.774 & 0.941 & 0.949 \\
FS & null & 500 & 0.997 & 7.973 & 0.947 & 0.262 & 0.917 & 0.932 \\
GRF & 40 & 500 & 1.000 & 7.400 & 0.939 & 0.320 & 0.920 & 0.919 \\
GRF & 40 & 700 & 1.000 & 7.596 & 0.943 & 0.235 & 0.931 & 0.934 \\
GRF & 120 & 500 & 1.000 & \(-4.088\) & 0.956 & 0.713 & 0.931 & 0.943 \\
GRF & null & 500 & 1.000 & 7.966 & 0.941 & 0.300 & 0.920 & 0.930 \\
DINA & 40 & 500 & 0.998 & 11.615 & 0.937 & 0.244 & 0.897 & 0.914 \\
DINA & 40 & 700 & 0.999 & 11.646 & 0.930 & 0.168 & 0.911 & 0.911 \\
DINA & 120 & 500 & 1.000 & \(-5.940\) & 0.965 & 0.705 & 0.937 & 0.950 \\
DINA & null & 500 & 0.994 & 13.800 & 0.931 & 0.233 & 0.893 & 0.910 \\

\end{longtable}

}

\emph{Reading.} On the binary design the field lower bound covers
\(\beta(\hat H)\) at 0.964 to 0.978 for forest search, 0.931 to 0.948
for GRF and 0.869 to 0.903 for DINA, and the unadjusted interval at
0.010 to 0.448; the field-s upper bound covers \(\beta(\hat H^c)\) at
0.908 to 0.950, and the pair covers jointly at 0.881 to 0.964. On the
continuous design the lower bound covers at 0.939 to 0.967, 0.939 to
0.956 and 0.930 to 0.965 in the same order, the unadjusted interval at
0.168 to 0.774, the upper bound at 0.893 to 0.941 and the pair at 0.910
to 0.950. On the binary design forest search's lower bound covers at or
above nominal, GRF's near it and DINA's below, the same ordering as on
the survival design of Section 5 of the main paper.

Both sets of campaigns were run on package builds that predate an
alignment of the multiplier re-selection's admission rule with the
screen's rounding; on the survival design of Section 5 of the main
paper, the same alignment changed no declaration or selected region and
moved three reported range endpoints by one replicate each.

\subsection{What is not recoverable}\label{sec-supp-notrecoverable}

Per-subject membership of the identified region is not recorded. No
classification summary finer than the four rates is therefore available
--- in particular, no agreement between two identifiers' regions on the
same replicate, which would answer whether they are finding the same
patients or different ones with similar rates. That question would need
a further campaign.

\section{Applications: further detail}\label{sec-supp-applications}

This section supports Section 6 of the main paper: the searches'
settings and funnels, the regions in detail, the classification
agreement on the breast cancer data, the ACTG 175 analyses in full, and
the quantities that support each bound. Numbers without an S prefix
refer to the main paper. Every table is drawn from the applications
extract at render.

\subsection{The searches in full}\label{sec-supp-app-searches}

The German Breast Cancer Study Group search enters seven covariates:
age, menopausal status, tumour size, grade, the number of positive lymph
nodes, and the progesterone and estrogen receptor concentrations in
fmol/l. The two receptors carry ten candidate cuts each, with the value
zero, no measured receptor, forced in as a cut of its own; the remaining
covariates take the package's default cuts. Of the 299 recurrences or
deaths, 94 are among the treated and 205 among the untreated, with 387
patients censored.

ACTG 175 is analysed on its zidovudine-plus-didanosine and didanosine
arms, 522 and 561 patients, 1,083 in all; every one of them has a
week-20 CD4 count, so the exclusion that would remove patients without
one removes nobody, and only two of the trial's four arms are analysed,
as in \citet{leonanderson2026}. The binary endpoint, the absence of CD4
improvement at week 20, is met by 430 patients; the whole-trial odds
ratio of 0.665 means the combination leaves fewer patients without
improvement. The continuous endpoint, the CD4 decline from baseline to
week 20, has a whole-trial mean difference of \(-27.6\), about
twenty-eight cells of decline spared, with a standard error of about
eight cells. Orienting every endpoint so that the adverse direction is
upward lets one search, one screen and one pair of bounds serve all
three outcome families without any change of sign at the point of
reading. Twelve covariates enter both searches: age, prior
antiretroviral therapy, weight, Karnofsky score, baseline CD4 and CD8
counts, and six binary indicators (haemophilia, homosexual activity,
injecting drug use, race, sex and symptomatic status). The binary search
carries ten candidate cuts on baseline CD4 and weight; the continuous
search carries ten on each of the six continuous covariates. On the CD4
scale the screening threshold and the per-split threshold are ten and
five cells. The specification for the continuous endpoint follows an
existing applied analysis of the same outcome, with the selection rule
as its one departure; that analysis uses \(\max_{\mathcal{A}} \beta\),
and the two rules select different regions on these data.
\citet{leonanderson2026} set \(\varepsilon\) at 0.10 on both trials.

The field's own Monte Carlo runs at 1,000 outer and 500 inner draws.
Beside the unadjusted analysis, the two-term estimate's two-sided
infinitesimal-jackknife interval is carried in
Table~\ref{tbl-supp-app-five} as a reference only, since Section 3.3
gives the structural reason the two-sided interval is the harder object.

\subsection{The candidate funnels}\label{sec-supp-app-funnels}

{

\begin{longtable}[]{@{}
  >{\raggedright\arraybackslash}p{(\linewidth - 10\tabcolsep) * \real{0.3000}}
  >{\raggedleft\arraybackslash}p{(\linewidth - 10\tabcolsep) * \real{0.1200}}
  >{\raggedleft\arraybackslash}p{(\linewidth - 10\tabcolsep) * \real{0.1300}}
  >{\raggedleft\arraybackslash}p{(\linewidth - 10\tabcolsep) * \real{0.1400}}
  >{\raggedleft\arraybackslash}p{(\linewidth - 10\tabcolsep) * \real{0.1500}}
  >{\raggedleft\arraybackslash}p{(\linewidth - 10\tabcolsep) * \real{0.1600}}@{}}

\caption{\label{tbl-supp-app-funnels}The candidate funnels and the
families over which each correction re-selects. The stages are not the
same across the identifiers, because the funnels are not the same
object: forest search enumerates and screens; GRF proposes a pool from
the covariate quantiles, independently of the outcome, and admits on the
hazard-ratio floor at re-ranking; DINA qualifies candidates on its
fitted surface and admits on the same floor. A dash marks a stage the
identifier does not have.}

\tabularnewline

\toprule\noalign{}
\begin{minipage}[b]{\linewidth}\raggedright
Stage
\end{minipage} & \begin{minipage}[b]{\linewidth}\raggedleft
GBSG, FS
\end{minipage} & \begin{minipage}[b]{\linewidth}\raggedleft
GBSG, GRF
\end{minipage} & \begin{minipage}[b]{\linewidth}\raggedleft
GBSG, DINA
\end{minipage} & \begin{minipage}[b]{\linewidth}\raggedleft
ACTG 175, binary
\end{minipage} & \begin{minipage}[b]{\linewidth}\raggedleft
ACTG 175, continuous
\end{minipage} \\
\midrule\noalign{}
\endhead
\bottomrule\noalign{}
\endlastfoot
Enumerated / proposed & 2,211 & 858 & 4,920 & 3,240 & 5,886 \\
Fittable & 1,410 & --- & --- & 2,584 & 4,702 \\
Qualifying on the surface & --- & --- & 30 & --- & --- \\
Clearing the screening threshold & 121 & 55 & 8 & 160 & 146 \\
Evaluated for consistency & --- & --- & --- & 137 & 131 \\
Passing the consistency screen & 9 & --- & --- & 5 & 8 \\
On the frontier & 3 & --- & --- & 4 & 6 \\
Re-selection family & 1,744 & 858 & 30 & 2,794 & 4,935 \\

\end{longtable}

}

On the continuous endpoint the 5,886 enumerated candidates pass through
three checks before the per-arm minimum: 5,631 pass the variance and
prevalence checks and 5,416 the redundancy check. GRF's restricted-mean
horizon is 62.3 months, and its pool of 858 is the same whatever the
endpoint.

\subsection{The breast cancer regions in
detail}\label{sec-supp-app-gbsg}

\begin{figure}

\centering{

\pandocbounded{\includegraphics[keepaspectratio,alt={Four panels. A: horizontal bars of the three identified regions by size, 75, 82 and 113 patients. B: for each region the unadjusted hazard ratio, the corrected estimate and the field lower bound on a log axis. C: the same for the complements with the field-s upper bound. D: the three lower bounds plotted against the three upper bounds. Reference lines at the whole-trial hazard ratio and at 0.75.}]{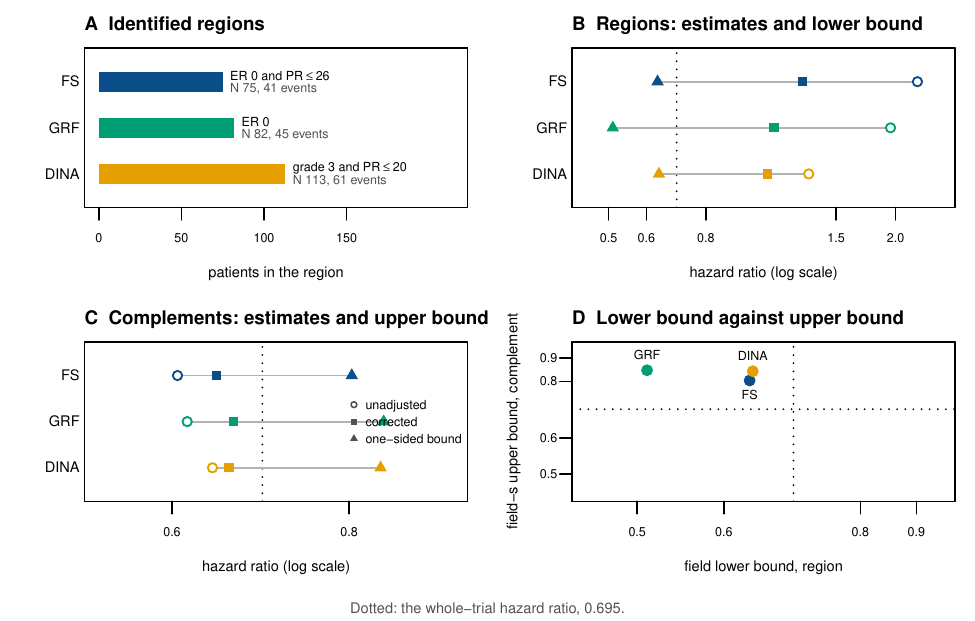}}

}

\caption{\label{fig-supp-app-gbsg}Selection-adjusted subgroup and
complement inference in the German Breast Cancer Study Group trial. A:
the regions forest search (FS), GRF and DINA return under the common
\(\max_{\mathcal{A}} N(\varepsilon)\) rule. B: in each region, the
unadjusted hazard ratio (open circle), the two-term corrected estimate
(filled square) and the nominal 95\% field lower bound on
\(\beta(\hat H)\) (filled triangle), the bound and the unadjusted
estimate joined by a segment that implies no interval. C: the same for
the complements, with the field-s upper bound on \(\beta(\hat H^c)\). D:
each region's field lower bound against its complement's field-s upper
bound. Dotted lines mark the whole-trial 0.695. All three corrected
region estimates lie above 0.90 with lower bounds below 1.0, suggestive
of limited benefit with a benefit not ruled out, and all three corrected
complement estimates lie below the whole-trial 0.695, a larger benefit
than the trial's; the upper bounds, all below one, are reported beside
them. The bounds target the working-model effects in the realized region
and its complement, subject to the coverage characterizations of Section
3.3 of the main paper.}

\end{figure}

Figure~\ref{fig-supp-app-gbsg} draws the three regions with their
estimates and bounds.

The frontier of Figure 1 offers three regions. The selected region ---
25 patients receiving hormonal therapy and 50 not, with 41 events ---
has an unadjusted hazard ratio of 2.222, within the \(\varepsilon\)
band, hazard ratios of 2.03 and above, on which the rule takes the
largest region. \(\max_{\mathcal{A}} N\) returns the 82 patients with no
measured estrogen receptor, a hazard ratio of 1.951, which contains the
selected region entirely and adds 7 patients whose progesterone receptor
exceeds 26 fmol/l; \(\max_{\mathcal{A}} \beta\), and with it
\(\min_{\mathcal{A}} N\), \(\min_{\mathcal{A}} N(\varepsilon)\) and the
unscreened \(\max_{\mathcal{S}} \beta\), returns the 61 with no measured
estrogen receptor and tumour size at most 35 mm, a hazard ratio of 2.54,
which shares 55 patients with the selected region, a Jaccard index of
0.68. The unadjusted elevation is concentrated where the regions
overlap: the 55 shared patients carry a hazard ratio of 2.89 (1.40,
5.94), and the cells outside the overlap are too small to fit.

GRF's region, the 82 with no measured estrogen receptor, has 26 treated
and 56 not, with 45 events, and its complement holds 88\% of the trial;
it is the GRF region \citet{leonanderson2026} report, reached here under
a different \(\varepsilon\) and a different floor, so a coincidence
rather than an anchor. DINA's native winner on its surface is grade 3
with progesterone receptor at most 102.5 fmol/l; its re-ranked region,
the 113 who are grade 3 with progesterone receptor at most 20 fmol/l,
has 31 treated and 82 not, with 61 events, at an unadjusted hazard ratio
of 1.315; its complement holds 84\% of the trial.

Forest search and GRF agree on 99.0\% of patients with a kappa of 0.950
and a Jaccard index of 0.915. DINA agrees with forest search on 82.8\%
of patients and with GRF on 81.8\%, but those percentages are carried by
the 526 patients every identifier leaves in the complement; the kappas,
0.277 and 0.256, and the Jaccard indices, 0.229 and 0.219, are the
informative figures. In all, 160 patients are selected by at least one
identifier and 526 by none. Three cells of the three-way partition are
empty by construction, because forest search's rule implies GRF's and,
among patients with no measured estrogen receptor, DINA's implies forest
search's.

{

\begin{longtable}[]{@{}
  >{\raggedright\arraybackslash}p{(\linewidth - 14\tabcolsep) * \real{0.1047}}
  >{\raggedleft\arraybackslash}p{(\linewidth - 14\tabcolsep) * \real{0.1628}}
  >{\raggedleft\arraybackslash}p{(\linewidth - 14\tabcolsep) * \real{0.1279}}
  >{\raggedleft\arraybackslash}p{(\linewidth - 14\tabcolsep) * \real{0.1395}}
  >{\raggedleft\arraybackslash}p{(\linewidth - 14\tabcolsep) * \real{0.1860}}
  >{\raggedleft\arraybackslash}p{(\linewidth - 14\tabcolsep) * \real{0.1163}}
  >{\raggedleft\arraybackslash}p{(\linewidth - 14\tabcolsep) * \real{0.0698}}
  >{\raggedleft\arraybackslash}p{(\linewidth - 14\tabcolsep) * \real{0.0930}}@{}}

\caption{\label{tbl-supp-app-pairwise}Classification agreement on the
breast cancer data, pairwise over the 686 patients.}

\tabularnewline

\toprule\noalign{}
\begin{minipage}[b]{\linewidth}\raggedright
Pair
\end{minipage} & \begin{minipage}[b]{\linewidth}\raggedleft
Both selected
\end{minipage} & \begin{minipage}[b]{\linewidth}\raggedleft
First only
\end{minipage} & \begin{minipage}[b]{\linewidth}\raggedleft
Second only
\end{minipage} & \begin{minipage}[b]{\linewidth}\raggedleft
Both complement
\end{minipage} & \begin{minipage}[b]{\linewidth}\raggedleft
Agreement
\end{minipage} & \begin{minipage}[b]{\linewidth}\raggedleft
Kappa
\end{minipage} & \begin{minipage}[b]{\linewidth}\raggedleft
Jaccard
\end{minipage} \\
\midrule\noalign{}
\endhead
\bottomrule\noalign{}
\endlastfoot
FS--GRF & 75 & 0 & 7 & 604 & 99.0\% & 0.950 & 0.915 \\
FS--DINA & 35 & 40 & 78 & 533 & 82.8\% & 0.277 & 0.229 \\
GRF--DINA & 35 & 47 & 78 & 526 & 81.8\% & 0.256 & 0.219 \\

\end{longtable}

}

{

\begin{longtable}[]{@{}
  >{\raggedright\arraybackslash}p{(\linewidth - 6\tabcolsep) * \real{0.2200}}
  >{\raggedleft\arraybackslash}p{(\linewidth - 6\tabcolsep) * \real{0.0800}}
  >{\raggedleft\arraybackslash}p{(\linewidth - 6\tabcolsep) * \real{0.1000}}
  >{\raggedright\arraybackslash}p{(\linewidth - 6\tabcolsep) * \real{0.6000}}@{}}

\caption{\label{tbl-supp-app-partition}The three-way partition of the
686 patients.}

\tabularnewline

\toprule\noalign{}
\begin{minipage}[b]{\linewidth}\raggedright
Region
\end{minipage} & \begin{minipage}[b]{\linewidth}\raggedleft
N
\end{minipage} & \begin{minipage}[b]{\linewidth}\raggedleft
Events
\end{minipage} & \begin{minipage}[b]{\linewidth}\raggedright
Who these patients are
\end{minipage} \\
\midrule\noalign{}
\endhead
\bottomrule\noalign{}
\endlastfoot
all three & 35 & 19 & grade 3 and estrogen receptor ≤ 0 fmol/l and
progesterone receptor ≤ 20 fmol/l \\
FS and GRF only & 40 & 22 & estrogen receptor ≤ 0 fmol/l and
progesterone receptor ≤ 26 fmol/l, but not both grade 3 and progesterone
receptor ≤ 20 fmol/l \\
FS and DINA only & 0 & 0 & empty by construction \\
GRF and DINA only & 0 & 0 & empty by construction \\
FS only & 0 & 0 & empty by construction \\
GRF only & 7 & 4 & estrogen receptor ≤ 0 fmol/l with progesterone
receptor above 26 fmol/l \\
DINA only & 78 & 42 & grade 3 and progesterone receptor ≤ 20 fmol/l,
with estrogen receptor above 0 fmol/l \\
flagged by none & 526 & 212 & none of the three selections \\

\end{longtable}

}

\subsection{The ACTG 175 analyses in full}\label{sec-supp-app-actg}

On the binary endpoint the selected region, the 72 patients weighing
more than 86 kg with baseline CD4 above 380 cells per cubic millimetre,
has 67.6\% of the treated showing no CD4 improvement against 36.8\% of
the controls, with 37 events. On this frontier
\(\max_{\mathcal{A}} N(\varepsilon)\) and \(\max_{\mathcal{A}} \beta\)
coincide, so \(\varepsilon\) does not decide the selection; the
next-largest passing candidate, baseline CD4 above 380 with weight above
81 kg, holds 117 patients at an odds ratio of 1.927 and lies outside the
band; four of the five passing candidates form the frontier and the
fifth is dominated. Where the selection is the same at
\(\varepsilon = 0.10\), the bounds move only in the third decimal with
\(\varepsilon\), the re-selection floor entering the field's draws. On
the continuous endpoint the selected region, the 79 men with baseline
CD4 above 507, has 40 on the combination and 39 on monotherapy, with a
consistency proportion of 0.91; their mean CD4 decline is 89.3 cells
under the combination against 18.2 under monotherapy. All eight passing
candidates are defined by a baseline-CD4 cut above 449 or 507 crossed
with age, weight, drug history or sex, so the elevation is carried by
the high-CD4 margin, itself a small and noisy stratum: baseline CD4
above 507 alone gives a mean difference of \(+48.5\) over 97 patients
(\(-18.2\), 115.1), and men alone \(-22.9\). The two regions are
selected on different endpoints of the same patients and are not the
same set; each leaves a complement of about 93\% of the trial.

{

\begin{longtable}[]{@{}
  >{\raggedright\arraybackslash}p{(\linewidth - 10\tabcolsep) * \real{0.2000}}
  >{\raggedleft\arraybackslash}p{(\linewidth - 10\tabcolsep) * \real{0.1600}}
  >{\raggedleft\arraybackslash}p{(\linewidth - 10\tabcolsep) * \real{0.1600}}
  >{\raggedleft\arraybackslash}p{(\linewidth - 10\tabcolsep) * \real{0.1600}}
  >{\raggedleft\arraybackslash}p{(\linewidth - 10\tabcolsep) * \real{0.1600}}
  >{\raggedleft\arraybackslash}p{(\linewidth - 10\tabcolsep) * \real{0.1600}}@{}}

\caption{\label{tbl-supp-app-five}The five analyses side by side. Each
column states its own effect scale; the three breast cancer columns
differ in the identifier alone and repeat Table 1 of the main paper.
Corrected rows carry the two-term estimate with its two-sided
infinitesimal-jackknife interval as a reference only. The field and
field-s rows are one-sided bounds at nominal 0.95; the Bonferroni pair
holds both at joint 0.95. Selection bias is on the log scale for the
ratio columns and in CD4 cells for the last.}

\tabularnewline

\toprule\noalign{}
\begin{minipage}[b]{\linewidth}\raggedright
\end{minipage} & \begin{minipage}[b]{\linewidth}\raggedleft
GBSG, FS (HR)
\end{minipage} & \begin{minipage}[b]{\linewidth}\raggedleft
GBSG, DINA (HR)
\end{minipage} & \begin{minipage}[b]{\linewidth}\raggedleft
GBSG, GRF (HR)
\end{minipage} & \begin{minipage}[b]{\linewidth}\raggedleft
ACTG 175, binary (OR)
\end{minipage} & \begin{minipage}[b]{\linewidth}\raggedleft
ACTG 175, continuous (MD, CD4 cells)
\end{minipage} \\
\midrule\noalign{}
\endhead
\bottomrule\noalign{}
\endlastfoot
Whole trial N (events) & 686 (299) & 686 (299) & 686 (299) & 1,083 (430)
& 1,083 \\
Whole-trial estimate, 95\% CI & 0.695 (0.544, 0.888) & 0.695 (0.544,
0.888) & 0.695 (0.544, 0.888) & 0.665 (0.520, 0.850) & −27.6 (−43.1,
−12.1) \\
Region & ER 0 and PR ≤ 26 & grade 3 and PR ≤ 20 & ER 0 & weight
\textgreater{} 86 kg and CD4 \textgreater{} 380 & CD4 \textgreater{} 507
and male \\
N (events) & 75 (41) & 113 (61) & 82 (45) & 72 (37) & 79 \\
Unadjusted, 95\% CI & 2.222 (1.169, 4.221) & 1.315 (0.746, 2.317) &
1.951 (1.037, 3.673) & 3.584 (1.315, 9.774) & 71.1 (−5.1, 147.3) \\
Corrected (two-term), IJ interval & 1.276 (0.570, 2.857) & 1.076 (0.433,
2.673) & 1.111 (0.486, 2.540) & 1.773 (0.604, 5.206) & 22.1 (−77.5,
121.6) \\
Field, lower bound & 0.634 & 0.638 & 0.510 & 0.686 & −38.3 \\
Complement N (events) & 611 (258) & 573 (238) & 604 (254) & 1,011 (393)
& 1,004 \\
Unadjusted, 95\% CI & 0.605 (0.465, 0.788) & 0.641 (0.489, 0.839) &
0.615 (0.472, 0.801) & 0.587 (0.454, 0.759) & −36.2 (−51.4, −20.9) \\
Corrected (two-term), IJ interval & 0.645 (0.402, 1.033) & 0.658 (0.394,
1.100) & 0.663 (0.411, 1.070) & 0.620 (0.372, 1.033) & −32.7 (−62.2,
−3.2) \\
Field-s, upper bound & 0.804 & 0.842 & 0.846 & 0.786 & −17.6 \\
Bonferroni pair & 0.509 / 0.840 & 0.572 / 0.887 & 0.404 / 0.874 & 0.597
/ 0.817 & −51.2 / −15.8 \\
Selection bias & 0.543 & 0.193 & 0.557 & 0.704 & 47.9 \\
Re-selection family & 1,744 & 30 & 858 & 2,794 & 4,935 \\

\end{longtable}

}

\emph{Reading the ACTG 175 estimates.} As in Section 6 of the main
paper, the corrected estimates carry the reading and the bounds are
reported beside them. On the binary endpoint the unadjusted odds ratio
in the region is 3.584 (1.315, 9.774), and the correction, over the
largest selection bias of the four ratio-scale analyses, 0.704 on the
log scale, brings it to 1.773: suggestive of limited benefit, but with a
field lower bound of 0.686, below 1.0, a benefit cannot be ruled out.
Outside the region the corrected estimate favours the combination
(Table~\ref{tbl-supp-app-five}); the field-s upper bound is 0.786, and
the Bonferroni pair's lower limit in the region is 0.597. The same shape
appears on the additive scale. The unadjusted mean difference in the
continuous region is 71.1 cells (\(-5.1\), 147.3); the correction
removes 47.9 cells of selection bias and leaves 22.1 on the adverse
side: suggestive of limited benefit, but with a field lower bound of
\(-38.3\), below zero, a benefit cannot be ruled out. Outside the region
the corrected estimate again favours the combination; the field-s upper
bound is \(-17.6\) cells, and the pair's lower limit in the region is
\(-51.2\). On the CD4 scale the screening thresholds sit well below the
trial's own effect and the floors are elevated above the claim
threshold, the case in which the control of Section 4 is conservative;
on the ratio scales the thresholds sit at the claim threshold, the case
in which Theorem 3 holds with the floors in force.

\emph{The common shape.} The five analyses return one reading in three
outcome families and on two trials. An unadjusted analysis of any of
these regions would report a substantial adverse effect; after
correction each region's estimate is much attenuated, suggestive of
limited benefit in every case, with a lower bound that does not rule a
benefit out. In every case the corrected estimate in the complement, the
part of the trial the search was not looking at, favours treatment, and
the complement bounds are reported beside it
(Table~\ref{tbl-supp-app-five}).

\subsection{Declaration at a protected
level}\label{sec-supp-app-declaration}

Section 6.2 of the main paper reads forest search's breast cancer
declaration against the calibrated screen;
Table~\ref{tbl-supp-app-declaration} reads forest search's declaration
on all three, the ACTG 175 endpoints at their own primary protected
levels --- an odds ratio of 0.75 on the binary endpoint and a mean
difference of ten cells per cubic millimetre in the combination's favour
on the continuous one, on grids of 0.70 to 0.85 and of five to fifteen
cells --- at \(\alpha = 0.10\), from the same 5,000 draws as the bounds.

\begin{longtable}[]{@{}
  >{\raggedright\arraybackslash}p{(\linewidth - 6\tabcolsep) * \real{0.4000}}
  >{\raggedright\arraybackslash}p{(\linewidth - 6\tabcolsep) * \real{0.2000}}
  >{\raggedright\arraybackslash}p{(\linewidth - 6\tabcolsep) * \real{0.2000}}
  >{\raggedright\arraybackslash}p{(\linewidth - 6\tabcolsep) * \real{0.2000}}@{}}
\caption{The three declarations read against the calibrated screen of
Section 4 of the main paper, at \(\alpha = 0.10\). The statistic \(T\)
is the region's standardized effect against the claim threshold; the
calibrated cutoff is equation (7) at the primary protected level; the
diagnostic is equation (8), the family-wise size of the default screen
under a uniform benefit of \(c_0\), read at the cutoff that screen
executes; the implied level is the screening level at which an exact
comparison would reproduce the calibrated cutoff, reported for
interpretation.}\label{tbl-supp-app-declaration}\tabularnewline
\toprule\noalign{}
\begin{minipage}[b]{\linewidth}\raggedright
\end{minipage} & \begin{minipage}[b]{\linewidth}\raggedright
GBSG, hazard ratio
\end{minipage} & \begin{minipage}[b]{\linewidth}\raggedright
ACTG 175 binary, odds ratio
\end{minipage} & \begin{minipage}[b]{\linewidth}\raggedright
ACTG 175 continuous, mean difference
\end{minipage} \\
\midrule\noalign{}
\endfirsthead
\toprule\noalign{}
\begin{minipage}[b]{\linewidth}\raggedright
\end{minipage} & \begin{minipage}[b]{\linewidth}\raggedright
GBSG, hazard ratio
\end{minipage} & \begin{minipage}[b]{\linewidth}\raggedright
ACTG 175 binary, odds ratio
\end{minipage} & \begin{minipage}[b]{\linewidth}\raggedright
ACTG 175 continuous, mean difference
\end{minipage} \\
\midrule\noalign{}
\endhead
\bottomrule\noalign{}
\endlastfoot
Region declared at the default level, \(N\) & 75 & 72 & 79 \\
\(T\) of the region / largest \(T\) in the family & 2.438 / 2.564 &
2.494 / 2.494 & 1.701 / 2.043 \\
Primary protected level \(c_0\) & HR 0.75 & OR 0.75 & MD \(-10\)
cells \\
Calibrated cutoff \(\hat\kappa_{0.10}(c_0)\) & 2.726 & 2.889 & 2.952 \\
Implied level \(2\Phi(\hat\kappa) - 1\) & 0.994 & 0.996 & 0.997 \\
Diagnostic at the default screen's cutoff,
\(\hat\alpha_{\mathrm{FW}}(c_0;\kappa)\) & 0.667 & 0.838 & 0.909 \\
Calibrated cutoff at the claim threshold & 3.503 & 3.611 & 3.625 \\
Diagnostic at the claim threshold & 0.974 & 0.989 & 0.997 \\
Calibrated rule at \(c_0\) & admits nothing & admits nothing & admits
nothing \\
\end{longtable}

In each trial the returned region's statistic sits below the calibrated
cutoff --- by 0.29, 0.40 and 1.25 --- and so does the largest statistic
in its family, so the calibrated screen admits no candidate; the same
holds at every protected level on each trial's grid and at the claim
threshold. Had the treatment benefited every patient by the protected
amount, the default screen would still have returned a harm subgroup on
0.667, 0.838 and 0.909 of such trials, on these families and designs. At
the claim threshold the diagnostic is a boundary size, 0.974, 0.989 and
0.997, the family-wise size of the default screen were every candidate
exactly at the claim threshold, and not any trial's error rate. The
implied levels, 0.994, 0.996 and 0.997, say what the calibration asks of
these trials: a screen far tighter than the default, which is what a
large overlapping family costs once the protection is placed at a
clinically meaningful benefit rather than at the claim threshold.

\subsection{The supporting quantities}\label{sec-supp-app-quantities}

{

\begin{longtable}[]{@{}
  >{\raggedright\arraybackslash}p{(\linewidth - 10\tabcolsep) * \real{0.3000}}
  >{\raggedleft\arraybackslash}p{(\linewidth - 10\tabcolsep) * \real{0.1400}}
  >{\raggedleft\arraybackslash}p{(\linewidth - 10\tabcolsep) * \real{0.1400}}
  >{\raggedleft\arraybackslash}p{(\linewidth - 10\tabcolsep) * \real{0.1400}}
  >{\raggedleft\arraybackslash}p{(\linewidth - 10\tabcolsep) * \real{0.1400}}
  >{\raggedleft\arraybackslash}p{(\linewidth - 10\tabcolsep) * \real{0.1400}}@{}}

\caption{\label{tbl-supp-app-diag}The quantities that support each
bound. The mean of the simulated selection root and the correlation of
the aligned roots are on each analysis's own working scale, the log
scale for the ratio columns and CD4 cells for the last.}

\tabularnewline

\toprule\noalign{}
\begin{minipage}[b]{\linewidth}\raggedright
\end{minipage} & \begin{minipage}[b]{\linewidth}\raggedleft
GBSG, FS
\end{minipage} & \begin{minipage}[b]{\linewidth}\raggedleft
GBSG, DINA
\end{minipage} & \begin{minipage}[b]{\linewidth}\raggedleft
GBSG, GRF
\end{minipage} & \begin{minipage}[b]{\linewidth}\raggedleft
ACTG 175, binary
\end{minipage} & \begin{minipage}[b]{\linewidth}\raggedleft
ACTG 175, continuous
\end{minipage} \\
\midrule\noalign{}
\endhead
\bottomrule\noalign{}
\endlastfoot
Mean of the simulated root, region & −0.022 & −0.002 & +0.015 & +0.106 &
+4.89 \\
Second-order estimate, region & 1.305 & 1.078 & 1.095 & 1.595 & 17.2 \\
Ordering bound \(\le\) second-order \(\le\) two-term & does not hold &
does not hold & holds & holds & holds \\
Second-order estimate, complement (field-s) & 0.647 & 0.663 & 0.674 &
0.640 & −31.1 \\
Scale ratio \(\rho^c\) & 1.002 & 1.013 & 0.985 & 0.988 & not computed \\
Bonferroni pair, simulated joint coverage & 0.9495 & 0.9499 & 0.9510 &
0.9490 & 0.9492 \\
Calibrated split \(\gamma\) & 0.025 & 0.025 & 0.025 & 0.025 & 0.025 \\
Correlation of the aligned roots & −0.026 & −0.020 & −0.013 & +0.054 &
−0.039 \\

\end{longtable}

}

Three of the diagnostics deserve a word. The field's second-order
estimate subtracts from the two-term estimate the mean of the simulated
selection root on the estimand's scale, where the bound subtracts a
quantile of the same law; the bound is not centred on it and no reading
depends on it. When that mean is positive the ordering bound \(\le\)
second-order \(\le\) two-term holds, as it does for GRF on the breast
cancer data and for both ACTG 175 analyses; it does not hold for forest
search and DINA there, where the mean is slightly negative (\(-0.022\)
and \(-0.002\) on the log scale) and the second-order step moves the
estimate back up, by 2.2\% and 0.2\%, after first corrections of 0.55
and 0.20. The scale ratio \(\rho^c\) (Proposition 3(iii)), the selected
complement's influence scale over the draw winners' mean scale, sits
between 0.985 and 1.013 wherever it is computed, so field-s and the
unstudentized complement bound agree on these data; on the continuous
endpoint it is not computed, and the two second-order estimates differ
by 0.02 cells. And the Bonferroni pair is not conservative on any of the
five: its simulated joint coverage lies between 0.9490 and 0.9510
against a nominal 0.95, the calibrated split of Section S1.5 settles at
\(\gamma = 0.025\), Bonferroni's own \(\alpha/2\), in every case, and
the aligned roots correlate at between \(-0.039\) and 0.054.

{

\begin{longtable}[]{@{}
  >{\raggedright\arraybackslash}p{(\linewidth - 10\tabcolsep) * \real{0.2500}}
  >{\raggedleft\arraybackslash}p{(\linewidth - 10\tabcolsep) * \real{0.1500}}
  >{\raggedleft\arraybackslash}p{(\linewidth - 10\tabcolsep) * \real{0.1500}}
  >{\raggedleft\arraybackslash}p{(\linewidth - 10\tabcolsep) * \real{0.1500}}
  >{\raggedleft\arraybackslash}p{(\linewidth - 10\tabcolsep) * \real{0.1500}}
  >{\raggedleft\arraybackslash}p{(\linewidth - 10\tabcolsep) * \real{0.1500}}@{}}

\caption{\label{tbl-supp-app-location}Where each bound sits. A region's
lower bound below the whole-trial value does not rule out the trial's
own benefit for that region; a complement's upper bound below one (or
zero) lies on the benefit side.}

\tabularnewline

\toprule\noalign{}
\begin{minipage}[b]{\linewidth}\raggedright
\end{minipage} & \begin{minipage}[b]{\linewidth}\raggedleft
Field lower, region
\end{minipage} & \begin{minipage}[b]{\linewidth}\raggedleft
Bonferroni lower, region
\end{minipage} & \begin{minipage}[b]{\linewidth}\raggedleft
Field-s upper, complement
\end{minipage} & \begin{minipage}[b]{\linewidth}\raggedleft
Bonferroni upper, complement
\end{minipage} & \begin{minipage}[b]{\linewidth}\raggedleft
Whole trial
\end{minipage} \\
\midrule\noalign{}
\endhead
\bottomrule\noalign{}
\endlastfoot
GBSG, FS (HR) & 0.634 & 0.509 & 0.804 & 0.840 & 0.695 \\
GBSG, DINA (HR) & 0.638 & 0.572 & 0.842 & 0.887 & 0.695 \\
GBSG, GRF (HR) & 0.510 & 0.404 & 0.846 & 0.874 & 0.695 \\
ACTG 175, binary (OR) & 0.686 & 0.597 & 0.786 & 0.817 & 0.665 \\
ACTG 175, continuous (MD, CD4 cells) & −38.3 & −51.2 & −17.6 & −15.8 &
−27.6 \\

\end{longtable}

}

\section{\texorpdfstring{Comparison with \citet{guohe2021} on a nested
continuum}{Comparison with @guohe2021 on a nested continuum}}\label{sec-guohe}

\citet{guohe2021} treat subgroup analysis in two steps: identify the
subgroup with the largest estimated effect by the argmax over a
prespecified candidate family, then bound the effect of the subgroup so
identified. Their post hoc design (their Section 5.2 and Table 7) runs
both steps on a nested continuum of cutpoints, where neighbouring
candidates differ by a single subject. It therefore shows what the
argmax identifies on such a family, and what inference must then
correct. We run the design exactly as published and set forest search's
post-selection bounds beside theirs.

\subsection{Setting}\label{sec-guohe-setting}

\textbf{Design.} Guo and He's Section 5.2 generates \(n = 400\) subjects
with \(W \sim \mathrm{Unif}[0, 80]\) and treatment
\(D \sim \mathrm{Bernoulli}(0.5)\) from the proportional hazards model
\(\lambda_0(t)\exp\{b(W)D\}\), with a Weibull(1, 1) baseline hazard,
about 40\% censoring, and \(b(w) = \beta_2\) for \(w \le 30\) and \(0\)
otherwise; \(\beta_2 \in \{0, 0.1, \ldots, 0.5\}\), with 2000 Monte
Carlo samples per value. The true harm region is \(H = \{W \le 30\}\),
with \(|H| \approx 150\). The candidates are \(S(c) = \{W \le c\}\) for
\(c \in [30, 60]\), and \(\beta(c)\) is the Cox log-hazard ratio of
treatment within \(S(c)\): a weighted average of \(b(\cdot)\) over
\([0, c]\) (their Section 5.2). Because an estimate changes only where
\(c\) passes an observed value of \(W\), we take the family to be the
distinct cuts at 30 and at every observed \(w \in (30, 60]\) --- nested,
cut at every observed value, about 151 candidates per replicate. This
discretisation is ours; Guo and He do not state one.

\textbf{Orientation and targets.} Guo and He take a larger \(\beta\) to
be better (their Section 2.1), and here \(\beta(c)\) is a log-hazard
ratio, so their selection \(\hat c = \arg\max_c \hat\beta(c)\) picks the
subgroup with the largest estimated hazard ratio. In forest search's
orientation that is the harm-side argmax, with no sign change. Write
\(\hat H = S(\hat c)\) for the selected subgroup and
\(\hat H^c = \{W > \hat c\}\) for its complement. Guo and He target the
best subgroup effect \(\gamma_{\max} = \sup_c \beta(c)\) (their
Algorithm 3 and Theorem 4) and the best selected subgroup effect
\(\gamma_s = \beta(\hat c)\) (their Theorem 5). For \(\beta_2 > 0\) the
best subgroup is \(S(30) = H\), so
\(\gamma_{\max} = \beta_2 = \log\theta^\dagger(H)\), where
\(\theta^\dagger(H)\) is the hazard ratio in \(H\), and
\(\gamma_s = \beta(\hat H)\), the true effect in the selected subgroup:
their two targets are forest search's two estimands. Because
\(H \subseteq \hat H\), \(\beta(\hat H) \le \beta_2\), with equality
only when \(\hat c = 30\).

\textbf{Truths.} The hazard ratio in \(H\) is
\(\theta^\dagger(H) = \exp(\beta_2) =\) 1.00, 1.11, 1.22, 1.35, 1.49 and
1.65 across the six cells. The complement's truth is exactly zero on the
log scale: \(\hat c \ge 30\), so \(\hat H^c\) lies inside
\(\{W > 30\}\), where \(b = 0\), and \(\beta(\hat H^c) = 0\) --- a
hazard ratio of 1.00 --- in every cell and replicate. At \(\beta_2 = 0\)
there is no harm region, and \(\beta(\hat H) = 0\) as well.

\textbf{Identifier.} The identifier is the pure argmax over the family,
with no consistency screen and no other filter, because that is Guo and
He's identifier. Removing forest search's screen reduces its selection
to this primitive, and the two targets then coincide. The unrestricted
argmax concentrates on the smallest admissible subgroups; here the
smallest admissible subgroup, \(S(30)\), is the true harm region itself,
so the caveat flatters identification on this design
(Section~\ref{sec-guohe-limits}). Guo and He note (their Section 6) that
when identification involves other considerations their bound serves as
a conservative approach to the subgroups so identified; that use is not
tested here.

\textbf{Guo and He's bound.} Their Algorithm 3 adds deterministic
offsets \(d(c) = (1 - n^{r - 1/2})\{\hat\gamma_{\max} - \hat\beta(c)\}\)
to every candidate, re-maximises over bootstrap samples, and subtracts a
bootstrap quantile of the re-centred maximum from \(\hat\gamma_{\max}\);
their bias-reduced estimate subtracts the bootstrap mean instead. The
tuning parameter \(r \in (0, 1/2)\) sets the offsets. We report
\(r = 1/12\); the other values in their Table 7 (\(1/3\), \(1/21\) and
\(1/30\)) are sensitivity only (Section~\ref{sec-guohe-detail}). We use
\(B = 2000\) bootstrap samples, our choice, since Guo and He state none
for their Section 5.

\textbf{Why Table 7 alone.} Table 7 is Guo and He's only continuum
design --- the only one whose subgroup is identified post hoc rather
than chosen from a predefined list --- and on its nested family
\(\hat H^c\) is never a candidate, which is what the complement
comparison of Section~\ref{sec-guohe-complement} turns on.

\textbf{Bounds compared.} On the selected subgroup there are four
one-sided 95\% lower bounds on \(\beta(\hat H)\): the unadjusted bound,
the bound Guo and He call naive, the standard within-subgroup Cox
analysis of \(\hat H\) as if it had been prespecified; Guo and He's
bound at \(r = 1/12\); MR (field); and MR (IJ two-term), the
conservative reference. Both MR bounds use multiplier resampling over
the fixed family: each of 5000 draws perturbs every candidate estimate
through one shared multiplier vector and re-applies the argmax. MR (IJ
two-term) centres on the two-term de-biased estimate
\(\tilde\beta(\hat H)\), which subtracts from \(\hat\beta(\hat H)\) the
resampling estimates of the selection optimism and of the same-draws
estimation term \citep{leon2024, leonanderson2026}, and uses its
infinitesimal-jackknife standard error. MR (field) inverts, around the
same \(\tilde\beta(\hat H)\), the distribution of a selection
discrepancy \(\Lambda^*\): multiplier perturbations of the
candidate-effect field, with the winner's entry set to its de-biased
value, are pushed through the re-selection map, with an inner Monte
Carlo estimating the selection drift. Its bound is
\(\tilde\beta(\hat H) - q_{0.95}(\Lambda^*)\), and its point estimate
\(\hat\beta_{\mathrm F}(\hat H)\) subtracts the mean of \(\Lambda^*\)
from \(\tilde\beta(\hat H)\) (1000 outer and 500 inner draws). On the
complement, the unadjusted, IJ two-term and field constructions give
one-sided 95\% upper bounds on \(\beta(\hat H^c)\), with \(\hat H^c\)
carried by the harm-side selection rather than selected; field-s is the
field for \(\hat H^c\) with each draw's reading studentized to the
selected complement's own scale. The Bonferroni pair sets the field
lower bound on \(\hat H\) and the field-s upper bound on \(\hat H^c\)
each at one-sided level 0.975, from the same aligned draws, for a joint
two-subgroup claim at level at least 0.95. Finally, \(\hat p(\hat H)\)
is the re-selection probability of the observed winner: the share of
multiplier draws whose argmax is \(\hat H\) itself. Full definitions are
in Section 3 of the main paper.

\textbf{Reading the bounds.} Bounds are read by their location on the
hazard-ratio scale against one and against the planted effect: a lower
bound on the hazard ratio in \(\hat H\) above one is on the harm side;
an upper bound on the hazard ratio in \(\hat H^c\) below one is on the
benefit side. The locations reported are where the bounds sit, not
claims.

Every cell has 2000 replicates, paired across methods on the same data
sets, so the Monte Carlo standard error of a coverage near 0.95 is about
0.005.

\subsection{Identification on a nested continuum}\label{sec-guohe-ident}

{

\begin{longtable}[]{@{}
  >{\raggedright\arraybackslash}p{(\linewidth - 12\tabcolsep) * \real{0.2174}}
  >{\raggedleft\arraybackslash}p{(\linewidth - 12\tabcolsep) * \real{0.1304}}
  >{\raggedleft\arraybackslash}p{(\linewidth - 12\tabcolsep) * \real{0.1304}}
  >{\raggedleft\arraybackslash}p{(\linewidth - 12\tabcolsep) * \real{0.1304}}
  >{\raggedleft\arraybackslash}p{(\linewidth - 12\tabcolsep) * \real{0.1304}}
  >{\raggedleft\arraybackslash}p{(\linewidth - 12\tabcolsep) * \real{0.1304}}
  >{\raggedleft\arraybackslash}p{(\linewidth - 12\tabcolsep) * \real{0.1304}}@{}}

\caption{\label{tbl-guohe-ident}What the argmax selects, by cell.
\(\hat c\) is the selected cutpoint. Specificity \((80 - \hat c)/50\)
and positive predictive value (PPV) \(30/\hat c\) are exact functions of
\(\hat c\); sensitivity and negative predictive value are identically 1
by nesting for \(\beta_2 > 0\) and are not tabulated. At \(\beta_2 = 0\)
there is no harm region, so specificity and PPV are shown as ---.
\(\lvert\hat H\rvert\) is the selected size, against
\(\lvert H\rvert \approx 150\). \(\beta(\hat H)\) is the true log-hazard
ratio in the selected subgroup; dilution is \(\beta_2\) minus its mean.
\(\hat p(\hat H)\) is the re-selection probability of the observed
winner. 2000 replicates per cell.}

\tabularnewline

\toprule\noalign{}
\begin{minipage}[b]{\linewidth}\raggedright
\end{minipage} & \begin{minipage}[b]{\linewidth}\raggedleft
\(\beta_2 = 0.0\)
\end{minipage} & \begin{minipage}[b]{\linewidth}\raggedleft
\(\beta_2 = 0.1\)
\end{minipage} & \begin{minipage}[b]{\linewidth}\raggedleft
\(\beta_2 = 0.2\)
\end{minipage} & \begin{minipage}[b]{\linewidth}\raggedleft
\(\beta_2 = 0.3\)
\end{minipage} & \begin{minipage}[b]{\linewidth}\raggedleft
\(\beta_2 = 0.4\)
\end{minipage} & \begin{minipage}[b]{\linewidth}\raggedleft
\(\beta_2 = 0.5\)
\end{minipage} \\
\midrule\noalign{}
\endhead
\bottomrule\noalign{}
\endlastfoot
\(\hat c\), median (IQR) & 40.0 (32.4, 52.4) & 36.3 (31.2, 48.7) & 33.5
(30.7, 42.1) & 32.3 (30.4, 37.5) & 31.4 (30.2, 35.3) & 30.9 (30.1,
33.5) \\
\(\hat c\), range & 30.0--60.0 & 30.0--60.0 & 30.0--60.0 & 30.0--60.0 &
30.0--60.0 & 30.0--59.9 \\
Mean specificity & --- & 0.794 & 0.848 & 0.890 & 0.912 & 0.939 \\
Mean PPV & --- & 0.788 & 0.836 & 0.874 & 0.898 & 0.924 \\
Mean \(\lvert\hat H\rvert\) (median) & 212.4 (202) & 201.8 (183) & 188.3
(171) & 177.8 (165) & 171.9 (161) & 165.3 (158) \\
Mean \(\beta(\hat H)\) & 0.000 & 0.082 & 0.169 & 0.261 & 0.358 &
0.459 \\
Dilution & 0.000 & 0.018 & 0.031 & 0.039 & 0.042 & 0.041 \\
\(\hat p(\hat H)\), median (IQR) & 0.196 (0.135, 0.276) & 0.209 (0.142,
0.299) & 0.229 (0.159, 0.324) & 0.244 (0.166, 0.347) & 0.272 (0.183,
0.393) & 0.309 (0.204, 0.442) \\

\end{longtable}

}

\begin{figure}

\centering{

\pandocbounded{\includegraphics[keepaspectratio]{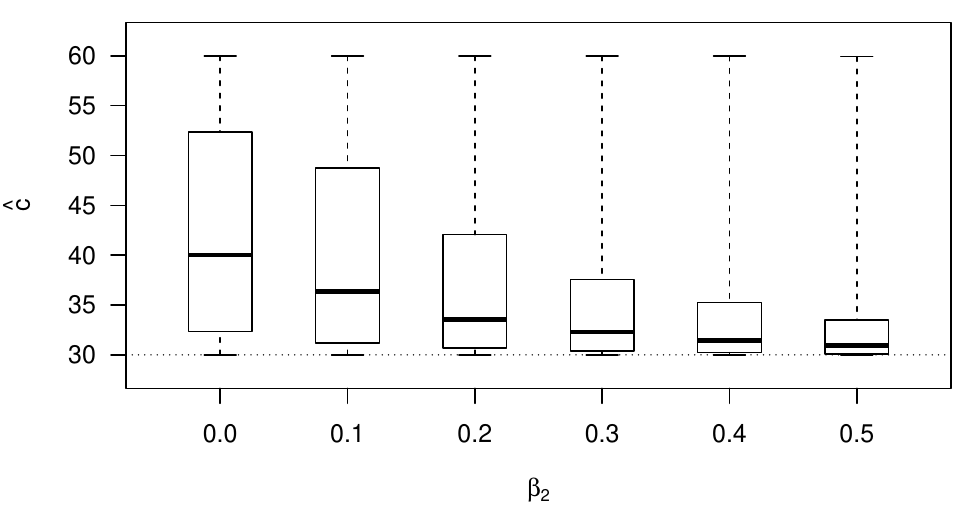}}

}

\caption{\label{fig-guohe-chat}Selected cutpoint \(\hat c\) by
\(\beta_2\), drawn from each cell's five-number summary (whiskers at the
minimum and maximum; 2000 replicates per cell). The dotted line is the
true boundary, \(c = 30\), which is also the family's smallest member.}

\end{figure}

For \(\beta_2 > 0\) every candidate contains \(H = S(30)\), so
\(H \subseteq \hat H\) in every replicate: sensitivity and negative
predictive value are identically 1 by nesting, not by merit, and
overshoot of \(\hat c\) beyond 30 is the only identification error.
Specificity, \((80 - \hat c)/50\), and positive predictive value,
\(30/\hat c\), are exact functions of \(\hat c\) and restate it
(Table~\ref{tbl-guohe-ident}). At \(\beta_2 = 0\) there is no harm
region and every selection is a false identification, so specificity and
positive predictive value are not defined as accuracy there; that cell
is the reference for the argmax's own pull toward the smallest
candidate. Its median \(\hat c\) is 40.0, with lower quartile 32.4, on a
family running from 30 to 60 --- and \(c = 30\) is where the true
boundary of every other cell sits (Figure~\ref{fig-guohe-chat}). As
\(\beta_2\) grows, \(\hat c\) concentrates on 30: its median falls from
36.3 at \(\beta_2 = 0.1\) to 30.9 at \(\beta_2 = 0.5\), and the mean
selected size \(|\hat H|\) from 202 to 165 (212 at the null), against
\(|H| \approx 150\).

Overshoot reaches the estimand. Because \(\beta(c)\) averages
\(b(\cdot)\) over \([0, c]\), \(\beta(\hat H)\) falls below \(\beta_2\)
whenever \(\hat c > 30\): the mean dilution \(\beta_2 - \beta(\hat H)\)
is 0.018 at \(\beta_2 = 0.1\) and about 0.04 from \(\beta_2 = 0.3\) on,
and the selected subgroup's hazard ratio --- the exponential of the mean
\(\beta(\hat H)\) --- is 1.09, 1.18, 1.30, 1.43 and 1.58, against
\(\theta^\dagger(H)\) from 1.11 to 1.65. Re-selection is spread thin
throughout: the median \(\hat p(\hat H)\) rises from 0.196 at the null
to 0.309 at \(\beta_2 = 0.5\) and stays low, because near-duplicate
candidates share the re-selection mass.

These figures describe what the argmax selects on this family. They do
not compare identifiers: FS, DINA and GRF search other families under
other rules.

{

\begin{longtable}[]{@{}
  >{\raggedright\arraybackslash}p{(\linewidth - 12\tabcolsep) * \real{0.2195}}
  >{\raggedleft\arraybackslash}p{(\linewidth - 12\tabcolsep) * \real{0.1301}}
  >{\raggedleft\arraybackslash}p{(\linewidth - 12\tabcolsep) * \real{0.1301}}
  >{\raggedleft\arraybackslash}p{(\linewidth - 12\tabcolsep) * \real{0.1301}}
  >{\raggedleft\arraybackslash}p{(\linewidth - 12\tabcolsep) * \real{0.1301}}
  >{\raggedleft\arraybackslash}p{(\linewidth - 12\tabcolsep) * \real{0.1301}}
  >{\raggedleft\arraybackslash}p{(\linewidth - 12\tabcolsep) * \real{0.1301}}@{}}

\caption{\label{tbl-guohe-ident-detail}Selection detail, by cell: the
share of replicates with \(\hat c = 30\); the median effective number of
competitors \(M_{\mathrm{eff}}\); and the marginal and error standard
deviations of the field's point estimate,
\(\mathrm{sd}{\hat\beta_{\mathrm F}(\hat H)}\) and
\(\mathrm{sd}{\hat\beta_{\mathrm F}(\hat H) - \beta(\hat H)}\).}

\tabularnewline

\toprule\noalign{}
\begin{minipage}[b]{\linewidth}\raggedright
\end{minipage} & \begin{minipage}[b]{\linewidth}\raggedleft
\(\beta_2 = 0.0\)
\end{minipage} & \begin{minipage}[b]{\linewidth}\raggedleft
\(\beta_2 = 0.1\)
\end{minipage} & \begin{minipage}[b]{\linewidth}\raggedleft
\(\beta_2 = 0.2\)
\end{minipage} & \begin{minipage}[b]{\linewidth}\raggedleft
\(\beta_2 = 0.3\)
\end{minipage} & \begin{minipage}[b]{\linewidth}\raggedleft
\(\beta_2 = 0.4\)
\end{minipage} & \begin{minipage}[b]{\linewidth}\raggedleft
\(\beta_2 = 0.5\)
\end{minipage} \\
\midrule\noalign{}
\endhead
\bottomrule\noalign{}
\endlastfoot
Share with \(\hat c = 30\) & 0.062 & 0.090 & 0.114 & 0.145 & 0.168 &
0.213 \\
\(M_{\mathrm{eff}}\), median & 2.12 & 2.13 & 2.14 & 2.16 & 2.20 &
2.23 \\
Marginal SD & 0.186 & 0.196 & 0.195 & 0.202 & 0.209 & 0.212 \\
Error SD & 0.186 & 0.191 & 0.188 & 0.193 & 0.197 & 0.202 \\

\end{longtable}

}

The argmax lands exactly on \(c = 30\) in 6\% of replicates at the null
and 21\% at \(\beta_2 = 0.5\) (Table~\ref{tbl-guohe-ident-detail}). The
effective number of competitors \(M_{\mathrm{eff}}\) --- the number of
independent standard normal variables whose expected maximum equals that
of the family's correlated null field --- has median 2.1 to 2.2 in every
cell: about 151 nested candidates compete like two independent ones. The
marginal standard deviation of the field's point estimate exceeds its
error standard deviation whenever \(\beta_2 > 0\), because
\(\beta(\hat H)\) itself moves with the selection.

\subsection{The selected subgroup's lower bound}\label{sec-guohe-lower}

\begin{figure}

\centering{

\pandocbounded{\includegraphics[keepaspectratio]{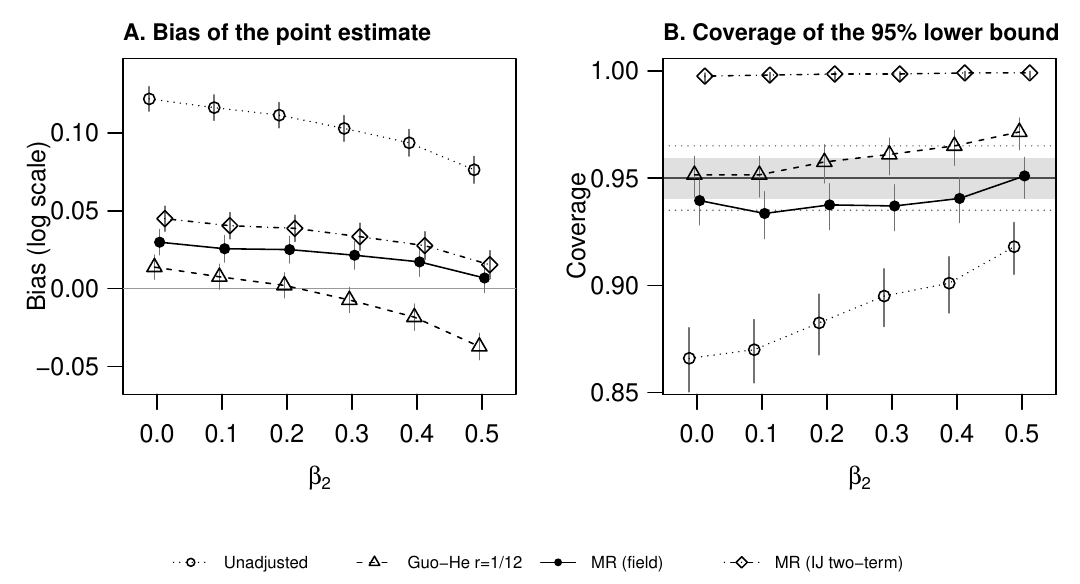}}

}

\caption{\label{fig-guohe-bias-cover}Bias and coverage for the four
one-sided 95\% lower bounds on \(\beta(\hat H)\), by cell. \textbf{A}:
mean bias of the point estimate against \(\beta(\hat H)\) on the log
scale, with bars at \(\pm 1.96\) Monte Carlo standard errors from the
empirical spread. \textbf{B}: coverage, with bars at the Wilson 95\%
interval for each estimate. In \textbf{B} the solid line is nominal
0.95; the shaded band is where a coverage estimate falls with
probability 0.95 when the true coverage is exactly nominal, and the
dotted lines widen that band for reading all 24 cell-by-method coverages
at once. A point inside the shading is indistinguishable from nominal;
one outside the dotted lines departs from it even allowing for
multiplicity. Points are offset horizontally so the bars separate. 2000
paired replicates per cell.}

\end{figure}%

The unadjusted bound covers at 0.866--0.918, below nominal in every cell
and furthest below it when \(\beta_2\) is small, as Guo and He found;
its point estimate carries much the largest bias throughout, \(+0.076\)
to \(+0.122\) (Figure~\ref{fig-guohe-bias-cover}). Guo and He's bound at
\(r = 1/12\) covers at 0.952--0.972 and MR (field) at 0.933--0.951. MR
(IJ two-term) covers at 0.998--0.999 with nearly twice their margin
(0.58--0.61).

The two corrections depart from nominal in opposite directions, and the
bands say by how much. Guo and He's bound sits above the pointwise band
in the three cells \(\beta_2 \ge 0.3\), and at \(\beta_2 = 0.5\) above
the simultaneous band as well: conservative, as their Section 2.5 leads
one to expect of a small \(r\). The field sits below the pointwise band
in the four cells \(\beta_2 \le 0.3\), and below the simultaneous band
only at \(\beta_2 = 0.1\), its lowest point at 0.933
(Section~\ref{sec-guohe-limits}). Their point estimates part the same
way: Guo and He's bias-reduced estimate crosses zero and over-corrects
as \(\beta_2\) grows, moving from \(+0.014\) to \(-0.037\), while the
field's retains a positive bias in every cell, \(+0.007\) to \(+0.030\).
The unadjusted bound is outside both bands in every cell, and the IJ
two-term bound above them in every cell.

Their published Table 7 is more conservative still than our replication,
at 0.961--0.972 for \(r = 1/12\); our replicated columns run about 0.01
below the published ones, a replication deficit carried here unchanged
(Table~\ref{tbl-guohe-rsens}). MR (field) reaches Guo and He's width on
this family --- margin 0.310--0.320 against their 0.305--0.306 --- so it
gets its coverage by correcting less, not by a narrower interval
(Table~\ref{tbl-guohe-lower}).

{

\begin{longtable}[]{@{}
  >{\raggedright\arraybackslash}p{(\linewidth - 12\tabcolsep) * \real{0.1724}}
  >{\raggedleft\arraybackslash}p{(\linewidth - 12\tabcolsep) * \real{0.1379}}
  >{\raggedleft\arraybackslash}p{(\linewidth - 12\tabcolsep) * \real{0.1379}}
  >{\raggedleft\arraybackslash}p{(\linewidth - 12\tabcolsep) * \real{0.1379}}
  >{\raggedleft\arraybackslash}p{(\linewidth - 12\tabcolsep) * \real{0.1379}}
  >{\raggedleft\arraybackslash}p{(\linewidth - 12\tabcolsep) * \real{0.1379}}
  >{\raggedleft\arraybackslash}p{(\linewidth - 12\tabcolsep) * \real{0.1379}}@{}}

\caption{\label{tbl-guohe-locations}Where each bound typically falls, on
the hazard-ratio scale: the exponential of the mean log-scale lower
bound. \(\theta^\dagger(H) = \exp(\beta_2)\) is the hazard ratio in
\(H\); HR in \(\hat H\) is the exponential of the mean
\(\beta(\hat H)\), the effect of the subgroup actually selected.
Coverage, margin and bias for the same bounds are in
Table~\ref{tbl-guohe-lower}.}

\tabularnewline

\toprule\noalign{}
\begin{minipage}[b]{\linewidth}\raggedright
\end{minipage} & \begin{minipage}[b]{\linewidth}\raggedleft
\(\beta_2 = 0.0\)
\end{minipage} & \begin{minipage}[b]{\linewidth}\raggedleft
\(\beta_2 = 0.1\)
\end{minipage} & \begin{minipage}[b]{\linewidth}\raggedleft
\(\beta_2 = 0.2\)
\end{minipage} & \begin{minipage}[b]{\linewidth}\raggedleft
\(\beta_2 = 0.3\)
\end{minipage} & \begin{minipage}[b]{\linewidth}\raggedleft
\(\beta_2 = 0.4\)
\end{minipage} & \begin{minipage}[b]{\linewidth}\raggedleft
\(\beta_2 = 0.5\)
\end{minipage} \\
\midrule\noalign{}
\endhead
\bottomrule\noalign{}
\endlastfoot
\(\theta^\dagger(H)\) & 1.00 & 1.11 & 1.22 & 1.35 & 1.49 & 1.65 \\
HR in \(\hat H\) & 1.00 & 1.09 & 1.18 & 1.30 & 1.43 & 1.58 \\
Unadjusted & 0.83 & 0.89 & 0.96 & 1.04 & 1.13 & 1.23 \\
Guo--He, \(r = 1/12\) & 0.75 & 0.81 & 0.87 & 0.95 & 1.03 & 1.12 \\
MR (field) & 0.76 & 0.82 & 0.89 & 0.97 & 1.06 & 1.16 \\
MR (IJ two-term) & 0.59 & 0.63 & 0.68 & 0.74 & 0.81 & 0.88 \\

\end{longtable}

}

Read against one, every typical lower bound lies below the selected
subgroup's hazard ratio in its cell (1.30, 1.43 and 1.58 from
\(\beta_2 = 0.3\) on). The unadjusted bound is the highest, at 1.23 for
\(\beta_2 = 0.5\), and it is the bound whose coverage is below nominal.
The typical Guo--He and field bounds are above one, on the harm side, at
\(\beta_2 = 0.4\) and 0.5 and below one in the other cells, the field a
little above Guo and He's in every cell (1.16 against 1.12 at
\(\beta_2 = 0.5\)). At \(\beta_2 = 0\) both sit at 0.75 and 0.76, below
one. The IJ two-term bound is below one in every cell (0.59 to 0.74
through \(\beta_2 = 0.3\)).

{

\begin{longtable}[]{@{}
  >{\raggedright\arraybackslash}p{(\linewidth - 14\tabcolsep) * \real{0.2197}}
  >{\raggedright\arraybackslash}p{(\linewidth - 14\tabcolsep) * \real{0.1212}}
  >{\raggedleft\arraybackslash}p{(\linewidth - 14\tabcolsep) * \real{0.1288}}
  >{\raggedleft\arraybackslash}p{(\linewidth - 14\tabcolsep) * \real{0.1136}}
  >{\raggedleft\arraybackslash}p{(\linewidth - 14\tabcolsep) * \real{0.0985}}
  >{\raggedleft\arraybackslash}p{(\linewidth - 14\tabcolsep) * \real{0.1212}}
  >{\raggedleft\arraybackslash}p{(\linewidth - 14\tabcolsep) * \real{0.1061}}
  >{\raggedleft\arraybackslash}p{(\linewidth - 14\tabcolsep) * \real{0.0909}}@{}}

\caption{\label{tbl-guohe-pre16}The sixteen replicated cells of Guo and
He's Tables 3--7 at \(r=1/12\): coverage of the one-sided 95\% lower
bound, mean margin and mean bias (log scale), for Guo and He's bound and
MR (field). 2000 paired replicates per cell.}

\tabularnewline

\toprule\noalign{}
\begin{minipage}[b]{\linewidth}\raggedright
Design
\end{minipage} & \begin{minipage}[b]{\linewidth}\raggedright
Cell
\end{minipage} & \begin{minipage}[b]{\linewidth}\raggedleft
Guo--He: coverage
\end{minipage} & \begin{minipage}[b]{\linewidth}\raggedleft
Guo--He: margin
\end{minipage} & \begin{minipage}[b]{\linewidth}\raggedleft
Guo--He: bias
\end{minipage} & \begin{minipage}[b]{\linewidth}\raggedleft
Field: coverage
\end{minipage} & \begin{minipage}[b]{\linewidth}\raggedleft
Field: margin
\end{minipage} & \begin{minipage}[b]{\linewidth}\raggedleft
Field: bias
\end{minipage} \\
\midrule\noalign{}
\endhead
\bottomrule\noalign{}
\endlastfoot
Two subgroups (Tables 3--5) & \(\beta_2 = 0.0\) & 0.946 & 0.265 & +0.003
& 0.945 & 0.302 & +0.010 \\
& \(\beta_2 = 0.1\) & 0.953 & 0.263 & -0.006 & 0.949 & 0.301 & +0.004 \\
& \(\beta_2 = 0.2\) & 0.948 & 0.262 & -0.017 & 0.945 & 0.304 & +0.003 \\
& \(\beta_2 = 0.3\) & 0.945 & 0.260 & -0.035 & 0.936 & 0.307 & -0.003 \\
& \(\beta_2 = 0.4\) & 0.950 & 0.260 & -0.057 & 0.936 & 0.309 & -0.012 \\
& \(\beta_2 = 0.5\) & 0.953 & 0.260 & -0.068 & 0.942 & 0.311 & -0.011 \\
\(k\) null subgroups (Table 6) & \(k = 2\) & 0.948 & 0.265 & +0.002 &
0.944 & 0.302 & +0.008 \\
& \(k = 6\) & 0.945 & 0.218 & +0.008 & 0.941 & 0.294 & +0.030 \\
& \(k = 10\) & 0.949 & 0.203 & +0.007 & 0.944 & 0.291 & +0.038 \\
& \(k = 12\) & 0.947 & 0.198 & +0.006 & 0.942 & 0.290 & +0.041 \\
Nested (Table 7) & \(\beta_2 = 0.0\) & 0.952 & 0.306 & +0.014 & 0.940 &
0.310 & +0.030 \\
& \(\beta_2 = 0.1\) & 0.952 & 0.305 & +0.008 & 0.933 & 0.310 & +0.026 \\
& \(\beta_2 = 0.2\) & 0.958 & 0.305 & +0.002 & 0.938 & 0.313 & +0.025 \\
& \(\beta_2 = 0.3\) & 0.961 & 0.305 & -0.007 & 0.937 & 0.314 & +0.021 \\
& \(\beta_2 = 0.4\) & 0.965 & 0.305 & -0.018 & 0.941 & 0.316 & +0.017 \\
& \(\beta_2 = 0.5\) & 0.972 & 0.306 & -0.037 & 0.951 & 0.320 & +0.007 \\

\end{longtable}

}

On the sixteen cells of Guo and He's Tables 3--7 that we replicate
(Table~\ref{tbl-guohe-pre16}), their bound at \(r = 1/12\) covers at
0.945--0.972 and MR (field) at 0.933--0.951. The field's margin stays at
0.290--0.320 in every design, whereas Guo and He's is 0.305--0.306 on
the nested family, 0.260--0.265 on the two-subgroup design, and down to
0.198 with twelve disjoint null subgroups: the field reaches their width
on the nested family only. Guo and He's bias-reduced estimate
over-corrects as separation grows (to \(-0.068\) on the two-subgroup
design), while the field's retained bias grows with the number of tied
null subgroups (to \(+0.041\) at twelve).

\subsection{The complement and the joint
claim}\label{sec-guohe-complement}

Guo and He's bound concerns the selected subgroup. The two sentences
that follow state what that leaves for its complement; the second holds
on this design's nested family, where \(\hat H^c\) is never a candidate.
It is not claimed for their two-subgroup design (their Section 5.1),
where the complement of the selected subgroup is the other candidate
(with two candidates, the argmin), nor for their MONET1 families (their
Section 4), which contain each subgroup and its complement.

\begin{quote}
Table 8 of \citet{guohe2021} reports an upper confidence bound on the
log-hazard-ratio scale; this is the identical one-sided lower-bound
construction under their Section 4 convention βᵢ = −log HR, applied to
the argmax-selected subgroup, and does not constitute an upper-bound
capability for a second, complementary subgroup.
\end{quote}

\begin{quote}
The Guo--He correction is defined through the maximum functional over
the supplied candidate family. The complement Ĥᶜ is not a member of that
family and is not the argmax of any functional, so no analogous
correction exists within their framework; the absence of a complement
bound, and hence of a joint two-subgroup claim, is structural rather
than an implementation gap.
\end{quote}

Because the complement's truth is exactly zero
(Section~\ref{sec-guohe-setting}), an upper bound on \(\beta(\hat H^c)\)
covers when it is at least 0, and nothing in
Table~\ref{tbl-guohe-complement} is scored against Guo and He, whose
framework does not define it.

\begin{figure}

\centering{

\pandocbounded{\includegraphics[keepaspectratio]{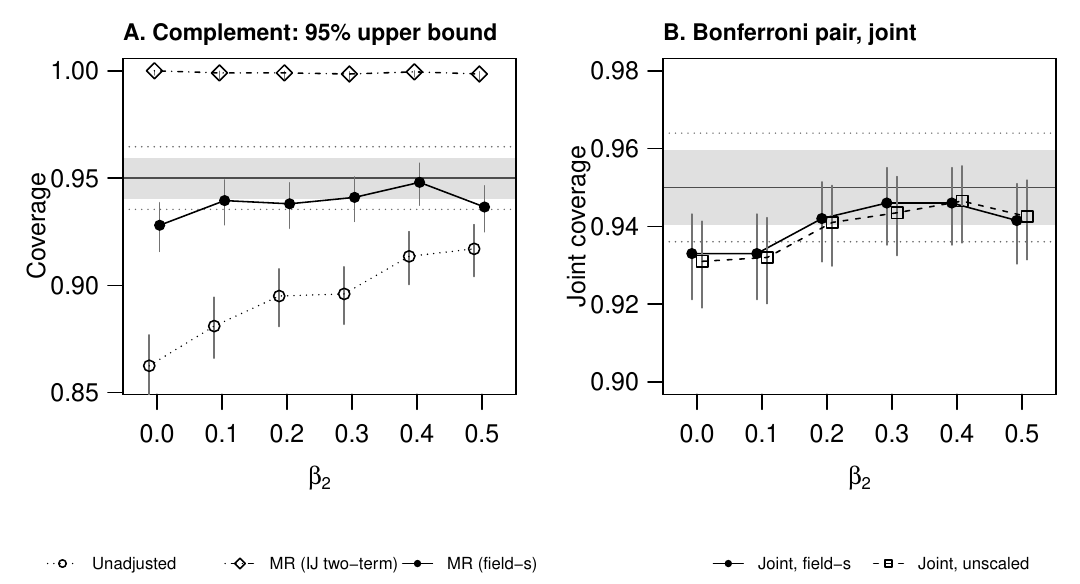}}

}

\caption{\label{fig-guohe-complement}Coverage on the second subgroup, by
cell, with bars at the Wilson 95\% interval. \textbf{A}: the one-sided
95\% upper bound on \(\beta(\hat H^c)\), against its truth of exactly 0.
\textbf{B}: the Bonferroni pair's joint coverage, with field-s on
\(\hat H^c\) and with the unscaled field beside it; note the narrower
vertical scale. Bands as in Figure~\ref{fig-guohe-bias-cover}: solid
nominal 0.95, shaded pointwise, dotted widened for reading each panel's
coverages at once. Guo and He have no counterpart to either panel. 2000
paired replicates per cell.}

\end{figure}%

{

\begin{longtable}[]{@{}
  >{\raggedright\arraybackslash}p{(\linewidth - 12\tabcolsep) * \real{0.3469}}
  >{\raggedleft\arraybackslash}p{(\linewidth - 12\tabcolsep) * \real{0.1088}}
  >{\raggedleft\arraybackslash}p{(\linewidth - 12\tabcolsep) * \real{0.1088}}
  >{\raggedleft\arraybackslash}p{(\linewidth - 12\tabcolsep) * \real{0.1088}}
  >{\raggedleft\arraybackslash}p{(\linewidth - 12\tabcolsep) * \real{0.1088}}
  >{\raggedleft\arraybackslash}p{(\linewidth - 12\tabcolsep) * \real{0.1088}}
  >{\raggedleft\arraybackslash}p{(\linewidth - 12\tabcolsep) * \real{0.1088}}@{}}

\caption{\label{tbl-guohe-complement}The complement and the joint claim,
by cell. Coverage of the one-sided 95\% upper bound on
\(\beta(\hat H^c)\) against its truth of exactly 0, with the typical
upper bound on the hazard-ratio scale (the exponential of the mean
log-scale bound); joint coverage of the Bonferroni pair (0.975 per side)
with field-s on \(\hat H^c\), and with the unscaled field beside it; the
correlation of the harm and complement draws under each scaling; and the
pair's typical bounds on the hazard-ratio scale. The unadjusted
complement bound uses the robust standard error, the unadjusted
harm-side bound the model-based Cox standard error; on the harm side the
two give unadjusted coverages within 0.005 of each other in every cell.
Guo and He have no counterpart to any row. Wilson intervals are in the
supplementary detail.}

\tabularnewline

\toprule\noalign{}
\begin{minipage}[b]{\linewidth}\raggedright
\end{minipage} & \begin{minipage}[b]{\linewidth}\raggedleft
\(\beta_2 = 0.0\)
\end{minipage} & \begin{minipage}[b]{\linewidth}\raggedleft
\(\beta_2 = 0.1\)
\end{minipage} & \begin{minipage}[b]{\linewidth}\raggedleft
\(\beta_2 = 0.2\)
\end{minipage} & \begin{minipage}[b]{\linewidth}\raggedleft
\(\beta_2 = 0.3\)
\end{minipage} & \begin{minipage}[b]{\linewidth}\raggedleft
\(\beta_2 = 0.4\)
\end{minipage} & \begin{minipage}[b]{\linewidth}\raggedleft
\(\beta_2 = 0.5\)
\end{minipage} \\
\midrule\noalign{}
\endhead
\bottomrule\noalign{}
\endlastfoot
Unadjusted: coverage & 0.863 & 0.881 & 0.895 & 0.896 & 0.913 & 0.917 \\
Unadjusted: upper bound (HR) & 1.24 & 1.24 & 1.24 & 1.24 & 1.26 &
1.26 \\
MR (IJ two-term): coverage & 1.000 & 0.999 & 0.999 & 0.999 & 1.000 &
0.999 \\
MR (IJ two-term): upper bound (HR) & 1.77 & 1.76 & 1.73 & 1.71 & 1.72 &
1.71 \\
MR (field-s): coverage & 0.928 & 0.940 & 0.938 & 0.941 & 0.948 &
0.936 \\
MR (field-s): upper bound (HR) & 1.34 & 1.34 & 1.33 & 1.31 & 1.33 &
1.32 \\
Joint coverage, field-s & 0.933 & 0.933 & 0.942 & 0.946 & 0.946 &
0.942 \\
Joint coverage, unscaled field & 0.931 & 0.932 & 0.941 & 0.944 & 0.947 &
0.943 \\
\(\mathrm{corr}(\Lambda^*, \Lambda^{*c})\), field-s & +0.015 & +0.020 &
+0.021 & +0.019 & +0.017 & +0.012 \\
\(\mathrm{corr}(\Lambda^*, \Lambda^{*c})\), unscaled & +0.018 & +0.021 &
+0.021 & +0.018 & +0.015 & +0.009 \\
Pair: lower bound on \(\hat H\) (HR) & 0.71 & 0.77 & 0.83 & 0.91 & 1.00
& 1.09 \\
Pair: upper bound on \(\hat H^c\) (HR) & 1.42 & 1.42 & 1.40 & 1.39 &
1.40 & 1.39 \\

\end{longtable}

}

The field-s upper bound covers at 0.928--0.948, with its low point in
the null cell, where the selection is loosest; that null cell is its
only one below the simultaneous band, and the rest sit at or just under
the pointwise band (Figure~\ref{fig-guohe-complement}). Its typical
location, 1.31 to 1.34, sits above one --- a typical bound is not on the
benefit side in \(\hat H^c\) --- so no benefit claim arises. The
unadjusted upper bound sits lower, at 1.24 to 1.26, but covers at only
0.863--0.917: selecting the most harmful subgroup pushes the
complement's estimate toward benefit, so an unadjusted upper bound is
too low. The IJ two-term bound covers at 0.999 or above, at 1.71 to
1.77.

The Bonferroni pair covers jointly at 0.933--0.946, and at 0.931--0.947
with the unscaled field on \(\hat H^c\) beside it. The two scalings
track each other cell for cell. Both fall just below the simultaneous
band in the two weakest cells, \(\beta_2 \le 0.1\), and sit within it
from \(\beta_2 = 0.2\) on, so the pair's lowest coverage is concentrated
where the selection is least settled. The correlation between the harm
and complement draws, \(\mathrm{corr}(\Lambda^*, \Lambda^{*c})\), is
\(+0.009\) to \(+0.021\) under either scaling, so a pair calibrated to
it coincides with Bonferroni. At 0.975 a side, the typical pair puts the
hazard ratio in \(\hat H\) at or above 0.71 to 1.09 across the cells and
the hazard ratio in \(\hat H^c\) at or below 1.39 to 1.42. The two
ranges overlap in every cell: at this sample size the typical joint
claim does not separate the two groups.

Two cautions apply. Joint coverage concerns the effects of the two
subgroups actually selected, \(\beta(\hat H)\) and \(\beta(\hat H^c)\),
not whether \(\hat H\) and \(\hat H^c\) partition the population
correctly. And because the complement's truth is a hazard ratio of 1.00
in every cell, a benefit claim for \(\hat H^c\) is never exercised.

\subsection{Limits}\label{sec-guohe-limits}

\begin{enumerate}
\def\labelenumi{\arabic{enumi}.}
\tightlist
\item
  \textbf{The field's low points are under-correction on a nested
  family.} Its harm-side coverage bottoms out at 0.933
  (\(\beta_2 = 0.1\)), with its point estimate biased upward in every
  cell, and the field-s complement bound bottoms out at 0.928 in the
  null cell. Both sit where the pick is least settled: re-selection is
  spread thin throughout, with median \(\hat p(\hat H)\) from 0.196 to
  0.309. Across Guo and He's replicated designs the nested family is the
  hardest case for the field's second-order correction, consistent with
  unequal candidate variances under nesting.
\item
  \textbf{Theory scope.} The design lies inside Guo and He's theory
  (their Section 3 continuum and Assumptions 4.1--4.3). Its roughly 151
  candidates, each differing from its neighbour by one subject, are the
  near-duplicate case that strains the no-tie condition (A5) on which
  the multiplier-resampling theory rests, so the field's calibration
  here is an empirical finding on this design, not a consequence of that
  theory. Condition (A5) is stated in Section S1.1.
\item
  \textbf{Identification is flattered by construction.} The true
  boundary is the family's smallest member, toward which the argmax is
  drawn even when there is no harm region; for \(\beta_2 > 0\)
  sensitivity and negative predictive value are identically 1, and
  overshoot is the only error.
\item
  \textbf{Forest search's screen and family construction are not
  exercised.} The identifier is the pure argmax over Guo and He's
  family. Forest search's screened identifier is evaluated in Section 5
  of the main paper.
\item
  \textbf{A benefit claim is untested.} The complement's truth is zero
  in every cell, so no cell holds a benefit for a bound to establish,
  and whether an upper bound on \(\hat H^c\) can support a benefit claim
  is not examined.
\item
  \textbf{One design, one sample size.} Everything is at \(n = 400\) on
  Guo and He's Section 5.2 design.
\item
  \textbf{Guo and He are not run on a screened family.} Their method is
  applied exactly as published, on the family their Section 5.2
  specifies; their Section 6 suggestion that it serves as a conservative
  approach after other identification is not tested, and no claim is
  made either way.
\item
  \textbf{No Adaptive column.} Guo and He's Table 7 reports none, and
  neither do we. In a pilot of our implementation of their
  cross-validated choice of \(r\) on this design (a two-value grid,
  \(r \in \{1/3, 1/12\}\)), changing only the seed offset of the
  adaptive procedure changed \(\hat r\) on 5 of 10 replicates, with the
  data and the selection unchanged, so an Adaptive column here would
  largely record the seed. Our implementation draws its cross-validation
  folds independently for each \(r\); Guo and He themselves call for a
  more reliable adaptive method (their Section 5.1).
\item
  \textbf{Everything is one-sided.} Every construction compared is a
  one-sided bound, so nothing here depends on the choice of a two-sided
  interval method.
\end{enumerate}

\subsection{Supplementary detail}\label{sec-guohe-detail}

{

\begin{longtable}[]{@{}
  >{\raggedleft\arraybackslash}p{(\linewidth - 14\tabcolsep) * \real{0.0855}}
  >{\raggedleft\arraybackslash}p{(\linewidth - 14\tabcolsep) * \real{0.1709}}
  >{\raggedleft\arraybackslash}p{(\linewidth - 14\tabcolsep) * \real{0.1282}}
  >{\raggedright\arraybackslash}p{(\linewidth - 14\tabcolsep) * \real{0.1624}}
  >{\raggedleft\arraybackslash}p{(\linewidth - 14\tabcolsep) * \real{0.1880}}
  >{\raggedleft\arraybackslash}p{(\linewidth - 14\tabcolsep) * \real{0.0598}}
  >{\raggedleft\arraybackslash}p{(\linewidth - 14\tabcolsep) * \real{0.0598}}
  >{\raggedleft\arraybackslash}p{(\linewidth - 14\tabcolsep) * \real{0.1453}}@{}}

\caption{\label{tbl-guohe-lower}One-sided 95\% lower bounds on
\(\beta(\hat H)\), by cell and method: coverage with its Wilson 95\%
interval; mean margin, estimate minus lower bound (log scale); mean bias
of the point estimate against \(\beta(\hat H)\) (log scale); and the
typical lower bound on the hazard-ratio scale, the exponential of the
mean log-scale bound. Point estimates: unadjusted \(\hat\beta(\hat H)\);
Guo and He's bias-reduced estimate; the field's
\(\hat\beta_{\mathrm F}(\hat H)\); the two-term \(\tilde\beta(\hat H)\).
\(\theta^\dagger(H) = \exp(\beta_2)\) is the hazard ratio in \(H\), and
HR in \(\hat H\) is the exponential of the mean \(\beta(\hat H)\). 2000
paired replicates per cell.}

\tabularnewline

\toprule\noalign{}
\begin{minipage}[b]{\linewidth}\raggedleft
\(\beta_2\)
\end{minipage} & \begin{minipage}[b]{\linewidth}\raggedleft
\(\theta^\dagger(H)\)
\end{minipage} & \begin{minipage}[b]{\linewidth}\raggedleft
HR in \(\hat H\)
\end{minipage} & \begin{minipage}[b]{\linewidth}\raggedright
Method
\end{minipage} & \begin{minipage}[b]{\linewidth}\raggedleft
Coverage (Wilson 95\%)
\end{minipage} & \begin{minipage}[b]{\linewidth}\raggedleft
Margin
\end{minipage} & \begin{minipage}[b]{\linewidth}\raggedleft
Bias
\end{minipage} & \begin{minipage}[b]{\linewidth}\raggedleft
Lower bound (HR)
\end{minipage} \\
\midrule\noalign{}
\endhead
\bottomrule\noalign{}
\endlastfoot
0.0 & 1.00 & 1.00 & Unadjusted & 0.866 (0.850, 0.880) & 0.305 & +0.122 &
0.83 \\
& & & Guo--He, \(r = 1/12\) & 0.952 (0.941, 0.960) & 0.306 & +0.014 &
0.75 \\
& & & MR (field) & 0.940 (0.928, 0.949) & 0.310 & +0.030 & 0.76 \\
& & & MR (IJ two-term) & 0.998 (0.994, 0.999) & 0.577 & +0.045 & 0.59 \\
0.1 & 1.11 & 1.09 & Unadjusted & 0.870 (0.855, 0.884) & 0.310 & +0.116 &
0.89 \\
& & & Guo--He, \(r = 1/12\) & 0.952 (0.941, 0.960) & 0.305 & +0.008 &
0.81 \\
& & & MR (field) & 0.933 (0.922, 0.944) & 0.310 & +0.026 & 0.82 \\
& & & MR (IJ two-term) & 0.998 (0.995, 0.999) & 0.582 & +0.040 & 0.63 \\
0.2 & 1.22 & 1.18 & Unadjusted & 0.882 (0.868, 0.896) & 0.317 & +0.111 &
0.96 \\
& & & Guo--He, \(r = 1/12\) & 0.958 (0.948, 0.965) & 0.305 & +0.002 &
0.87 \\
& & & MR (field) & 0.938 (0.926, 0.947) & 0.313 & +0.025 & 0.89 \\
& & & MR (IJ two-term) & 0.999 (0.996, 0.999) & 0.591 & +0.039 & 0.68 \\
0.3 & 1.35 & 1.30 & Unadjusted & 0.895 (0.881, 0.908) & 0.323 & +0.103 &
1.04 \\
& & & Guo--He, \(r = 1/12\) & 0.961 (0.952, 0.969) & 0.305 & -0.007 &
0.95 \\
& & & MR (field) & 0.937 (0.925, 0.947) & 0.314 & +0.021 & 0.97 \\
& & & MR (IJ two-term) & 0.999 (0.996, 0.999) & 0.598 & +0.033 & 0.74 \\
0.4 & 1.49 & 1.43 & Unadjusted & 0.901 (0.887, 0.913) & 0.325 & +0.094 &
1.13 \\
& & & Guo--He, \(r = 1/12\) & 0.965 (0.956, 0.972) & 0.305 & -0.018 &
1.03 \\
& & & MR (field) & 0.941 (0.929, 0.950) & 0.316 & +0.017 & 1.06 \\
& & & MR (IJ two-term) & 0.999 (0.996, 1.000) & 0.601 & +0.028 & 0.81 \\
0.5 & 1.65 & 1.58 & Unadjusted & 0.918 (0.905, 0.929) & 0.329 & +0.076 &
1.23 \\
& & & Guo--He, \(r = 1/12\) & 0.972 (0.963, 0.978) & 0.306 & -0.037 &
1.12 \\
& & & MR (field) & 0.951 (0.941, 0.960) & 0.320 & +0.007 & 1.16 \\
& & & MR (IJ two-term) & 0.999 (0.996, 1.000) & 0.608 & +0.015 & 0.88 \\

\end{longtable}

}

{

\begin{longtable}[]{@{}
  >{\raggedleft\arraybackslash}p{(\linewidth - 12\tabcolsep) * \real{0.0794}}
  >{\raggedright\arraybackslash}p{(\linewidth - 12\tabcolsep) * \real{0.0873}}
  >{\raggedleft\arraybackslash}p{(\linewidth - 12\tabcolsep) * \real{0.1667}}
  >{\raggedleft\arraybackslash}p{(\linewidth - 12\tabcolsep) * \real{0.1667}}
  >{\raggedleft\arraybackslash}p{(\linewidth - 12\tabcolsep) * \real{0.1667}}
  >{\raggedleft\arraybackslash}p{(\linewidth - 12\tabcolsep) * \real{0.1667}}
  >{\raggedleft\arraybackslash}p{(\linewidth - 12\tabcolsep) * \real{0.1667}}@{}}

\caption{\label{tbl-guohe-rsens}Coverage of the one-sided 95\% lower
bound on \(\beta(\hat H)\) at every \(r\) of Guo and He's Table 7, and
for the unadjusted bound: our replication, with its Wilson 95\%
interval, beside the published value (Guo and He 2021, Table 7).}

\tabularnewline

\toprule\noalign{}
\begin{minipage}[b]{\linewidth}\raggedleft
\(\beta_2\)
\end{minipage} & \begin{minipage}[b]{\linewidth}\raggedright
Source
\end{minipage} & \begin{minipage}[b]{\linewidth}\raggedleft
\(r = 1/3\)
\end{minipage} & \begin{minipage}[b]{\linewidth}\raggedleft
\(r = 1/12\)
\end{minipage} & \begin{minipage}[b]{\linewidth}\raggedleft
\(r = 1/21\)
\end{minipage} & \begin{minipage}[b]{\linewidth}\raggedleft
\(r = 1/30\)
\end{minipage} & \begin{minipage}[b]{\linewidth}\raggedleft
Unadjusted
\end{minipage} \\
\midrule\noalign{}
\endhead
\bottomrule\noalign{}
\endlastfoot
0.0 & Replicated & 0.935 (0.924, 0.945) & 0.952 (0.941, 0.960) & 0.953
(0.942, 0.961) & 0.954 (0.943, 0.962) & 0.866 (0.850, 0.880) \\
& Published & 0.947 & 0.961 & 0.962 & 0.962 & 0.872 \\
0.1 & Replicated & 0.940 (0.928, 0.949) & 0.952 (0.941, 0.960) & 0.954
(0.944, 0.962) & 0.954 (0.944, 0.962) & 0.870 (0.855, 0.884) \\
& Published & 0.960 & 0.972 & 0.972 & 0.972 & 0.879 \\
0.2 & Replicated & 0.945 (0.934, 0.954) & 0.958 (0.948, 0.965) & 0.958
(0.948, 0.966) & 0.958 (0.948, 0.966) & 0.882 (0.868, 0.896) \\
& Published & 0.958 & 0.966 & 0.967 & 0.967 & 0.890 \\
0.3 & Replicated & 0.951 (0.941, 0.960) & 0.961 (0.952, 0.969) & 0.961
(0.952, 0.969) & 0.962 (0.952, 0.969) & 0.895 (0.881, 0.908) \\
& Published & 0.959 & 0.969 & 0.970 & 0.970 & 0.895 \\
0.4 & Replicated & 0.954 (0.943, 0.962) & 0.965 (0.956, 0.972) & 0.967
(0.958, 0.974) & 0.967 (0.958, 0.974) & 0.901 (0.887, 0.913) \\
& Published & 0.962 & 0.968 & 0.968 & 0.968 & 0.906 \\
0.5 & Replicated & 0.961 (0.952, 0.969) & 0.972 (0.963, 0.978) & 0.973
(0.964, 0.979) & 0.973 (0.965, 0.979) & 0.918 (0.905, 0.929) \\
& Published & 0.964 & 0.972 & 0.973 & 0.973 & 0.901 \\

\end{longtable}

}

{

\begin{longtable}[]{@{}
  >{\raggedleft\arraybackslash}p{(\linewidth - 8\tabcolsep) * \real{0.1429}}
  >{\raggedleft\arraybackslash}p{(\linewidth - 8\tabcolsep) * \real{0.2143}}
  >{\raggedleft\arraybackslash}p{(\linewidth - 8\tabcolsep) * \real{0.2143}}
  >{\raggedleft\arraybackslash}p{(\linewidth - 8\tabcolsep) * \real{0.2143}}
  >{\raggedleft\arraybackslash}p{(\linewidth - 8\tabcolsep) * \real{0.2143}}@{}}

\caption{\label{tbl-guohe-rmargin}Mean margin and mean bias of the point
estimate (log scale), shown as margin / bias, for Guo and He's bound at
every \(r\) of their Table 7.}

\tabularnewline

\toprule\noalign{}
\begin{minipage}[b]{\linewidth}\raggedleft
\(\beta_2\)
\end{minipage} & \begin{minipage}[b]{\linewidth}\raggedleft
\(r = 1/3\)
\end{minipage} & \begin{minipage}[b]{\linewidth}\raggedleft
\(r = 1/12\)
\end{minipage} & \begin{minipage}[b]{\linewidth}\raggedleft
\(r = 1/21\)
\end{minipage} & \begin{minipage}[b]{\linewidth}\raggedleft
\(r = 1/30\)
\end{minipage} \\
\midrule\noalign{}
\endhead
\bottomrule\noalign{}
\endlastfoot
0.0 & 0.307 / +0.040 & 0.306 / +0.014 & 0.306 / +0.012 & 0.306 /
+0.011 \\
0.1 & 0.308 / +0.034 & 0.305 / +0.008 & 0.305 / +0.006 & 0.305 /
+0.005 \\
0.2 & 0.311 / +0.030 & 0.305 / +0.002 & 0.305 / +0.000 & 0.305 /
-0.000 \\
0.3 & 0.313 / +0.022 & 0.305 / -0.007 & 0.304 / -0.009 & 0.304 /
-0.010 \\
0.4 & 0.315 / +0.014 & 0.305 / -0.018 & 0.304 / -0.021 & 0.304 /
-0.021 \\
0.5 & 0.319 / -0.002 & 0.306 / -0.037 & 0.305 / -0.040 & 0.305 /
-0.041 \\

\end{longtable}

}

{

\begin{longtable}[]{@{}
  >{\raggedleft\arraybackslash}p{(\linewidth - 10\tabcolsep) * \real{0.0870}}
  >{\raggedleft\arraybackslash}p{(\linewidth - 10\tabcolsep) * \real{0.1826}}
  >{\raggedleft\arraybackslash}p{(\linewidth - 10\tabcolsep) * \real{0.1826}}
  >{\raggedleft\arraybackslash}p{(\linewidth - 10\tabcolsep) * \real{0.1826}}
  >{\raggedleft\arraybackslash}p{(\linewidth - 10\tabcolsep) * \real{0.1826}}
  >{\raggedleft\arraybackslash}p{(\linewidth - 10\tabcolsep) * \real{0.1826}}@{}}

\caption{\label{tbl-guohe-wilson}Wilson 95\% intervals for the coverages
of Table~\ref{tbl-guohe-complement}: the three one-sided 95\% upper
bounds on \(\beta(\hat H^c)\) and the two joint coverages of the
Bonferroni pair.}

\tabularnewline

\toprule\noalign{}
\begin{minipage}[b]{\linewidth}\raggedleft
\(\beta_2\)
\end{minipage} & \begin{minipage}[b]{\linewidth}\raggedleft
Unadjusted
\end{minipage} & \begin{minipage}[b]{\linewidth}\raggedleft
MR (IJ two-term)
\end{minipage} & \begin{minipage}[b]{\linewidth}\raggedleft
MR (field-s)
\end{minipage} & \begin{minipage}[b]{\linewidth}\raggedleft
Joint, field-s
\end{minipage} & \begin{minipage}[b]{\linewidth}\raggedleft
Joint, unscaled
\end{minipage} \\
\midrule\noalign{}
\endhead
\bottomrule\noalign{}
\endlastfoot
0.0 & 0.863 (0.847, 0.877) & 1.000 (0.998, 1.000) & 0.928 (0.916, 0.939)
& 0.933 (0.921, 0.943) & 0.931 (0.919, 0.941) \\
0.1 & 0.881 (0.866, 0.894) & 0.999 (0.996, 1.000) & 0.940 (0.928, 0.949)
& 0.933 (0.921, 0.943) & 0.932 (0.920, 0.942) \\
0.2 & 0.895 (0.881, 0.908) & 0.999 (0.996, 1.000) & 0.938 (0.927, 0.948)
& 0.942 (0.931, 0.951) & 0.941 (0.930, 0.951) \\
0.3 & 0.896 (0.882, 0.909) & 0.999 (0.996, 0.999) & 0.941 (0.930, 0.951)
& 0.946 (0.935, 0.955) & 0.944 (0.933, 0.953) \\
0.4 & 0.913 (0.900, 0.925) & 1.000 (0.997, 1.000) & 0.948 (0.937, 0.957)
& 0.946 (0.935, 0.955) & 0.947 (0.936, 0.956) \\
0.5 & 0.917 (0.904, 0.928) & 0.999 (0.996, 0.999) & 0.936 (0.925, 0.946)
& 0.942 (0.930, 0.951) & 0.943 (0.931, 0.952) \\

\end{longtable}

}

\section{Details of the identifiers}\label{sec-supp-identifiers}

Section 2.1 of the main paper defines the three identifiers as
admissible families \(\mathcal{A}_{\mathrm{FS}}\),
\(\mathcal{A}_{\mathrm{GRF}}\) and \(\mathcal{A}_{\mathrm{DINA}}\) of
one candidate space \(\mathcal{S}\); the table below records the floor
each places on the statistic it proposes with and the floor on
\(\hat\beta(g)\) at re-ranking. This section gives the definitions in
full: how each identifier generates its candidates, the size and event
requirements it applies, its floors, and the family over which
multiplier resampling re-selects. Every quantity below is a setting,
stated at the value this paper uses. Throughout, \(n_{\min}\) is the
eligibility minimum on subgroup size, 60 by default, and \(c\) is the
screening threshold on \(\hat\beta(g)\), \(\log 1.0\) for a ratio
measure.

\textbf{Size and events.} The size minimum reaches all three identifiers
through \(\mathcal{S}\). The event minimum does not: forest search
requires a minimum number of events in each arm of every candidate it
fits, 10 by default, and neither model-based identifier imposes one. The
asymmetry is deliberate. Forest search enumerates \(\mathcal{S}\)
exhaustively and fits the working model in every member, so a coarse
guard against candidates too sparse in one arm to fit stably is worth
its cost there. GRF and DINA propose regions from a fitted surface, and
each proposal is fit once more at re-ranking, where the refit decides
estimability: on the survival path it returns nothing for a region with
fewer than two events, arms pooled, or without a finite coefficient, and
on the generalized-linear paths for a region with fewer than three
subjects in either arm, a fit that does not converge, or a coefficient
that is not finite. A region the refit cannot score cannot be selected.
The two size tests also differ at the boundary, forest search keeping
candidates larger than \(n_{\min}\) and the other two candidates of at
least \(n_{\min}\), a difference of one subject that Section 2.1 of the
main paper notes.

\textbf{The constructions as algorithms.} The two model-based
identifiers turn a fitted surface into candidates of the same
interpretable form as forest search's; the steps are these.

\emph{DINA.} Input: a time-to-event outcome, treatment and covariates;
\(n_{\min}\), \(c\), depth two.

\begin{enumerate}
\def\labelenumi{\arabic{enumi}.}
\tightlist
\item
  On each half of a random split, fit the propensity and per-arm working
  models on one half and the difference in natural parameters, linear in
  the covariates, on the other; average the two coefficient vectors.
\item
  Evaluate the fitted effect for every patient, on the log-hazard scale,
  positive for harm.
\item
  List single cuts of each covariate at every distinct value, and
  two-cut conjunctions of different covariates on each one's interior
  deciles, or at every distinct value of one with at most nine, each cut
  in either direction; keep regions of at least \(n_{\min}\).
\item
  Keep those whose mean fitted effect is at or above \(c\).
\item
  Refit each kept region by the standard analysis for \(\hat\beta(g)\).
  The qualifying candidates enter re-ranking on \(\hat\beta(g)\) against
  \(c\), and the family multiplier resampling re-selects over is the
  regions that cleared size and the surface floor, under the screening
  threshold.
\end{enumerate}

\emph{GRF.} Input: a time-to-event outcome, treatment and covariates;
\(n_{\min}\), \(c\), depth two.

\begin{enumerate}
\def\labelenumi{\arabic{enumi}.}
\tightlist
\item
  Fit a causal survival forest, with treatment probability one half, for
  the difference in restricted mean survival to 0.8 of the smaller arm's
  largest event time.
\item
  Compute each patient's doubly robust scores under treatment and under
  control.
\item
  List single cuts at each covariate's interior deciles and two-cut
  conjunctions of different covariates at their quintile boundaries,
  each cut in either direction; keep the regions of at least
  \(n_{\min}\) patients, short of the whole trial.
\item
  Score each region by its doubly robust contrast, control less treated,
  positive for harm; if none is at or above 0, GRF proposes nothing.
\item
  Otherwise refit every listed region by the standard analysis for
  \(\hat\beta(g)\). The qualifying candidates enter re-ranking on
  \(\hat\beta(g)\) against \(c\), and the family multiplier resampling
  re-selects over is every size-eligible region of its grid, under the
  screening threshold.
\end{enumerate}

\textbf{Survival outcomes.} Table~\ref{tbl-supp-identifiers} sets the
three identifiers side by side for the hazard ratio, the case of the
operating characteristics of Section 5 of the main paper and of the GBSG
analysis.

\begin{longtable}[]{@{}
  >{\raggedright\arraybackslash}p{(\linewidth - 6\tabcolsep) * \real{0.1094}}
  >{\raggedright\arraybackslash}p{(\linewidth - 6\tabcolsep) * \real{0.2969}}
  >{\raggedright\arraybackslash}p{(\linewidth - 6\tabcolsep) * \real{0.2969}}
  >{\raggedright\arraybackslash}p{(\linewidth - 6\tabcolsep) * \real{0.2969}}@{}}
\caption{The three identifiers for a survival outcome. Every entry is a
setting at this paper's default. \(n_{\min}\) is the eligibility minimum
on size and \(c\) the screening threshold on
\(\hat\beta(g)\).}\label{tbl-supp-identifiers}\tabularnewline
\toprule\noalign{}
\begin{minipage}[b]{\linewidth}\raggedright
\end{minipage} & \begin{minipage}[b]{\linewidth}\raggedright
Forest search
\end{minipage} & \begin{minipage}[b]{\linewidth}\raggedright
DINA
\end{minipage} & \begin{minipage}[b]{\linewidth}\raggedright
GRF
\end{minipage} \\
\midrule\noalign{}
\endfirsthead
\toprule\noalign{}
\begin{minipage}[b]{\linewidth}\raggedright
\end{minipage} & \begin{minipage}[b]{\linewidth}\raggedright
Forest search
\end{minipage} & \begin{minipage}[b]{\linewidth}\raggedright
DINA
\end{minipage} & \begin{minipage}[b]{\linewidth}\raggedright
GRF
\end{minipage} \\
\midrule\noalign{}
\endhead
\bottomrule\noalign{}
\endlastfoot
Candidates proposed & every conjunction of at most two cuts of the
candidate cut set, enumerated exhaustively & single cuts at every
distinct value of each covariate, in either direction, and two-cut
conjunctions on an interior-decile grid, or at every distinct value of a
covariate with at most nine & single cuts on an interior-decile grid and
two-cut conjunctions on a quintile grid, in either direction \\
Size & more than \(n_{\min}\) & at least \(n_{\min}\) & at least
\(n_{\min}\) \\
Events, per arm & at least 10 in each arm & none & none \\
Floor at proposal & \(\hat\beta(g) > c\), the Cox coefficient in the
candidate & region mean of the cross-fit effect surface at or above
\(c\), on the same log-hazard scale & DR-score contrast at or above 0,
applied to the region GRF selects on its own statistic, so it decides
whether GRF proposes anything rather than which regions are carried
forward \\
Floor at re-ranking & not required & \(\hat\beta(g) \ge c\), over the
regions that cleared size and the surface floor &
\(\hat\beta(g) \ge c\), over every size-eligible region of its grid \\
Replicability & \(p_{\mathrm{cons}}(g) \ge 0.90\) & none & none \\
Family re-selected by multiplier resampling & every conjunction of at
most two cuts of size at least \(n_{\min}\), under the screening
threshold and the consistency screen; the per-arm event minimum is not
re-applied, so this family contains the one the search ranked & the
regions that cleared size and the surface floor, under the screening
threshold & every size-eligible region of its grid, under the screening
threshold \\
\end{longtable}

The GBSG search, drawn in Figure 1 of the main paper, makes the rules
concrete. Nine candidates pass the screen, all of them variants of one
region, and the rules separate only in how far that region is trimmed:
the largest admissible subgroup carries the weakest effect, the
\(\varepsilon\) band excludes it in favour of a subgroup of 75, and the
four remaining rules agree on the smallest and strongest of the three.
The spread of the frontier --- from HR 1.95 at 82 subjects to 2.54 at 61
--- is the size-versus-effect trade the choice of rule settles. The
simulations of Section 5 set the screening threshold at a hazard ratio
of 0.90 for all three identifiers; the applications use a hazard ratio
of one. DINA's frontier caps (\texttt{max\_subgroups},
\texttt{max\_per\_covariate}) trim only a displayed table of single
cuts; its family is every qualifying region, and the caps bind in no
analysis here.

\textbf{Binary outcomes.} A binary outcome is coded with \(Y = 1\) the
adverse event, so that a positive effect is harm, as for the hazard
ratio. The effect \(\hat\beta(g)\) is the log odds ratio, estimated by
logistic regression within the region, as in \citet{leonanderson2026};
the risk difference and the risk ratio are estimated the same way when
chosen instead. Table~\ref{tbl-supp-identifiers-binary} gives the
requirements and floors. This paper's binary analyses use the odds
ratio.

\begin{longtable}[]{@{}
  >{\raggedright\arraybackslash}p{(\linewidth - 6\tabcolsep) * \real{0.1094}}
  >{\raggedright\arraybackslash}p{(\linewidth - 6\tabcolsep) * \real{0.2969}}
  >{\raggedright\arraybackslash}p{(\linewidth - 6\tabcolsep) * \real{0.2969}}
  >{\raggedright\arraybackslash}p{(\linewidth - 6\tabcolsep) * \real{0.2969}}@{}}
\caption{The three identifiers for a binary outcome on the odds-ratio
scale.}\label{tbl-supp-identifiers-binary}\tabularnewline
\toprule\noalign{}
\begin{minipage}[b]{\linewidth}\raggedright
\end{minipage} & \begin{minipage}[b]{\linewidth}\raggedright
Forest search
\end{minipage} & \begin{minipage}[b]{\linewidth}\raggedright
DINA
\end{minipage} & \begin{minipage}[b]{\linewidth}\raggedright
GRF
\end{minipage} \\
\midrule\noalign{}
\endfirsthead
\toprule\noalign{}
\begin{minipage}[b]{\linewidth}\raggedright
\end{minipage} & \begin{minipage}[b]{\linewidth}\raggedright
Forest search
\end{minipage} & \begin{minipage}[b]{\linewidth}\raggedright
DINA
\end{minipage} & \begin{minipage}[b]{\linewidth}\raggedright
GRF
\end{minipage} \\
\midrule\noalign{}
\endhead
\bottomrule\noalign{}
\endlastfoot
Size & more than \(n_{\min}\) & at least \(n_{\min}\) & at least
\(n_{\min}\) \\
Events, per arm & at least 10 subjects with \(Y = 1\) in each arm & none
& none \\
Floor at proposal & \(\hat\beta(g) > c\) on the log-odds scale & region
mean of the cross-fit log-odds surface at or above \(c\) & DR-score
contrast, a difference in event probability, at or above 0, applied as
in the survival case \\
Floor at re-ranking & not required & \(\hat\beta(g) \ge c\) &
\(\hat\beta(g) \ge c\) \\
The within-region fit & at proposal & at re-ranking: at least three
subjects in each arm, convergence and a finite coefficient & at
re-ranking, as for DINA \\
\end{longtable}

\textbf{Continuous and count outcomes.} No event requirement applies to
any identifier; the size minimum alone does, with the within-region fit
at re-ranking requiring three subjects in each arm. The floors are the
same threshold on the mean-difference scale for a continuous outcome,
and on the log-rate scale for a count, DINA's surface being fit on that
scale; GRF's forest is fit to the count on the log scale by default.


\end{document}